\documentclass[review,3p,12pt]{elsarticle}

\usepackage{amssymb}
\usepackage{amsmath}
\usepackage[ruled,linesnumbered]{algorithm2e}
\usepackage{setspace}
\usepackage{siunitx}
\usepackage{lineno}

\usepackage{multirow}
\usepackage{booktabs}
\usepackage{overpic}
\usepackage[colorlinks,linkcolor=blue,anchorcolor=blue,citecolor=blue]{hyperref}
\usepackage{xurl}
\usepackage{etoolbox}
\usepackage{silence}
\usepackage{sectsty}
\usepackage[dvipsnames]{xcolor}
\usepackage{multibib}
\usepackage{longtable}            
\usepackage{rotating}
\usepackage{array}
\usepackage{pdflscape}         
\usepackage{array}
\usepackage[table]{xcolor}
\newcites{S}{References}
\makeatletter
\providecommand*\@extra@b@citeb{}
\patchcmd{\NAT@wrout}{\bibcite{#5}}{\bibcite{#5\@extra@b@citeb}}{}{\PackageWarning{main}{NAT@wrout patch failed}}
\let\mf@bibliographyS\bibliographyS
\renewcommand\bibliographyS[1]{{\def\@extra@b@citeb{@S}\mf@bibliographyS{#1}}}
\AtBeginDocument{%
  \def\mb@@citex[#1][#2]#3{%
    \global\let\@save@auxout\@auxout
    \let\@auxout\@newciteauxhandle
    \def\@extra@b@citeb{@S}%
    \std@@citex[#1][#2]{#3}%
    \def\@extra@b@citeb{}%
    \let\@auxout\@save@auxout
    \let\@citex\std@@citex}}
\makeatother
\makeatletter
\patchcmd{\logtheLineNumber}{\@auxout}{\@mainaux}{}{\PackageWarning{main}{lineno patch failed}}
\makeatother

\newcommand{\papertitle}{MechFoundry: An agentic loop that poses and performs engineering analyses and optimization of real-world objects}

\journal{}

\begin{document}

\newgeometry{left=1.5cm,right=1.5cm,top=2cm,bottom=2cm}

\begin{frontmatter}


  \title{\papertitle}

  \pdfstringdefDisableCommands{%
    \def\corref#1{}%
  }

  \author[cornell]{Xue-Ling Luo}
  \author[cornell]{Marco Conte}
  \author[ubc]{Peter Yichen Chen}
  \author[cornell]{Nikolaos Bouklas\corref{cor1}}
  \ead{nb589@cornell.edu}

  \cortext[cor1]{Corresponding author.}
  
  \affiliation[cornell]{addressline={Cornell University},
              city={Ithaca},
              state={NY},
              country={USA}}

  \affiliation[ubc]{addressline={University of British Columbia},
  city={Vancouver},
  country={Canada}
  }

  \begin{abstract}
    AI agents can automate software engineering because a compiler or a test suite checks that the code does what the user or the agent intended. In engineering analysis, however, the steps that precede the computation have few easily verifiable checks: modeling a design or an object, deciding how it is used, where supports and loads are applied, and what material each part is made of. A human expert therefore makes these decisions, carries out the steps, and judges the result, one object at a time. We present MechFoundry, an agentic loop coupled with bespoke software that automates structural analyses of real-world objects without a human in the loop. From an object's lexical description, it generates a shape, poses a loading case relevant to how the object is used, assigns a material to each part, solves the nonlinear structural response, and improves the design against the loads it carries. A differentiable GPU solver built on an adaptive octree that need not conform to the surface enables the transition from shape generation to simulation and optimization. Across 50 everyday and engineering objects, the agent posed and solved a loading case in 6.5 minutes and for US\$1.3 per object on average. Without a test suite to check against, we had language-model reviewers grade each session, as a first indication of where the agent's engineering judgment holds and where it fails.
\end{abstract}



\end{frontmatter}



\section{Introduction}

AI agents now carry out substantial software engineering, writing, running and revising code with limited supervision. They work in part because software supplies its own verification: the code compiles and runs, and the unit tests pass or fail. For engineering analysis, finite element method (FEM) solvers automated the computation of how an object deforms decades ago: given a geometry, a mesh, a set of supports and loads, and a material model, a solver returns the response without a human in the loop. The computation itself can be checked, but the steps that precede it have neither a compiler nor a test suite: how the real object is used, where boundary conditions for loads and supports are applied, and what material model and properties each of its parts should have. A human expert therefore makes these decisions, carries out the steps, and judges the result, which is why structural analysis of a real-world object still takes hours to days~\cite{owen2026survey}, and why it is still carried out one object at a time. 

The need for expert actions at these steps shapes engineering practice. Because setting up the first meaningful simulation of a new object is expensive, analysis and redesign usually come late, after the key design decisions have been made. Many objects are never analyzed at all, because the conventional workflow starts from a parametric computer-aided design (CAD) model built in a narrow design space, and everyday hardware, legacy parts and free-form real-world shapes often have no such model. Design optimization, which could in principle improve any of them, is sensitive to the initial design~\cite{sigmund2013topology}, particularly for shape and level-set methods that cannot form new topological features~\cite{allaire2004structural,wang2003level}. Good starting points for free-form objects are scarce, and multistart strategies need many of them~\cite{papadopoulos2021computing}.

Recent work approaches this problem from two directions, and neither closes the gap. Language-model agents now plan and run experiments in chemistry and materials science~\cite{boiko2023autonomous,szymanski2023autonomous}, and in mechanics they drive finite element workflows, writing meshing and solver scripts for commercial and open-source codes~\cite{ni2024mechagents,mudur2025feabench,qi2025feagpt}. These studies mostly analyze structures whose geometry is parametric, given in CAD or specified by detailed instructions or a knowledge base, and are benchmarked against reference solutions built manually or analytically, so the agent is asked to operate a solver rather than to decide what to solve. In the other direction, generative models turn text, sketches or images into free-form shapes in seconds to minutes~\cite{li2025triposg,hunyuan3d2025hunyuan3d,khan2024text2cad,guan2026cad}, and recent work makes such shapes ready for simulation, mostly as rigid assets annotated with scale, joints and affordances, or optimized for stability under gravity~\cite{cao2026physx,chen2024atlas3d,ranawaka2026simfoundry}. Work that simulates generated deformable objects loads them mainly by gravity or grasping, or analyzes them linearly on voxel grids~\cite{chen2026diaggen,guo2024physically,zhan2025physiopt}. In none of it does the system infer the loading case from how the object is used, solve its nonlinear response, and feed that response back into the design to enable optimization.

Closing that gap for AI agents on a generated free-form shape is obstructed at every step. Generated surfaces contain disconnected pieces and self-intersections, which conventional tetrahedral mesh generators reject and robust meshers accept only within a user-set envelope~\cite{hu2020fast}, and conforming hexahedral meshes are harder still to produce automatically~\cite{schneider2022large}. A generated surface carries no CAD faces on which supports and loads can be placed, so an agent is left testing raw coordinates, such as finding all points with height $z<0.05$ to assign a base support, which demands costly visual reasoning and trial and error for complex topology. The surface is not organized into parts, which limits per-part material assignment. And any change of shape during optimization requires a new mesh, which can fail again.

We present MechFoundry, an agentic loop coupled with bespoke software that carries a real-world object from its name to a nonlinear structural analysis without a human in the loop, and on to an improved design (Fig.~\ref{fig:overview}). An AI agent operator, a coding agent working from generic engineering instructions, answers every approval gate of the workflow in place of a user: it writes a case brief stating how the object is used, where it is held, what loads it and how large each load is, and inspects each rendered result before approving it. AI subagents propose supports, loads and materials on the segmented parts of the generated surface. Each shape is carried as a signed distance field on an adaptive octree that need not conform to it, on which the diffuse Nitsche method~\cite{nguyen2018diffuse} applies supports and loads through a thin diffused layer and solves the nonlinear elastostatic problem on graphics processing units (GPUs)~\cite{macklin2022warp}. Because the solver is differentiable, the same loop continues into shape and topology optimization, redistributing material to optimize performance. We run the loop on 50 everyday and engineering objects from brief descriptions alone, compare its solutions against body-fitted finite elements, and have language-model reviewers grade its geometry, supports and loads, and materials, as a first indication of where the agent's engineering judgment holds and where it fails.

\section{Results}
\label{sec:results}

\subsection*{A workflow an AI agent can run autonomously}

\begin{figure}[t]
  \centering
  \includegraphics[width=\linewidth]{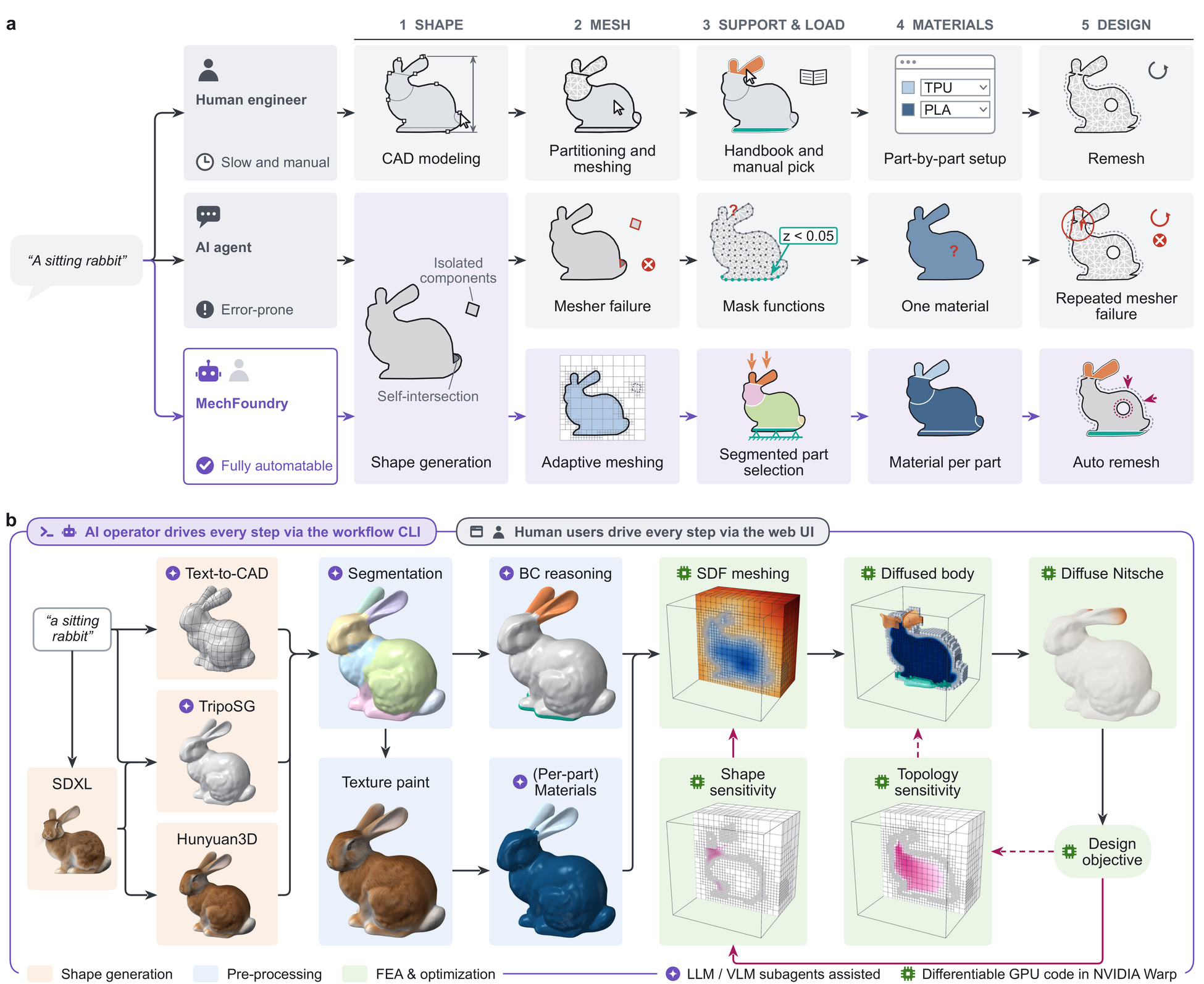}
  \caption{\textbf{MechFoundry carries an object from a text description to a posed and solved loading case, and can then improve its design.} \textbf{a}, The five steps of a deformation simulation for ``a sitting rabbit'', carried out by a human engineer (top), by a general-purpose AI agent writing its own scripts (middle, illustrative only: a schematic of how generated shapes can break conventional tools) and by MechFoundry (bottom). \textbf{b}, The MechFoundry workflow, taking the Stanford bunny for illustration. Shape generation (orange): a text-to-CAD model, TripoSG, or Hunyuan3D-2.1 with images generated by Stable Diffusion XL (SDXL). Pre-processing (blue): the surface is segmented into parts that receive supports and loads (boundary condition, BC), and materials. Finite element analysis (FEA) and optimization (green): the signed distance field (SDF) of the surface defines an adaptive octree carrying a diffused body, namely a phase field, solved by the diffuse Nitsche method. Sensitivities with respect to the SDF (shape) and to the phase field (topology, optional with dashed arrows) update the geometry. Star: steps that AI subagents (large language or vision-language models, LLM/VLM) can assist. Chip: differentiable GPU code in NVIDIA Warp. A person drives the workflow through the web user interface (UI), or the AI agent operator through the command line interface (CLI).}
  \label{fig:overview}
\end{figure}

Simulating how an object deforms, conventionally with finite element analysis, takes five steps (Fig.~\ref{fig:overview}a): building its shape, dividing it into a mesh of small elements, specifying where it is supported and loaded, assigning its materials and, when the design is to be improved, changing the shape and repeating the analysis. A human engineer carries out each step manually in CAD and finite element software, requiring hours or even days (Fig.~\ref{fig:overview}a, top row), and a general-purpose AI agent writing scripts for the same tools meets the obstacles of generated shapes described in the Introduction (Fig.~\ref{fig:overview}a, middle row). MechFoundry replaces each step with one that works on generated shapes and that a language agent can reason about and drive (Fig.~\ref{fig:overview}a, bottom row), with nonlinear solutions and optimization built in.

\paragraph{Shape.} A text prompt, a rough sketch or an image conditions a pre-trained 3D generator (Fig.~\ref{fig:overview}b, orange), which is TripoSG~\cite{li2025triposg} by default. Alternatively, a text-to-CAD model~\cite{khan2024text2cad,guan2026cad} returns a CAD solid, or an image drawn from the text by Stable Diffusion XL (SDXL)~\cite{podell2024sdxl} conditions TripoSG or Hunyuan3D-2.1~\cite{hunyuan3d2025hunyuan3d}, or the user uploads a surface. Small disconnected pieces of a generated shape are removed so that one body is kept, and the surface is scaled to its real-world size (Methods, ``Implicit geometry'', and Supplementary Section~\ref{sec:si-geometry}).

\paragraph{Supports, loads and materials.} The surface is segmented into parts by clustering the per-facet features from a pre-trained network, PartField~\cite{liu2025partfield}, or, for a CAD solid, by its faces (Fig.~\ref{fig:overview}b, blue). This allows AI agents to perform targeted selections on the surface with less reasoning and trial and error. From the AI agent operator's reasoning or the user's written description of how the object is held and loaded, an AI subagent writes a short program for each support or load region, which selects surface facets by combining operations such as picking a part, a position or a facing direction. The AI subagent also proposes the load values and the material constants, one material per part when the parts differ in material (Methods, ``Boundary-region identification'' and ``Language-model-assisted workflow'').

\paragraph{Mesh and solve.} The surface is turned into its signed distance field (SDF), whose value is the shortest distance from a spatial location to the surface, and whose sign is decided by the generalized winding number~\cite{jacobson2013robust}, negative inside and positive outside. From this field, an adaptive octree is built on the GPU, refined near the surface and coarse farther from the surface (Fig.~\ref{fig:overview}b, green). Because the octree cells do not need to follow the surface and the winding number tolerates defects on the surface, the octree can also be built for surfaces that a conventional tetrahedral mesher may reject: three of the 50 generated surfaces of Fig.~\ref{fig:agentic} and Supplementary Section~\ref{si:opus-results} intersect themselves, so the mesher failed to generate conventional tetrahedral volumetric meshes for them, whereas MechFoundry completed them. To sharpen the representation, the SDF is filtered to obtain a phase field on the octree, taking a value of one inside the body and zero outside, with a thin transition layer at the surface called the diffused zero level set. Supports and loads act through this transition layer, and the supports are imposed weakly on it by the diffuse Nitsche method~\cite{nguyen2018diffuse}. The solver computes the solution of a nonlinear elastostatic problem utilizing a compressible neo-Hookean material and is written as differentiable GPU code in NVIDIA Warp~\cite{macklin2022warp} (Methods, ``Diffuse Nitsche finite element analysis''). The diffuse Nitsche method is readily reusable for all types of standard constitutive and damage theories encountered in solid mechanics.


\paragraph{Design.} Because the solver is differentiable, an approximate gradient of a design objective with respect to the signed distance field can be obtained from one additional linear solve by the adjoint method, and moving the surface along this gradient enables shape optimization (Methods, ``Sensitivity analysis'' and ``Shape optimization''). The gradient vanishes away from the surface, so a second design variable mirroring the phase field lets topological features emerge, enabling topology optimization (Methods, ``Topology optimization''). Both of these optimization procedures are easily interchangeable by following the computational graph backward to a different endpoint. By default, the surface and the octree are extracted again at every iteration, and the support and load regions are restored on each new surface, so they stay in place while the shape changes. The optimization examples in this work use one material and are loaded by tractions and body forces only, without prescribed non-zero displacements.

\paragraph{Operation of the workflow.} The steps run in a given order and wait at six approval gates, stages at which the workflow waits for the user, such as the boundary regions and the material. There, the proposals of the AI subagents (DeepSeek-V4.1-Flash~\cite{xu2026deepseek} by default) take effect only after the user accepts or edits them. A person answers the gates in a web interface (Supplementary Section~\ref{sec:si-webui}), or, alternatively, an AI agent operator, a coding agent (Claude Code with Opus~5.5 in Fig.~\ref{fig:agentic}, or others), answers them through a command line interface in place of the person (Methods, ``Operation by a coding agent''). Working from an operator skill, a set of generic instructions, reference material and scripts that frame its role as that of an engineer, the operator first writes a case brief stating how the real object is used, where it is held, what loads it experiences and how large each load is. From the brief, a script computes estimates of the peak stress and deflection, treating the loaded member as a cantilever. The operator then looks at each rendered picture of the shape, the segmented or selected regions, and the solution, before approving each step, compares the solve with the estimates and closes with a verdict on whether the case it ran is the one its instruction describes (Supplementary Fig.~\ref{fig:si-operator-loop}). The pre-trained models are used as released, and no model is trained or fine-tuned in this work.

We evaluate MechFoundry in three settings: agent-operated simulations of 50 objects, each from a few words of description (Fig.~\ref{fig:agentic}, Supplementary Fig.~\ref{fig:si-gallery} and Supplementary Section~\ref{si:opus-results}), variants of one object that differ in shape, per-part material properties, or supports and loads (Fig.~\ref{fig:showcase}), and finally shape and topology optimization of generated shapes under the loads they carry (Fig.~\ref{fig:optimization}).

\clearpage

\subsection*{From an object's name to a solved loading case in minutes}

\begin{figure}[htp]
  \centering
  \includegraphics[width=\linewidth]{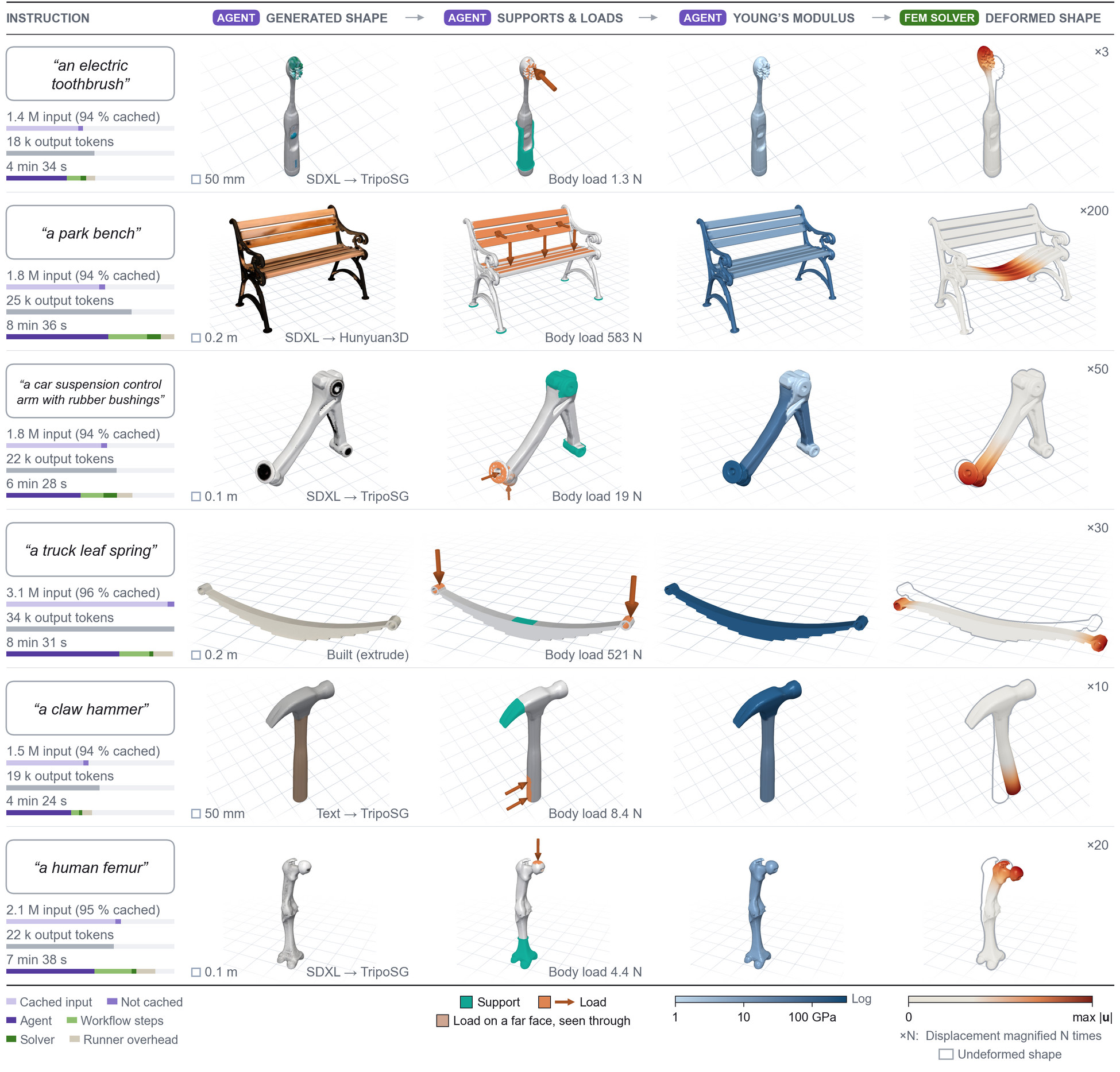}
  \caption{\textbf{The AI agent operator poses and solves loading cases from a few words of description.} Six of the 50 sessions autonomously run by Claude Code with Opus~5.5 (Supplementary Section~\ref{si:opus-results}), one per row. AGENT marks the steps that the operator decides, and FEM SOLVER the diffuse Nitsche solve that the operator reviews. Generated shape: the simulated body on a floor grid with the cell size at the lower left, and its generation route at the lower right (Built (extrude): a profile extruded by the operator). Supports (teal), loads (orange, with arrows along the load) and the weight of the body (body load). Young's modulus of each part, on a logarithmic scale, indicates the stiffness of the material. Parts with different colors also have different Poisson's ratios and densities. Deformed shape: the displacement magnitude $|\mathbf{u}|$, magnified by the factor at the upper right, with the undeformed outline in gray. Left column: input tokens with their cached share, output tokens and wall-clock time of the session, split into the operator (Agent), the workflow steps, the solver and the runner overhead. Runner overhead includes process starting, GPU selection, output rendering and other steps that cannot be classified into the other three categories. }
  \label{fig:agentic}
\end{figure}

We first demonstrate that the AI agent operator can take an object from its lexical description to a meaningful loading case and its solution. We chose 50 everyday and engineering objects (Supplementary Fig.~\ref{fig:si-gallery}) in 10 categories, five per category: household items, furniture, toys, food, animals, vehicle parts, machines and tools, medical and anatomy items, buildings, and landmarks. Fig.~\ref{fig:agentic} shows six of the 50 sessions, and Supplementary Section~\ref{si:opus-results} shows all of them. The operator, Claude Code with Opus~5.5, received only a brief lexical description of each object, such as ``a park bench'' (full prompt in Supplementary Section~\ref{sec:si-prompt}). Before any computation started, within the checks of its operator skill, the operator planned how the object is used, where it is held, what loads it and how large each load is, its real size, and its materials. 

After planning, the operator chose the route that produced each shape (Fig.~\ref{fig:agentic}, second column). Four of the six shapes came from an SDXL drawing passed to TripoSG or Hunyuan3D-2.1, one came from TripoSG directly, and the operator built the leaf spring as an extruded profile. The 3D surfaces undergo segmentation, which allows the operator to easily pose loading cases according to how each object is used (Fig.~\ref{fig:agentic}, third column). For instance, the electric toothbrush is held static at the grip and loaded on the brush head, and the park bench is loaded on the seat and the backrest and restrained at its four feet. The loading values are translated from realistic conditions, for example a weight or a force that a person applies, and each body carries its self-weight. The operator then assigns material properties (Fig.~\ref{fig:agentic}, fourth column) that can be different for different parts of an object. For example, in the case of the hammer, steel properties are applied to the head while wood properties are applied to the handle. This is followed by a GPU-accelerated finite element solver that produces deformation solutions (Fig.~\ref{fig:agentic}, last column). 

Each of the six sessions shown in Fig.~\ref{fig:agentic} took 4.4--8.6\,min of wall-clock time and 1.4--3.1 million input tokens, of which 94--96\% were reused from earlier steps at a lower price (cached), and 18--34 thousand output tokens (Fig.~\ref{fig:agentic}, first column). Most of the time was spent on operator reasoning and writing, and the workflow steps and the solver took the rest. All 50 sessions took 5\,h 23\,min, 6.5\,min per object on average, and cost a total of US\$65.29 at API price for the operator's model calls (Supplementary Table~\ref{tab:S_model_evaluation_astra}), excluding the AI subagent calls that normally cost less than US\$0.02 per object (Supplementary Section~\ref{sec:si-agents}).

We gave the same 50 instructions to the lower-cost Sonnet~5.5 and Haiku~5.5 models of the same model family (Supplementary Section~\ref{sec:si-model-compare}). Sonnet~5.5 cost less than half as much as Opus~5.5 (US\$26.22 against US\$65.29) and took less time (3\,h 56\,min against 5\,h 23\,min), whereas Haiku~5.5 cost US\$7.44 and took 5\,h 39\,min, but did not complete 4 sessions. Two panels of language-model reviewers, one running three Claude Fable~5.1 agents, from the same model family as the operators, and the other one running three GPT-6-Astra agents, graded the geometry, the supports and loads, and the materials of each session. The two panels agreed slightly to fairly (Cohen's kappa 0.19--0.25). Both panels graded the geometry, supports and loads, and materials as high in 15, 26 and 18 objects respectively from 50 Sonnet~5.5 sessions, 15, 24 and 19 objects from Opus~5.5 sessions, and 10, 20 and 15 objects from Haiku~5.5 sessions. Of the 12 materials that both graded low, 10 were in objects with soft parts (Supplementary Section~\ref{sec:si-model-eval}). Opus~5.5 received the fewest low grades for geometry, whereas for the materials and the supports and loads, Sonnet~5.5 and Opus~5.5 were rated similarly, although Sonnet~5.5 cost much less. 

Because the octree does not follow the surface, the diffuse Nitsche solution can differ from a conventional finite element analysis on a mesh that follows the surface. We therefore solved each of the 50 sessions again with conventional finite elements on three linear tetrahedral meshes that follow the surface and are discretized differently (Supplementary Section~\ref{si:error-comparison}). Tetrahedra were used because a mesh that follows a generated surface is far easier to autonomously build from tetrahedra than from hexahedra~\cite{schneider2022large}. Linear tetrahedra, however, are known to be stiffer than trilinear hexahedra in bending and to lock more readily~\cite{schneider2022large}, so part of the discrepancy can come from a different choice of element type. Three surfaces intersect themselves and were rejected by the tetrahedral mesher, leaving 47 objects to be compared. The comparison used double precision, a tighter Newton tolerance, one averaged material per object, and supports and loads posed as the diffuse Nitsche method poses them. At the default finest octree level of 7 (Methods, ``Adaptive octree extraction''), the median displacement discrepancy between the two solutions, measured as the relative $L^2$ norm after removal of a rigid-body motion, was 4.6--6.6\% against three tetrahedral meshes. One octree level finer, the discrepancy fell to 2.7--3.0\%. Thus, for a typical object, the two methods predict displacement fields that differ by a few percent. Their displacement gradients differ more: in the $H^1$ norm, which additionally measures the gradient, the median discrepancy was 16--19\% at level 7 and 8.1--8.4\% at level 8. The diffuse Nitsche solutions were slightly stiffer, with a median ratio of their compliance to the linear tetrahedral one of 0.91 at level 7 and 0.96 at level 8. At level 8, preparing the octree, the boundary conditions and the solver took a median of 2.2\,s, against 16--44\,s for preparing the tetrahedral meshes. The solves took a median of 7.6\,s for the diffuse Nitsche method and 7.9--21\,s on the tetrahedral meshes. Regarding the larger discrepancies, for 14 of the 47 objects at octree level 7 and for 10 at level 8, mostly objects with thin walls, narrow gaps or small supports near the resolution of the octree, the $L^2$ discrepancy exceeded 10\% against at least one tetrahedral mesh. 

\subsection*{Varying shape, materials, supports and loads}

\begin{figure}[htp]
  \centering
  \includegraphics[width=\linewidth]{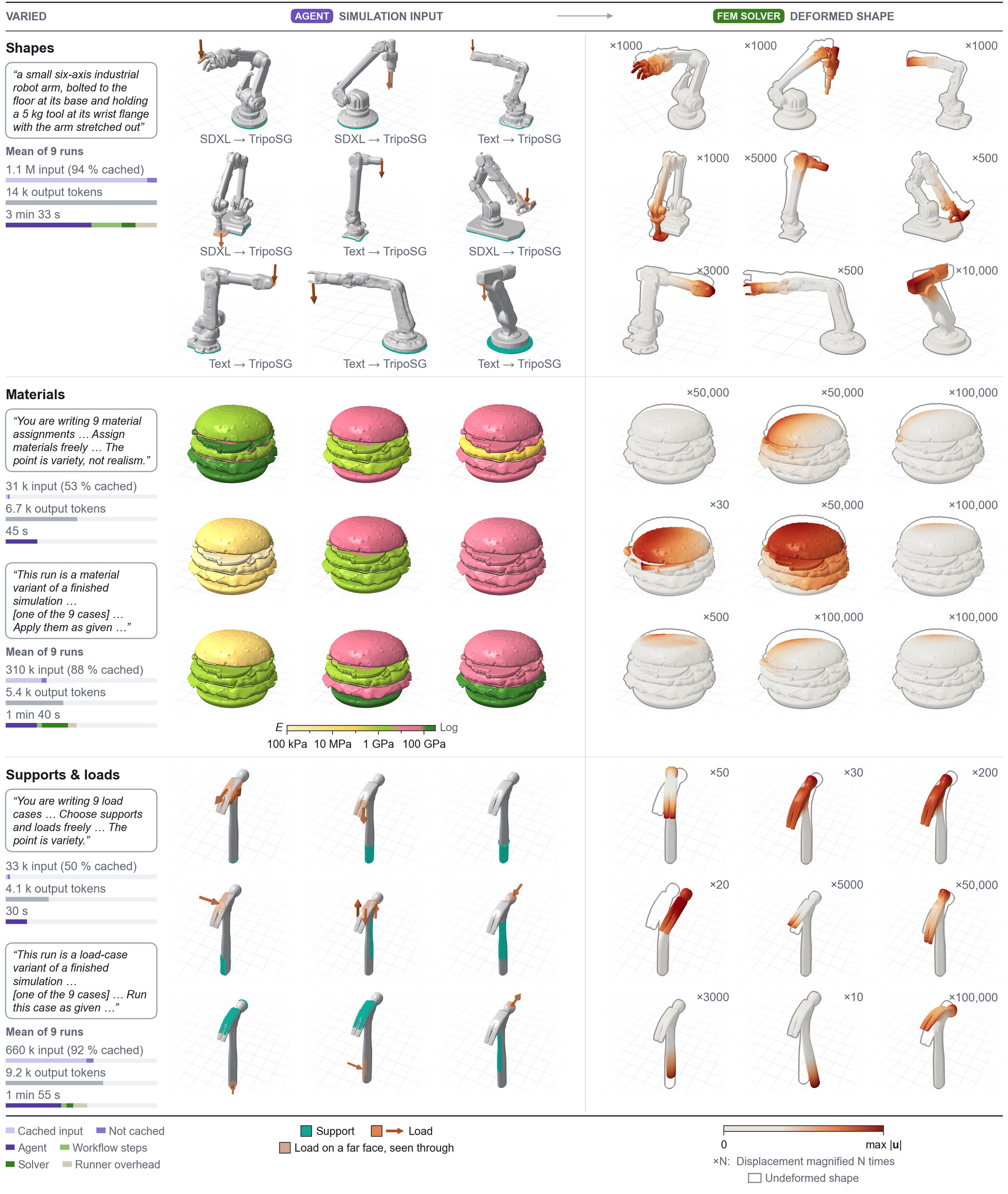}
  \caption{\textbf{The AI agent operator generates varied simulations of one object.} Each row varies one input over nine sessions by Sonnet~5.5 and shows the instruction and the mean cost of a session (left), the input set by the operator (AGENT, middle) and the deformed shape (FEM SOLVER, right). \textbf{Shapes}: nine robot arms from one instruction, differing in the generation seed and labeled at the bottom with the generation route that the operator chose. \textbf{Materials}: nine material assignments for the parts of a hamburger, written in one request (upper left text block and cost bars) and each written assignment applied and solved (lower left text block and cost bars). \textbf{Supports and loads}: nine load cases for a claw hammer, written in one request (upper left) and each run realizes one of the load cases (lower left), with the cost bars as for the hamburger. Colors, deformed shapes, and cost bars as in Fig.~\ref{fig:agentic}.}
  \label{fig:showcase}
\end{figure}

Comparing design options early and optimizing from several starting points or with adaptive sampling both need many variants of one object that differ in shape, materials, supports and loads. So do the training datasets of learned surrogates that can speed up both~\cite{yang2021deep,hong2025deepjeb}. With the AI agent operator, the same workflow can produce such variations without a person (Fig.~\ref{fig:showcase}). This autonomy can be achieved either by repeating a session with a different generation random seed or by asking the operator to write many cases and then running each. The prompt for varying shapes is in Supplementary Section~\ref{sec:si-prompt}, and the other prompts are in the released code. All variations are run with Sonnet~5.5. 

\paragraph{Shapes.} We gave nine operator sessions the same instruction, ``a small six-axis industrial robot arm, bolted to the floor at its base and holding a 5 kg tool at its wrist flange with the arm stretched out'', with a different generation random seed for each (Fig.~\ref{fig:showcase}, top). The operator chose the generation route of each independently, and consequently the nine robot arms differ in their proportions, joints and bases. In each, the operator placed the support on the base and the same load at the wrist, as the instruction states, and the solved deflections differ among the arms. A session took 3\,min 33\,s and 1.1 million input tokens (94\% cached) on average.

\paragraph{Materials.} For a hamburger, for example, one request asked the operator to write nine material assignments, with each segmented part of the object free to take any material, so that the cases vary as widely as possible (Fig.~\ref{fig:showcase}, middle). The assigned Young's moduli are shown on a scale from 100\,kPa to 100\,GPa. Writing the nine assignments took 45\,s, and applying each assignment to the finished simulation and solving it took 1\,min 40\,s on average. Under the same supports and loads, the nine burgers deform in different patterns, depending on which layers are soft.

\paragraph{Supports and loads.} For the claw hammer, a request likewise asked for nine load cases, with the supports and loads chosen freely (Fig.~\ref{fig:showcase}, bottom). The nine cases hold the hammer at different parts and load it at different places and in different directions. Writing the nine cases took 30\,s, and each run took 1\,min 55\,s on average.

In summary, the variations use one description to produce simulations with different geometry, material maps, boundary regions and full-field displacements. Each further variation costs one more session, which is cheaper than a fresh session starting from a minimal description. 

\clearpage

\subsection*{Shape and topology optimization}

\begin{figure}[htp]
  \centering
  \includegraphics[width=0.98\linewidth, trim=10mm 0mm 10mm 0mm, clip]{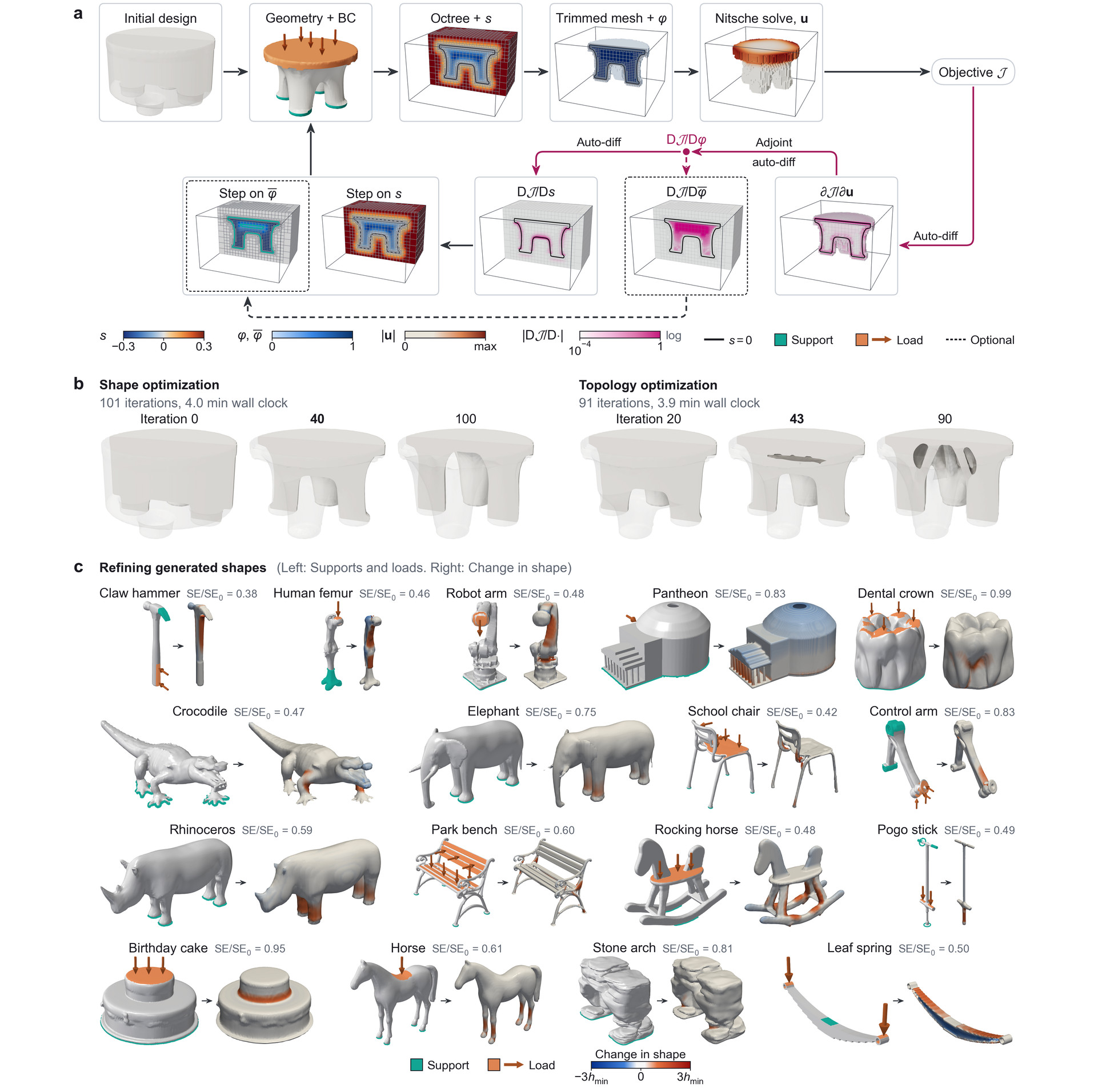}
  \caption{\textbf{Gradient-based optimization moves material to where generated objects carry their loads.} \textbf{a}, One iteration for a table loaded on its top (orange) and held at its four feet (teal). Forward (black arrows in the first row): the signed distance field $s$, the phase field $\varphi$, the diffuse Nitsche solve for the displacement $\mathbf{u}$ and the objective $\mathcal{J}$. Backward (magenta arrows): automatic differentiation and the adjoint method give the sensitivity $\mathrm{D}\mathcal{J}/\mathrm{D}s$, which updates $s$. Optionally (dashed), the interior sensitivity updates a hidden phase field $\bar{\varphi}$, which opens new topological features. Sensitivity magnitudes are on a logarithmic scale. The black line is the surface $s=0$. \textbf{b}, Shape optimization (left, 101 iterations, 4.0\,min) and topology optimization (right, 91 iterations, 3.9\,min) of the table toward half of its initial volume while minimizing compliance, with the near half translucent to reveal the profile. Iteration 43 of topology optimization is shown in \textbf{a}. \textbf{c}, Refinement of 17 generated bodies from the agent-operated sessions under their own supports and loads (left of each pair), with growth of the volume penalized. Right of each pair: the signed distance of the refined surface from the generated one, from $-3\,h_{\min}$ (blue: material removed) to $3\,h_{\min}$ (red: material added), with $h_{\min}$ the finest octree cell size. SE/SE$_0$: strain energy relative to the initial generated body, where lower is stiffer for the same load. Times are on one NVIDIA RTX A6000 GPU.}
  \label{fig:optimization}
\end{figure}

MechFoundry can then improve a design against the loads it carries. A generated shape resembles the object it names, but nothing in its generation ensures that it carries those loads well. It can, however, serve as a realistic starting point for optimization, whose result depends on the design it starts from. Because the diffuse Nitsche solver is differentiable, the shape can be changed to optimize for its load case while keeping the support and load regions in place (Fig.~\ref{fig:optimization}a). Each iteration solves the deformation, evaluates the objective, i.e., the total strain energy under a penalty on volume above a target (Methods, ``Shape optimization''), and computes its sensitivity with respect to the signed distance field by the adjoint method. A gradient descent step moves the signed distance field, after which the surface and the octree are extracted again and the support and load regions are restored (Methods, ``Sensitivity analysis'' and ``Shape optimization''). For topology optimization, the sensitivity with respect to a hidden phase field also allows for the emergence of new topological features inside the solid (Methods, ``Topology optimization'').

We first optimized a rough initial design of a table under a volume penalty targeting half of its initial volume, with fixed supports at its four feet and a distributed load on its top surface (Fig.~\ref{fig:optimization}b). Shape optimization carved the underside into four legs under the top in 101 iterations and 4.0\,min. For topology optimization, with the hidden phase field enabled from iteration 35, the first cavity-like feature opened inside the solid at iteration 43 and reached the surface by iteration 80, leaving the underside with a more complex topology after 91 iterations and 3.9\,min (Supplementary Figs.~\ref{fig:si-shapeopt} and~\ref{fig:hiddenphi}). Because half of the material was removed, the total strain energy, a measure of energy stored in the material due to deformation, ended at 1.62 times its initial value in both runs, at a volume fraction of approximately 0.5, and opening cavities gave a different topology at a similar stiffness (Supplementary Figs.~\ref{fig:si-so-objective} and~\ref{fig:si-hp-objective}).

We then applied shape optimization to bodies from the 50 agent-operated sessions to improve each generated design under the loads that its session posed, and show 17 of them in Fig.~\ref{fig:optimization}c, each under the supports and loads of its session. The design objective is now the total strain energy under a penalty on volume above the initial volume, aiming at a stiffer structure under load control by redistributing the volume. Bodies with a different material per part used one material instead, i.e., the volume-weighted mean of their parts (Supplementary Section~\ref{sec:si-shapeopt}); concurrent multi-material optimization is possible~\cite{luo2026towards} but beyond the scope of this work. In these optimization cases, each run stopped after 40 updates or before the surface moved more than $1.5\,h_{\min}$ (99th percentile) or $3\,h_{\min}$ (anywhere) from the generated one, where $h_{\min}$ is the size of the finest octree cell, so that the object stays recognizably the same (Supplementary Section~\ref{sec:si-shapeopt}). With growth of the volume penalized, the refinement reinforced the load-carrying members (e.g., the lower legs of the elephant, the rhinoceros and the horse, the shaft of the femur and the legs of the school chair), and removed material from regions that carry little load. Under a fixed load, which is the case for all examples shown, a lower strain energy means a stiffer body. The strain energy at the last accepted shape fell to 0.38--0.99 times its initial value, below 0.5 for 7 of the 17 bodies, for example the claw hammer (0.38), the school chair (0.42) and the femur (0.46), whereas the dental crown changed little (0.99).

\section{Discussion}
\label{sec:conclusions}

MechFoundry lets an AI agent operator decide and act at the steps that precede an engineering analysis, from how an object is used, held and loaded to what material it is made of, and then turn a few words of description into a nonlinear simulation of a deformable object, with every approval gate of the workflow answered by the operator. Generated shapes are simulated on GPUs by the diffuse Nitsche method on an adaptive octree of their signed distance field, a grid that need not conform to the surface, and AI subagents can place supports, loads and materials through the segmented parts of each shape. In our tests, the operator took each of the 50 everyday and engineering objects from its name to a solved loading case in minutes (Supplementary Fig.~\ref{fig:si-gallery}), and the same workflow produced variations of shape, materials, and supports and loads from one description. For a typical object, the displacements differed from those of body-fitted finite elements by a few percent. Gradient-based optimization made generated shapes stiffer under their loads.

The conventional finite element analysis steps that take an expert hours to days for a real-world object, from modeling and meshing to placing supports and loads, ran without a human in the loop. The procedure can be repeated across objects, shapes, materials and loads at the cost of one session each, about 6.5\,min and US\$1.3 of operator model calls per object with Opus~5.5, or less than half that cost with Sonnet~5.5, which received similar grades for the supports, loads, and materials. Engineers could therefore analyze a design while it is still being decided, and analyze free-form objects that have no CAD model at all.

The current workflow has limits. The solver is static, with dead loads and without contact, so it describes how an object carries its loads, whereas motion and contact with other bodies are not yet modeled. Solutions from the diffuse treatment of the surface converge at about first order with mesh refinement, strains (displacement gradients) are less accurate than displacements, and features near the octree resolution, such as thin walls and small supports, show larger discrepancies from body-fitted solutions. We did not evaluate the simulations with human experts or with a more rigorous protocol for language-model evaluation. Either would need a separate study that first defines what makes a finite element simulation of an everyday object sound, from its loading case and materials to its numerical settings.

Contact with friction and fluid--structure interaction would let the workflow pose loading cases in which objects touch, are grasped or are surrounded by fluid, which robot environments need, and dynamic analysis would extend static responses to incorporate inertial effects. A session takes minutes, so optimization can start from many generated candidates of one object and explore designs from multiple starting points. Further steps include incorporation of material models such as plasticity and viscoelasticity, as well as neural representations of constitutive laws~\cite{fuhg2023modular,fuhg2025review}, which can help in rapid identification of unknown materials. An additional direction is the enforcement of manufacturability constraints on the designed objects. Beyond engineering design, the same simulations could serve as data. Learned surrogates of mechanical response~\cite{yang2021deep,lu2021learning} need training sets that vary shapes, loads and materials~\cite{karniadakis2021physics}, whereas existing simulation datasets vary them only within predefined parametrizations~\cite{hong2025deepjeb,lejeune2020mechanical,hendriks2025wallpaper}. Virtual environments for training robots~\cite{hwangbo2019learning} need deformable objects with known mechanical response~\cite{yin2021modeling}, digital twins need model building that scales~\cite{niederer2021scaling}, and world models~\cite{hafner2025mastering}, which still violate basic physical laws~\cite{motamed2026generative}, could learn physical responses from such data. Therefore, running the workflow over large lists of objects, and training surrogates and world models on the resulting simulations, is the next, broader objective of this effort.

\section*{Methods}

MechFoundry takes a text prompt, a scribble, a single image or an uploaded surface to a mechanically analyzed and, on request, optimized solid object. The surface is scaled to physical size and segmented, boundary regions corresponding to parts are selected on the segments, and the deformation is solved on an adaptive octree by the diffuse Nitsche method, without a body-fitted volumetric mesh. AI subagents, requests to a multimodal language model, propose inputs at approval gates, stages at which the workflow waits for the user, and an AI agent operator can answer the gates in place of a person. The pre-trained components are used as released, and no deep learning model needs to be trained or fine-tuned in this work. The computation runs on one NVIDIA RTX A6000 GPU and an AMD Ryzen Threadripper 3970X CPU, implemented in NVIDIA Warp~\cite{macklin2022warp} and PyTorch~\cite{paszke2019pytorch}, and results use the default configuration given in the Supplementary Information, apart from a few settings in the optimization examples. Further details are in the Supplementary Methods.

\subsection*{Implicit geometry}

\paragraph{Signed distance representation.} The body is carried by a scalar field $s(\mathbf{X})$, negative inside and positive outside, whose zero level set is the surface. When $|s|$ equals the distance to the surface, $s$ is a signed distance field and satisfies the eikonal condition $|\nabla s|=1$, which the diffuse Nitsche method~\cite{nguyen2018diffuse} assumes.

\paragraph{3D shape synthesis by TripoSG and alternatives.} By default, the shape comes from TripoSG~\cite{li2025triposg}, whose rectified-flow transformer~\cite{liu2022flow} samples a latent code conditioned on an image, or on a scribble or text through its scribble model. Its decoder maps the code to a truncated signed distance field, which saturates away from the surface and only approximates a distance near it. Therefore, it serves only to extract the surface. Alternatively, Text2CAD~\cite{khan2024text2cad} and CAD-Coder~\cite{guan2026cad} turn a text prompt into a CAD solid, Hunyuan3D-2.1~\cite{hunyuan3d2025hunyuan3d} turns an image into a colored surface, and Stable Diffusion XL~\cite{podell2024sdxl} can draw the input image from text.

\paragraph{Adaptive octree extraction.} An adaptively refined octree, a grid of cubic cells subdivided only where needed, is built in the manner of multiresolution isosurface extraction~\cite{mescheder2019occupancy}, by subdividing the cells near the surface, keeping neighboring cells within one level of each other~\cite{sundar2008bottom,burstedde2011p4est} and querying the field at the new nodes only. For TripoSG, a first extraction queries the truncated signed distance field and is followed by marching cubes~\cite{lorensen1987marching} to extract a surface. Every surface is then scaled uniformly to an approved physical length and becomes the source surface. A second extraction, on the exact signed distance to the source surface, with its sign from the generalized winding number~\cite{jacobson2013robust}, builds the octree mesh for the analysis, with finest cell width $h_{\min}=0.015\,L_{\max}$ at the default octree level 7, where $L_{\max}$ is the longest side of the bounding box of the scaled source surface. The nodes of a fine cell that are not nodes of a neighboring coarse cell are hanging nodes~\cite{burstedde2011p4est}, whose values are averaged from the coarse cell's nodes.

\paragraph{Signed distance reinitialization.} The discretized signed distance field cannot represent fine details (sharp edges, thin walls, notches, etc.) and, interpolated trilinearly, no longer satisfies the eikonal condition. Therefore, we extract a reconstructed surface from it by marching cubes and re-evaluate the exact signed distance to that surface at every node. Thus, the solution is that of the reconstructed surface, on which sharp edges are rounded at the scale of $h_{\min}$, thinner features are not represented faithfully, and members a few $h_{\min}$ thick are not precisely resolved in bending.

\subsection*{Boundary-region identification}

In finite element frameworks, a boundary region is conventionally located by interactive selection or by a geometric test. Neither is viable on a non-conforming octree mesh for AI subagents that work from images and text. Therefore, we find the region on an explicit surface in three steps. First, the source surface is segmented, by default into five segments by clustering~\cite{ward1963hierarchical} the per-facet features of the pre-trained part-feature network PartField~\cite{liu2025partfield}. A surface from a text-to-CAD model keeps its CAD faces as segments. Second, a selector program, written and visually checked by an AI subagent, builds each region as a set of facets from 25 primitive selectors, for example by position, normal direction, segment or flood fill, combined by Boolean operations. Third, a quadrature point of the octree mesh belongs to the region when its distance to the nearest selected facet is below $0.75\,h_{\min}$. The marked points form the indicator $M_D^{i}$ of the $i$-th Dirichlet (support) region, $M_N^{i}$ of the $i$-th Neumann (load) region, or $M^{i}$ for either. A support prescribes one scalar displacement value, on all three components or on chosen ones, and a load prescribes a constant traction. A body force $\mathbf{b}$ may also be added. All loads are dead loads, which do not follow the deformation, and the analysis is static.

\subsection*{Diffuse Nitsche finite element analysis}

\paragraph{Phase field and diffused zero level set.} Because the octree cells do not conform to the surface, we follow the diffuse-domain approach~\cite{li2009solving} and the diffuse Nitsche method~\cite{nguyen2018diffuse} and replace the body $\Omega$ by a phase field, taking a value of unity in the interior and zero in the exterior, $\varphi=\tfrac12\bigl(1-\tanh(s/\varepsilon)\bigr)$ with $\varepsilon=\tfrac12h_{\min}$ in the transition layer near the boundary. Its thin transition layer is the diffused zero level set. Elements whose nodal values all satisfy $\varphi<10^{-3}$ are removed, leaving the analysis (trimmed) mesh $\mathcal{K}$, and its elements with a node at $|s|<3.45\,\varepsilon$, where $10^{-3}<\varphi<1-10^{-3}$, form the diffuse boundary band $\mathcal{B}$. Integrals over the body $\Omega$ and over its surface in region $i$, $\partial\Omega^{i}$, become volume integrals over the octree,
\begin{equation}
  \int_{\Omega}(\cdot)\,\mathrm{d}V\approx\int_{\mathcal{K}}(\cdot)\,w(\varphi)\,\mathrm{d}V,\qquad
  \int_{\partial\Omega^{i}}(\cdot)\,\mathrm{d}A\approx\int_{\mathcal{B}}(\cdot)\,|\nabla\varphi|\,M^{i}\,\mathrm{d}V,
  \label{eq:m-measures}
\end{equation}
with the material weight $w(\varphi)=(1-\varphi_0)\varphi+\varphi_0$, $\varphi_0=10^{-4}$.

\paragraph{Weak form.} The solid is a compressible neo-Hookean material~\cite{bonet2008nonlinear} with strain energy density $W(\mathbf{F})$ defined in Supplementary Methods, deformation gradient $\mathbf{F}=\mathbf{I}+\nabla\mathbf{u}$ and first Piola--Kirchhoff stress $\mathbf{P}$. Dirichlet conditions are imposed weakly by a symmetric Nitsche method~\cite{nitsche1971variationsprinzip}. The trilinear displacement $\mathbf{u}$ satisfies $R(\mathbf{u},\mathbf{v})=0$ for an arbitrary test displacement $\mathbf{v}$ in the same space, with
\begin{align}
  R(\mathbf{u},\mathbf{v}) ={}&
  \int_{\mathcal{K}} w(\varphi)\,\bigl(\mathbf{P}\!:\!\nabla\mathbf{v}-\mathbf{b}\cdot\mathbf{v}\bigr)\,\mathrm{d}V
  -\sum_i\int_{\mathcal{B}}\chi\,M_N^{i}\,(\bar{\mathbf{t}}^{i}\cdot\mathbf{v})\,|\nabla\varphi|\,\mathrm{d}V
  \nonumber\\
  &+\sum_i\int_{\mathcal{B}}\chi\,M_D^{i}\Bigl[
      (\mathbf{P}\nabla\varphi)\cdot\mathbf{v}
      +\bigl(\mathbb{A}[\nabla\mathbf{v}]\nabla\varphi\bigr)\cdot(\mathbf{u}-\bar{\mathbf{u}}^{i})
  +\beta\,(\mathbf{u}-\bar{\mathbf{u}}^{i})\cdot\mathbf{v}\,|\nabla\varphi|
    \Bigr]\mathrm{d}V ,
  \label{eq:m-residual}
\end{align}
where $\mathbb{A}[\mathbf{H}]=(\partial\mathbf{P}/\partial\mathbf{F})\!:\!\mathbf{H}$ is the material tangent, $\chi(\mathbf{X})$ is one where $\varphi>\varphi_0$ and zero elsewhere, $\bar{\mathbf{u}}^{i}$ and $\bar{\mathbf{t}}^{i}$ are the prescribed displacement and traction, and $\beta(\mathbf{X})$ is the penalty. The bracket holds the consistency term (the boundary traction on the Dirichlet region), the symmetry term (its transpose, which makes the tangent symmetric) and the penalty term.

\paragraph{Load correction and penalty stabilization.} Because $M_N^{i}$ also reaches the diffused zero level set of the adjacent, unloaded face, we replace the weight $|\nabla\varphi|$ of the traction term by the positive part of $-\nabla\varphi\cdot\mathbf{n}_f^{i}$, with $\mathbf{n}_f^{i}$ the region's mean normal, and rescale each traction once so that the region's total force equals the traction times its projected area. 

The penalty term must dominate the consistency and symmetry terms. Following the stability argument of Ref.~\cite{nguyen2018diffuse}, derived there for the scalar Poisson problem, the penalty factor $\beta$ of each band element is $5$ times the largest eigenvalue of a generalized eigenvalue problem~\cite{embar2010imposing} that bounds its normal-traction energy by its strain energy, with a floor. Computed with the small-strain elasticity tensor for efficiency, the solution of $\beta$ only estimates the bound that coercivity needs.

\paragraph{Newton solution and linear solver.} All elements carry a $2\!\times\!2\!\times\!2$ Gauss rule. Newton's method is applied in a single load step, without line search in our results, until the residual norm falls to $10^{-2}$ of its initial value by default or below an absolute floor. The tangent omits the second derivative of $\mathbf{P}$ in the symmetry term, so the iteration is quasi-Newton. The linear systems are solved in either single precision or double precision by conjugate gradients with an algebraic-multigrid preconditioner (NVIDIA AmgX~\cite{naumov2015amgx}, through JAX-AMG~\cite{liu2026jax}). Because $\varepsilon$ is tied to $h_{\min}$, the modeling error of the diffuse-domain approximation is refined together with the discretization error. In a preliminary comparison on one problem, the displacement error converges at roughly first order, against roughly second order for body-fitted finite elements (Supplementary Methods).

\subsection*{Sensitivity analysis}

Optimization minimizes an objective $\mathcal{J}(\mathbf{U},\mathbf{m})$ subject to $\mathbf{R}(\mathbf{U},\mathbf{m})=\mathbf{0}$, where $\mathbf{U}$ and $\mathbf{m}$ are the nodal values of the displacement $\mathbf{u}$ and of the signed distance field $s$, and $\mathbf{R}$ is the residual vector, $R(\mathbf{u},\mathbf{v})=\mathbf{V}^{\mathsf{T}}\mathbf{R}$ in Eq.~\eqref{eq:m-residual} with $\mathbf{V}$ the nodal values of $\mathbf{v}$. Finite differences would cost one nonlinear solve per design variable, prohibitive for a nodal design space. Therefore, the derivative is obtained by the adjoint state method~\cite{allaire2004structural}, at the cost of one extra linear solve with the same symmetric tangent for the adjoint state $\boldsymbol{\lambda}$,
\begin{equation}
  \frac{\mathrm{D}\mathcal{J}}{\mathrm{D}\mathbf{m}}
  =\frac{\partial\mathcal{J}}{\partial \mathbf{m}}
  +\boldsymbol{\lambda}^{\mathsf{T}}\frac{\partial\mathbf{R}}{\partial \mathbf{m}},
  \qquad
  \Bigl(\frac{\partial\mathbf{R}}{\partial\mathbf{U}}\Bigr)^{\!\mathsf{T}}\boldsymbol{\lambda}
  =-\Bigl(\frac{\partial\mathcal{J}}{\partial\mathbf{U}}\Bigr)^{\!\mathsf{T}}.
  \label{eq:m-adjoint}
\end{equation}
Reverse-mode automatic differentiation over a chain-rule recording \textit{tape} gives the explicit derivatives, and a custom node on the tape solves for $\boldsymbol{\lambda}$ with the same preconditioned conjugate-gradient iteration and stopping rule as each Newton step. Outside the boundary band $\mathcal{B}$, the slope $\mathrm{d}\varphi/\mathrm{d}s=-2\varphi(1-\varphi)/\varepsilon$ nearly vanishes, so the chain from $s$ to $\varphi$ passes almost no gradient back. Thus, we keep it off the tape there to save time, and $s$ carries no stiffness sensitivity in the interior. The result is approximate, because the forward state is converged only to the Newton tolerance, the tangent is quasi-Newton, and quantities rebuilt off the tape, such as the penalty, contribute no sensitivity. A central-difference test agrees within $1\%$ at nodes farther than $\varepsilon$ from the surface and within $15\%$ closer to it, and the sensitivity is used as an inexact search direction.

\subsection*{Shape optimization}

As in level-set structural optimization~\cite{wang2003level,allaire2004structural}, the design variable for shape optimization is $s$. We seek a stiff design whose volume does not exceed a target fraction of the initial volume. Therefore, we minimize the normalized strain energy under a one-sided volume penalty,
\begin{equation}
  \mathcal{J}
  =\frac{\mathrm{SE}}{\mathrm{SE}_0}
  +w_{\mathrm{vol}}\,\mathrm{softplus}_k\!\Bigl(\frac{V/V_0-f_{\mathrm{target}}}{1-f_{\mathrm{target}}}\Bigr),
  \label{eq:m-objective}
\end{equation}
with $\mathrm{SE}=\int w(\varphi)W\,\mathrm{d}V$ the strain energy, $V=\int\varphi\,\mathrm{d}V$ the volume, $\mathrm{SE}_0$ and $V_0$ their first-iteration values, $w_{\mathrm{vol}}=5$, $f_{\mathrm{target}}=0.5$ and a smoothed hinge $\operatorname{softplus}_k$, $k=500$. Under traction loading, this is compliance minimization~\cite{bendsoe1989optimal} in the linear regime. With self-weight, the load also changes with the design. Under prescribed displacements, the strain energy falls as material is removed, so the optimization examples in this work are loaded by tractions and body forces, without prescribed non-zero displacements. Because $\mathrm{D}\mathcal{J}/\mathrm{D}s$ is meaningful only near the zero level set, the shape changes by moving its boundary. We update $s$ by plain gradient descent with the gradient set to zero near the boundary condition regions. A gradient descent step does not preserve the eikonal condition, so every iteration the field $s$ is reinitialized from a newly extracted surface and a fresh octree is extracted. The re-extracted surface re-creates the facets of the boundary condition regions inexactly and can leave sharp edges, so optionally the patch of facets around each region is saved at the first iteration and stitched back into each new surface with a smoothed junction~\cite{fuchs1977optimal,taubin1995signal}. The optimization examples in Fig.~\ref{fig:optimization}c replace the volume term by a one-sided penalty on volume above $V_0$, use the gradient mask only and skip the per-iteration reinitialization (details in the ``Refinement of generated bodies'' paragraph of Supplementary Section~\ref{sec:si-shapeopt}), whereas the table design problem in Fig.~\ref{fig:optimization}b uses both the gradient mask and stitching.

\subsection*{Topology optimization}

A moving boundary can merge surfaces, pinch off a neck or open a through-hole, but it cannot force a new topological feature to emerge inside the solid, a known limitation of level-set methods~\cite{xia2019stable}, because the sensitivity with respect to $s$ vanishes there. Therefore, we add a second nodal design variable, the hidden phase field $\bar{\varphi}$, which mirrors $\varphi$ and evolves where $s$ does not. At node $i$, with step $\eta_\varphi$,
\begin{equation}
  \bar{\varphi}_i \leftarrow
  \begin{cases}
    \tfrac12\bigl(1-\tanh(s_i/\varepsilon)\bigr), & |s_i|<3\varepsilon,\\[2pt]
    \bar{\varphi}_i-\eta_\varphi\,g_i, & |s_i|\ge3\varepsilon,
  \end{cases}
  \label{eq:m-hiddenphi}
\end{equation}
clamped to $[0,1]$, while $s$ receives its gradient only within $3\varepsilon$ of the surface. The solve uses only $\varphi$ and the tape does not record $\bar{\varphi}$, so the signal $g_i$ is the gradient of $\mathcal{J}$ with respect to $\varphi$, taken as that with respect to $\bar{\varphi}$. With the interior weight $w(\varphi)$ recorded on the tape, in the form of the solid isotropic material with penalization interpolation~\cite{bendsoe1989optimal} with exponent one, the same backward pass returns $g_i\simeq\partial\mathcal{J}/\partial\varphi_i+\boldsymbol{\lambda}^{\mathsf{T}}\partial\mathbf{R}/\partial\varphi_i$ without an additional adjoint solve. In $g_i$, the explicit derivative term is non-negative and alone could only remove material, whereas the adjoint term carries the structural response and can keep material. A void is admitted in an element whose nodes all lie at $s<-3\varepsilon$ and include one with $\bar{\varphi}<\tfrac12$. There, $\bar{\varphi}$ is mapped back to a signed distance, $\varepsilon\operatorname{atanh}(1-2\bar{\varphi})$, before the surface re-extraction, and the cavity then evolves with $s$ updating, as an ordinary part of the surface.

\subsection*{Language-model-assisted workflow}

A deterministic state machine, built with LangGraph~\cite{wang2024agent}, runs the stages in order and makes AI subagent requests at several of them. Each workflow execution is called a session. The requests go by default to DeepSeek-V4.1-Flash~\cite{xu2026deepseek} with thinking mode off. Each request carries no memory of earlier requests, uses zero temperature and must answer in a fixed data format. The proposals are applied only after the user, a person in the web interface or the AI agent operator in the command line interface, accepts or edits them at one of six approval gates: geometry input, physical scale, boundary regions, boundary values, material, and solver and optimizer settings. At the boundary-region gate, the AI subagent writes one selector program per region from the user's description of the supports and loads, rendered views of the segments, and a summary of each segment. Its instructions tell it to propose only the regions and values that the description states, and the user previews the regions before approving them. At the physical-scale, boundary-value and material gates, it proposes the length, the values and the (per-part) material constants respectively, choosing Young's modulus and Poisson's ratio, from which the shear and bulk moduli of the neo-Hookean model follow, against a reference table. A typical session uses 10--50 thousand uncached input tokens and 1--2 thousand output tokens for subagents of DeepSeek-V4.1-Flash, which costs less than US\$0.02 per session.

\subsection*{Operation by a coding agent}

\paragraph{AI agent operator and case brief.} The AI subagent proposes only what the user's description states. Therefore, to run the workflow from only an object name, description, image or surface mesh, an AI agent operator poses a loading case meaningful for the real object, chooses the numerical settings and looks at each rendered picture before approving the step that made it. The AI agent operator is a coding agent, Claude Code (with Opus~5.5, Sonnet~5.5, or Haiku~5.5, accessed in October 2026) in this work, working from an operator skill that consists of generic instructions, reference material and scripts that frame its role as that of an engineer in computer-aided engineering. It first writes a case brief on how the real object is used, the actions on it, where it is held, and its real size, mass and materials, each value marked as given or decided. Each load is sized from the mechanism that sets it, for example a mass times the gravitational acceleration or, for a force that a person applies, a table of 11 human actions in the operator's reference material. A script supplied with the skill, namely the driver, computes from the brief the support reactions and cantilever estimates of the peak stress and deflection, which are upper bounds for a member held at both ends.

\paragraph{Driver checks and operator's visual checks.} The brief from the AI agent operator can propose a case that is not of interest or that the workflow cannot faithfully or robustly complete. Therefore, the driver first checks that the operator selects the most severe case, and that the proposed supports constrain degrees of freedom as a real contact does and do not leave the object unbalanced. It also warns of a predicted deflection over $30\%$ of the largest extent or a stress at $90\%$ of the material strength. The operator then corrects the load, size or section, or keeps the case unchanged, and no case is refused for its large deflection or stress. The driver also checks the generated shape against the brief's size, wall thickness and mass, and rejects a body whose mass is off while untried generation routes remain. It stops wherever a rendered picture, for example of the geometry, the boundary regions or the solution fields, must be judged by the AI agent operator. A rejected shape is replaced by the next generation route in a fixed order, from the generators to a public-domain or CC~BY photograph and, normally as a last resort, a body built by the operator from the brief's dimensions.

\section*{Statements and Declarations}

\noindent \textbf{Acknowledgments.} The authors declare no funding for this work. X.-L. L. acknowledges Kandice Lu (Cornell University) for feedback on the web user interface and relevant discussions. 

\noindent \textbf{Code availability.} All code is open-sourced on GitHub (\url{https://github.com/LuoXueling/MechFoundry}).

\noindent \textbf{Data availability.} Raw data of experiments is released on Zenodo (\url{https://zenodo.org/records/23511564}). 

\noindent \textbf{Competing interests.} The authors declare no competing interests.

\begingroup\hbadness=1500 

\endgroup


\clearpage
\newgeometry{left=1.5cm,right=1.5cm,top=2cm,bottom=2cm}
\nolinenumbers

\makeatletter
\def\@blstr{1}
\def\@listI{\leftmargin\leftmargini \topsep\z@ \partopsep\z@ \parsep\z@ \itemsep\z@}
\let\@listi\@listI
\def\@listii{\leftmargin\leftmarginii \labelwidth\leftmarginii
  \advance\labelwidth-\labelsep \topsep\z@ \partopsep\z@ \parsep\z@ \itemsep\z@}
\def\@listiii{\leftmargin\leftmarginiii \labelwidth\leftmarginiii
  \advance\labelwidth-\labelsep \topsep\z@ \partopsep\z@ \parsep\z@ \itemsep\z@}
\setlength{\@sep}{0pt}
\setlength{\@@sep}{0pt}
\makeatother
\singlespacing
\setlength{\parskip}{0pt plus 1pt}
\setlength{\emergencystretch}{1.5em}

\setcounter{page}{1}
\setcounter{section}{0}
\setcounter{subsection}{0}
\setcounter{subsubsection}{0}
\setcounter{figure}{0}
\setcounter{table}{0}
\setcounter{equation}{0}
\setcounter{algocf}{0}
\setcounter{footnote}{0}
\renewcommand{\thepage}{S\arabic{page}}
\renewcommand{\thesection}{S\arabic{section}}
\renewcommand{\thefigure}{S\arabic{figure}}
\renewcommand{\thetable}{S\arabic{table}}
\renewcommand{\theequation}{S\arabic{equation}}
\renewcommand{\thealgocf}{S\arabic{algocf}}
\renewcommand{\theHsection}{S\arabic{section}}
\renewcommand{\theHfigure}{S\arabic{figure}}
\renewcommand{\theHtable}{S\arabic{table}}
\renewcommand{\theHequation}{S\arabic{equation}}
\renewcommand{\theHalgocf}{S\arabic{algocf}}
\renewcommand{\citenumfont}[1]{#1}
\renewcommand{\bibnumfmt}[1]{[#1]}

\makeatletter
\newcount\SI@k \newcount\SI@n \newcount\SI@last
\newcount\SI@fs \newcount\SI@fe \newcount\SI@ts \newcount\SI@te
\let\SI@section\section
\newcommand{\SI@record}[1]{%
  \immediate\write\@auxout{\string\expandafter\string\gdef\string\csname\space SI@sec@\the\value{section}\string\endcsname{{\unexpanded{#1}}{\the\value{figure}}{\the\value{table}}}}}
\RenewDocumentCommand{\section}{s o m}{%
  \IfBooleanTF{#1}{\SI@section*{#3}}{%
    \IfNoValueTF{#2}{\SI@section{#3}}{\SI@section[#2]{#3}}%
    \SI@record{#3}}}
\AtEndDocument{\immediate\write\@auxout{\string\gdef\string\SI@end{{\the\value{section}}{\the\value{figure}}{\the\value{table}}}}}
\def\SI@getstart#1#2#3{\def\SI@title{#1}\SI@fs=#2\relax\SI@ts=#3\relax}
\def\SI@getnext#1#2#3{\SI@fe=#2\relax\SI@te=#3\relax}
\def\SI@getcount#1#2#3{\SI@last=#1\relax\advance\SI@last\@ne}
\newcommand{\SI@range}[4]{%
  \ifnum#4>#3\relax
    \advance#3\@ne
    \ifnum#3=#4\relax #1~S\the#3\else #2~S\the#3--S\the#4\fi
  \fi}
\newcommand{\SIcontents}{%
  \@ifundefined{SI@end}{\item Run LaTeX again to list the sections.}{%
    \expandafter\SI@getcount\SI@end
    \SI@k=\@ne
    \@whilenum\SI@k<\SI@last\do{%
      \expandafter\expandafter\expandafter\SI@getstart\csname SI@sec@\the\SI@k\endcsname
      \SI@n=\SI@k \advance\SI@n\@ne
      \@ifundefined{SI@sec@\the\SI@n}%
        {\expandafter\SI@getnext\SI@end}%
        {\expandafter\expandafter\expandafter\SI@getnext\csname SI@sec@\the\SI@n\endcsname}%
      \item \hyperlink{section.S\the\SI@k}{S\the\SI@k\quad\SI@title}%
      \ifnum\SI@fe>\SI@fs\relax\ (\SI@range{Figure}{Figures}\SI@fs\SI@fe\ifnum\SI@te>\SI@ts\relax; \SI@range{Table}{Tables}\SI@ts\SI@te\fi)%
      \else\ifnum\SI@te>\SI@ts\relax\ (\SI@range{Table}{Tables}\SI@ts\SI@te)\fi\fi
      \advance\SI@k\@ne}}}
\makeatother

\pdfbookmark[0]{Supplementary Information}{supplementary-information}
\begin{nolinenumbers}
\begin{center}
  \thispagestyle{plain}
  \vspace*{-0.5cm}
  {\Large\bfseries Supplementary Information for\par}
  \vspace{1em}
  {\Large \papertitle\par}
  \vspace{2em}
  {\large Xue-Ling Luo\textsuperscript{a}, Marco Conte\textsuperscript{a}, Peter Yichen Chen\textsuperscript{b}, and Nikolaos Bouklas\textsuperscript{a,*}\par}
  \vspace{1em}
  {\itshape \textsuperscript{a}Cornell University, Ithaca, NY, USA\par \textsuperscript{b}University of British Columbia, Vancouver, Canada\par}
  \vspace{1em}
  {\small \textsuperscript{*}Corresponding author. E-mail: nb589@cornell.edu\par}
\end{center}

\vspace{1em}
\noindent\textbf{This file includes:}
\begin{itemize}
  \SIcontents
  \item Supplementary References
\end{itemize}

\noindent\begin{minipage}{\linewidth}
  \centering
  \expandafter\def\csname @captype\endcsname{figure}
  \includegraphics[width=\linewidth]{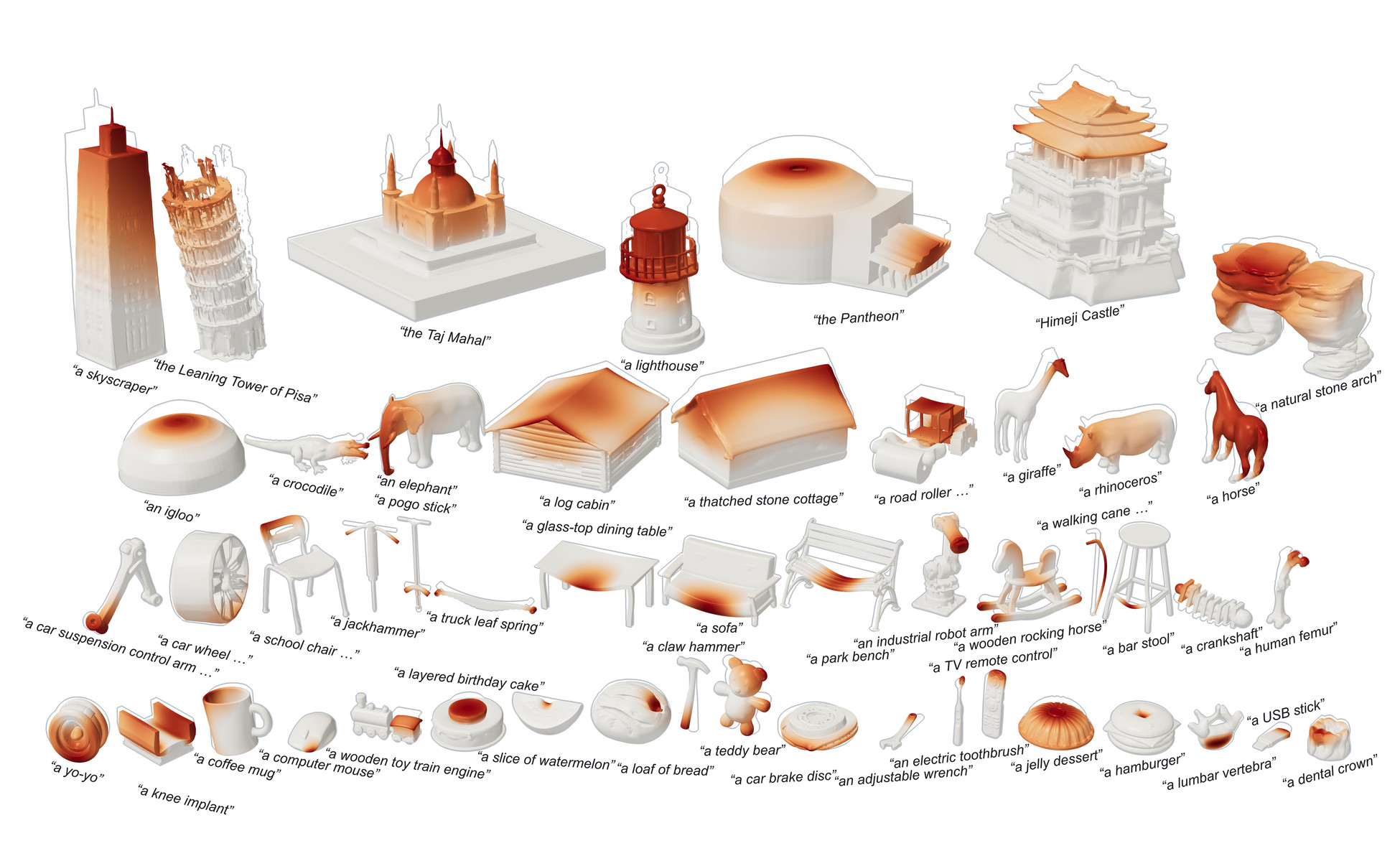}
  \caption{\textbf{The 50 objects simulated by the AI agent operator} (Claude Code with Opus~5.5), each labeled with its instruction (Section~\ref{si:opus-results}). Color shows the displacement magnitude on the deformed shape, with the undeformed outline in gray.}
  \label{fig:si-gallery}
\end{minipage}
\end{nolinenumbers}

\clearpage
\newpage


\section{Supplementary Methods}
\label{sec:si-methods}

\subsection{Overview}
\label{sec:si-overview}

\begin{figure}[tb]
  \centering
  \includegraphics[width=\linewidth]{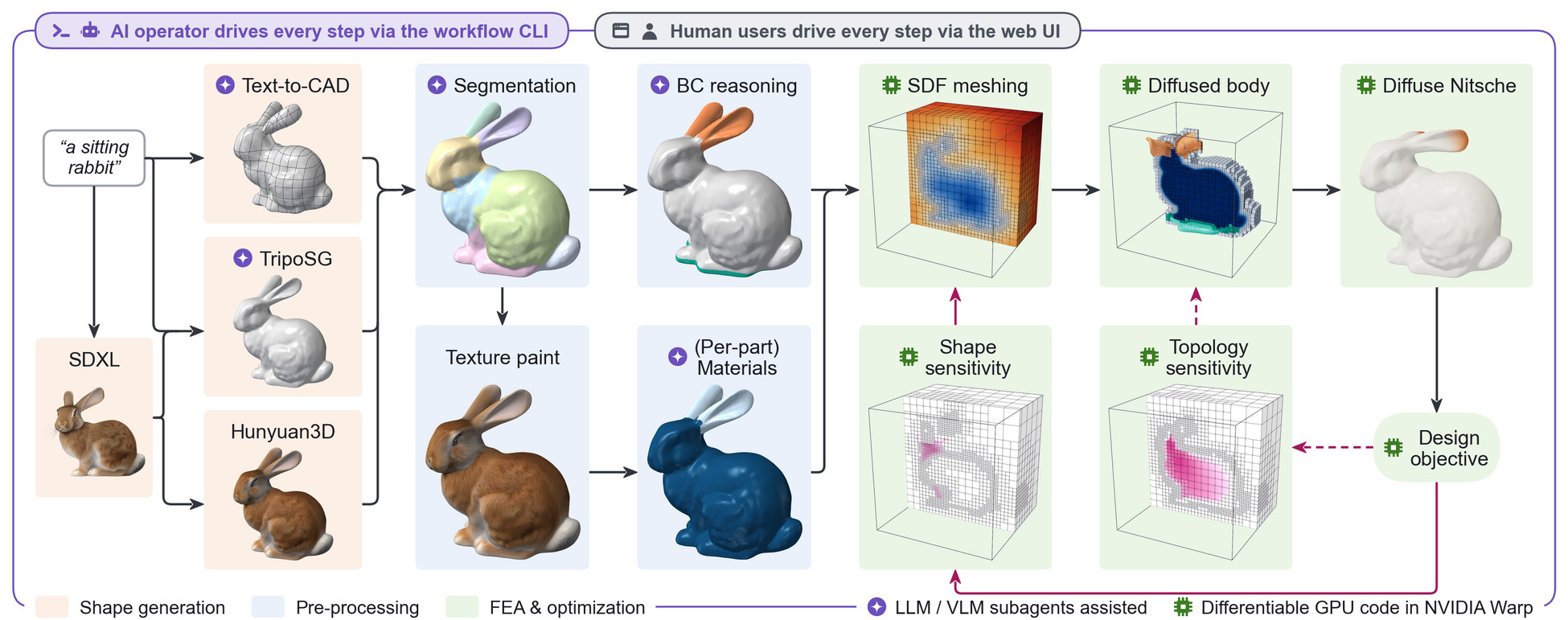}
  \caption{\textbf{Workflow overview.} A text prompt, optionally first drawn into an image by SDXL, is turned into a surface by TripoSG, Hunyuan3D-2.1 or a text-to-CAD model (Text2CAD or CAD-Coder) (orange: shape generation, Sec.~\ref{sec:si-geometry}). The surface is then segmented, and boundary condition regions are reasoned on the segments (Sec.~\ref{sec:si-bcregions}). Optionally, the surface is textured and per-part (heterogeneous) materials are assigned (Sec.~\ref{sec:si-nitsche}) (blue: pre-processing). The surface defines a signed distance field (SDF) $s$, from which an adaptive octree is extracted. The octree is trimmed to the body, the phase field $\varphi$ is evaluated on it, and the diffuse Nitsche solve returns the deformation and the objective $\mathcal{J}$ (green: finite element analysis and optimization, Sec.~\ref{sec:si-nitsche}). In shape optimization, the sensitivity $\mathrm{D}\mathcal{J}/\mathrm{D}s$ updates the signed distance field, from which the octree is rebuilt (solid arrows, Secs.~\ref{sec:si-sensitivity} and \ref{sec:si-shapeopt}). In topology optimization, the sensitivity $\mathrm{D}\mathcal{J}/\mathrm{D}\varphi$ also updates the hidden phase field $\bar{\varphi}$ (dashed arrows, Sec.~\ref{sec:si-hiddenphi}). Steps marked with a star can be assisted by an AI subagent (Sec.~\ref{sec:si-agents}), and steps marked with a chip run as differentiable GPU code in NVIDIA Warp. The whole workflow is driven either by a user through the web interface (Sec.~\ref{sec:si-agents}) or by the AI agent operator through the command line (Sec.~\ref{sec:si-operator}).}
  \label{fig:si-overview}
\end{figure}

MechFoundry takes a text prompt, a scribble with or without that prompt, or a single image to a completed structural analysis or, on request, an optimized solid (Fig.~\ref{fig:si-overview}). AI subagents (requests to a multimodal language model, Sec.~\ref{sec:si-agents}) assist the user throughout, and an AI agent operator can run the whole workflow without a person (Sec.~\ref{sec:si-operator}). All of this is implemented over a bespoke software base, designed to allow the agent to seamlessly pivot among required tasks (Sec.~\ref{sec:si-implementation} gives the implementation details and lists the components that we developed and those used as released). The input (a text prompt, a scribble or an image) conditions a 3D shape generator, from which a surface is extracted and scaled to physical dimensions. The user may instead upload a surface (Sec.~\ref{sec:si-geometry}). The surface is then segmented (Sec.~\ref{sec:si-bcregions}), an AI subagent turns the user's written description of the supports and loads into boundary regions on the segments (Sec.~\ref{sec:si-agents}) and defines part-specific materials (Sec.~\ref{sec:si-nitsche}), and the user reviews and approves the regions, their prescribed values and the material constants in the web interface (Sec.~\ref{sec:si-agents}). The computation then works on a signed distance field, the shortest distance to the scaled surface, negative inside and positive outside, and an adaptive octree is constructed and refined directly from the signed distance field around its zero level set, without a body-fitted (conforming) volumetric mesh (Sec.~\ref{sec:si-geometry}). All boundary conditions are imposed weakly; essential (Dirichlet) boundary conditions are imposed through Nitsche terms on the diffused zero level set, the thin transition layer of a phase field that is one inside the body and zero outside, and a single diffuse Nitsche solve returns the predicted deformation (Sec.~\ref{sec:si-nitsche}) with accompanying stress and strain fields. When the shape is optimized the solve is repeated, and an approximate sensitivity of the mechanical response with respect to the nodal signed distance field and the phase field is computed by the adjoint method with automatic differentiation (Sec.~\ref{sec:si-sensitivity}). The surface moves along that signed distance sensitivity while the boundary condition regions are kept in place (Sec.~\ref{sec:si-shapeopt}), and a hidden phase field lets voids open inside the solid (Sec.~\ref{sec:si-hiddenphi}).

\subsection{Implicit geometry}
\label{sec:si-geometry}

\paragraph{Signed distance representation.} The geometry is carried by a scalar field $s(\mathbf{X})$ that is negative inside the body and positive outside, and its zero level set $\partial\Omega=\{s=0\}$ is the surface. When $|s|$ equals the distance to the surface, $s$ is a signed distance field (SDF), and it satisfies the eikonal condition $|\nabla s|=1$. The diffuse Nitsche method (Sec.~\ref{sec:si-nitsche}) assumes this condition when it sets the width of the phase-field profile and evaluates surface integrals over the diffused zero level set. Given a watertight explicit 3D surface, the exact signed distance can be computed efficiently: a bounding volume hierarchy (a tree of nested boxes that makes the search for the nearest facet fast) gives the distance, and an inside-outside test (generalized winding number~\citeS{jacobson2013robust}) gives its sign.  

\paragraph{3D shape synthesis by TripoSG.} The default source of a 3D shape is TripoSG~\citeS{li2025triposg}, a pre-trained generator that produces 3D shapes from latent codes. Its image-conditioned model is used for an image, whether uploaded or generated from the text description. Its guidance-distilled scribble model is used for a scribble, with or without text, and also for text alone. The input serves only as conditioning for its rectified-flow transformer~\citeS{liu2022flow}, which samples a latent code $\Theta$. The latent space is that of TripoSG's variational autoencoder. Its encoder, which maps an explicit 3D surface into a latent code, serves mainly for training, whereas generation uses the decoder. From the latent code, its decoder evaluates a field at arbitrary points of a normalized domain,
\begin{equation}
  \tilde{s}(\overline{\mathbf{X}})
  =\mathcal{D}\bigl(\overline{\mathbf{X}};\Theta\bigr)
  =T\bigl(2\sigma(\ell)-1\bigr),
  \qquad
  \overline{\mathbf{X}}\in[-1,1]^{3},\quad T=10^{-2},
  \label{eq:truncation}
\end{equation}
where $\ell(\overline{\mathbf{X}};\Theta)$ is the network output, $\sigma$ the logistic function and $T$ the truncation bound. The decoded field, though similar to a signed distance field, is bounded to $[-T,T]$ and saturates away from the surface. Therefore, it is called a truncated signed distance field, meaningful only within a band near the surface. The field only approximates a distance: it satisfies the eikonal condition only approximately, even within the band, and it lives in normalized coordinates. Therefore, it does not directly serve as the input of the diffuse Nitsche method that works on the exact distance. However, the field represents the zero level set and can be used to extract the surface by adaptive octree extraction (see the ``Adaptive octree extraction'' paragraph below) and marching cubes.

\paragraph{Text-to-CAD generation.} Alternatively, one of two text-to-CAD models turns a text prompt into a CAD solid. Text2CAD~\citeS{khan2024text2cad} encodes the prompt with a frozen pre-trained BERT encoder~\citeS{devlin2019bert} and one adaptive transformer layer, then decodes, one token at a time, a quantized sequence of sketch-and-extrude operations. Its sequence processor builds that sequence into a boundary representation (B-rep) with Open CASCADE. CAD-Coder~\citeS{guan2026cad} is a language model fine-tuned on pairs of text and CadQuery scripts, and then further trained by reinforcement learning against a reward on geometry (Chamfer distance) and output format. It writes a CadQuery script. Both models are trained on descriptions from the Text2CAD corpus, which state geometry, ranging from abstract shapes to explicit sketch coordinates and extrusions. The text prompt needs to be detailed and specific, in the form each model follows best: one short geometric sentence for Text2CAD, and a step-by-step construction with explicit sketch coordinates for CAD-Coder. An AI subagent can help write it in that form (Sec.~\ref{sec:si-agents}). The solids that either model returns are tessellated face by face. Of the connected components, the one enclosing the largest volume is kept, and a result that does not bound a closed, positive volume is rejected. If the part has fewer facets than a budget of $2\times10^{4}$, its long facet edges are bisected to a target edge length chosen so that the refined mesh stays around the budget. The tessellated part becomes the \textit{source surface}, as an upload does, and each facet keeps the index of the B-rep face it tessellates, which can be directly used for boundary condition region identification (Sec.~\ref{sec:si-bcregions}). Besides these two text-to-CAD models, the AI agent operator (Sec.~\ref{sec:si-operator}) also has a tool we developed that directly builds simple CAD models (extrusions, revolutions and frames), which is much lighter than Text2CAD and CAD-Coder. 

\paragraph{Textured image-to-3D generation.} Optionally, Hunyuan3D-2.1~\citeS{hunyuan3d2025hunyuan3d} produces a colored surface from an image in two stages. The shape stage, a flow-matching diffusion transformer, generates a mesh from the image, after the image's background has been removed (unless it is already transparent). By default, this stage takes 30 sampling steps and decodes the shape at an octree resolution of $384$ with the accelerated FlashVDM decoder~\citeS{lai2025unleashingvecsetdiffusionmodel}. The mesh is then cleaned in several steps: floating fragments and degenerate facets are removed, the mesh is reduced to a budget of $4\times10^{4}$ facets by default, the connected component enclosing the largest volume is kept, and single-triangle holes are filled. A result that is still not closed afterwards is rejected. The paint stage, a multiview diffusion model, then textures that same surface from the image (six views at $512^2$ pixels) without remeshing it, and the albedo texture is sampled back onto the surface's own vertices as vertex colors. The colored surface becomes the source surface, as on the text-to-CAD path.

\paragraph{Image input drawn from text.} Either image-conditioned generator (TripoSG and Hunyuan3D-2.1) can also take an image that the workflow draws from the text description, so that the user need not supply one. The drawing uses the base model of Stable Diffusion XL (SDXL)~\citeS{podell2024sdxl}, a text-to-image diffusion model, with a publicly released fine-tune of its variational autoencoder that is stable in half precision (SDXL-VAE). The drawing is sampled from the text description, with a fixed suffix asking for a single whole object on a plain background and with a fixed negative prompt, at $1024^{2}$ pixels with $30$ denoising steps and classifier-free guidance~\citeS{ho2022classifier} of $7.0$. The drawing is then cut out from its background with a mask from the RMBG-1.4 background-removal network, the same network that TripoSG applies to uploaded images without a transparent background. Finally, the drawing is sent to TripoSG's image model or to Hunyuan3D-2.1 to generate the source surface.

\paragraph{Surface painting.} Hunyuan3D-2.1's paint stage can also color a surface that has no color, whether supplied by the user, generated by TripoSG or produced by a text-to-CAD model. The reference image is either supplied by the user or drawn by SDXL from a text description. In the latter case, the drawing is conditioned on the surface's depth through a depth ControlNet~\citeS{zhang2023adding}, a network that adds spatial conditioning to a diffusion model. Only vertex colors are added, and the geometry is unchanged. A surface of more than $4\times10^{4}$ facets is painted through a reduced copy, and the colors are carried to the surface's vertices by nearest sample. Painting is optional and can be applied at two points: before the surface is first read and scaled, so that every later rendering for AI subagents carries the colors, or before assigning a spatially varying material (Sec.~\ref{sec:si-nitsche}). 

\begin{figure}[tb]
  \centering
  \includegraphics[width=\linewidth]{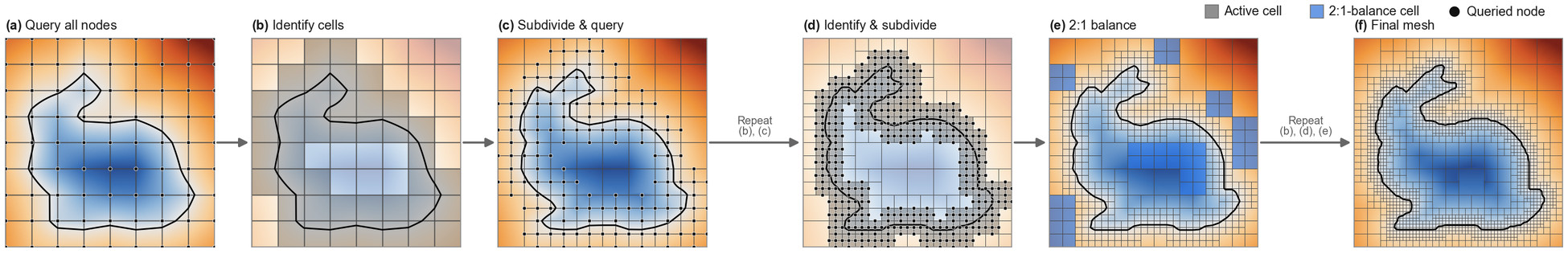}
  \caption{\textbf{Schematic of adaptive octree extraction in two dimensions.} The planar analogue of the octree (a quadtree) is built on a section of the signed distance field, from level $3$ to $6$. \textbf{a}, The field is queried at all nodes of the initial grid. \textbf{b}, Cells whose corner values change sign or lie in the narrow band around the surface are identified (gray). \textbf{c}, They are subdivided, and the field is queried at the new nodes only (black dots). Steps b and c repeat at the next level. \textbf{d}, Identification and subdivision at the next level, before balancing. \textbf{e}, The $2\!:\!1$ balance splits further cells (blue) so that neighboring cells differ by at most one level. Steps b, d and e repeat up to the final level. \textbf{f}, The final mesh: the finest cells wrap the surface, and the cells far from it stay coarse.}
  \label{fig:si-octree-steps}
\end{figure}

\paragraph{Adaptive octree extraction.} An adaptively refined octree is a grid of cubic cells that is subdivided only where needed. It is built from either the truncated signed distance field from TripoSG or the signed distance field computed through the bounding volume hierarchy on a given explicit surface, as long as the field preserves the zero level set of the underlying surface. The octree is built from level $l_{\mathrm{init}}$ to $l_{\mathrm{final}}$ inside a cubic region by identify-subdivide-query cycles that we developed in PyTorch, in the manner of multiresolution isosurface extraction~\citeS{mescheder2019occupancy} (Fig.~\ref{fig:si-octree-steps}). A cubic cell at level $l$ has a side length of $1/2^l$ of the cubic region's side length. At each level, three steps are taken. First, a cell of that level is marked to be subdivided into eight equal cells when its corner values change sign or any corner satisfies $|s|<0.9\,\tau$, where the width parameter $\tau$ sets a narrow band about the surface. Second, $2\!:\!1$ balance, the rule that neighboring cells differ by at most one level, is restored~\citeS{sundar2008bottom,burstedde2011p4est} by a downward ripple process: splitting a cell at level $l-1$ to eight leaf cells at level $l$ forces the split of every coarser 26-neighboring leaf (every leaf sharing a face, an edge or a corner with it) to level $l-1$. Third, the field is queried at the new nodes only, through the decoder of TripoSG or the bounding volume hierarchy. A dense background grid of $(2^{l_{\mathrm{final}}}+1)^{3}$ nodes holds the queried values on octree nodes and is completed by nearest-neighbor upsampling.

\begin{figure}[tb]
  \centering
  \includegraphics[width=\linewidth]{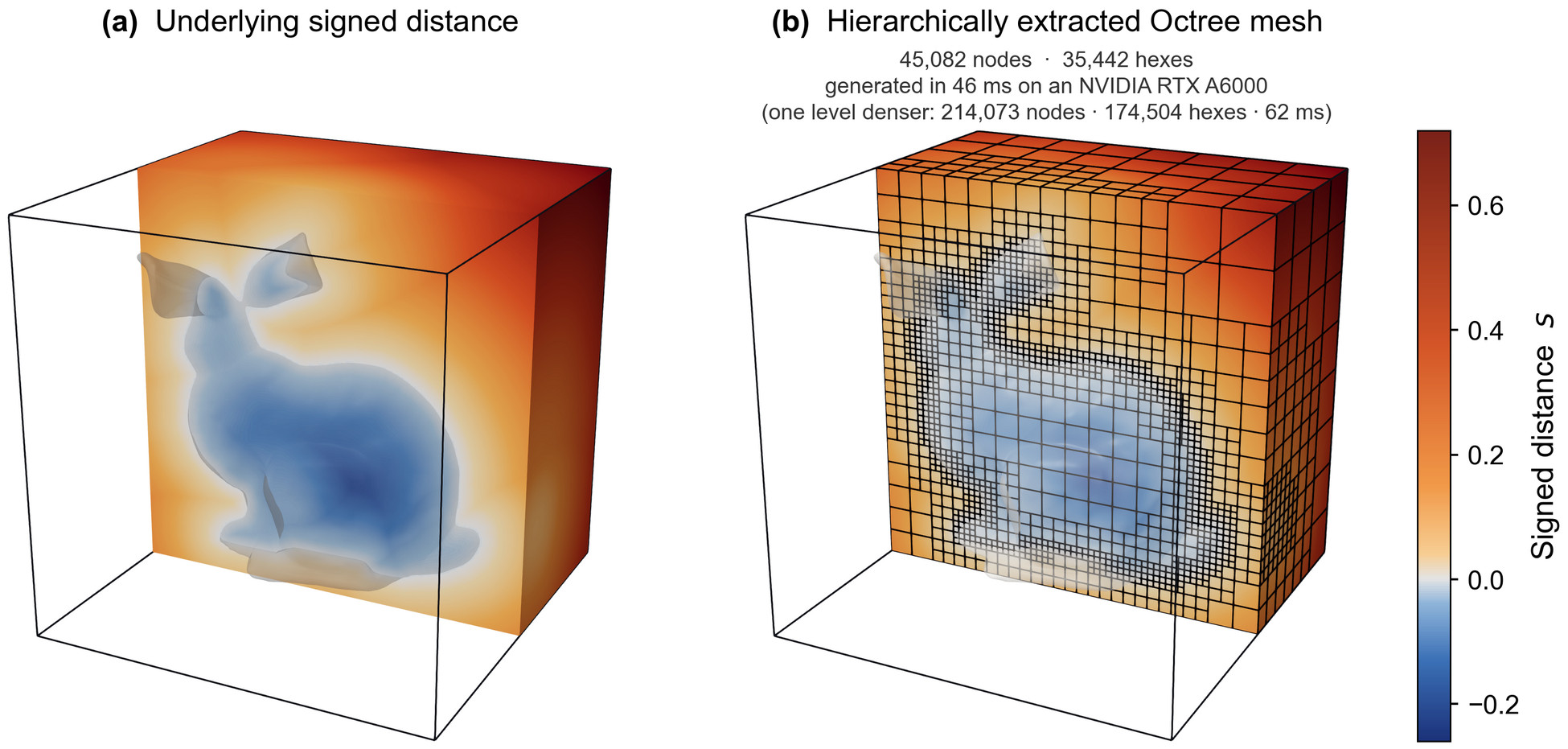}
  \caption{\textbf{From a signed distance field to an adaptive octree.} \textbf{a}, The signed distance field to an explicit surface, on a section through the cubic region enclosing the body (translucent): negative inside, zero on the surface and positive outside. \textbf{b}, The octree extracted from that field (see the ``Adaptive octree extraction'' paragraph), on the same section with its cell edges drawn. Cells are subdivided only where the field changes sign or lies in the narrow band around the surface, so the finest cells wrap the surface and the cells far from it stay coarse. This octree is built from level $3$ to $6$. Its size and build time are given in the panel.}
  \label{fig:si-sdfoctree}
\end{figure}

For TripoSG, the extraction runs twice, both times from $l_{\mathrm{init}}$ to $l_{\mathrm{final}}$, but on different fields and in different coordinates: (i) on the truncated signed distance field of the generator to extract the source surface, and (ii) on the signed distance to the scaled source surface to build the analysis octree mesh used later. The first extraction queries the TripoSG decoder on a cubic domain of $[-1,1]^3$ with $\tau=T$, so that the band criterion selects the part of the truncated signed distance field that has not saturated. The first extraction outputs the background grid of truncated signed distance values. This grid is filtered (see the ``Spurious-component removal for TripoSG'' paragraph below), and marching cubes~\citeS{lorensen1987marching} extracts the source surface from it. The source surface is then scaled (see the ``Physical scaling'' paragraph below) for the second extraction (see the ``Octree extraction with the scaled surface'' paragraph below), which outputs both the octree mesh and the dense background grid (Fig.~\ref{fig:si-sdfoctree}). For a user-supplied surface or a source surface from Hunyuan3D-2.1 or text-to-CAD models, only scaling and one (the second) extraction are performed. 

\begin{figure}[tb]
  \centering
  \includegraphics[width=0.8\linewidth]{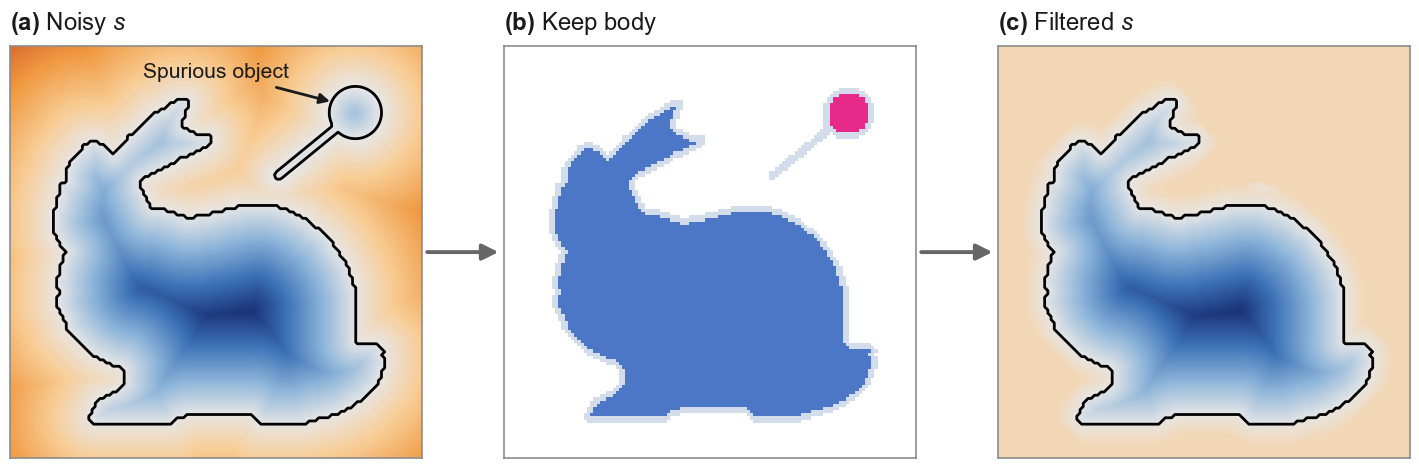}
  \caption{\textbf{Schematic of spurious-component removal in two dimensions.} \textbf{a}, A generated truncated signed distance field $s$ with a spurious object detached from the body. \textbf{b}, The interior regions are labeled as separate components. The body is kept (blue, with its one-cell margin in light blue) and the spurious component is removed (magenta). \textbf{c}, The filtered field: outside the kept component, it is set to its maximum value.}
  \label{fig:si-spurious}
\end{figure}

\paragraph{Spurious-component removal for TripoSG.} Generated fields from TripoSG can contain small disconnected negative regions, which would otherwise become isolated islands of material. To remove them, we label the isolated components of the interior voxels (those with negative truncated signed distance) on the dense background grid separately and keep only one of them (Fig.~\ref{fig:si-spurious}). By default, this is the largest component by volume. However, if the component whose centroid lies closest to the domain center is a different one and holds at least $20\%$ of the volume of the largest component, that one is kept instead. The kept component is dilated by a one-voxel margin, and the field is set to $\max\tilde{s}$ everywhere outside the dilated component. 

\begin{figure}[tb]
  \centering
  \includegraphics[width=0.55\linewidth]{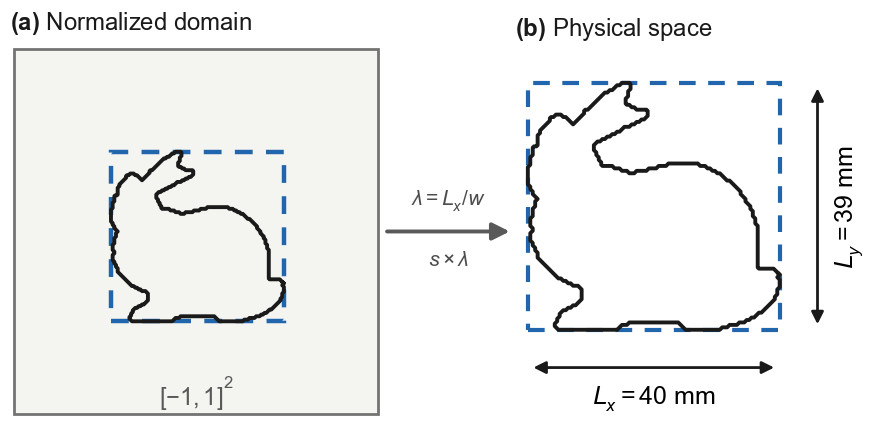}
  \caption{\textbf{Physical scaling, shown in two dimensions.} \textbf{a}, The source surface in the normalized domain $[-1,1]^2$, with its bounding box (dashed) of width $w$ along the nominated axis $x$. \textbf{b}, The same surface after uniform scaling by $\lambda=L_x/w$, so that its extent along $x$ equals the nominated length, here $L_x=40$~mm. The other extent follows ($L_y=39$~mm), and signed distance values scale by the same factor $\lambda$.}
  \label{fig:si-scaling}
\end{figure}

\paragraph{Physical scaling.} The source surface is scaled uniformly so that its extent along a nominated axis equals a length (Fig.~\ref{fig:si-scaling}) entered by the user or proposed by an AI subagent (Sec.~\ref{sec:si-agents}). The scale is computed against bounds snapped to the finest lattice of the analysis octree at level $l_{\mathrm{final}}$. Therefore, the realized extent matches the nominated value to within one finest-cell width $h_{\min}$. 

\paragraph{Octree extraction with the scaled surface.} Using the same adaptive octree extraction code we developed, introduced above, the second octree extraction builds the octree mesh on which the mechanical deformation will be solved. Its field is the signed distance to the scaled source surface, evaluated on the GPU through a bounding volume hierarchy. The octree is extracted in a cubic region of side $1.92\,L_{\max}$, where $L_{\max}$ is the longest side of the bounding box of the scaled source surface, which is centered in the region. Specifically, the octree is translated to align its lattice with the surface's bounding box padded by $0.15\,L_{\max}+h_{\min}/2$, and the octree is then clipped to that padded box, so that at level $l_{\mathrm{final}}$ every finest leaf has side $h_{\min}=1.92\,L_{\max}/2^{l_{\mathrm{final}}}$. The width parameter $\tau$ for octree extraction is $h_{\min}$, and the band criterion is therefore $|s|<0.9\,h_{\min}$. In addition to the general extraction steps (described in the above ``Adaptive octree extraction'' paragraph), every cell containing a vertex of the scaled source surface is split down to level $l_{\mathrm{final}}$. An uploaded surface must therefore be tessellated finely enough to carry a vertex in each finest-level octree cell it crosses. A TripoSG surface meets this requirement, because its marching-cubes vertices are spaced at about $h_{\min}$. A text-to-CAD part's surface is refined by edge bisection so that it also meets the requirement (see the ``Text-to-CAD generation'' paragraph above).

\begin{figure}[tb]
  \centering
  \includegraphics[width=0.8\linewidth]{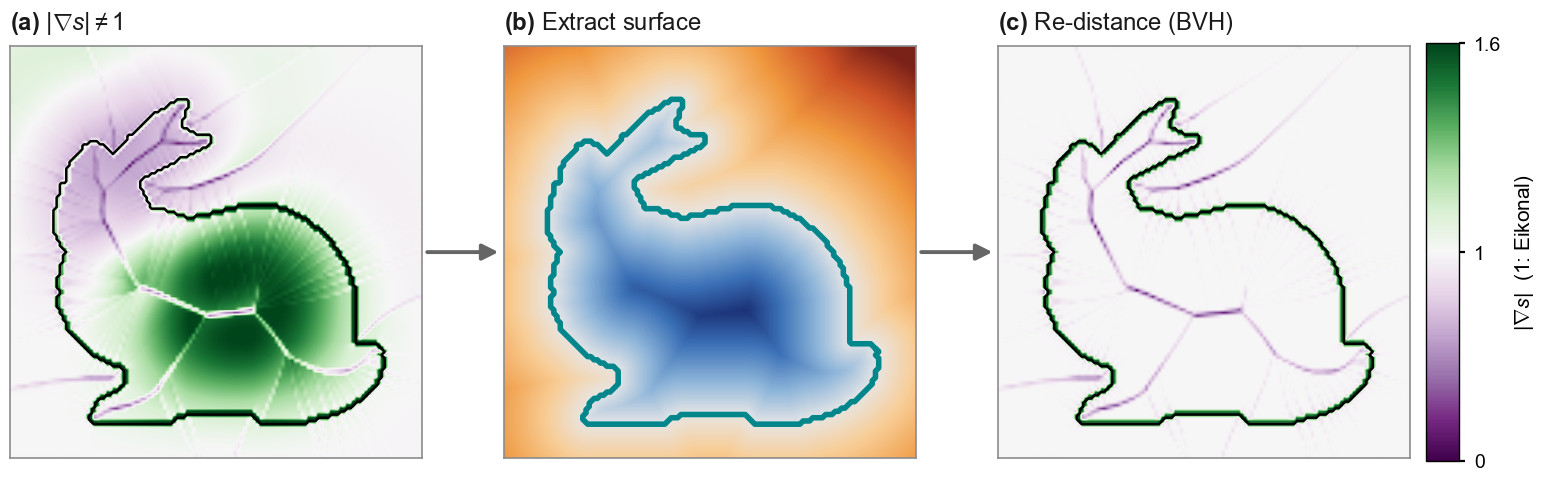}
  \caption{\textbf{Schematic of signed distance reinitialization in two dimensions.} \textbf{a}, A field that is not a true distance (here the exact distance multiplied by a smooth synthetic distortion for illustration), shown as $|\nabla s|$ (green above one, purple below one). \textbf{b}, Its zero level set is extracted as an explicit surface by marching cubes. \textbf{c}, The exact signed distance to that surface, computed through the bounding volume hierarchy (BVH), restores $|\nabla s|=1$ everywhere except along the medial axis (thin purple lines), where the gradient of a distance is undefined.}
  \label{fig:si-reinit}
\end{figure}

\paragraph{Signed distance reinitialization.} The queried nodal distance field values on the dense background grid (or the corresponding octree nodes) cannot represent fine details (sharp edges, thin walls, notches, etc.) of the original surface due to the resolution of discretization, and no longer satisfy the eikonal condition $|\nabla s|\approx 1$ on trilinear finite elements, which the diffuse Nitsche method assumes. Therefore, before the diffuse Nitsche solve, we reinitialize the signed distance field values obtained during octree extraction from the source surface based on a reconstructed surface that the discretization preserves. Reinitialization has three steps (Fig.~\ref{fig:si-reinit}): (i) marching cubes extracts a new surface from the dense background grid, (ii) the exact signed distance to that surface is re-evaluated at every grid node through the bounding volume hierarchy, with its sign given by the generalized winding number~\citeS{jacobson2013robust}, and (iii) the octree reads its nodal values from the result to be used for the diffuse Nitsche solve. Thus, the mechanical solution is for the reconstructed surface: sharp edges are rounded at the scale of $h_{\min}$, features thinner than about $h_{\min}$ are not represented faithfully, and members only a few $h_{\min}$ thick, though represented, are spanned by one or two trilinear elements and are poorly resolved in bending.

\paragraph{Hanging nodes.} Where a fine octree cell neighbors a coarser one, the fine cell has nodes at the middle of the coarse cell's edges and faces that the coarse cell does not have (Fig.~\ref{fig:si-octree-steps}f). These are called hanging nodes~\citeS{burstedde2011p4est}. A hanging node carries no degrees of freedom of its own and is treated by a multi-point constraint. Simply put, its value is the average of the two end nodes of the coarse edge it lies on, or of the four corner nodes of the coarse face it lies on. Therefore, any field stored on the octree, such as the displacement solved later (Sec.~\ref{sec:si-nitsche}), stays continuous across the boundary between coarse and fine cells.

\subsection{Boundary-region identification}
\label{sec:si-bcregions}

A boundary condition has two parts: the region of the surface it acts on, and the value it prescribes there. Conventionally in finite element frameworks, the region is located by partitioning the surface interactively or by writing a geometric test (a predicate) by hand. In our workflow, boundary conditions are defined on a non-conforming octree mesh, on which interactive selection and hand-written geometric tests are not viable for AI subagents that work from images and text. Therefore, we find the region on an explicit surface in three steps. First, the source surface (Sec.~\ref{sec:si-geometry}) is segmented into parts. Second, a selector program builds each boundary condition region as a set of surface facets (triangles), selected from those segments by geometric and topological tests. Third, each region is mapped onto the octree mesh (Sec.~\ref{sec:si-geometry}), where it turns into quadrature point indicators and enters the boundary term integrals in the weak form of the diffuse Nitsche method (Sec.~\ref{sec:si-nitsche}).

\begin{figure}[tb]
  \centering
  \begin{minipage}[t]{0.25\linewidth}\centering
    \includegraphics[width=\linewidth]{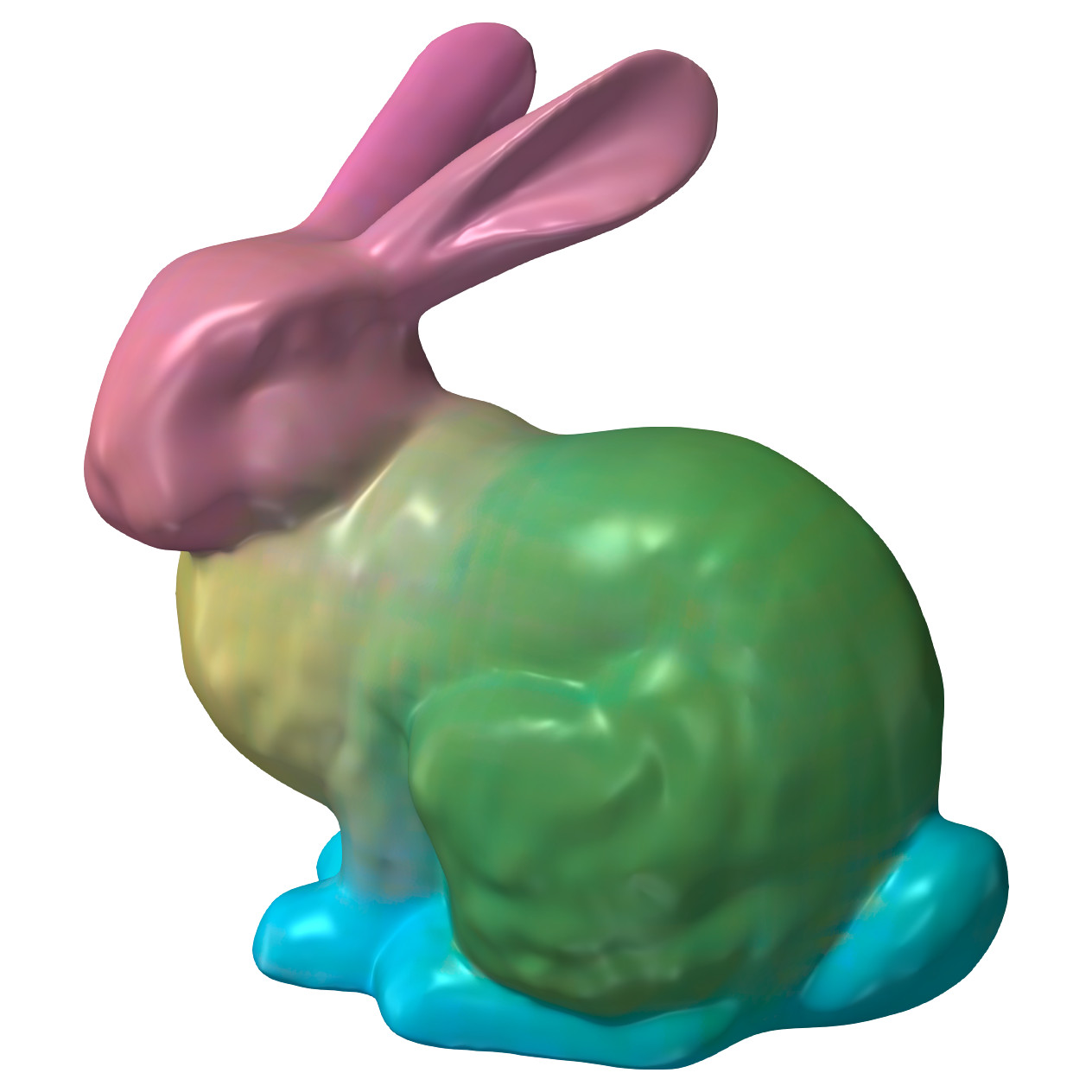}\\
    \textbf{a}
  \end{minipage}
  \begin{minipage}[t]{0.25\linewidth}\centering
    \includegraphics[width=\linewidth]{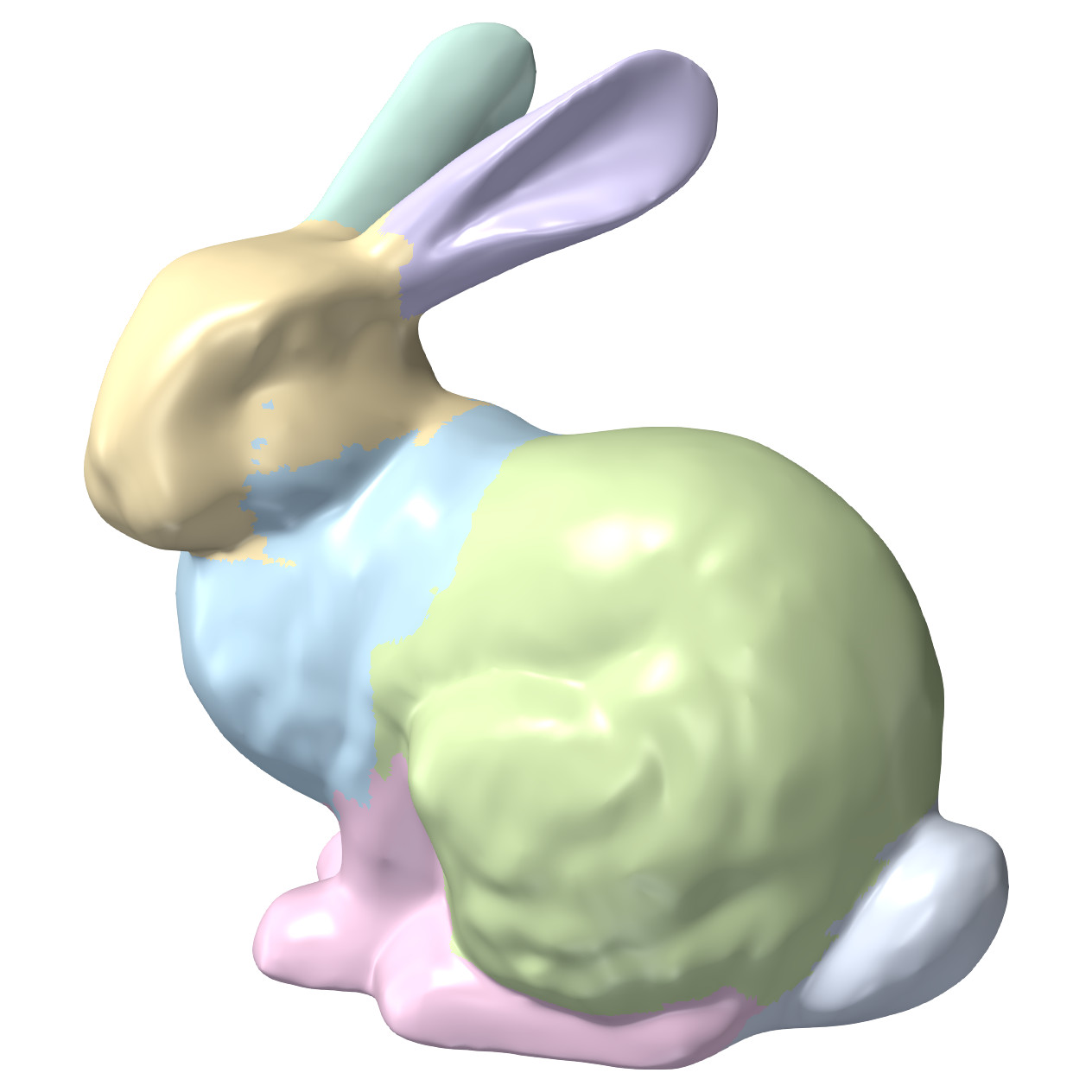}\\
    \textbf{b}
  \end{minipage}
  \caption{\textbf{Pre-trained part segmentation.} \textbf{a}, The per-facet PartField descriptor ($448$ channels, normalized to unit length) of a $161\,568$-facet bunny, projected onto its three leading principal components, which are mapped to lightness plus the two chroma axes of a perceptual color space. These principal components carry $45.7\%$ of the descriptor variance. Facets with similar descriptors take similar colors. \textbf{b}, The segmentation obtained by clustering that descriptor: $8$ clusters, chosen for this example as the smallest number that gives each ear its own part (the default is $5$). }
  \label{fig:si-partfield}
\end{figure}

\paragraph{Pre-trained part segmentation model: PartField.} By default, a source surface that does not come from a text-to-CAD model is segmented with the pre-trained part-feature network PartField~\citeS{liu2025partfield} and its released weights. PartField takes $10^{5}$ surface samples, drawn with probability proportional to area, and predicts a triplane feature field over the shape (features stored on three axis-aligned planes). Each facet is then given a descriptor: the mean of the 448-channel part feature over $500$ random points on the facet, normalized to unit length (Fig.~\ref{fig:si-partfield}a). These descriptors are clustered in two stages: K-means first groups them into at most $512$ small clusters, and Ward's hierarchical clustering~\citeS{ward1963hierarchical} then repeatedly merges the two clusters whose merger least increases the spread of the descriptors within clusters. The hierarchy is cut at five clusters by default (Fig.~\ref{fig:si-partfield}b). Because the clustering uses only the descriptors and ignores which facets are neighbors, a segment can be disconnected.

\begin{figure}[tb]
  \centering
  \includegraphics[width=\linewidth]{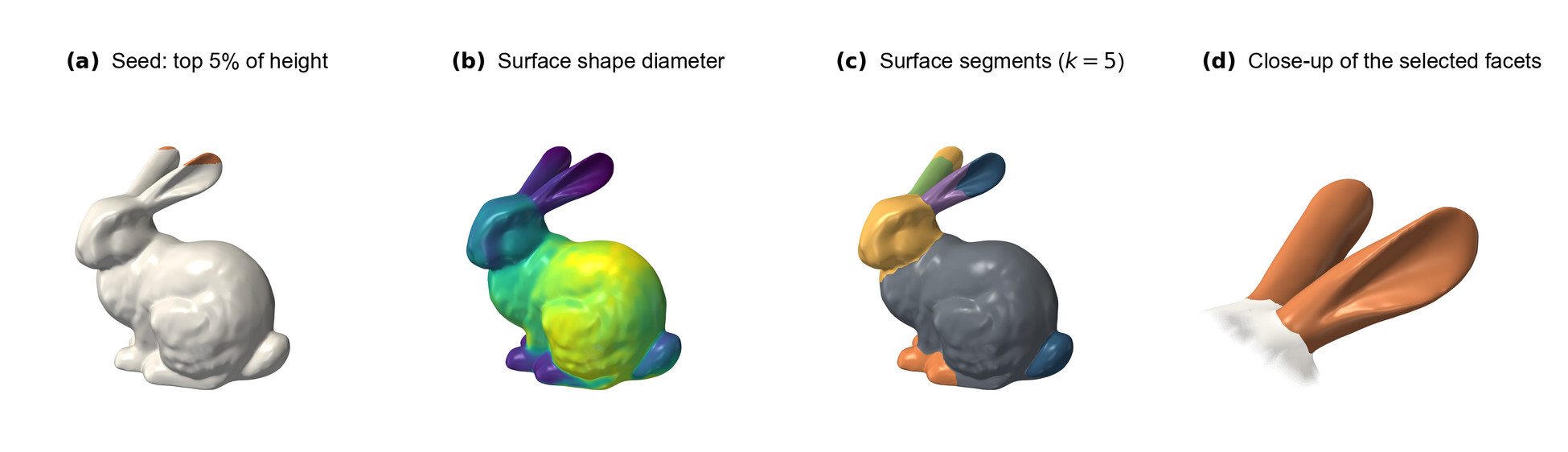}
  \caption{\textbf{Building a boundary region from a shape-diameter segmentation.} \textbf{a}, A seed: the facets in the top $5\%$ of the object's height, which land on both ear tips ($2\,264$ facets in two components). \textbf{b}, The shape diameter, the length of rays cast inward from each facet, measures local thickness and separates the slender ears and feet (small shape diameter) from the body (large shape diameter). \textbf{c}, Facets are grouped on that measure by a Gaussian mixture with $k=5$ components and cut into connected segments by a graph cut over facet adjacency, giving $8$ segments. \textbf{d}, The selected region: the segments containing the seed together with their adjacent segments, $24\,710$ of the surface's $161\,568$ facets. }
  \label{fig:si-bcregions}
\end{figure}

\paragraph{Shape-diameter segmentation.} Alternatively, segments follow the shape diameter function~\citeS{shapira2008consistent}, a per-facet measure of local thickness (Fig.~\ref{fig:si-bcregions}b), computed by our GPU reimplementation of the segmentation pipeline of the Computational Geometry Algorithms Library (CGAL)~\citeS{alliez2016cgal}. For each facet, a cone of rays (by default $25$ rays within $120^\circ$) is cast inward, and the ray lengths, cleaned of outliers and smoothed, give the local shape diameter. A mixture of five Gaussian distributions, fitted to the logarithm of the diameters, defines five classes. A graph cut~\citeS{boykov2001fast} then assigns each facet a class, favoring the classes that fit its diameter while discouraging neighboring facets from taking different classes (smoothing weight $\lambda=0.25$). A border along a concave crease costs less than one along a convex crease, so borders tend to follow creases. Finally, the connected components of each class become distinct segments (Fig.~\ref{fig:si-bcregions}c).

\paragraph{CAD-face segmentation.} A source surface from either text-to-CAD model (Sec.~\ref{sec:si-geometry}) already carries its own partition, because each facet keeps the index of the boundary-representation (B-rep) face it tessellates. Physical scaling moves only the vertices and keeps the facet order. Therefore, the segment indices survive it. For such surfaces, this partition is the default segmentation: one segment per B-rep face, labeled with the face's surface type (plane, cylinder, cone, sphere, torus, B-spline and others). 

\begin{figure}[tb]
  \centering
  \includegraphics[width=\linewidth]{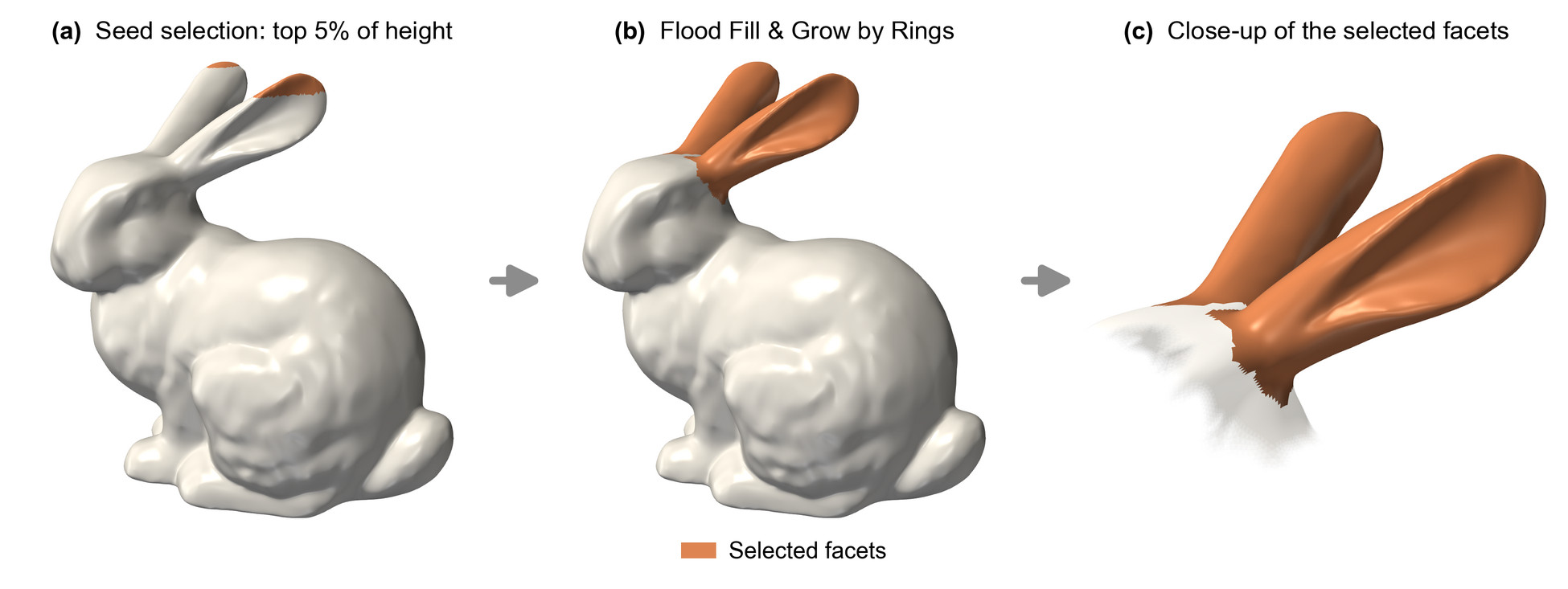}
  \caption{\textbf{Building the same region by seed and flood fill, without a segmentation.} \textbf{a}, The seed of Fig.~\ref{fig:si-bcregions}a. \textbf{b}, A flood fill gated on the angle between neighboring facet normals, followed by growth by facet rings ($14$ rings on each ear), follows each ear and stops at its base, selecting $26\,024$ facets. \textbf{c}, A close-up of the selected facets. Compared with the selection in Fig.~\ref{fig:si-bcregions}, this achieves an intersection-over-union of $0.944$, which measures the agreement between the two routes.}
  \label{fig:si-bcmanual}
\end{figure}

\paragraph{Selector programs.} A selector program builds a region from 25 primitive selectors that we developed, each of which selects facets by a simple test. They fall into five groups: position (for example, the facets inside a box or within a top percentile of height), orientation and curvature (for example, the facets whose normal points along a given direction, or the flat facets), segment selection (whole segments by their identifier, or the shape-diameter segment that best overlaps a given starting set of facets, called a seed), topological operations (for example, flood fill, which spreads a set across neighboring facets while the angle between their normals stays small, and ring growth, which adds the facets adjacent to a set), and Boolean combination (intersection, union, subtraction, complement). Figure~\ref{fig:si-bcregions} shows a region built from a seed and the shape-diameter segments, and Fig.~\ref{fig:si-bcmanual} shows the same region built from the same seed without any segment selection, by flood fill and ring growth. The AI subagent looks at the rendered views of the surface and descriptions of each segment (written by another AI subagent call if segmentation is performed), writes these programs from the user's description of the supports and loads (Sec.~\ref{sec:si-agents}) and visually examines the selection, which can be much easier with segments provided by either of the surface segmentation approaches.

\paragraph{Regions on the analysis octree mesh.}
Each approved region is turned into a set of facets. Through our implementation, the selected facets then mark the quadrature points (the integration points of each element) of the octree mesh that belong to the region: a quadrature point belongs to the region when its distance to the nearest selected facet, computed through a bounding volume hierarchy, is below the match tolerance $\delta=0.75\,h_{\min}$, where $h_{\min}$ is the side length of the finest octree cell (Sec.~\ref{sec:si-geometry}). The marked points form the region's indicator. We write $M_D^{i}$ and $M_N^{i}$ for the indicator of the $i$-th Dirichlet (support) and Neumann (load) region respectively, and $M^{i}$ for either. The indicator marks points on both sides of the surface. The diffuse Nitsche method (Sec.~\ref{sec:si-nitsche}) then weights them so that only the thin layer around the surface contributes to boundary term integrals.

\paragraph{Supported boundary conditions and loads.} Each Dirichlet (support) region prescribes one scalar displacement value. By default, all three displacement components take that value: zero is a clamp, and a nonzero value moves the region by the same amount along $x$, $y$ and $z$. Alternatively, the value applies only to chosen components (see the ``Component-wise Dirichlet boundary conditions'' paragraph of Sec.~\ref{sec:si-nitsche}). At least one Dirichlet region is required to prevent free rigid-body motion, which would make the system singular. Each Neumann (load) region prescribes a constant traction vector. A uniform body force $\mathbf{b}$ and a uniform acceleration $\mathbf{a}$, such as gravity, may be added. The acceleration acts only as the body force $\rho\mathbf{a}$, where $\rho$ is the density, and the analysis remains static. All loads are dead loads in the reference configuration: they do not follow the deformation, and tractions are given per unit undeformed area. Where regions of the same type overlap, their prescribed values are averaged as a simplified treatment.

\subsection{Diffuse Nitsche analysis}
\label{sec:si-nitsche}

\begin{figure}[tb]
  \centering
  \includegraphics[width=\linewidth]{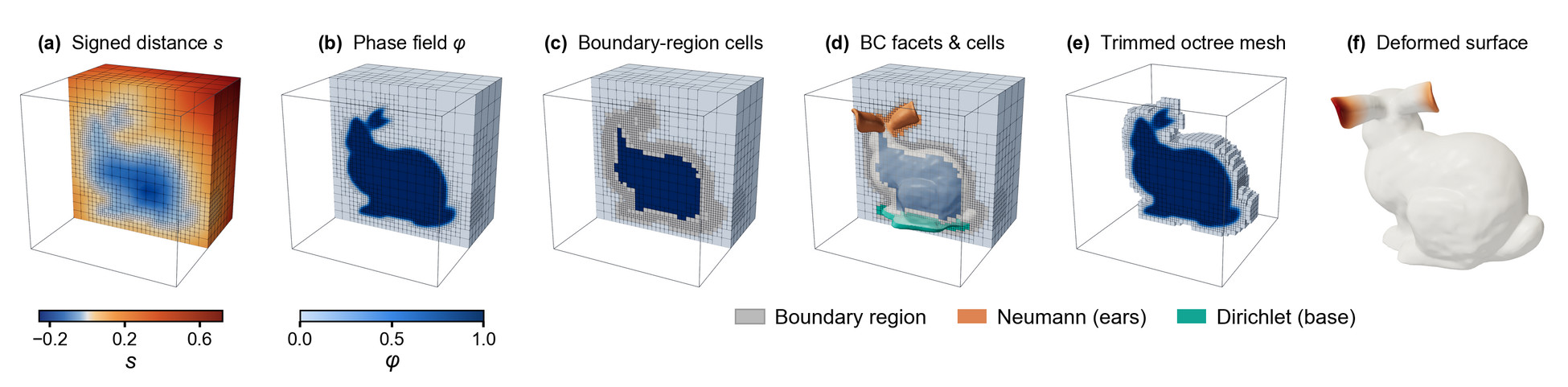}
  \caption{\textbf{The diffused zero level set and the regions the weak form is integrated on}, all on one octree mesh cut on a mid-plane, with the cell edges drawn. \textbf{a}, The signed distance field $s$ carried on the octree. Away from the surface it is stored at the coarse resolution the mesh provides there. \textbf{b}, The phase field $\varphi=\tfrac12(1-\tanh(s/\varepsilon))$ with $\varepsilon=h_{\min}/2$: one inside, zero outside, with a thin transition layer. \textbf{c}, The diffuse boundary band $\mathcal{B}$: every cell with a node at which $10^{-3}<\varphi<1-10^{-3}$ (gray), equivalently $|s|<3.45\,\varepsilon$. Because the octree is graded toward the surface, the band holds $28\,130$ of the $35\,442$ cells. \textbf{d}, The embedded surface (translucent) with a Neumann (traction) region on the ears (orange) and a Dirichlet (prescribed-displacement) region on the base (teal), painted on the facets that define each region and on the cells that carry it ($2\,581$ cells for the ears and $3\,386$ for the base, at $\delta=0.75\,h_{\min}$). \textbf{e}, The trimmed mesh handed to the solver, after cells with $\varphi<10^{-3}$ at all nodes are removed: $29\,549$ of the $35\,442$ cells remain. \textbf{f}, The deformed surface under a downward traction on the ears, colored by displacement magnitude. The octree is graded from level $3$ to $6$ ($h_{\min}=0.018$ on a domain of side $1.16$), and the ear region is the seeded flood-fill selection of Fig.~\ref{fig:si-bcmanual}.}
  \label{fig:si-nitsche}
\end{figure}

\paragraph{Phase field and diffused zero level set.} Following the diffuse-domain approach~\citeS{li2009solving} and the diffuse Nitsche method~\citeS{nguyen2018diffuse}, the sharp domain is replaced by a phase field (Fig.~\ref{fig:si-nitsche}b), a smooth function that is one inside the body and zero outside. Its thin transition layer at the surface, on which the boundary terms are integrated, is the diffused zero level set. The phase field is computed from the reinitialized signed distance field on the octree nodes (Sec.~\ref{sec:si-geometry}),
\begin{equation}
  \varphi(\mathbf{X})=\tfrac12\Bigl(1-\tanh\tfrac{s(\mathbf{X})}{\varepsilon}\Bigr),
  \qquad
  \varepsilon=\tfrac12 h_{\min},
  \label{eq:phasefield}
\end{equation}
where $h_{\min}$ is the width of the finest octree cell. Elements whose eight nodal phase-field values $\varphi_n=\varphi(s_n)$ all satisfy $\varphi_n<10^{-3}$ are considered to lie outside the body and are removed (Fig.~\ref{fig:si-nitsche}e). The remaining elements form the analysis mesh $\mathcal{K}$, namely the trimmed octree. The full octree of Sec.~\ref{sec:si-geometry} before this removal is called the untrimmed octree. The elements of $\mathcal{K}$ with a node in $10^{-3}<\varphi_n<1-10^{-3}$ (equivalently, a node at $|s|<3.45\,\varepsilon$) form the diffuse boundary band $\mathcal{B}\subset\mathcal{K}$, the set of elements covered by the diffused zero level set (Fig.~\ref{fig:si-nitsche}c). Because the reinitialized $s$ satisfies the eikonal condition almost exactly (Sec.~\ref{sec:si-geometry}), the transition layer has a width of order $\varepsilon$, and the surface measure below integrates approximately to the surface area. At each quadrature point of $\mathcal{B}$, $\varphi$ and its gradient are evaluated in closed form from the interpolated $s$, with $\nabla s$ renormalized to unit length:
\begin{equation}
  |\nabla\varphi|=\frac{2\varphi(1-\varphi)}{\varepsilon},
  \qquad
  \mathbf{n}=-\frac{\nabla\varphi}{|\nabla\varphi|}.
\end{equation}
Integrals over the sharp domain $\Omega$ and over its boundary $\partial \Omega$ are then approximated by weighted volume integrals over the trimmed octree $\mathcal{K}$,
\begin{equation}
  \int_{\Omega}(\cdot)\,\mathrm{d}V\approx\int_{\mathcal{K}}(\cdot)\,w(\varphi)\,\mathrm{d}V,
  \qquad
  \int_{\partial\Omega^{i}}(\cdot)\,\mathrm{d}A\approx\int_{\mathcal{B}}(\cdot)\,|\nabla\varphi|\,M^{i}\,\mathrm{d}V,
  \label{eq:measures}
\end{equation}
with the regularized material weight $w(\varphi)=(1-\varphi_0)\varphi+\varphi_0$, $\varphi_0=10^{-4}$, and $M^{i}$ the quadrature-point indicator of the approved boundary region $i$ (Sec.~\ref{sec:si-bcregions} and Fig.~\ref{fig:si-nitsche}d). The floor $\varphi_0$ prevents a singular stiffness far from the diffused boundary.

\paragraph{Neo-Hookean constitutive model.} The solid is an isotropic, compressible neo-Hookean material~\citeS{bonet2008nonlinear}. Its strain energy density is
\begin{equation}
  W(\mathbf{F})=\frac{\mu}{2}\bigl(J^{-2/3}I_1-3\bigr)+\frac{\kappa}{2}(\ln J)^2,
  \qquad
  I_1=\mathbf{F}\!:\!\mathbf{F},\quad J=\det\mathbf{F},
\end{equation}
with $\mathbf{F}=\mathbf{I}+\nabla\mathbf{u}$ the deformation gradient, $\mu=E/\bigl[2(1+\nu)\bigr]$ the shear modulus and $\kappa=E/\bigl[3(1-2\nu)\bigr]$ the bulk modulus, where $E$ is Young's modulus and $\nu$ Poisson's ratio. The first Piola--Kirchhoff stress follows as
\begin{equation}
  \mathbf{P}=\mu J^{-2/3}\mathbf{F}
  +\Bigl(\kappa\ln J-\tfrac{\mu}{3}J^{-2/3}I_1\Bigr)\mathbf{F}^{-\mathsf{T}}.
\end{equation}
The trilinear element is fully integrated, without a treatment of volumetric locking. Therefore, a nearly incompressible material ($\nu$ close to $\tfrac12$, as for rubbers and silicone) is expected to respond over-stiffly.

\paragraph{Spatially varying material.} By default, $E$, $\nu$ and the density $\rho$ are uniform. On request, all three vary in space instead. They are then given on a $64^3$ grid over the surface's bounding cube by one of the two sources below, both of which read the surface's colors (see the ``Surface painting'' paragraph of Sec.~\ref{sec:si-geometry}). The values on the $64^3$ grid are assigned to the octree nodes it covers, the three quantities are interpolated trilinearly within each octree cell to its quadrature points, and at each quadrature point $\mu$ and $\kappa$ are computed from the interpolated $E$ and $\nu$. The body load becomes $\mathbf{b}+\rho(\mathbf{X})\,\mathbf{a}$. Once the user approves it, the field is used in forward solves.

\emph{Predicted field.} The pre-trained material predictor VoMP~\citeS{dagli2025vomp} renders the scaled surface, or its painted copy, from 150 views, aggregates image features per voxel, and decodes each voxel's material latent code to $(E,\nu,\rho)$. For computational efficiency, the views are rasterized rather than path-traced via Blender as in the released VoMP pipeline, and the view encoder runs in half precision. The rasterized views closely match the path-traced ones, but the predicted fields may differ noticeably, meaning the prediction can be sensitive to the way the views are rendered. This approach is incorporated in the workflow but is not used in our agentic results. 

\emph{Material per part.} Alternatively, the AI subagent, a multimodal language model (by default DeepSeek-V4.1-Flash~\citeS{xu2026deepseek}), assigns one homogeneous, isotropic material to each part (Sec.~\ref{sec:si-agents}). The parts are the segments that defined the boundary regions (Sec.~\ref{sec:si-bcregions}). The multimodal language model receives one image of six views of the surface in its own colors from texture, above the same six views with parts in different colors, together with text of each part's color, area fraction, position and dominant facing direction, short descriptions of the object and its parts that it has written from rendered views for boundary region identification (Sec.~\ref{sec:si-bcregions}), and the user's description of the material. Its instructions contain a reference table of ranges of Young's modulus and Poisson's ratio for $32$ typical materials and rules for reading a material from appearance, and a material that the user names is taken as given. Values outside broad physical ranges are rejected, which catches unit slips. To fill the $64^3$ material grid, sample points are generated on the surface, and every cell that holds a part of the surface, or is enclosed by such cells, takes the part of its nearest sample. The user can edit the values per part before approval.

\paragraph{Nitsche weak form.} Essential (Dirichlet) boundary conditions are imposed weakly by a symmetric Nitsche method~\citeS{nitsche1971variationsprinzip} on the diffused zero level set. Together with Eq.~\eqref{eq:measures}, this is the diffuse Nitsche method~\citeS{nguyen2018diffuse}. Mechanical equilibrium is stated in weak form via the Galerkin method: the discrete displacement $\mathbf{u}$ in its trilinear finite element space is sought such that $R(\mathbf{u},\mathbf{v})=0$ for an arbitrary test displacement $\mathbf{v}$ in the same trilinear space. Because $R$ is linear with respect to $\mathbf{v}$, it can be written as $R=\mathbf{V}^{\mathsf{T}}\mathbf{R}$, with $\mathbf{V}$ the nodal values of $\mathbf{v}$ and $\mathbf{R}$ the residual vector, one entry per nodal degree of freedom. 

The notation is as follows. $\mathbb{A}[\mathbf{H}]=(\partial\mathbf{P}/\partial\mathbf{F})\!:\!\mathbf{H}$ is the material tangent applied to a tensor $\mathbf{H}$. The gate $\chi(\mathbf{X})$ equals one where $\varphi>\varphi_0=10^{-4}$ and zero elsewhere. $M_D^{i}$ and $M_N^{i}$ are the indicators of the Dirichlet and Neumann regions (Sec.~\ref{sec:si-bcregions}), and $\mathbf{n}_f^{i}$ is the normalized (unweighted) mean of $\mathbf{n}$ over the quadrature points of Neumann region $i$. $\bar{\mathbf{u}}^{i}$ and $\bar{\mathbf{t}}^{i}$ are the prescribed displacement and traction on a boundary condition region $i$, and $\beta(\mathbf{X})$ is the penalty (see the ``Penalty stabilization'' paragraph below). $\langle x\rangle_{+}=\max(x,0)$. 

The assembled residual is
\begin{align}
  R(\mathbf{u},\mathbf{v}) =\;&
  \int_{\mathcal{K}} w(\varphi)\,\mathbf{P}\!:\!\nabla\mathbf{v}\,\mathrm{d}V
  \;-\;\int_{\mathcal{K}} w(\varphi)\,\mathbf{b}\cdot\mathbf{v}\,\mathrm{d}V
  \nonumber\\
  &-\sum_i\int_{\mathcal{B}}\chi\,M_N^{i}\,
     (\bar{\mathbf{t}}^{i}\cdot\mathbf{v})\,
     |\nabla\varphi|\,\mathrm{d}V 
  \nonumber\\
  &+\sum_i\int_{\mathcal{B}}\chi\,M_D^{i}\Bigl[
      (\mathbf{P}\nabla\varphi)\cdot\mathbf{v}
      +\bigl(\mathbb{A}[\nabla\mathbf{v}]\nabla\varphi\bigr)\cdot(\mathbf{u}-\bar{\mathbf{u}}^{i})
      +\beta\,(\mathbf{u}-\bar{\mathbf{u}}^{i})\cdot\mathbf{v}\,|\nabla\varphi|
    \Bigr]\mathrm{d}V .
  \label{eq:residual}
\end{align}
The first line is the internal and body-force work. The second line is the applied traction. The three bracketed terms of the third line are the Nitsche terms, namely consistency, symmetry, and penalty terms. The consistency term, $(\mathbf{P}\nabla\varphi)\cdot\mathbf{v}$, is the boundary traction on the surface where Dirichlet boundary conditions are applied. In a body-fitted Galerkin finite element method with strongly imposed Dirichlet conditions, $\mathbf{v}=\mathbf{0}$ on the Dirichlet boundary, so this boundary-traction term vanishes. The symmetry term is the derivative of the consistency term's stress in the direction $\mathbf{v}$, contracted with $\mathbf{u}-\bar{\mathbf{u}}^{i}$, so that $R(\mathbf{u},\mathbf{v})$ is the first variation of a discrete potential and the tangent is symmetric. The penalty term drives $\mathbf{u}$ toward $\bar{\mathbf{u}}^{i}$. 

\paragraph{Component-wise Dirichlet boundary conditions.} Equation~\eqref{eq:residual} shows the full-vector Dirichlet case. In our implementation, a Dirichlet condition on a subset of the components carries a diagonal projector $\boldsymbol{\psi}_D^{i}\in\{0,1\}^3$, which replaces $\mathbf{v}$ by $\boldsymbol{\psi}_D^{i}\odot\mathbf{v}$ and $\mathbf{u}-\bar{\mathbf{u}}^{i}$ by $\boldsymbol{\psi}_D^{i}\odot(\mathbf{u}-\bar{\mathbf{u}}^{i})$ in all three bracketed terms, while $\mathbb{A}[\nabla\mathbf{v}]$ remains unchanged. Because the projector is diagonal, the tangent stays symmetric, and the unconstrained components are left free.

\paragraph{Load correction.} \textit{Correction of loaded region:} The Neumann region indicator, $M_N^{i}$, takes every quadrature point close to the region's facets (Sec.~\ref{sec:si-bcregions}), including points in the diffused band of the adjacent face, which is not selected as the Neumann region and is not supposed to be loaded. In the traction term of Eq.~\eqref{eq:residual}, with the original weight $|\nabla\varphi|$, such points would be loaded and the region would receive extra total force along its edges. Therefore, we replace the surface measure $|\nabla\varphi|$ from Eq.~\eqref{eq:measures} by modified weights $\langle-\nabla\varphi\cdot\mathbf{n}_f^{i}\rangle_{+}$, i.e., the component of $-\nabla\varphi$ along the mean normal $\mathbf{n}_f^{i}$ of the Neumann region kept only where it is positive. With the projection, their weight is smaller, because their $-\nabla\varphi$ turns away from $\mathbf{n}_f^{i}$. Although this modified weight alleviates the extra traction on unwanted regions, the corrected traction on a curved surface is distributed differently from the standard Neumann traction, which applies $\bar{\mathbf{t}}^i$ uniformly per unit area: more traction is applied where the surface faces along $\mathbf{n}_f^{i}$. The distribution is unchanged on a flat surface. This treatment also introduces deviation in the total force, which is addressed next. 

\textit{Correction of total force:} With the diffuse Nitsche method in our implementation, the total force deviates from the intended total force $\bar{\mathbf{t}}^{i}A^{i}$, where $A^{i}$ is the area of the selected region projected onto the plane normal to $\mathbf{n}_f^{i}$, for two reasons, even before the loaded region correction above. First, the region indicator $M_N^{i}$ reaches only the match tolerance (Sec.~\ref{sec:si-bcregions}), so it covers only part of the diffused zero level set (about $90\%$ of it at a match tolerance of $0.75\,h_{\min}$), which may be corrected by an enlarged indicator but at the risk of loading more unwanted regions and a longer computation. Second, the accuracy of a $2\!\times\!2\!\times\!2$ Gauss quadrature over the boundary band $\mathcal{B}$ depends on where the surface falls between the quadrature points, which may be corrected by higher-order quadrature rules at the cost of computation time. In our tests, these two effects together delivered between $88\%$ and $99.5\%$ of the intended force, depending on how the surface was aligned with the grid. We therefore scale the traction $\bar{\mathbf{t}}^i$ of each Neumann region $i$ by $A^{i}/W^{i}$. Here $W^{i}$ is the integral of $\langle-\nabla\varphi\cdot\mathbf{n}_f^{i}\rangle_{+}$ over $M_N^{i}$ with the quadrature and gate $\chi$ in Eq.~\eqref{eq:residual}. The loaded area $A^{i}$ is computed on the source (not diffused) surface as the sum of the region's facet areas, each multiplied by the positive part of the cosine between the facet normal and the region's area-weighted mean facet normal, which approximates $\mathbf{n}_f^i$. The scale factors are computed once at the first assembly of the solve. As a result, the total force on each region is $\bar{\mathbf{t}}^{i}A^{i}$, independent of the boundary band width, the mesh-surface alignment, and the loaded region correction.

\paragraph{Penalty stabilization.}
The penalty $\beta$ is a per-element field chosen so that it dominates the consistency and symmetry terms. The requirement is coercivity: the quadratic form of the discrete tangent must stay non-negative for every test displacement, so that each Newton system has a solution. The empirical choice, a penalty proportional to $E/h$ with $h$ the element size, assumes that the boundary area and the element volume of a cut element scale together. That assumption fails on an element that the surface clips at a corner, where the boundary measure $|\nabla\varphi|$ is large relative to the volume measure $\varphi$. Therefore, we compute the penalty per element instead. The quadratic form of the tangent follows from Eq.~\eqref{eq:residual}: the tangent is the derivative of $R(\mathbf{u},\mathbf{v})$ with respect to $\mathbf{u}$ in a direction $\delta\mathbf{u}$, and its quadratic form sets $\delta\mathbf{u}=\mathbf{v}$. The body-force and traction terms do not depend on $\mathbf{u}$ and drop out from the quadratic form. The internal work term gives $\nabla\mathbf{v}\!:\!\mathbb{A}[\nabla\mathbf{v}]\,w(\varphi)$, with $w(\varphi)\approx\varphi$. The consistency term gives $(\mathbb{A}[\nabla\mathbf{v}]\nabla\varphi)\cdot\mathbf{v}$, and the symmetry term, without the second derivative of $\mathbf{P}$, gives the same expression. With $\nabla\varphi=-|\nabla\varphi|\,\mathbf{n}$, the two together give $-2(\mathbb{A}[\nabla\mathbf{v}]\,\mathbf{n})\cdot\mathbf{v}\,|\nabla\varphi|$. The penalty term gives $\beta\,|\mathbf{v}|^{2}\,|\nabla\varphi|$. Therefore, on an element $\Omega_e$ in $\mathcal{B}$, for every test displacement $\mathbf{v}$ (dropping the gate $\chi$ and the region indicators $M^i$ for brevity), the coercivity requirement reads
\begin{equation}
  \int_{\Omega_e}\nabla\mathbf{v}\!:\!\mathbb{A}[\nabla\mathbf{v}]\,\varphi\,\mathrm{d}V
  +\int_{\Omega_e}\beta\,|\mathbf{v}|^{2}\,|\nabla\varphi|\,\mathrm{d}V
  -2\int_{\Omega_e}\bigl(\mathbb{A}[\nabla\mathbf{v}]\,\mathbf{n}\bigr)\cdot\mathbf{v}\,|\nabla\varphi|\,\mathrm{d}V
  \;\ge\;0,
  \label{eq:coercivity}
\end{equation}
where the first term is the internal work, the second the penalty, and the third collects the consistency and symmetry terms. The third term has indefinite sign and therefore threatens the bound. We follow the stability argument of Ref.~\citeS{nguyen2018diffuse}, derived there for the scalar Poisson problem (for example, steady heat conduction), but bound only the normal traction rather than the full stress, which gives a smaller penalty. On each element, there exists a constant $C_1$ such that the integrated squared traction is bounded by the internal energy,
\begin{equation}
  \int_{\Omega_e}\bigl\|\mathbb{A}[\nabla\mathbf{v}]\,\mathbf{n}\bigr\|^{2}\,|\nabla\varphi|\,\mathrm{d}V
  \;\le\;C_1\int_{\Omega_e}\nabla\mathbf{v}\!:\!\mathbb{A}[\nabla\mathbf{v}]\,\varphi\,\mathrm{d}V .
  \label{eq:tracebound}
\end{equation}
Then, Eq.~\eqref{eq:coercivity} holds whenever $\beta\ge C_1$: with $a=\|\mathbb{A}[\nabla\mathbf{v}]\,\mathbf{n}\|$ and $b=|\mathbf{v}|$, the cross term ($2ab$) is bounded by $2ab\le a^{2}/\beta+\beta b^{2}$, where the integrated $a^{2}/\beta$ part is bounded by the internal energy (the first term in Eq.~\eqref{eq:coercivity}) through Eq.~\eqref{eq:tracebound}, and the $\beta b^{2}$ part by the penalty (the second term in Eq.~\eqref{eq:coercivity}). The smallest such constant $C_1$ is found as follows. First, we write $\mathbf{v}=\mathbf{N}_e\mathbf{c}$, with $\mathbf{N}_e$ the element's shape functions and $\mathbf{c}$ the nodal values. Eq.~\eqref{eq:tracebound} then becomes $\mathbf{c}^{\mathsf{T}}\mathbf{A}_e\mathbf{c}\le C_1\,\mathbf{c}^{\mathsf{T}}\mathbf{B}_e\mathbf{c}$ ($\mathbf{A}_e$ and $\mathbf{B}_e$ shown below), so $C_1$ is the largest eigenvalue of a generalized eigenvalue problem~\citeS{embar2010imposing,nguyen2018diffuse}, solved per element in the boundary band $\mathcal{B}$,
\begin{equation}
  \mathbf{A}_e\mathbf{c}=\Lambda\,\mathbf{B}_e\mathbf{c},
  \qquad
  \beta_e=\max\Bigl(s_\beta\,\Lambda_{\max}^{(e)},\;0.2\,\frac{\lambda_L+2\mu}{h_{\min}}\Bigr),
  \qquad s_\beta=5,
  \label{eq:beta}
\end{equation}
where $\lambda_L=\kappa-\tfrac{2}{3}\mu$ is the first Lam\'e parameter, $s_\beta$ a safety factor, $\mathbf{A}_e$ the boundary traction energy matrix, built from the traction $\mathbf{T}_e$ that each nodal displacement produces on the diffused zero level set, and $\mathbf{B}_e$ the regularized form, defined below, of the internal strain energy matrix $\mathbf{B}_e^{0}$:
\begin{equation}
  \mathbf{A}_e=\int_{\Omega_e}\mathbf{T}_e^{\mathsf{T}}\mathbf{T}_e\,|\nabla\varphi|\,\mathrm{d}V,
  \quad
  \mathbf{T}_e=\bigl(\mathbb{C}\!:\!\nabla\mathbf{N}_e\bigr)\mathbf{n},
  \quad
  \mathbf{B}_e^{0}=\int_{\Omega_e}(\nabla\mathbf{N}_e)^{\mathsf{T}}\mathbb{C}\,(\nabla\mathbf{N}_e)\,\varphi\,\mathrm{d}V,
  \label{eq:betamat}
\end{equation}
where $\mathbb{C}=2\mu\,\mathbb{I}_{\mathrm{sym}}+\lambda_L\,\mathbf{I}\otimes\mathbf{I}$ is the small-strain elasticity tensor (the material tangent at zero deformation), and both matrices are integrated with the element's $2\!\times\!2\!\times\!2$ Gauss rule. Because the constant $\mathbb{C}$ replaces the configuration-dependent tangent $\partial\mathbf{P}/\partial\mathbf{F}$, the $24\!\times\!24$ eigenproblems are solved only once per mesh, at the reference configuration. The energy matrix is regularized as $\mathbf{B}_e=\mathbf{B}_e^{0}+\gamma_e\mathbf{I}$, with $\gamma_e=\max\bigl(|\operatorname{tr}\mathbf{B}_e^{0}|/24,\, (\lambda_L+2\mu)\,h_{\min}\,10^{-6}\bigr)$ to ensure its positive-definiteness. With a spatially varying material, $\mathbb{C}$, $\lambda_L$ and $\mu$ take reference values from the median $E$ and median $\nu$ of the material grid, and both terms of Eq.~\eqref{eq:beta} are multiplied by the ratio of the element's $\lambda_L+2\mu$ to its reference value computed from the medians. Because of the regularization and the small-strain tangent simplification, $\Lambda_{\max}^{(e)}$ only estimates the bound needed for coercivity, and the safety factor $s_\beta$ and the floor in Eq.~\eqref{eq:beta} are empirical safeguards that we add.

\begin{figure}[tb]
  \centering
  \includegraphics[width=\linewidth]{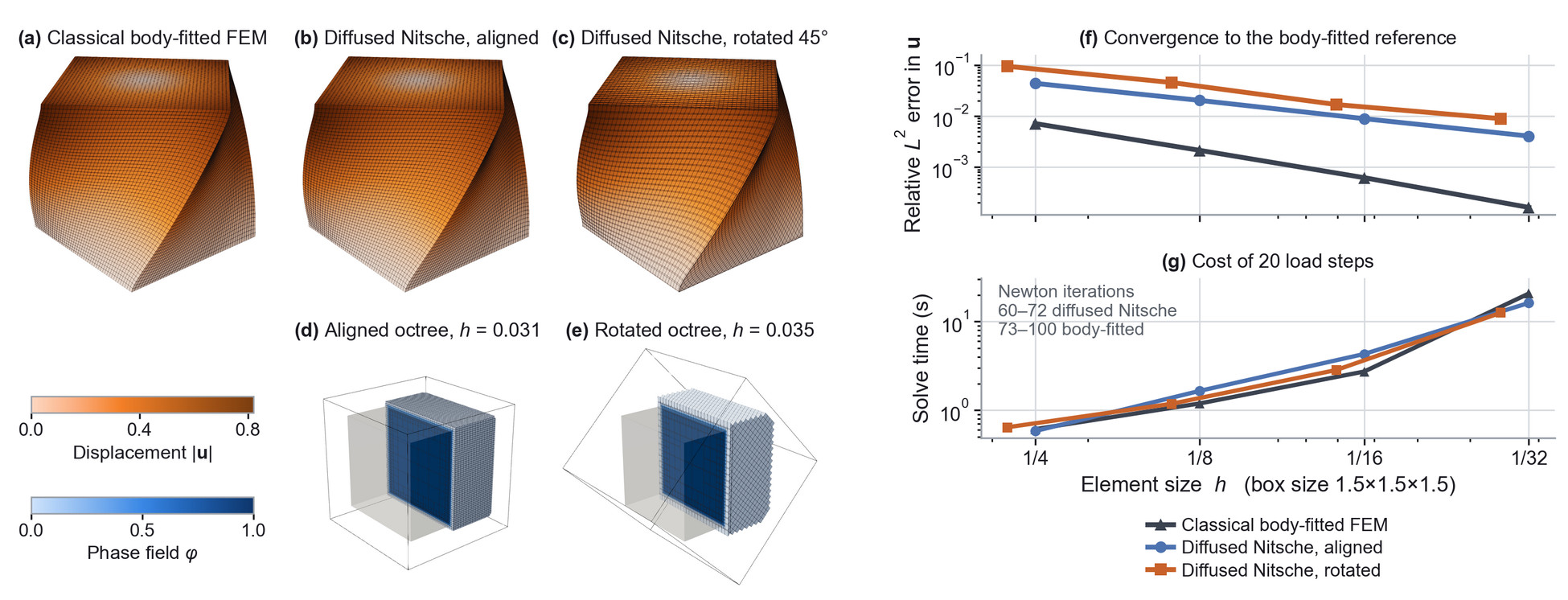}
  \caption{\textbf{Accuracy of the diffuse Nitsche method against body-fitted finite elements.} A block of side $1.5$ is clamped on one face and twisted by $45^\circ$ about its axis on the opposite face (neo-Hookean, $E=10$, $\nu=0.3$) and solved three ways. \textbf{a--c}, The deformed block colored by displacement magnitude on one scale: a body-fitted uniform hexahedral mesh, the block embedded in an octree whose cells align with its faces, and the block rotated by $45^\circ$ inside the octree so that no cell face conforms to it. \textbf{d},\textbf{e}, The two discretizations at level $6$: the background domain (wireframe), the geometry (gray, translucent) and the cells kept after trimming, colored by $\varphi$. \textbf{f}, Relative $L^2$ error in displacement against a body-fitted $96^3$ reference, evaluated on a shared $64^3$ lattice, for octrees of level $3$ to $6$ (finest cell size $h_{\min}=1/4$ to $1/32$ for the aligned block, $12.5\%$ larger for the rotated one) and body-fitted meshes (of the same element size $h=h_{\min}$ but uniform), with least-squares slopes of $1.15$ (aligned), $1.17$ (rotated) and $1.82$ (body-fitted). \textbf{g}, Wall time of the same runs. The study runs at single precision, a relative tolerance of $2\times10^{-5}$, an absolute floor of $10^{-8}$, $20$ load steps, a Newton iteration cap of $50$ and line search. The Nitsche penalty inflates the initial residual of a prescribed twist by orders of magnitude. }
  \label{fig:si-convergence}
\end{figure}

\paragraph{Element quadrature.} All elements, the diffuse boundary band included, carry a trilinear basis and a $2\!\times\!2\!\times\!2$ Gauss rule. The $\tanh$ profile of the phase field of width $\varepsilon=h_{\min}/2$ is therefore sampled at two points in the finest cell in each direction. The resulting quadrature error is part of the discretization error measured in Fig.~\ref{fig:si-convergence}. Although we implemented the adaptively refined quadrature used in Ref.~\citeS{nguyen2018diffuse}, adapted there from the finite cell method~\citeS{parvizian2007finite}, it is not used, because it costs more and gains less accuracy than mesh refinement. 

\paragraph{Newton solution and linear solver.} The nonlinear system is solved by Newton's method in a single load step, without line search in our results. By default, the iteration stops when the residual norm has fallen to $10^{-2}$ of its initial value, or below an absolute floor of $10^{-5}$ (in the selected force unit), within at most $20$ iterations. The tangent stiffness matrix omits the second derivative of $\mathbf{P}$ in the Nitsche symmetry term for efficiency. Therefore, the iteration is quasi-Newton. The omitted term is symmetric, so the tangent remains symmetric. Single and double precision are both available. Double precision, the default, has a better convergence rate but costs several times more, especially on a GPU with limited double-precision capacity. Each linear system is solved by conjugate gradients preconditioned by an aggregation algebraic-multigrid V-cycle (NVIDIA's AmgX~\citeS{naumov2015amgx}, via the JAX-AMG Python interface~\citeS{liu2026jax}, by default, or the \texttt{pyamgx} Python package) to a relative residual reduction of $5\times10^{-6}$ within at most $1000$ iterations. Single precision bounds the smallest strain that the solve can resolve: at the default relative tolerance, a solve whose peak strain lies below about $1.2\times10^{-4}$ tends to stall at a fixed residual. The user interface therefore roughly estimates the peak strain from the approved loads and stiffness and warns when the estimate lies below that value. 

\paragraph{Preliminary comparison against body-fitted finite elements.} We implemented the components above in one framework in PyTorch and NVIDIA Warp (Sec.~\ref{sec:si-implementation}) and measured the discretization error of the diffuse Nitsche method and compared it against body-fitted finite elements on one problem, a block twisted by $45^\circ$ via Dirichlet boundary conditions, over four refinement levels against a body-fitted reference (Fig.~\ref{fig:si-convergence}). Because $\varepsilon$ is tied to the smallest element size of the octree mesh, the modeling error of the diffuse-domain approximation is refined together with the discretization error. At the finest level (cell size $h=1/32$ for a block of side $1.5$), the relative $L^2$ displacement error is $4.06\times10^{-3}$ with the block aligned to the octree (Fig.~\ref{fig:si-convergence}b and d) and $8.91\times10^{-3}$ with the block rotated by $45^\circ$ (Fig.~\ref{fig:si-convergence}c and e), against $1.62\times10^{-4}$ for a body-fitted mesh of the same $h$ (Fig.~\ref{fig:si-convergence}a). The least-squares slopes over the four levels are $1.15$, $1.17$ and $1.82$, respectively. On this problem, the diffuse Nitsche method converges at roughly first order, against roughly second order for body-fitted finite elements. Rotating the block so that no cell face conforms to it about doubles the error and keeps the order.

\subsection{Sensitivity analysis}
\label{sec:si-sensitivity}

\paragraph{Adjoint state method.} Optimizing the shape minimizes a scalar objective $\mathcal{J}(\mathbf{U},\mathbf{m})$ of the deformed state, such as the strain energy, and needs its derivative with respect to each design variable. Finite differences would cost one nonlinear solve per design variable, which is prohibitively expensive for a high-dimensional design space such as the nodal signed distance field. Therefore, the derivative is obtained by the adjoint state method~\citeS{allaire2004structural}, at the cost of one extra linear solve. The optimization is a constrained problem of the form
\begin{equation}
  \min_{\mathbf{m}}\ \mathcal{J}(\mathbf{U},\mathbf{m})
  \qquad\text{subject to}\qquad
  \mathbf{R}(\mathbf{U},\mathbf{m})=\mathbf{0},
  \label{eq:constrained}
\end{equation}
where $\mathbf{U}$ and $\mathbf{m}$ are the nodal values of the displacement $\mathbf{u}$ and of the signed distance field $s$, and $\mathbf{R}$ is the residual vector of the ``Nitsche weak form'' paragraph (Sec.~\ref{sec:si-nitsche}). The equilibrium defines the state implicitly as a function $\mathbf{U}(\mathbf{m})$. Therefore, the total derivative of the objective is
\begin{equation}
  \frac{\mathrm{D}\mathcal{J}}{\mathrm{D}\mathbf{m}}
  =\frac{\partial\mathcal{J}}{\partial \mathbf{m}}
  +\frac{\partial\mathcal{J}}{\partial\mathbf{U}}\,\frac{\mathrm{D}\mathbf{U}}{\mathrm{D}\mathbf{m}}.
  \label{eq:totalderiv}
\end{equation}
The two partial derivatives of $\mathcal{J}$ follow from automatic differentiation (the chain rule applied to each operation of the code) of its explicit expression. The factor $\mathrm{D}\mathbf{U}/\mathrm{D}\mathbf{m}$ does not, because $\mathbf{U}$ is the output of a nonlinear solve. Instead, differentiating $\mathbf{R}(\mathbf{U}(\mathbf{m}),\mathbf{m})=\mathbf{0}$ with respect to $\mathbf{m}$ gives
\begin{equation}
  \frac{\partial\mathbf{R}}{\partial\mathbf{U}}\,\frac{\mathrm{D}\mathbf{U}}{\mathrm{D}\mathbf{m}}
  +\frac{\partial\mathbf{R}}{\partial \mathbf{m}}=\mathbf{0}
  \quad\Longrightarrow\quad
  \frac{\mathrm{D}\mathbf{U}}{\mathrm{D}\mathbf{m}}
  =-\Bigl(\frac{\partial\mathbf{R}}{\partial\mathbf{U}}\Bigr)^{-1}\frac{\partial\mathbf{R}}{\partial \mathbf{m}},
  \label{eq:ift}
\end{equation}
which is the implicit function theorem. Here $\partial\mathbf{R}/\partial\mathbf{U}$ is the tangent stiffness matrix, which Newton's method already assembles in approximate form (see the ``Newton solution and linear solver'' paragraph of Sec.~\ref{sec:si-nitsche}). Substituting Eq.~\eqref{eq:ift} into the second term of Eq.~\eqref{eq:totalderiv}, the product $(\partial\mathcal{J}/\partial\mathbf{U})(\partial\mathbf{R}/\partial\mathbf{U})^{-1}$ appears. It is a row vector that does not depend on which entry of $\mathbf{m}$ is differentiated, so it is computed once. Its negative, written $\boldsymbol{\lambda}^{\mathsf{T}}$, is called the adjoint state, and
\begin{equation}
  \frac{\mathrm{D}\mathcal{J}}{\mathrm{D}\mathbf{m}}
  =\frac{\partial\mathcal{J}}{\partial \mathbf{m}}
  +\boldsymbol{\lambda}^{\mathsf{T}}\frac{\partial\mathbf{R}}{\partial \mathbf{m}},
  \qquad
  \Bigl(\frac{\partial\mathbf{R}}{\partial\mathbf{U}}\Bigr)^{\!\mathsf{T}}\boldsymbol{\lambda}
  =-\Bigl(\frac{\partial\mathcal{J}}{\partial\mathbf{U}}\Bigr)^{\!\mathsf{T}}.
  \label{eq:adjoint}
\end{equation}
One linear solve with the transposed tangent therefore yields the gradient with respect to all entries of $\mathbf{m}$ at once. The tangent is symmetric (Sec.~\ref{sec:si-nitsche}), so the adjoint system is solved with the same tangent.

\paragraph{Hybrid automatic differentiation and adjoint solve.} Equation~\eqref{eq:adjoint} is evaluated by reverse-mode automatic differentiation over a recorded ``tape'' (the record of every operation in the forward computation, replayed backward to obtain derivatives), on which the explicit derivatives $\partial\mathcal{J}/\partial \mathbf{m}$ and $\partial\mathbf{R}/\partial \mathbf{m}$ of Eq.~\eqref{eq:adjoint}, including the chain $s\rightarrow\varphi\rightarrow\mathbf{R}$ (from the signed distance field through the phase field of Eq.~\eqref{eq:phasefield} to the residual), are supplied automatically. We register the adjoint solve, which automatic differentiation cannot supply, as a custom node on the tape: it solves for $\boldsymbol{\lambda}$ from $\partial\mathcal{J}/\partial\mathbf{U}$, so one backward pass returns the whole gradient without forming $\mathrm{D}\mathbf{U}/\mathrm{D}\mathbf{m}$. On the tape, the signed distance field enters the residual only through the diffuse boundary band $\mathcal{B}$, whose phase field $\varphi$ and gradient $\nabla\varphi$ are recomputed from the nodal distance $s$ at every solve. Outside the boundary band, $\varphi$ lies within $10^{-3}$ of zero or one, where the slope $\mathrm{d}\varphi/\mathrm{d}s=-2\varphi(1-\varphi)/\varepsilon$ of Eq.~\eqref{eq:phasefield} nearly vanishes, so the chain from $s$ to $\varphi$ passes almost no gradient back. Therefore, to save the time of recording and back-propagating the interior, the material weight $w(\varphi)$ of Eq.~\eqref{eq:measures} is evaluated there from the nodal values and included in the quadrature weights without being recorded on the tape. The signed distance field then carries no stiffness sensitivity in the interior. The design field lives on the untrimmed octree and the solve on the analysis mesh $\mathcal{K}$ (Sec.~\ref{sec:si-nitsche}). The field is copied to the analysis mesh before the solve, and the gradient is scattered back to the untrimmed octree after it.

\paragraph{Accuracy of the sensitivity.} The result is an approximate sensitivity, for three reasons. First, the forward state satisfies $\mathbf{R}=\mathbf{0}$ only to the Newton tolerance, whereas Eq.~\eqref{eq:adjoint} assumes the equality exactly. Second, the adjoint system is solved with the quasi-Newton tangent assembled at the converged state, which omits the second derivative of $\mathbf{P}$ in the Nitsche symmetry term (Sec.~\ref{sec:si-nitsche}). It is solved by the same preconditioned conjugate-gradient iteration, with the same stopping rule (a nominal relative reduction of $5\times10^{-6}$), as each Newton step. Third, the derivative $\partial\mathbf{R}/\partial \mathbf{m}$ is incomplete: the penalty $\beta$, the quadrature weights, the band membership, the mean normals $\mathbf{n}_f^{i}$ of the Neumann regions, the trimming and the hanging-node constraints all depend on the geometry, but they are rebuilt without the tape recording, so their variation with the design is not captured. In addition, at hanging nodes (Sec.~\ref{sec:si-geometry}) the gradient is replaced by values interpolated from the coarse-cell nodes, which do not receive the hanging nodes' own contributions as the exact adjoint would require. A central-difference test checks the derivative at three nodes of a fixed, uniform hexahedral mesh of $16\times8\times8$ elements without hanging nodes, in double precision and at tight Newton tolerances, reaching agreement within 1\% at nodes farther than $\varepsilon$ from the surface and within 15\% at nodes closer to it. The sensitivity is therefore used as an inexact search direction.

\subsection{Shape optimization}
\label{sec:si-shapeopt}

\paragraph{Objective.} The design variable is the nodal signed distance field $s$ on the untrimmed octree (Sec.~\ref{sec:si-nitsche}), as in level-set structural optimization~\citeS{wang2003level,allaire2004structural}. We seek a stiff design that uses at most a target fraction of the initial volume. Therefore, we minimize the normalized strain energy under a one-sided volume penalty,
\begin{equation}
  \mathcal{J}
  =\frac{\mathrm{SE}}{\mathrm{SE}_0}
  +w_{\mathrm{vol}}\,\mathrm{softplus}_k\!\Bigl(\frac{\hat{V}-f_{\mathrm{target}}}{\Delta_0}\Bigr),
  \qquad
  \hat{V}=\frac{V}{V_0},
  \label{eq:objective}
\end{equation}
with $w_{\mathrm{vol}}=5$, $f_{\mathrm{target}}=0.5$ by default, and $\operatorname{softplus}_k(x)=k^{-1}\log\bigl(1+e^{kx}\bigr)$ at $k=500$, a smoothed hinge, so that only volume above the target is penalized. Here $\mathrm{SE}=\int w(\varphi)\,W(\mathbf{F})\,\mathrm{d}V$ is the strain energy and $V=\int\varphi\,\mathrm{d}V$ is the volume. $\mathrm{SE}_0$ and $V_0$ are the first-iteration values of $\mathrm{SE}$ and $V$, and $\Delta_0=\left|1-f_{\mathrm{target}}\right|$, all held constant and not differentiated. Under traction loading, this objective is compliance minimization~\citeS{bendsoe1989optimal} in the linear limit. With self-weight, the load also changes with the design. Under prescribed displacements, however, the strain energy falls as material is removed, and the one-sided volume penalty does not oppose that. Therefore, the optimization examples are loaded by tractions and body forces, without prescribed non-zero displacements.

\paragraph{Gradient-descent update.} The volume requirement enters only through the penalty of Eq.~\eqref{eq:objective}, so equilibrium (Eq.~\eqref{eq:constrained}) is the only constraint. Its sensitivity $\mathrm{D}\mathcal{J}/\mathrm{D}s$ is meaningful only near the zero level set, because $\partial\varphi/\partial s$ decays on the scale of $\varepsilon$ and the stiffness outside the diffuse boundary band is not differentiated (see the ``Hybrid automatic differentiation and adjoint solve'' paragraph of Sec.~\ref{sec:si-sensitivity}). Therefore, this formulation changes the shape by only moving its boundary. Before each step, the gradient is set to zero at nodes absent from the trimmed mesh and at all nodes of boundary band elements that touch any of the boundary condition regions, so the optimizer cannot move the boundary regions directly (see the ``Boundary-region preservation'' paragraph below for additional steps to exactly freeze boundary condition regions). Updates are plain gradient descent at a learning rate of $0.01$ by default, with no bound on the step size and no smoothing of the gradient, for up to $101$ iterations by default. The loop stops earlier when two successive relative reductions of $\mathcal{J}$ both lie between zero and $10^{-9}$. The learning rate is dimensional: $\mathcal{J}$ is dimensionless and $s$ is a length, so it carries units of length squared. A suitable value therefore depends on the model's extent, unit system and mesh resolution. Optionally, the learning rate is instead set at every update to $0.5\,\varepsilon/\max|\mathrm{D}\mathcal{J}/\mathrm{D}s|$ over the free nodes, so that each step changes $s$ by at most $0.5\,\varepsilon$ (the fraction $0.5$ is the default). 

\begin{figure}[tb]
  \centering
  \includegraphics[width=\linewidth]{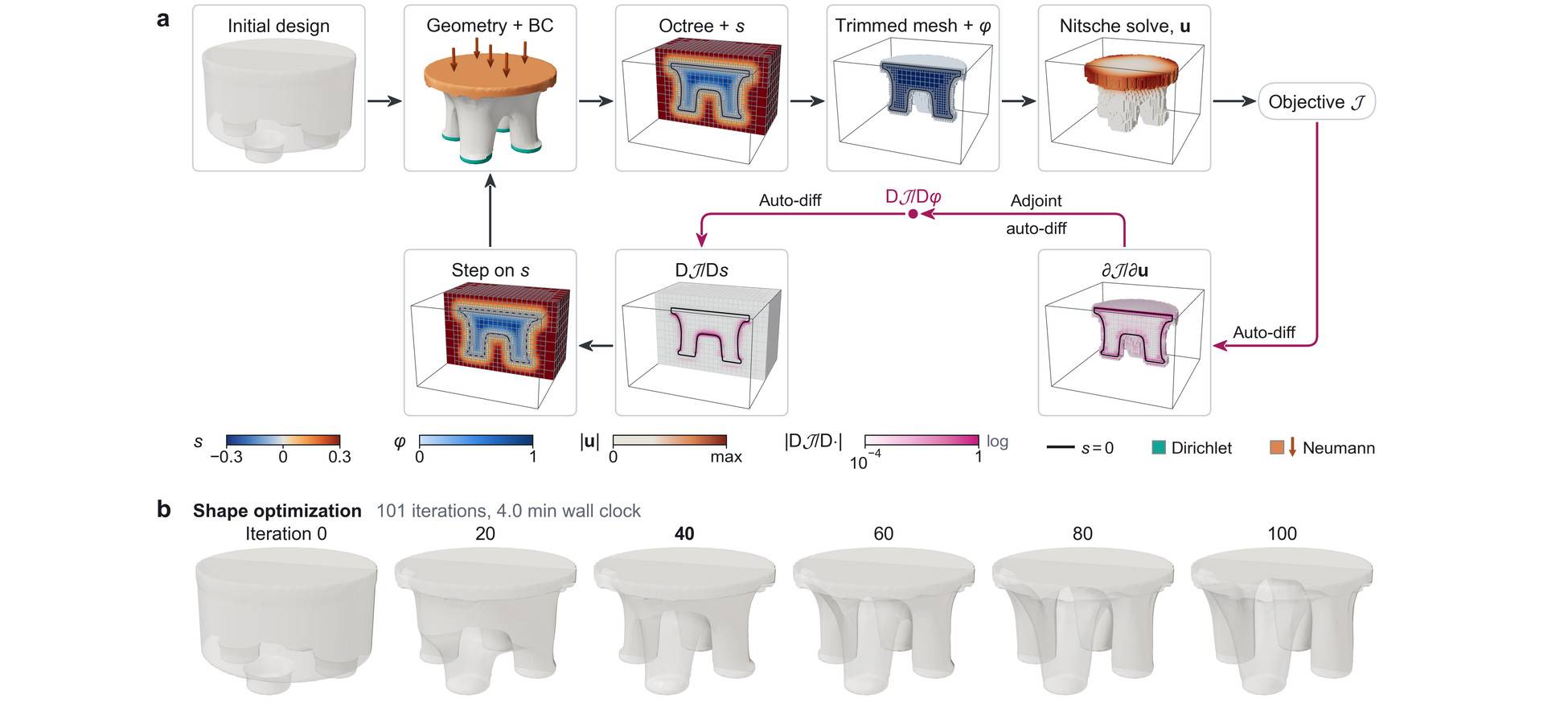}
  \caption{\textbf{Shape optimization of an exemplary table problem from a rough initial design.} \textbf{a}, One iteration, shown at iteration $40$. Forward (top row, black arrows): starting from the initial design, the current surface carries its Dirichlet (teal, the four feet) and Neumann (orange, the top) regions. Its signed distance field $s$ on the octree defines the phase field $\varphi$ on the trimmed mesh (Eq.~\eqref{eq:phasefield}). The diffuse Nitsche solve returns the displacement $\mathbf{u}$, from which the objective $\mathcal{J}$ is evaluated (Eq.~\eqref{eq:objective}). Backward (bottom row, magenta arrows, right to left, Sec.~\ref{sec:si-sensitivity}): automatic differentiation gives $\partial\mathcal{J}/\partial\mathbf{U}$, the adjoint state method with automatic differentiation gives $\mathrm{D}\mathcal{J}/\mathrm{D}\varphi$, and automatic differentiation of Eq.~\eqref{eq:phasefield} gives $\mathrm{D}\mathcal{J}/\mathrm{D}s$, which is confined to a band around the surface and is frozen as zero on the boundary-condition regions. Each sensitivity is shown as its magnitude divided by its own 99th percentile, on a log scale. A gradient step on $s$ at a learning rate of $8$ moves the zero level set from its previous position to the new one, and the geometry is rebuilt with the boundary-condition patches stitched back (``Boundary-region preservation'' paragraph of Sec.~\ref{sec:si-shapeopt}). This example uses both the gradient mask and stitching, which leads to the design for the next iteration. \textbf{b}, The design over the $101$ iterations ($4.0$~min on an NVIDIA RTX A6000), cut in half with the near half translucent. Iteration $40$ is the one shown in a. }
  \label{fig:si-shapeopt}
\end{figure}

\paragraph{Design iteration.} Each iteration runs the diffuse Nitsche solve on the current analysis mesh (Sec.~\ref{sec:si-nitsche}), evaluates the sensitivity of Eq.~\eqref{eq:adjoint} and takes the descent step on $s$ (Fig.~\ref{fig:si-shapeopt}a, for an exemplary table design problem). The step does not preserve the eikonal condition, on which the diffuse Nitsche method relies for the width of the phase-field profile and for the surface measure $|\nabla\varphi|$ (Sec.~\ref{sec:si-nitsche}). It also creates a strong discontinuity near the frozen boundary condition regions. Therefore, the geometry is rebuilt (see the next paragraph) and $s$ is reinitialized every iteration from the steps of Sec.~\ref{sec:si-geometry}, but in a different order. For a single forward solve, the octree is extracted first and then reads the reinitialized values, whereas in design iterations, the field is reinitialized first and the octree is extracted afterwards, from the new surface, so that its refinement follows the moved boundary.

\paragraph{Geometry rebuild.} The rebuild has five steps. (i) The updated nodal field on the untrimmed mesh is scattered from the octree onto the dense background grid ($129^3$ nodes at the default depths $l_{\mathrm{init}}=4$ to $l_{\mathrm{final}}=7$). The other grid nodes keep their previous values, and on the boundary regions the field is frozen (held at its initial values). (ii) Marching cubes extracts a new surface from the grid, and, when stitching is on, the boundary patches saved at the first iteration are restored into it (Fig.~\ref{fig:si-shapeopt}a, see the ``Boundary-region preservation'' paragraph below). (iii) As in the ``Signed distance reinitialization'' paragraph of Sec.~\ref{sec:si-geometry}, the exact signed distance to that surface is recomputed at the grid nodes. (iv) As in the ``Octree extraction with the scaled surface'' paragraph, but against the reconstructed surface instead of the source surface, a fresh octree is extracted from level $l_{\mathrm{init}}$ to $l_{\mathrm{final}}$, with every cell that contains a vertex of the surface forced to the finest level. (v) The octree reads its nodal values from the reinitialized grid, and the analysis mesh is trimmed from it (Sec.~\ref{sec:si-nitsche}). Finally, the displacement is interpolated from the old mesh onto the new one as the initial guess of the next Newton solve. The traction scale factors $A^{i}/W^{i}$ of the ``Load correction'' paragraph (Sec.~\ref{sec:si-nitsche}) are computed at the first iteration and kept, while the region where traction is applied is preserved, so that the load does not jump between iterations. The rebuild is a small part of an iteration: in a separate timed run of the problem of Fig.~\ref{fig:si-shapeopt} at the default depths ($98\,296$ nodes in the untrimmed octree), one iteration took a median of $2.28$~s on an NVIDIA RTX A6000, of which the forward solve took $1.35$~s, the adjoint $0.47$~s and the rebuild $0.31$~s. Further efficiency improvement can be achieved by constructing octree meshes with a wider refined band and reusing the previous octree mesh while it still resolves the boundary band, which avoids rebuilding the mesh and thus the AmgX hierarchy at every design iteration. 

\begin{figure}[tb]
  \centering
  \includegraphics[width=0.6\linewidth]{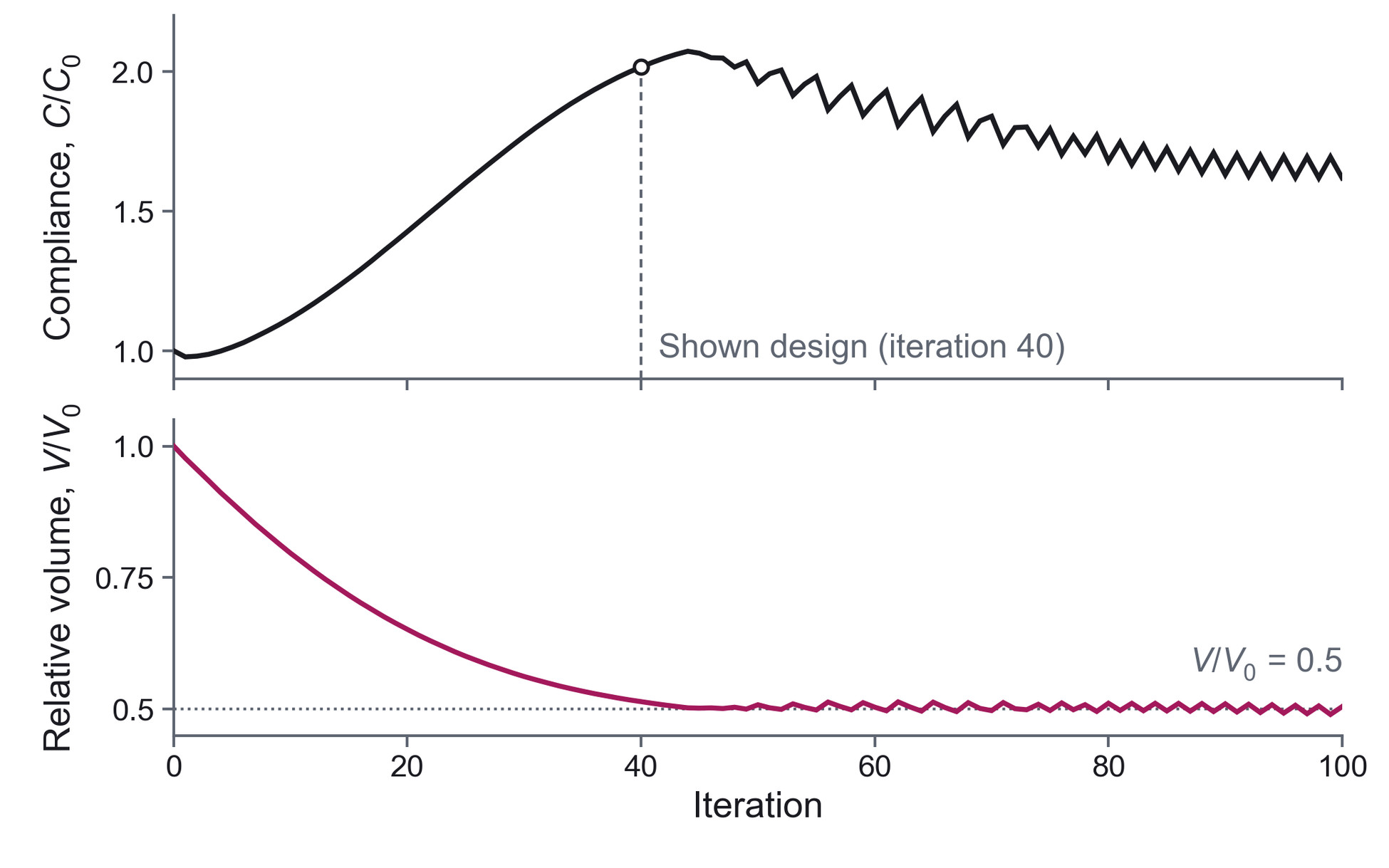}
  \caption{\textbf{Strain energy and volume over the run of Fig.~\ref{fig:si-shapeopt}.} Strain energy normalized by its initial value (top plot, labeled as compliance) and the volume fraction $V/V_0$ (bottom plot) over $101$ iterations. The volume penalty of Eq.~\eqref{eq:objective} acts only above the target of $0.5$ (dotted). While the design is above target, the optimizer removes material, and the strain energy of the loaded part rises. The volume reaches the target near iteration $45$, and the strain energy peaks at $2.07$ times its initial value at iteration $44$. Thereafter, the boundary redistributes material at nearly fixed volume and the strain energy falls, to $1.62$ at iteration $100$ with the volume fraction at $0.505$. The dashed line marks iteration $40$, the design shown in Fig.~\ref{fig:si-shapeopt}. }
  \label{fig:si-so-objective}
\end{figure}

\paragraph{Behavior of the exemplary run.} In the traction-loaded table design run of Fig.~\ref{fig:si-shapeopt}b, the strain energy rises while the penalty removes material down to the target volume, and then falls as the boundary redistributes material at nearly fixed volume (Fig.~\ref{fig:si-so-objective}).

\paragraph{Boundary-region preservation.}
The regions that carry boundary conditions must stay in place while the rest of the surface moves. A simple approach is using boundary condition indicators $M^{i}$, computed from the selected facets before the first iteration (see the ``Regions on the analysis octree mesh'' paragraph of Sec.~\ref{sec:si-bcregions}), as the mask of the gradient update, which keeps the optimizer from moving the regions. However, the surface is re-extracted every iteration from an $s$ field that has changed around the frozen region, so the region's facets are re-created and not exactly the same as the initial facets. This approach can also leave a non-smooth surface, with sharp edges, around the boundary condition regions. 

\begin{figure}[tb]
  \centering
  \includegraphics[width=\linewidth]{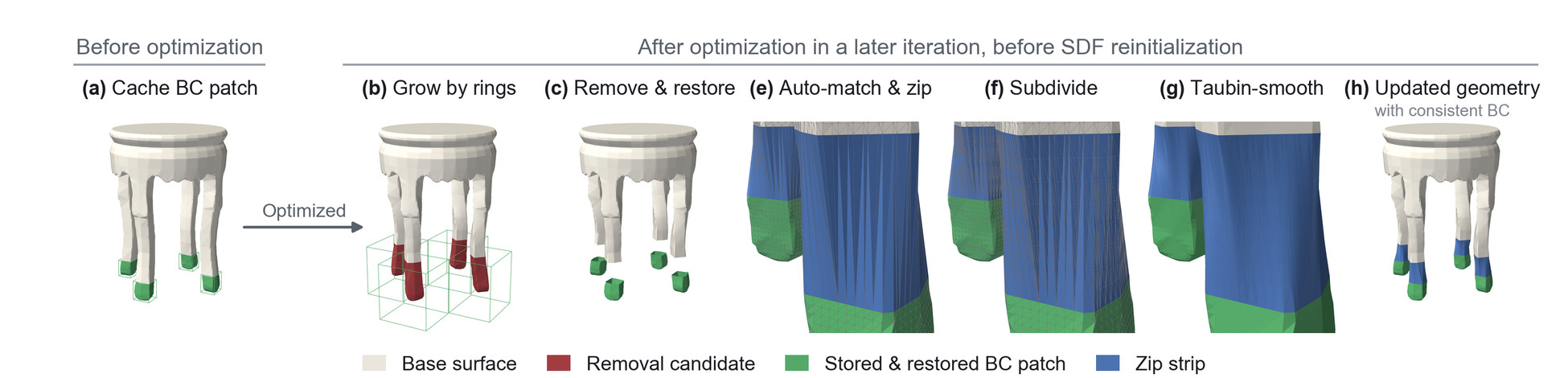}
  \caption{\textbf{Schematic of preserving boundary-condition regions across a shape update.} The saved patch carrying each region (left) is re-inserted into the surface extracted from the updated field. The neighborhood of the updated surface is grown by facet adjacency until it is one connected piece whose removal leaves a manifold remainder exposing at least as many open loops as the patch. Surplus loops are merged, and each remaining loop is bridged to a patch loop by a dynamic program that minimizes the total squared edge length. Long bridge edges are subdivided and the new facets relaxed by Taubin smoothing. As a result, the region's facets are restored exactly in the reconstructed surface and are joined continuously to the moving surface. The update shown is a prescribed twist of the initial surface standing in for a design step.}
  \label{fig:bczip}
\end{figure}

\begin{figure}[tb]
  \centering
  \includegraphics[width=0.8\linewidth]{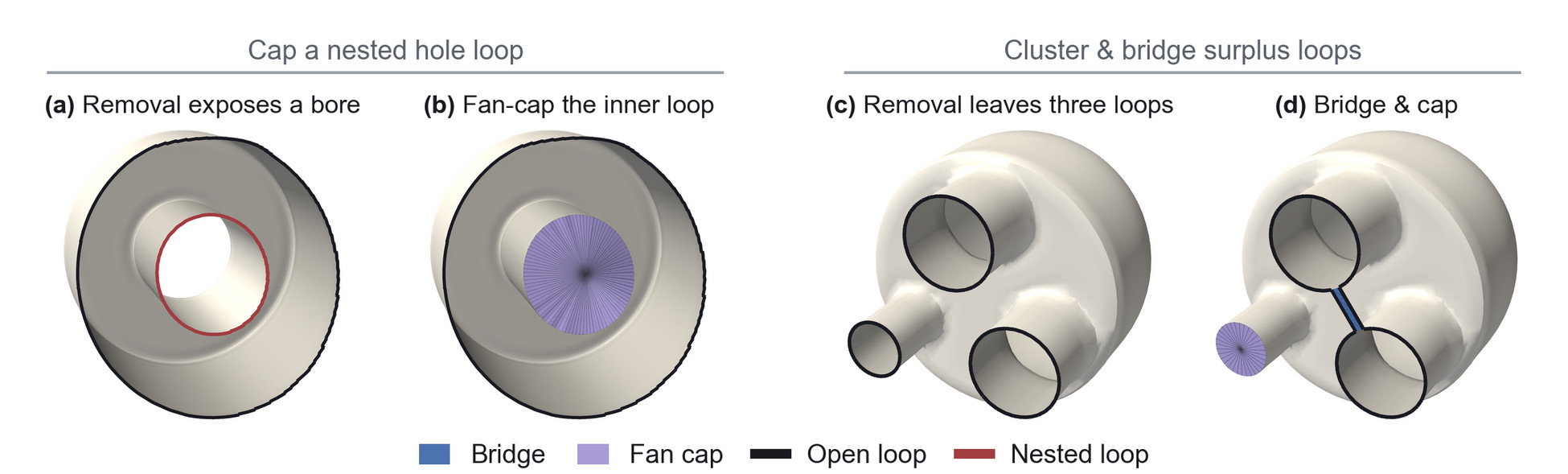}
  \caption{\textbf{The two topological branches of the patch restoration.} Each pair shows the surface after the neighborhood of a boundary region has been removed (left) and after the branch has acted (right), on a body built on purpose to reach that branch. \textbf{a},\textbf{b}, Removal around a clamped end face exposes a through-bore, so a second hole loop (red) lies nested inside the outer one (black). The nested loop has no counterpart on the saved patch and is sealed by a triangle fan ($112$ triangles), leaving one loop. \textbf{c},\textbf{d}, Removal around a region spanning three struts leaves three hole loops for a patch with one. The two nearest are joined by a bridge strip ($8$ triangles, blue) and the remaining loop is capped ($32$ triangles), leaving one loop to be bridged to the patch.}
  \label{fig:si-bczipx}
\end{figure}

Optionally, we restore the surface patches in the boundary condition regions explicitly (Fig.~\ref{fig:bczip}). At the first iteration, the surface facets whose three vertices all lie within twice the background-grid spacing $h_{\min}$ of a frozen node, widened by two facet rings, are saved as a patch to be restored later for each connected boundary condition region. At each iteration after a new surface is rebuilt, the facets neighboring the surface patches are removed: starting from the facets on the new surface with a vertex in the patch's bounding box enlarged by a factor of $1.1$, the removed facets are grown by adjacency, up to $80$ rings, until they form one connected piece whose removal leaves a manifold remainder (the remainder's open edges forming simple closed chains, i.e., loops), with at least as many open boundary loops (not counting loops nested inside another) as the restored patch. Extra loops that the removal exposes are closed by triangle fans or joined by bridge strips (Fig.~\ref{fig:si-bczipx}). Each remainder loop is joined to its closest patch loop by a strip of new triangles chosen by an optimal contour tiling after Fuchs, Kedem and Uselton~\citeS{fuchs1977optimal}, here a dynamic program that minimizes the total squared edge length. Finally, the strip is relaxed by Taubin smoothing~\citeS{taubin1995signal} with the patch and the remainder surface held fixed. As a result, each region's facets are restored exactly in the reconstructed surface from which the next signed distance field is computed. As a side effect, the junction between the patch and the moving surface is continuous, though not tangent-continuous.

\paragraph{Refinement of generated bodies.}
Each of the 50 bodies of the AI agent operator sessions (Sec.~\ref{si:opus-results}) is refined by a short shape optimization of its signed distance field $s$. Fig.~\ref{fig:optimization}c of the main text shows 17 of them that have observable and meaningful refinement. The aim is to smooth small generation defects and slightly thicken weak members while keeping each object recognizable. The objective keeps the strain energy of Eq.~\eqref{eq:objective} but replaces its volume term by a one-sided penalty on growth above the initial volume,
\begin{equation}
  \mathcal{J}
  =\frac{\mathrm{SE}}{\mathrm{SE}_0}
  +w_{\mathrm{vol}}\,\mathrm{softplus}_k\!\Bigl(\frac{V}{V_0}-1\Bigr),
  \label{eq:si-refine-objective}
\end{equation}
with $w_{\mathrm{vol}}=5$ and $k=500$ as in Eq.~\eqref{eq:objective}. The penalty is one-sided and does not necessarily hold the volume at $V_0$. The boundary regions and the nodes outside the trimmed analysis mesh are frozen by the gradient mask alone, without patch restoration. Because the bodies span four orders of magnitude in size, the learning rate is fixed per body at the first update, so that the largest nodal change of $s$ is $0.5\,h_{\min}$, and then held in later iterations. Neither $s$ nor the octree is rebuilt between iterations, because on members a few cells thick reinitialization moves the surface by a fraction of a cell even with a zero step. The surface $\Gamma_k$ at iteration $k$ is the marching-cubes zero level set of the current $s$. With $d(\mathbf{x},\Gamma)=\min_{\mathbf{y}\in\Gamma}\lVert\mathbf{x}-\mathbf{y}\rVert$ the distance from a point $\mathbf{x}$ to a surface $\Gamma$, let $P_{99}(\Gamma_a\to\Gamma_b)$ be the 99th percentile of $d(\mathbf{x}_i,\Gamma_b)$ over the vertices $\mathbf{x}_i$ of $\Gamma_a$. The change of $\Gamma_k$ from the generated (initial) surface $\Gamma_0$ is measured in both directions by the percentile and the maximum distance,
\begin{equation}
  D_{99}=\max\bigl(P_{99}(\Gamma_k\to\Gamma_0),\,P_{99}(\Gamma_0\to\Gamma_k)\bigr),
  \qquad
  D_{\max}=\max\Bigl(\max_{\mathbf{x}_i\in\Gamma_k}d(\mathbf{x}_i,\Gamma_0),\,\max_{\mathbf{x}_i\in\Gamma_0}d(\mathbf{x}_i,\Gamma_k)\Bigr).
  \label{eq:si-refine}
\end{equation}
A shape is accepted while $D_{99}\le1.5\,h_{\min}$ and $D_{\max}\le3\,h_{\min}$ (Eq.~\eqref{eq:si-refine}). The percentile allows a neck to thicken locally but stops a change spread over the body, and the maximum catches an erased small feature such as a tusk. A run stops before the first shape that exceeds either bound, or after 40 accepted updates. Shape optimization currently supports only uniform materials, so a body with a different material per part uses the volume-weighted mean of its parts' $E$, $\nu$ and $\rho$, with the part volumes taken from the labeling that defines the per-part material field (Sec.~\ref{sec:si-nitsche}). Concurrent multi-material optimization is possible~\citeS{luo2026towards} but beyond the scope of this work.

\subsection{Topology optimization}
\label{sec:si-hiddenphi}

\begin{figure}[tb]
  \centering
  \includegraphics[width=\linewidth]{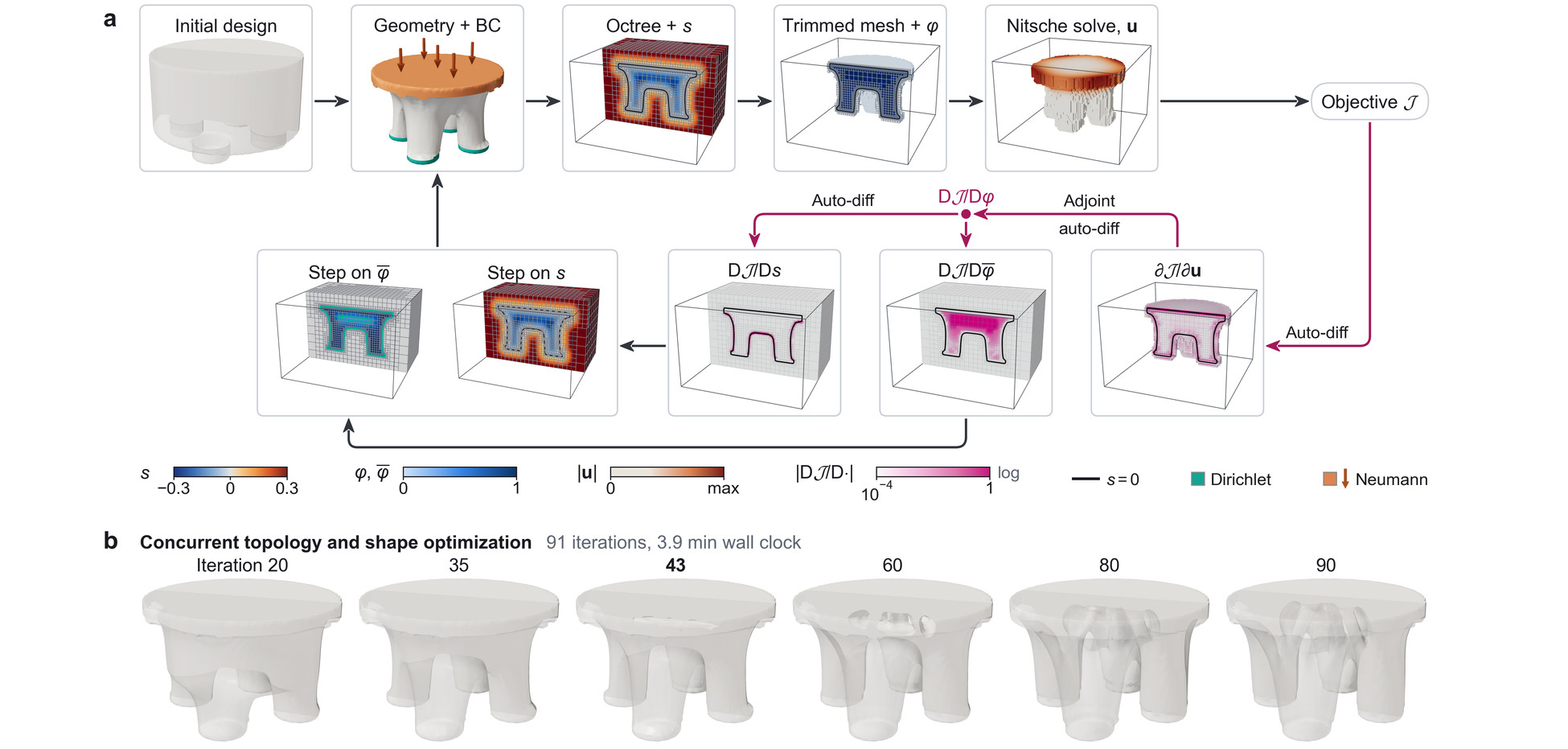}
  \caption{\textbf{Topology optimization of the same exemplary table design problem in Fig.~\ref{fig:si-shapeopt}.} \textbf{a}, One iteration, shown at iteration $43$, whose update carves the first cavity. The backward pass now also gives the signal $g$ of Eq.~\eqref{eq:shadowgrad} (labeled $\mathrm{D}\mathcal{J}/\mathrm{D}\bar{\varphi}$ in the panel), the gradient with respect to $\varphi$ in the interior, used as that with respect to $\bar{\varphi}$, while $\mathrm{D}\mathcal{J}/\mathrm{D}s$ is kept only within the band around the surface. The hidden phase field $\bar{\varphi}$ is updated by a step in the interior (``Step on $\bar{\varphi}$'', teal contour $\bar{\varphi}=\tfrac12$, gray outside the body), and the closed interior region where $\bar{\varphi}<\tfrac12$ is admitted as a new cavity and is reflected in $s$ when the geometry is rebuilt. The solve uses only the phase field $\varphi$ computed from $s$ instead of the hidden phase field $\bar{\varphi}$ being updated. \textbf{b}, The design over the $91$ iterations ($3.9$~min). The hidden phase field is enabled at iteration $35$. The bold frame shows the surface after iteration $43$'s update, with the first cavity seen through the translucent near half, and the cavities open to the surface by iteration $80$.}
  \label{fig:hiddenphi}
\end{figure}

\paragraph{Interior voids.} The shape optimization of Sec.~\ref{sec:si-shapeopt} changes topology only through moving boundaries. Because the surface is re-extracted at every iteration, a moving boundary can merge surfaces, pinch off a neck or open a through-hole. It cannot, however, open a cavity deep inside the solid, a known limitation of level-set methods~\citeS{xia2019stable}, because the adjoint sensitivity with respect to the signed distance field vanishes away from the diffused zero level set (Sec.~\ref{sec:si-sensitivity}). Therefore, we add a second design variable: a nodal field $\bar{\varphi}$ (one in material, zero in void), called the \textit{hidden} phase field, which mirrors the phase field $\varphi$ of Eq.~\eqref{eq:phasefield} and can evolve where the signed distance field does not (Fig.~\ref{fig:hiddenphi}). The phase field $\varphi$ is built only from the signed distance field (Eq.~\eqref{eq:phasefield}) and is the actual input to the residual vector $\mathbf{R}$, while the design variable $\bar{\varphi}$ changes the geometry only when a void is admitted as $\bar{\varphi}$ updates. The signed distance field remains the geometric design variable and receives its gradient within $3\varepsilon$ of the surface. Farther away, its gradient is set to zero, and the hidden phase field receives the update there.

\paragraph{Hidden phase field update.} The hidden phase field is kept on the nodes of the untrimmed octree, which also carries the signed distance field, and is initialized at the first iteration from the phase field of the current geometry (Eq.~\eqref{eq:phasefield}). At each iteration, it is updated node by node by
\begin{equation}
  \bar{\varphi}_i \;\leftarrow\;
  \begin{cases}
    \tfrac12\bigl(1-\tanh(s_i/\varepsilon)\bigr),
      & |s_i| < 3\varepsilon \quad\text{(near the surface)},\\[4pt]
    \bar{\varphi}_i - \eta_\varphi\,g_i,
      & |s_i| \ge 3\varepsilon,
  \end{cases}
  \label{eq:hiddenphi}
\end{equation}
then clamped to $[0,1]$. At hanging nodes, its values are replaced by the average over the end nodes of the coarse edge or the corner nodes of the coarse face, as in the ``Hanging nodes'' paragraph of Sec.~\ref{sec:si-geometry}. Within $3\varepsilon$ of the surface, the first line makes the hidden phase field follow the phase field of the current geometry (Eq.~\eqref{eq:phasefield}), whereas outside it, the second line lets the field descend with step $\eta_\varphi$ along the sensitivity $g_i$ defined next.

\paragraph{Sensitivity signal.} Because the solve uses only $\varphi$, the objective has no derivative with respect to $\bar{\varphi}$. Since $\bar{\varphi}$ mirrors $\varphi$, the gradient of $\mathcal{J}$ with respect to $\varphi$ at the nodes farther than $3\varepsilon$ from the surface is used in its place. For this, unlike in Sec.~\ref{sec:si-sensitivity}, the interior residual records its weight $w(\varphi)$ of Eq.~\eqref{eq:measures} on the tape as a function of the nodal $\varphi$,
\begin{equation}
  R_{\mathrm{int}}(\mathbf{v})=\int_{\mathcal{K}\setminus\mathcal{B}}
  w(\varphi)\,\mathbf{P}\!:\!\nabla\mathbf{v}\;\mathrm{d}V,
  \label{eq:simpresidual}
\end{equation}
which, with the linear weight $w(\varphi)$, has the form of the solid isotropic material with penalization interpolation~\citeS{bendsoe1989optimal} with exponent one, while the chain from $s$ to $\varphi$ stays off the tape in the interior, because it passes almost no gradient back there (Sec.~\ref{sec:si-sensitivity}). The body-force term carries the same weight, and the volume $\int\varphi\,\mathrm{d}V$ of the objective also depends on the nodal $\varphi$. Therefore, the backward pass of Sec.~\ref{sec:si-sensitivity} returns, without an additional solve, the analogue of Eq.~\eqref{eq:adjoint} for $\varphi$,
\begin{equation}
  g_i \;\simeq\;
  \frac{\partial\mathcal{J}}{\partial\varphi_i}
   +\boldsymbol{\lambda}^{\mathsf{T}}\frac{\partial\mathbf{R}}{\partial\varphi_i},
  \label{eq:shadowgrad}
\end{equation}
where $\boldsymbol{\lambda}$ is the adjoint already computed for the signed distance field. This gradient with respect to $\varphi$ is taken as the gradient with respect to $\bar{\varphi}$ in Eq.~\eqref{eq:hiddenphi}.

\paragraph{Void nucleation.} The explicit term of Eq.~\eqref{eq:shadowgrad}, $\partial\mathcal{J}/\partial\varphi_i$, is non-negative, because at fixed displacement more material stores more energy and adds volume. Without the adjoint term, it could only lower $\bar{\varphi}$. The adjoint term, $\boldsymbol{\lambda}^{\mathsf{T}}\partial\mathbf{R}/\partial\varphi_i$, which carries the structural response, can have either sign, so the signal can also keep material where the structure needs it. A void is admitted in an element when all eight of its nodes lie in the deep solid, $s<-3\varepsilon$ (so that the void does not re-cut the existing boundary, which the signed distance field update already moves), and at least one of them carries $\bar{\varphi}<\tfrac12$. In addition, a candidate void thinner than the phase-field profile can resolve is screened out: for each node in the deep solid with $\bar{\varphi}<\tfrac12$, the extent of the region $\bar{\varphi}<\tfrac12$ is measured along each coordinate axis over the node's adjacent element edges, and if the smallest of the three extents is below $2\varepsilon$, the node's $\bar{\varphi}$ is reset just above $\tfrac12$. When this screen would leave no element to admit, its resets are reverted, so that the values below $\tfrac12$ keep accumulating, and no void is admitted in that iteration.

On an iteration that admits a void (at least one element passes the thin-void screen), the hidden phase field is interpolated trilinearly onto the background grid nodes inside the admitted elements and is then mapped back to a distance by inverting Eq.~\eqref{eq:phasefield}, $\bar{s}=\varepsilon\operatorname{atanh}(1-2\bar{\varphi})$, and $\bar{s}$ overwrites the signed distance field $s$ at the octree nodes and at those grid nodes in step (i) of the geometry rebuild (Sec.~\ref{sec:si-shapeopt}). Signed distance reinitialization then proceeds as in steps (ii)--(iii) of the geometry rebuild: marching cubes extracts the reconstructed surface, which now carries both the cavity from the $\bar{\varphi}$ update and the boundary moved by the $s$ update, the generalized winding number classifies the region inside the cavity as exterior, and the signed distance is recomputed. Thereafter, the cavity is an ordinary part of the surface and evolves with the signed distance field under the shape optimization of Sec.~\ref{sec:si-shapeopt}. 

Because the octree is coarse in the deep solid before the void nucleates, the nodes there at which $\bar{\varphi}$ has fallen below $1/2$ are recorded, whether or not a void was admitted there, and at every later rebuild the octree is refined around each recorded location, so that the $\bar{\varphi}$ field is tracked there directly. After a void is admitted, the hidden phase field is reset at every node from the reinitialized signed distance field. On iterations that admit no void, $s$ rather than $\bar{s}$ is written onto the grid, so that decreases of $\bar{\varphi}$ that have not led to admission accumulate over iterations without carving partial voids.

\begin{figure}[tb]
  \centering
  \includegraphics[width=0.6\linewidth]{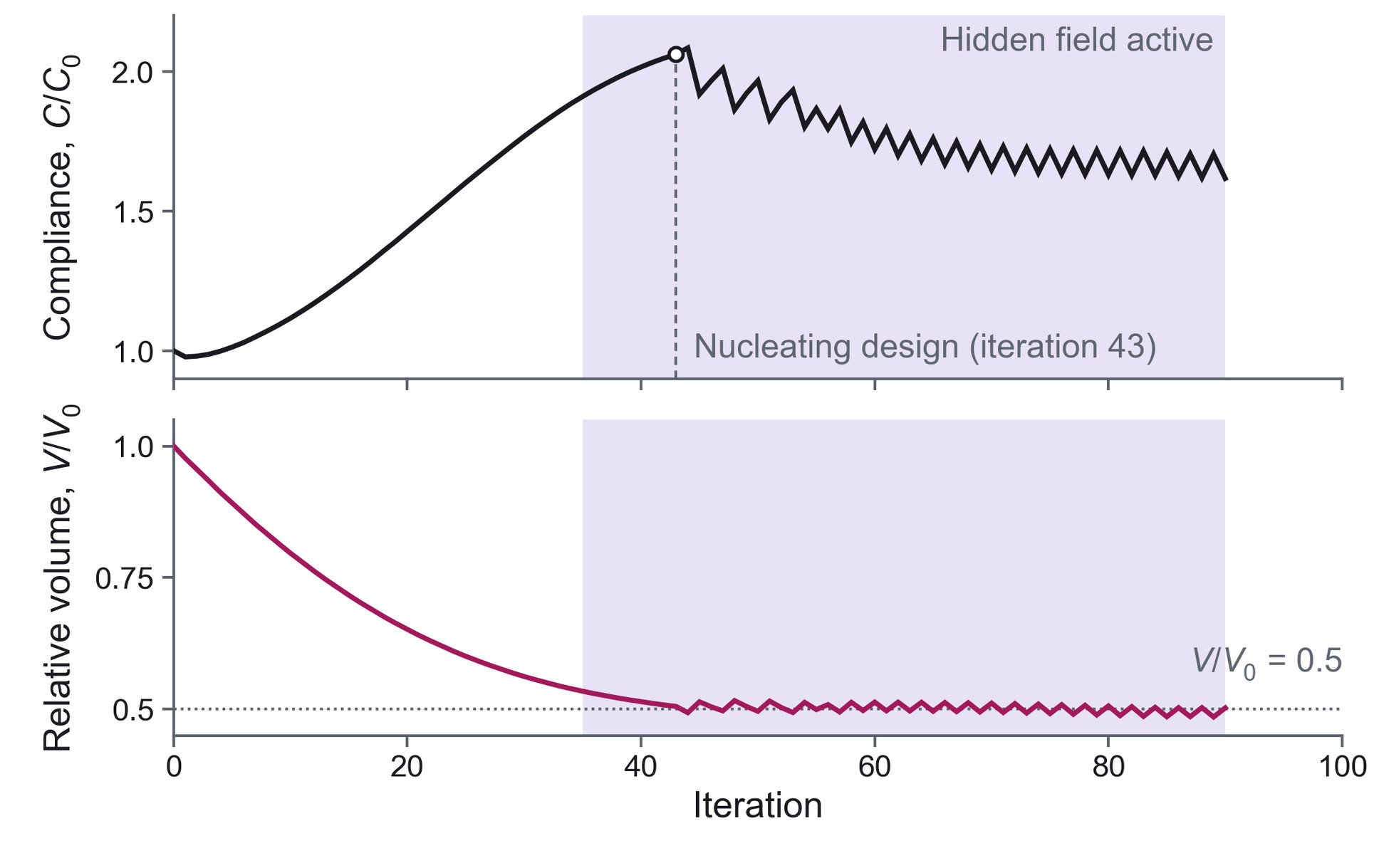}
  \caption{\textbf{Strain energy and volume for the hidden-phase-field run of Fig.~\ref{fig:hiddenphi}.} The interior branch of Eq.~\eqref{eq:hiddenphi} is enabled at iteration $35$ (purple line and shading), when boundary motion has brought the volume fraction to $0.53$. The first cavity is admitted at iteration $43$ (dashed), with the volume fraction at $0.51$. The strain energy peaks at $2.08$ times its initial value at iteration $44$ and ends at $1.62$, with the volume fraction at $0.502$. Up to iteration $35$, the curves coincide with those of Fig.~\ref{fig:si-so-objective}.}
  \label{fig:si-hp-objective}
\end{figure}

\paragraph{Behavior of the exemplary run.} The run shown in this section (Fig.~\ref{fig:hiddenphi}) differs from the previous table design problem (Fig.~\ref{fig:si-shapeopt}) in adding the hidden phase field as a second design variable, in a hidden-phase-field step $\eta_\varphi=12$ (the signed distance field keeps the step of $8$ of Sec.~\ref{sec:si-shapeopt}), in $91$ iterations, and in enabling the hidden-phase-field update of Eq.~\eqref{eq:hiddenphi} only from iteration $35$. Other settings are consistent with the run of Fig.~\ref{fig:si-shapeopt}. Up to iteration $35$, the run is therefore the shape optimization of Sec.~\ref{sec:si-shapeopt}. The first cavity is admitted at iteration $43$ (Fig.~\ref{fig:hiddenphi}b). It then grows and merges with the exterior surface, and the underside of the table ends with a more complex topology. Figure~\ref{fig:si-hp-objective} shows the objective over the run.

\subsection{Language-model-assisted workflow}
\label{sec:si-agents}

\paragraph{Session, state machine and approval gates.} A deterministic state machine, which we designed and built with LangGraph~\citeS{wang2024agent}, runs the workflow as a sequence of processing stages, in the order of the preceding sections. Each workflow execution is called a session. At several stages it sends a request to the AI subagent, a multimodal language model. Each request carries no memory of earlier requests, uses zero temperature, and its answer must fit a fixed data format, which the workflow reads. The AI subagent's proposals are applied only after the user (a person or the AI agent operator) accepts or edits them at an approval gate, a stage at which the workflow waits for the user. There are six approval gates: geometry input, physical scale, boundary regions, boundary values, material, and solver and optimizer settings. An optional gate offers painting of a surface without color (see the ``Surface painting'' paragraph of Sec.~\ref{sec:si-geometry}). The user sets the generation and segmentation settings, the octree depths, the solver tolerances, the optimizer settings and the run mode at their gates. 

\paragraph{Descriptions of rendered views.} The segment descriptions used for writing boundary region selector programs (Sec.~\ref{sec:si-bcregions}) come from an AI subagent request, separate from the program writing request, made before any region is proposed. This request receives rendered views of the object, or of its segments, and returns a short description. The rendered views come with a legend that gives each segment's color, position, area fraction, dominant facing and, for CAD faces, surface type. The descriptions are context for the requests that follow, so that those requests know what each segment stands for without examining the rendered views again.

\paragraph{Boundary regions from the user's description.} The boundary-region gate asks the user to describe the supports and the applied-load regions in words. The AI subagent turns this description into the selector programs of Sec.~\ref{sec:si-bcregions}, one per region, based on the segmented source surface. It works from a compact summary: the overall bounding box, the segmentation method and its parameters, the descriptions of rendered views above and each segment's identifier, area fraction, bounds, centroid and the fractions of its facets whose normals lie within $45^\circ$ of each of the six axis directions. Its instructions tell it to propose only the regions and values that the user's description states. The user then previews the resulting regions and approves them.

\paragraph{Boundary-condition values.} At the boundary-condition value gate, the AI subagent proposes the boundary-condition values (Sec.~\ref{sec:si-bcregions}) from the user's description, and this request also receives the user's description of the supports and loads from the previous boundary-region gate. Load directions are proposed in the frame of the source surface, with $+y$ as up and $+z$ as the front of the object. A load is returned either as a traction or as a total force, and gravity as an acceleration, separately from a body force per unit volume. A total force is divided by the region's loaded area $A^{i}$ on the scaled source surface (see the ``Load correction'' paragraph of Sec.~\ref{sec:si-nitsche}).

\paragraph{Physical scale and material values.} At the physical-scale gate and the material gate, the AI subagent likewise proposes, from a description that the user types there, the length used in physical scaling (Sec.~\ref{sec:si-geometry}) and the material constants (Sec.~\ref{sec:si-nitsche}). Elastic constants are chosen against the reference table of that section and converted into the session's units. At the material gate, it can instead propose one material per part from renderings of the parts (Sec.~\ref{sec:si-nitsche}).

\paragraph{Revision of user text inputs.} For TripoSG, Text2CAD and CAD-Coder, whose input is the user's text, the user can have an AI subagent revise a draft into the form that the generator follows best (see the ``Text-to-CAD generation'' paragraph of Sec.~\ref{sec:si-geometry}).

\paragraph{Cost of AI subagents.} A typical session uses 10--50 thousand uncached input tokens and 1--2 thousand output tokens of DeepSeek-V4.1-Flash with thinking mode off, depending on the number of retries and on the complexity of the segments and boundary condition regions.

\paragraph{Interface and checkpoints.} We implement two interfaces for the workflow: a command line interface for the AI agent operator and a web interface (Streamlit) for human users. Both drive the state machine, show its approval gates (as forms in the web interface) and run background tasks. The web interface also keeps a persistent view of the current geometry, the segmented boundary regions and the solution fields. A checkpoint is written at every stage. Therefore, a session can be resumed, or returned to any earlier gate and re-decided, which discards the steps that followed.

\begin{table}[tb]
  \centering
  \footnotesize
  \caption{Default configuration. The generation block is for the default generator (TripoSG). The optimization block applies only when that mode is selected.}
  \label{tab:defaults}
  \smallskip
  \begin{tabular}{@{}lll@{}}
    \toprule
    & Quantity & Default \\
    \midrule
    Generation & inference steps & $16$ (text, scribble), $50$ (image) \\
               & guidance scale & $7.0$ (image), none (guidance-distilled \\
               & & text and scribble model) \\
               & prompt / scribble conditioning weight & $1.0$ / $0.4$ (scribble), \\
               & & $1.0$ / $0$ (text) \\
               & random seed & $42$ \\
               & image drawn from text (optional) & seed $0$, drawn by SDXL at $1024^{2}$, \\
               & & $30$ steps, guidance $7.0$ (fixed) \\
    \midrule
    Segmentation & method & CAD faces (text-to-CAD surface), otherwise \\
               & & PartField (shape diameter on request) \\
               & clusters & $5$ \\
               & shape diameter: rays / cone / & $25$ / $120^{\circ}$ / $0.25$ \\
               & smoothing weight & \\
               & match tolerance $\delta$ & $0.75\,h_{\min}$ \\
    \midrule
    Discretization & octree depth & $4\rightarrow7$ \\
               & phase-field width $\varepsilon$ & $h_{\min}/2$ \\
               & element / quadrature & trilinear hex, $2\!\times\!2\!\times\!2$ Gauss \\
               & & (all elements, band included) \\
               & precision & double (solve), half (generator network, \\
               & & octree-extraction sampling) \\
    \midrule
    Material & $E$, $\nu$ & user-entered or approved \\
               & (form default) & $10^{6}$ in the selected units, $0.3$ \\
               & $\rho$ (uniform) & unset, required only with an acceleration \\
               & spatial field & VoMP or per part, forward solves only \\
    \midrule
    Solve & Newton tolerance, rel.\ / abs. & $10^{-2}$ / $10^{-5}$ (force unit) \\
               & Newton iteration cap & $20$ \\
               & load steps / line search & $1$ / disabled \\
               & linear solver & AMG-preconditioned CG, $5\times10^{-6}$, \\
               & & cap $1000$ iterations \\
               & AMG interface & JAX-AMG (default) or \texttt{pyamgx} \\
               & penalty safety factor $s_\beta$ & $5$ \\
               & correction of total force ($A^{i}/W^{i}$) & on \\
    \midrule
    Optimization & optimizer / learning rate & gradient descent / $0.01$ \\
               & automatic learning rate & off, step $0.5\,\varepsilon$ when on \\
               & iterations / stopping tolerance & $101$ / $10^{-9}$ \\
               & $w_{\mathrm{vol}}$ / target fraction & $5$ / $0.5$ \\
               & boundary-patch dilation & $2$ facet rings \\
               & reinitialization & every iteration, with \\
               & & boundary-region preservation \\
               & boundary-region preservation & gradient mask, patch restoration \\
               & & optional (Fig.~\ref{fig:bczip}) \\
    \bottomrule
  \end{tabular}
\end{table}

\subsection{Operation by a coding agent}
\label{sec:si-operator}

\paragraph{AI agent operator and operator skill.} The workflow of the preceding sections waits at each approval gate until the user answers it, and the AI subagent proposes only the regions and values that the user's description states (see the ``Boundary regions from the user's description'' paragraph of Sec.~\ref{sec:si-agents}). Therefore, to run the workflow from only the name of an object, a description, an image or a surface mesh, an AI agent operator poses a loading case that is meaningful for the real object, chooses the numerical settings and looks at each rendered picture before it approves the step that made it. The operator is a coding agent, Claude Code in this work, which runs a language model with shell access to execute commands. It is separate from the AI subagent of Sec.~\ref{sec:si-agents} and runs the whole workflow end to end through the command line interface instead of the web interface (Sec.~\ref{sec:si-agents}). It works from an \emph{operator skill} that we developed, a set of written instructions, reference material and scripts that frame its role as that of an engineer in computer-aided engineering. The operator drives the same state machine and answers its six approval gates (Sec.~\ref{sec:si-agents}), and the optional painting gate, in place of the user. Its answers enter the state machine as data and go through the same validation as a person's answers would, and it reports the result with the evidence for it.

\begin{figure}[tb]
  \centering
  \includegraphics[width=\linewidth]{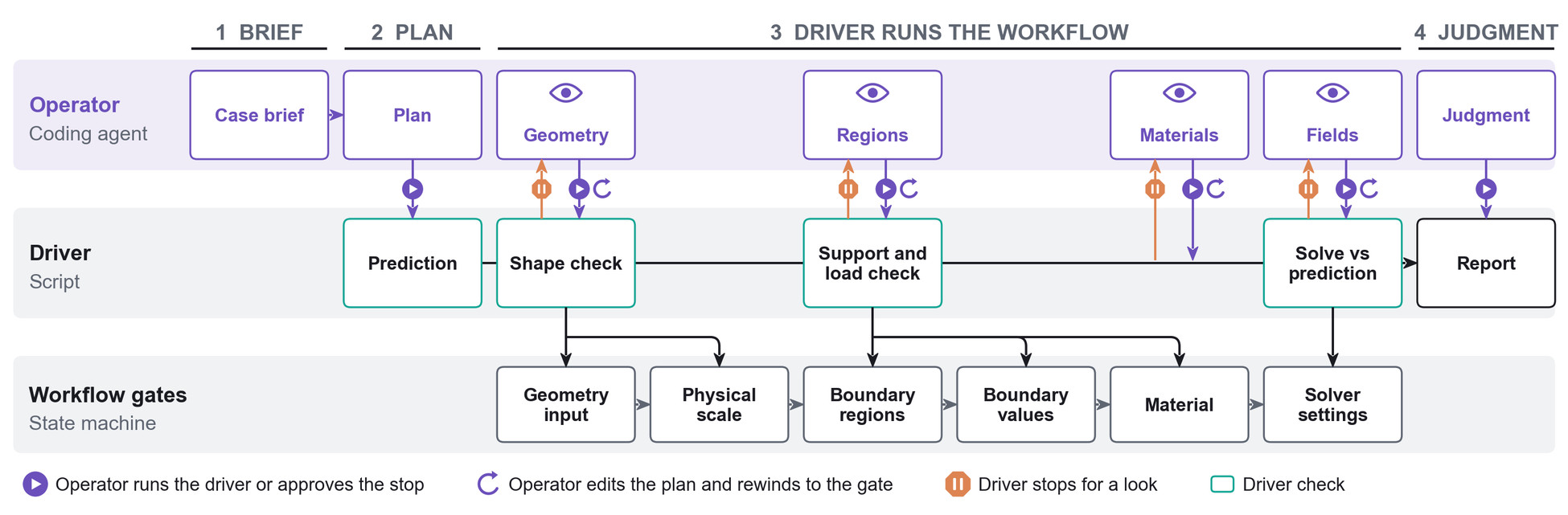}
  \caption{\textbf{Operator loop.} The operator (top row) writes the case brief and the plan, runs the driver, and writes the final judgment. The driver (middle row) predicts the response from the brief, checks the shape, the supports and loads, and the solve against the prediction, and sends the plan's answers to the workflow gates (bottom row). At each stop, the operator looks at the rendered pictures and either approves the stop by running the driver again or edits the plan, which rewinds the session to that stop's gate.}
  \label{fig:si-operator-loop}
\end{figure}

\paragraph{Four-step procedure.} The operator works in four steps (Fig.~\ref{fig:si-operator-loop}). First, before anything is generated or solved, it writes a \emph{case brief}, which states the loading case and the reasoning behind it. Second, it transcribes the brief into a \emph{plan}, a file with one answer per gate. Third, it runs the \emph{driver}, a script we developed that checks the plan's answers and sends them to the state machine. The driver pauses at a \emph{stop} wherever a picture must be looked at: a drawing, a photograph or a painted surface when one is made, the geometry, the boundary regions, the per-part materials when they are used, and the solution fields. At each stop, the driver names the pictures for the operator to look at and the sections of the reference material that apply to the stop. Two branches follow: running the driver again approves the stop, and editing the plan instead rewinds the session to that stop's gate and resends from there (see the ``Interface and checkpoints'' paragraph of Sec.~\ref{sec:si-agents}). Finally, the operator writes a judgment of the result, which the driver adds to the session's report.

\paragraph{Case brief.} The case brief poses the loading case that a user would otherwise state at the gates. How much the operator decides depends on what the instruction gives. From a name only, it decides how the object is used, where it is held, the loads and their sizes, the object's size, its material and the question the analysis answers. From a more detailed instruction, it keeps what was given and decides the rest, which for a full case is only the conversion of units. It reasons from how the real object is used, and unless the user asks otherwise, the question is the deflection and the pattern of stress. A case beyond static elasticity, for example one with contact between bodies, a moving load or plastic yielding, is run as a simplified static elastic case, and the report says what was simplified. Each value in the brief is marked as given by the user or decided by the operator.

The brief has four parts. \emph{Use} is one sentence on who uses the object and how, followed by a \emph{use table} with one row per action on the object in each state of its use that loads it (in service, in handling, in assembly and in storage). Each row gives the action's type (an applied force, a weight, or an action of the surroundings), where its size comes from, its direction, the member (a structural part of the object) it enters and the supports that hold the object meanwhile. One of the states is named as the one to solve. \emph{Contacts} lists each support, where it is, which displacement components it fixes and why, and its type. A grip, a socket or a fastening can transmit a moment as well as a force, whereas a rest or a hang transmits a force only. Each support is normally imposed as a prescribed displacement (see the ``Supported boundary conditions and loads'' paragraph of Sec.~\ref{sec:si-bcregions}). \emph{Load} gives, for each load, the face through which it enters, its direction, and its size, which is derived from the mechanism that sets it, either a mass times the gravitational acceleration, a pressure times an area, or a quoted clause of a design code. A force that a person applies, when nothing in the instruction sets it, is taken from a table of 11 human actions in the reference material. For example, a whole-hand grip is 250--500\,N, and the driver refuses a size outside the row's range. \emph{Size and material} gives the real object's mass and extents, its thinnest member, the cross-section of the member the solve loads most, the members each load passes through on its way to the supports (the \emph{load path}), the Young's modulus of the governing material, and the material of each visibly distinct part.

\paragraph{Prediction from the brief.} The driver fetches the brief's structured lines and computes the theoretical predictions from them, so the operator does no hand calculation. From the use table, the driver computes the moment of each action, the force times its lever arm, at each member of the load path between where the action enters and the supports, and the reactions at the supports. At the most heavily loaded cross-section, it then estimates (i) the stress as $M(d/2)/I+F/A_{\mathrm{s}}$, from the bending moment $M$, the axial force $F$, the section's second moment of area $I$, its area $A_{\mathrm{s}}$ and its depth $d$, and (ii) the deflection as $FL^{3}/(3EI)$ and the axial shortening for an action that acts mainly along the member, summed over the actions and treating the member as a cantilever of length $L$ from its nearest support. These two estimates, which are upper bounds for a member held at both ends, form the \emph{prediction} with which the solve is later compared.

\paragraph{Driver checks.} A brief can propose a use case that is not of interest or that the solver cannot accurately or robustly solve. Therefore, the driver also checks the case at four points. First, we prefer the use state where members are loaded as heavily as possible to reflect the critical loading conditions. Therefore, before anything is generated, when a use state loads some member more than one third more heavily than the one selected by the operator, the driver requires that state to be solved instead, unless the heavier state's loading is one the current solver cannot apply, such as a pressure inside a closed cavity. 

Second, a proposed support should not constrain more than its real contact does, and the proposed loads should not leave the object unbalanced on its proposed supports. Therefore, at the approval of the boundary regions, the driver checks that each proposed support holds only what its specified type allows. For example, a grip-type support holds along at least half the length it grips, and a rest-type support that needs restraint against sliding receives the fewest extra fixed components that stop the sliding as a simplification for friction. 

Third, the driver's prediction assumes small deflections, and an everyday object neither deflects by a large fraction of its length nor works near the strength of its material. A case past either limit usually means that the load or the size in the brief is wrong. Therefore, the driver points out such a case, for example a predicted deflection over $30\%$ of the object's largest extent or a predicted stress at $90\%$ of the material strength. The operator either corrects the load, the size or the section, or keeps the case. No case is refused for its deflection or stress. 

Fourth, a generated shape can differ from the real object. Therefore, the driver checks the shape against the brief: the largest extent, and the smallest when a wall or the loaded section sets it, must match the real ones within a factor of 1.5, and a body whose load-bearing wall is more than three times the real thickness, or whose mass is more than five times the real mass, is regarded as another structure. Among these rules, the mass rule refuses a generated body while another way of generating the shape remains to be tried, and otherwise the verdict is that the brief's case could not be run. A member thinner than two of the smallest octree cells ($2h_{\min}$) is flagged as too stiff in bending at that resolution, and the operator may keep the body or refine the octree by one level. 

\paragraph{Operator's visual checks.} The driver's checks work on numbers, whereas whether a picture shows the intended object and regions is judged by the operator. At each stop, the operator reads the rendering, which shows six views of the body, labeled by the direction they are seen from. At the geometry stop, it checks (i) the body is recognizably the object asked for, (ii) the member the load enters and the member the use state loads most are present and attached, (iii) the body is one piece standing as in the use state, (iv) members meant to be separate are not fused, and (v) its size and thinnest member are those of the brief. It rejects a body that is the wrong object, several objects or only part of one object, or whose primary member is missing (a missing secondary member is accepted and declared absent). In the rendering of the boundary regions, it checks (i) each support lies on its contact region and each load on the face the brief names, (ii) each load region's loaded area $A^{i}$ (see the ``Load correction'' paragraph of Sec.~\ref{sec:si-nitsche}) is at least $80\%$ of its facet area, so that the region faces mostly one way, (iii) no load region is smaller than 50 facets or three cells across, and (iv) no region covers more than $40\%$ of the entire surface without reason. A failing region is revised first by a new segmentation (Sec.~\ref{sec:si-bcregions}) and then by a new description for writing the selector programs. At the solve, it checks (i) the supports do not move, (ii) the deformed shape and the peak stress are where the brief predicted, and (iii) the fields vary smoothly.

\paragraph{Shape generation routes.} A generator can return a shape that fails the driver checks or the operator's visual checks, so the operator has several \emph{routes}, i.e., ways of obtaining the shape. From the type of object, the brief lists the routes that can make its shape, out of those of Sec.~\ref{sec:si-geometry}, and the driver restricts the generation routes to a fixed order: TripoSG from text, an SDXL drawing from the description as the image input of TripoSG, Hunyuan3D-2.1 from that same drawing, and CAD-Coder for joined prismatic parts. Bodies from TripoSG and CAD-Coder count against a budget of two generated bodies. The description is at most about 20 words, naming the material, the object and two or three defining features. If no generated body is accepted, the driver searches public collections for a public-domain or CC~BY photograph of the object. The operator judges the candidates and either picks one for shape generation, or rejects them all with a reason. As a last resort, the operator builds the body from dimensions with the tool for simple CAD models (see the ``Text-to-CAD generation'' paragraph of Sec.~\ref{sec:si-geometry}): a box, a cylinder, a profile extruded along one axis with through-holes, a frame of bars and plates joined into one watertight body, or a body of revolution. However, building comes first when the brief gives one of two reasons: a real member thinner than $1/50$ of the largest extent (a sheet, shell or lattice, which the generators may fill solid), or a load path through a single member whose real cross-section was stated by the brief. 

\paragraph{Attempt limits.} The driver counts each attempt at each stage, and refuses one past its limit (Table~\ref{tab:operator-limits}). The operator's instructions add that each retry makes one change (the words or the seed, the regions or the segmentation, or one setting). At a limit, or at an error that no change of plan can fix, the operator writes a failure report and stops. The report gives the stage, each attempt and what it showed, the last error and what a human user could try next.

\begin{table}[tb]
  \centering
  \footnotesize
  \caption{Attempt limits per operator session, first attempts included.}
  \label{tab:operator-limits}
  \smallskip
  \begin{tabular}{@{}ll@{}}
    \toprule
    Stage & Limit \\
    \midrule
    TripoSG generation from text & $1$, or $2$ when no watertight surface \\
                                 & was returned \\
    SDXL drawings & $3$ \\
    Hunyuan3D-2.1 generations & $2$ \\
    Photograph searches & $1$, or $2$ when the first found nothing \\
    CAD-Coder generations & $3$, within the two generated bodies \\
    Built or uploaded bodies & $3$ \\
    Region previews / region revisions & $4$ / $4$ \\
    Re-segmentations & $3$ \\
    Per-part material valuations & $3$ \\
    Solves & $5$ \\
    \bottomrule
  \end{tabular}
\end{table}

\paragraph{Report and comparison with the prediction.} An inspection script collects the evidence of a run into a report: the geometry and resolution, the boundary regions and their applied loads, the Newton convergence of the diffuse Nitsche solve, the peak displacement and strain against plausibility thresholds, and the peak stress, compared with a yield strength when one is supplied. The driver states the agreement between the solve and its prediction as the ratio of the solved to the predicted value, for the peak displacement and for the peak surface stress away from the boundary regions. A ratio of at most 2 counts as agreement, with no lower limit, because the driver's estimate is an upper bound. A solve two to five times the estimate is a disagreement, which the operator must explain from the units, the size or face of a load, a contact or the resolution. A larger ratio marks a wrong case, which is corrected and solved again, or reported in the final judgment with the ratio and whether its cause was found.

\paragraph{Final judgment.} The operator's final judgment opens with one of five verdicts: the brief's case was run, a variant of it was run, a simplified case was run, the brief's case could not be run, or the governing load cannot be applied and a secondary case was run. It then lists the deviations from the brief's case (another object, a support on another part or holding other components, a load on another face or in another direction), the actually run case with each decided value, the results with their units, the prediction against the result, what each support holds and how that overstates the real contact, and the caveats. The driver requires a verdict and a list of deviations and accepts ``the brief's case was run'' only when it finds no simplification: the body is the real one, no member is absent, every member is covered by the solved state, no contact is simplified and the materials are unchanged. 

\subsection{Implementation and defaults}
\label{sec:si-implementation}

We developed the workflow, including the differentiable diffuse Nitsche solver, octree extraction, boundary-region selection, sensitivity analysis, shape and topology optimization, the state machine with its interfaces, and the operator skill and driver. The pre-trained models (TripoSG, Hunyuan3D-2.1, Text2CAD, CAD-Coder, SDXL, RMBG-1.4, PartField, VoMP, DeepSeek-V4.1-Flash and Claude Code) and libraries (AmgX, LangGraph and TetGen) are used as released, and methods from earlier work are cited where they are used.

Field evaluation, octree construction, segmentation, assembly, quadrature and the linear algebra run on the GPU (NVIDIA RTX A6000), implemented in PyTorch~\citeS{paszke2019pytorch} and NVIDIA Warp~1.17~\citeS{macklin2022warp}, with the linear solves in AmgX~\citeS{naumov2015amgx} via the Python interface JAX-AMG~\citeS{liu2026jax} or \texttt{pyamgx}. Orchestration and the sequential steps of boundary-region preservation (``Boundary-region preservation'' paragraph of Sec.~\ref{sec:si-shapeopt}) run on the CPU (AMD Ryzen Threadripper 3970X). No network is trained or fine-tuned in this work. Results use the default configuration of Table~\ref{tab:defaults}, apart from a few solver and patch settings in the optimization examples and the refinement settings described in the ``Refinement of generated bodies'' paragraph above.

\clearpage
\newpage

\section{Prompt template used for agentic simulations}
\label{sec:si-prompt}

\noindent \textbf{Simulate based on abstract ideas}

{\itshape
\begin{quote}
You are working in the repository [PATH TO THE CODE REPOSITORY] (run commands from that root). Read the skill at [PATH TO THE SKILL] and follow it exactly; read what the skill and the runner point you to, no more. Do not edit any file in the skill folder or any source code.

The user's whole instruction is: "[NAME OF THE OBJECT]". Always paint the generated shape with the texture step so the body has realistic colours: the result goes into a paper figure.

Carry the task through to the end as the skill says. When you are done (or stopped at a limit), reply briefly with: the session id; the verdict line of operator/judgement.md (do not repeat the judgement: it is read from the file); and, in a few lines, any place where the skill was unclear, where you had to guess, or where you deviated from it.
\end{quote}
}

\noindent \textbf{Generate different shapes}

The prompt for the industrial robot arm example is the same for all agents. The instruction is unchanged, each run sets a different generation seed as an environment variable, and the operator may choose a different generation route. Compared to the prompt with only abstract ideas, the prompt here also specifies boundary conditions. 

{\itshape
\begin{quote}
You are working in the repository [PATH TO THE CODE REPOSITORY] (run commands from that root). Read the skill at [PATH TO THE SKILL] and follow it exactly; read what the skill and the runner point you to, no more. Do not edit any file in the skill folder or any source code.

The user's whole instruction is: "a small six-axis industrial robot arm, bolted to the floor at its base and holding a 5 kg tool at its wrist flange with the arm stretched out". Always paint the generated shape with the texture step so the body has realistic colours: the result goes into a paper figure.

Carry the task through to the end as the skill says. When you are done (or stopped at a limit), reply briefly with: the session id; the verdict line of operator/judgement.md (do not repeat the judgement: it is read from the file); and, in a few lines, any place where the skill was unclear, where you had to guess, or where you deviated from it.
\end{quote}
}

The prompts for writing different materials or boundary conditions for the same simulation, and for running a simulation with each of them, are lengthy and can be found in the released repository.

\newpage

\section{Results on 50 objects in 10 different categories by Opus 5.5 agents}
\label{si:opus-results}

\begin{figure}[hp]
    \centering
    \includegraphics[width=0.7\linewidth]{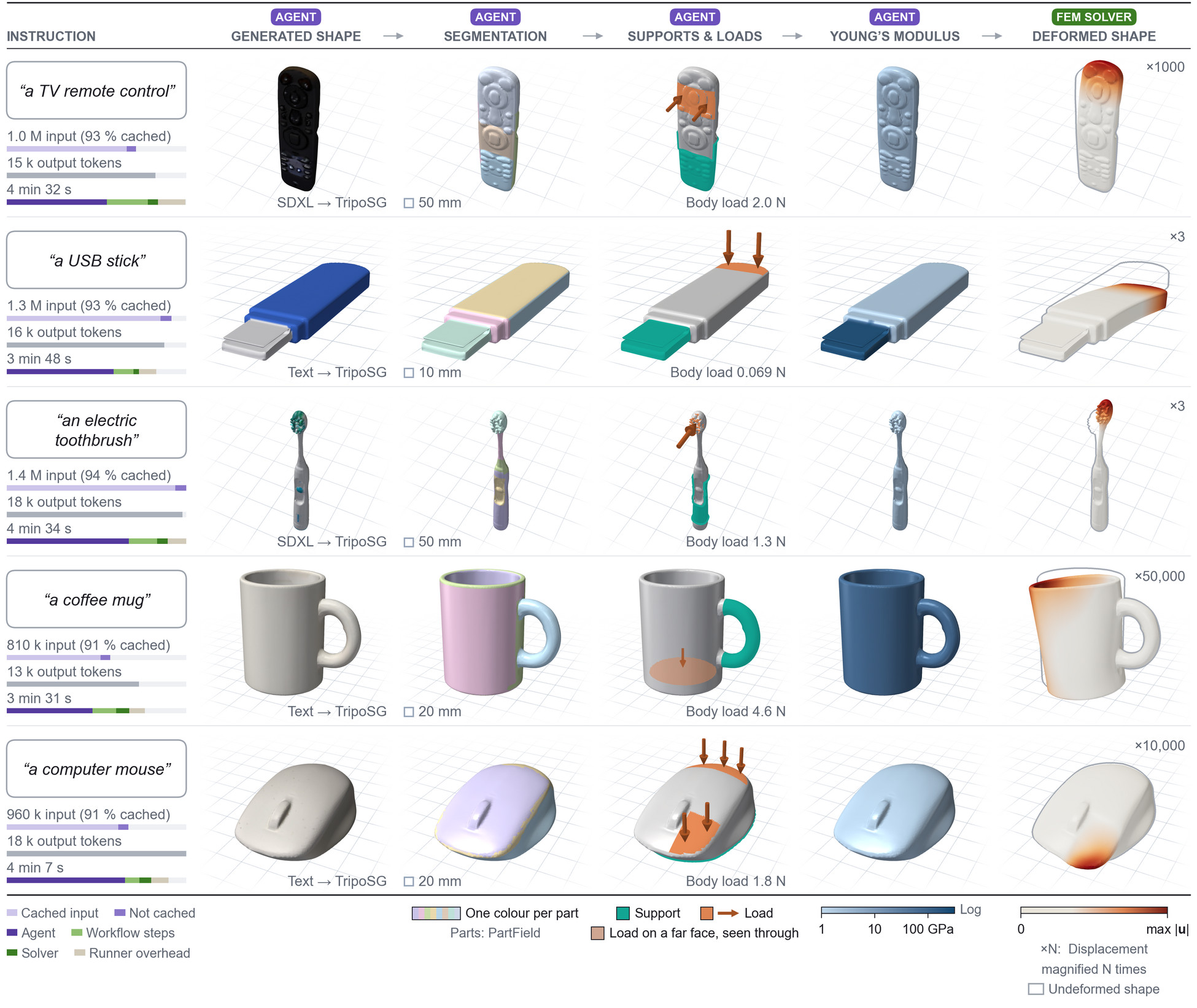}
    \caption{Settings and results on five household items, by Claude Opus 5.5.}
\end{figure}

\begin{figure}[hp]
    \centering
    \includegraphics[width=0.7\linewidth]{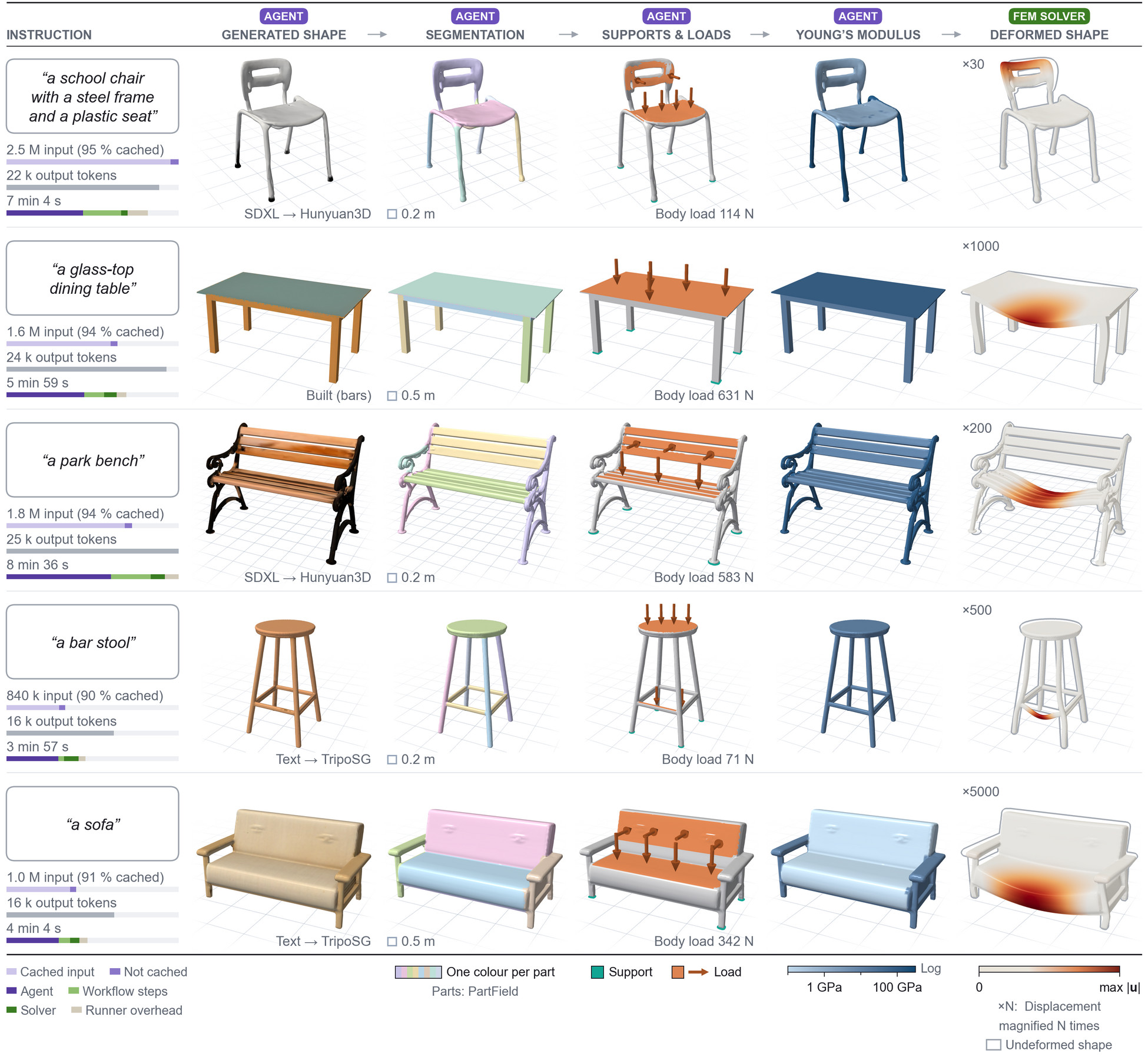}
    \caption{Settings and results on five furniture items, by Claude Opus 5.5.}
\end{figure}

\begin{figure}[hp]
    \centering
    \includegraphics[width=0.7\linewidth]{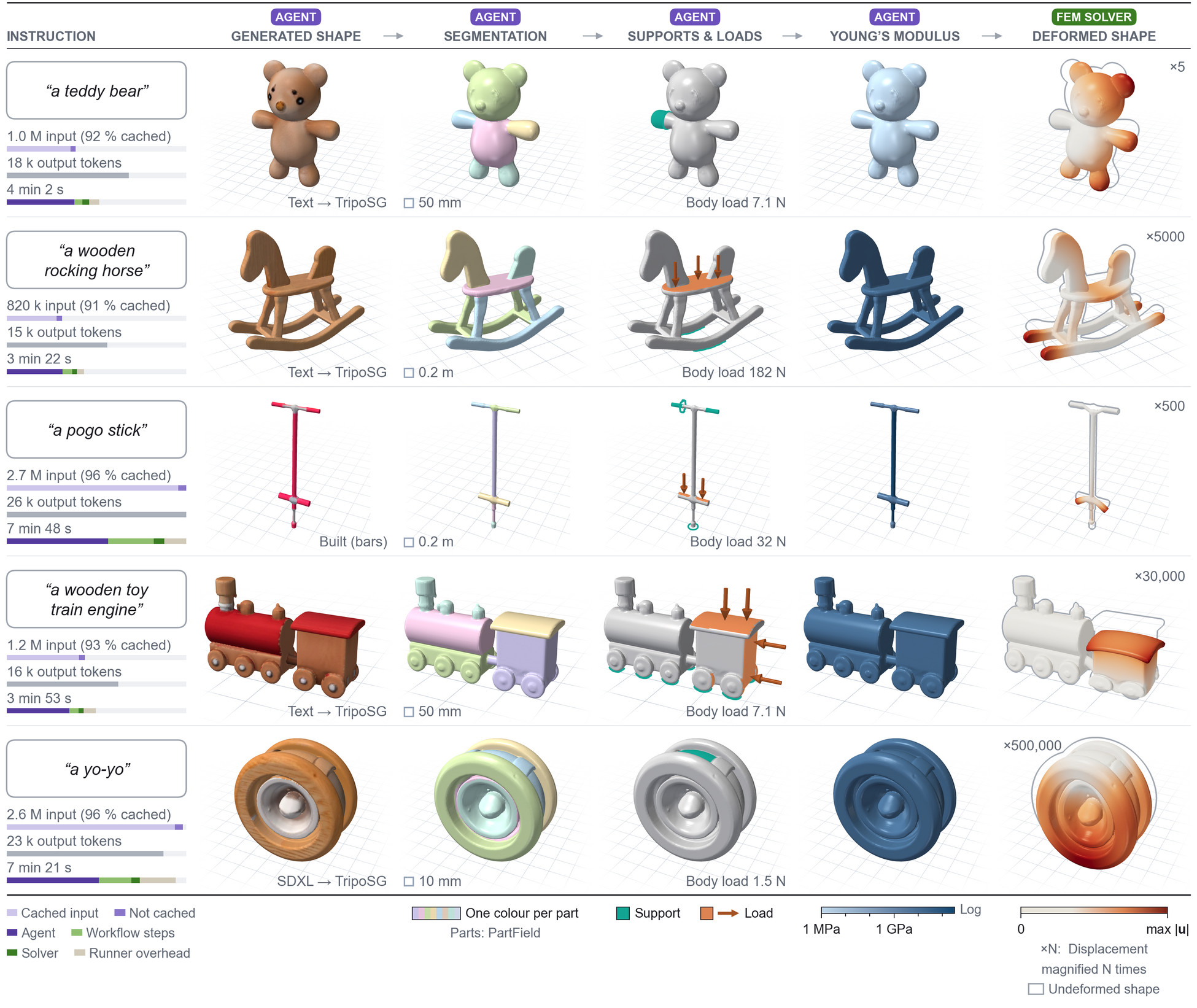}
    \caption{Settings and results on five toys, by Claude Opus 5.5.}
\end{figure}

\begin{figure}[hp]
    \centering
    \includegraphics[width=0.7\linewidth]{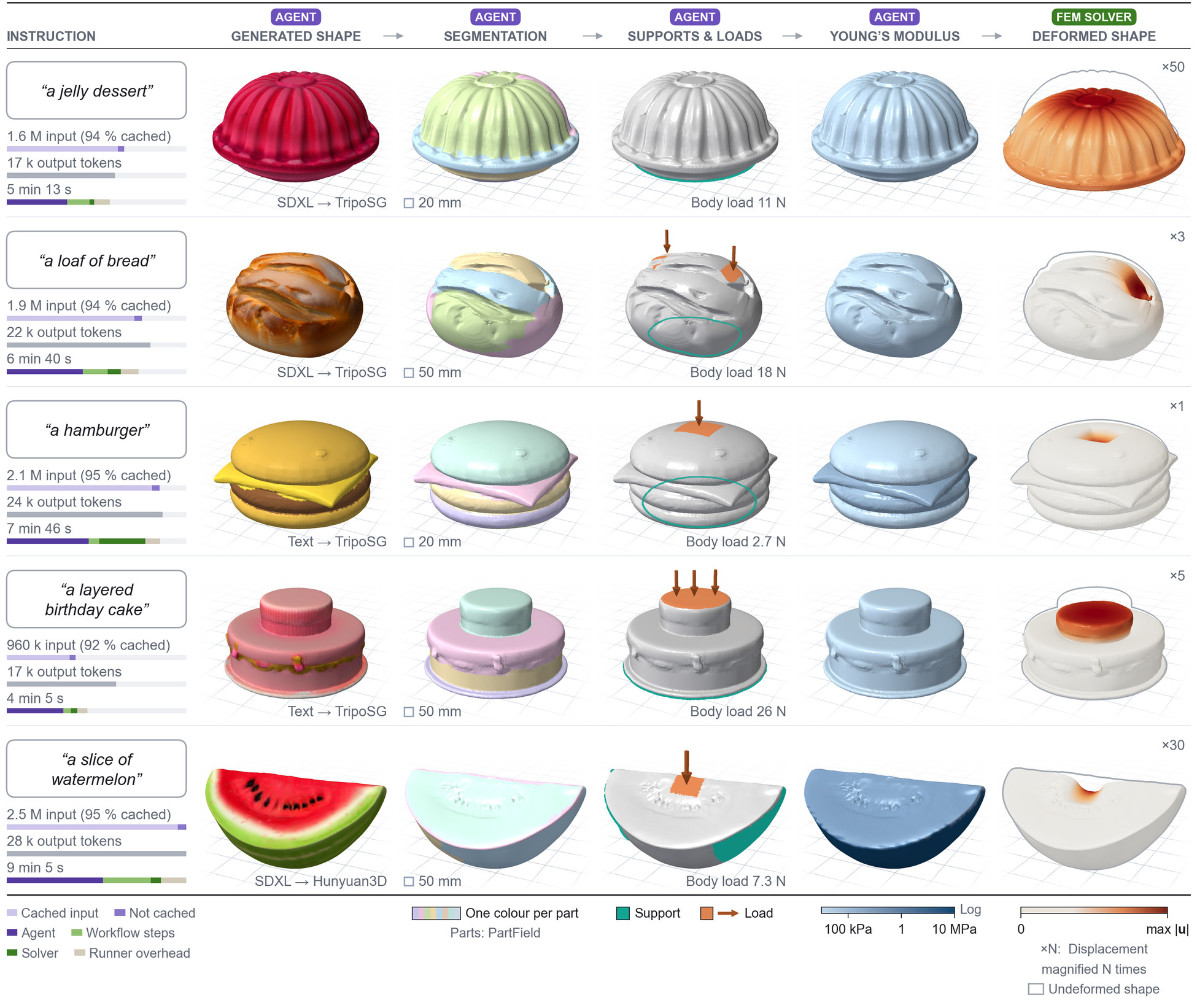}
    \caption{Settings and results on five food items, by Claude Opus 5.5.}
\end{figure}

\begin{figure}[hp]
    \centering
    \includegraphics[width=0.7\linewidth]{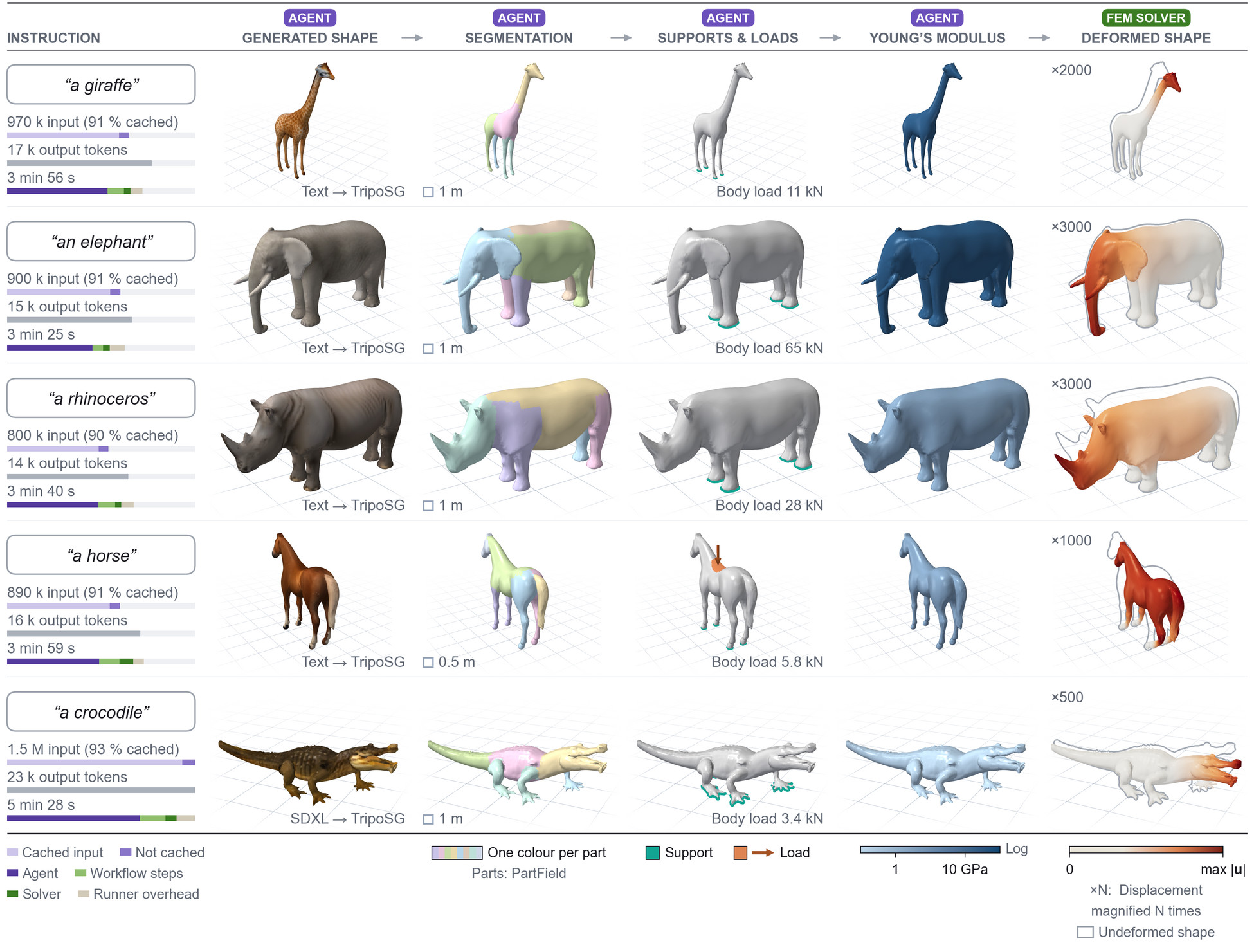}
    \caption{Settings and results on five animals, by Claude Opus 5.5.}
\end{figure}

\begin{figure}[hp]
    \centering
    \includegraphics[width=0.7\linewidth]{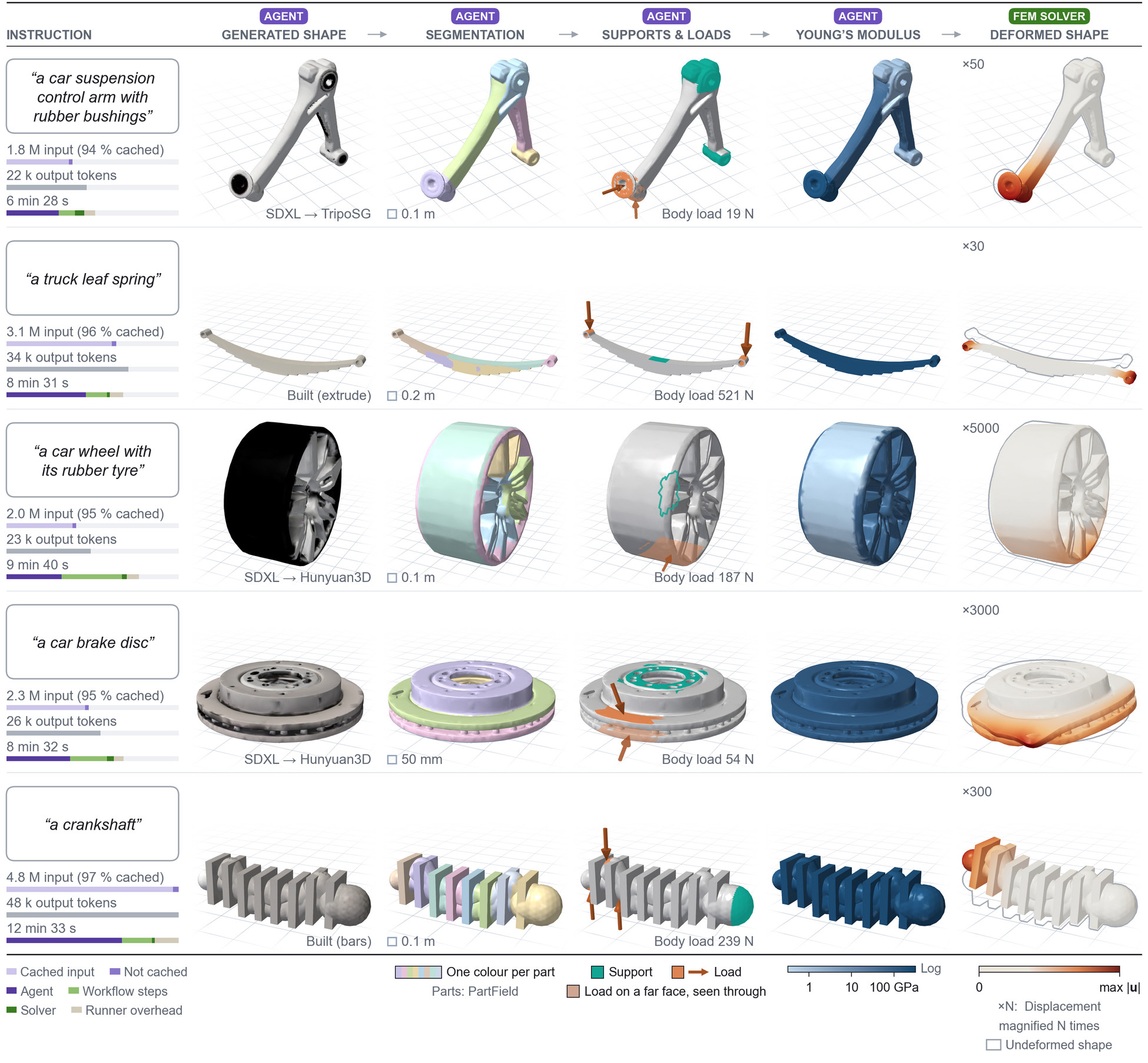}
    \caption{Settings and results on five vehicle parts, by Claude Opus 5.5.}
\end{figure}

\begin{figure}[hp]
    \centering
    \includegraphics[width=0.7\linewidth]{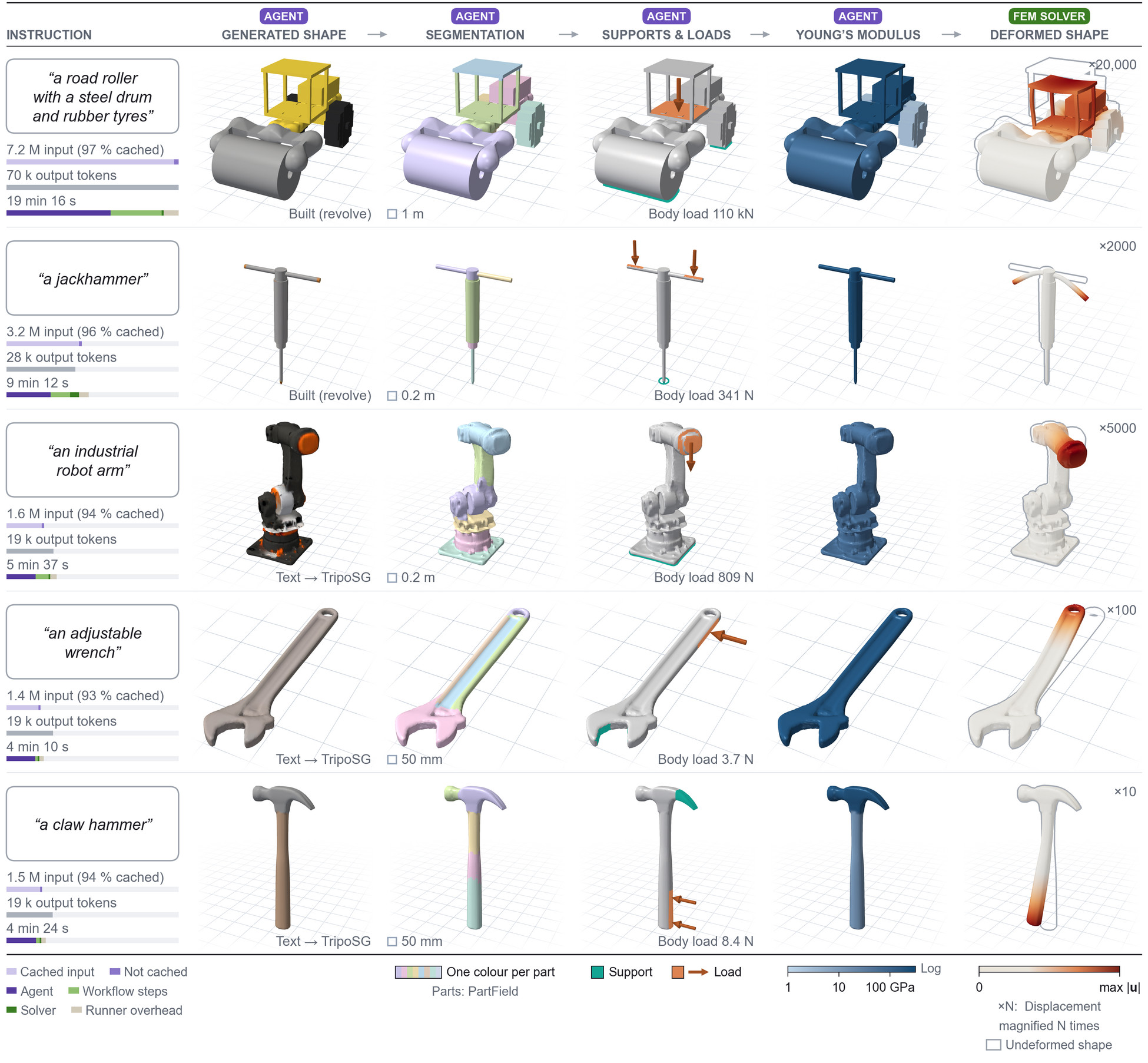}
    \caption{Settings and results on five machines or tools, by Claude Opus 5.5.}
\end{figure}

\begin{figure}[hp]
    \centering
    \includegraphics[width=0.7\linewidth]{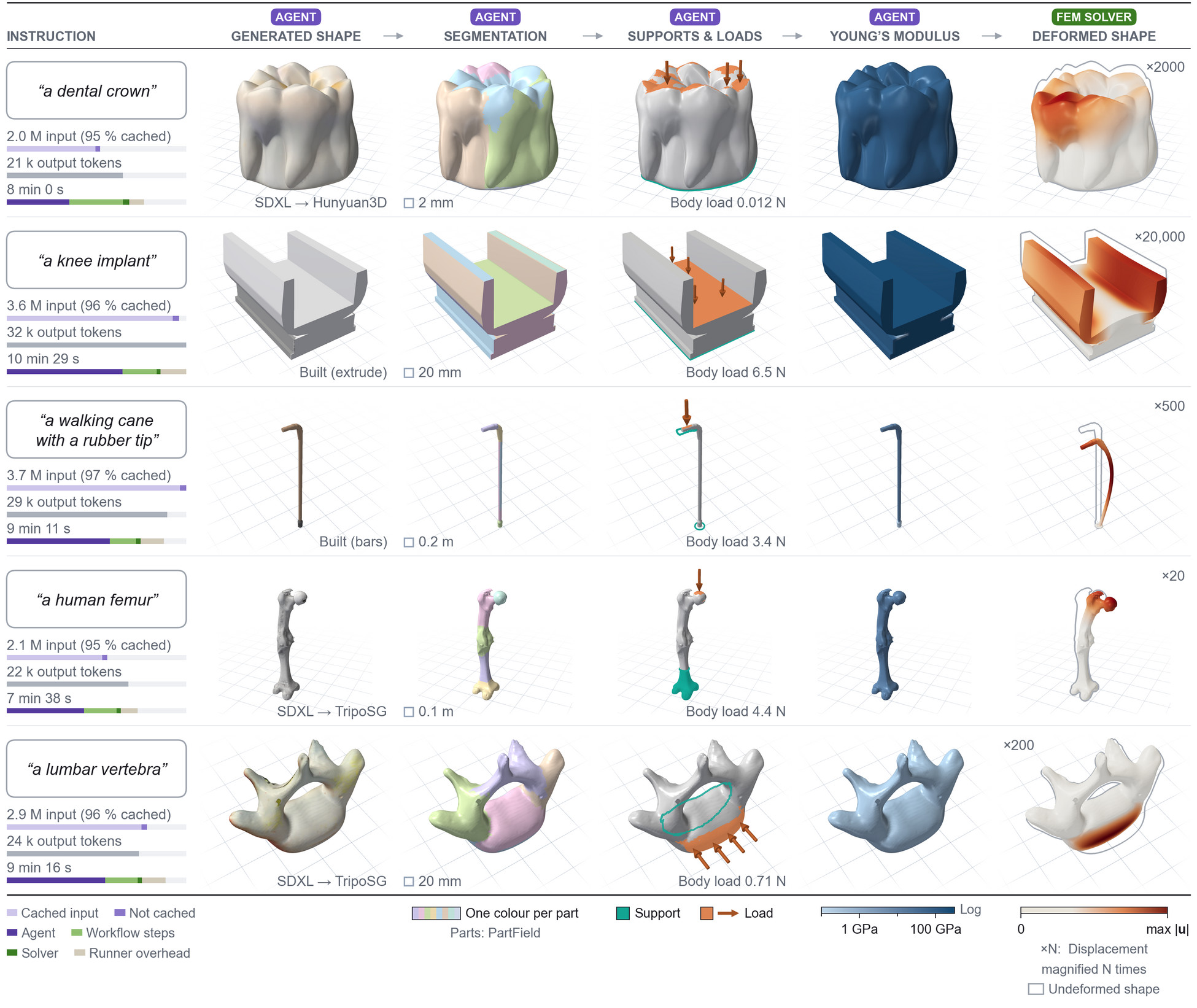}
    \caption{Settings and results on five medical and anatomy items, by Claude Opus 5.5.}
\end{figure}

\begin{figure}[hp]
    \centering
    \includegraphics[width=0.7\linewidth]{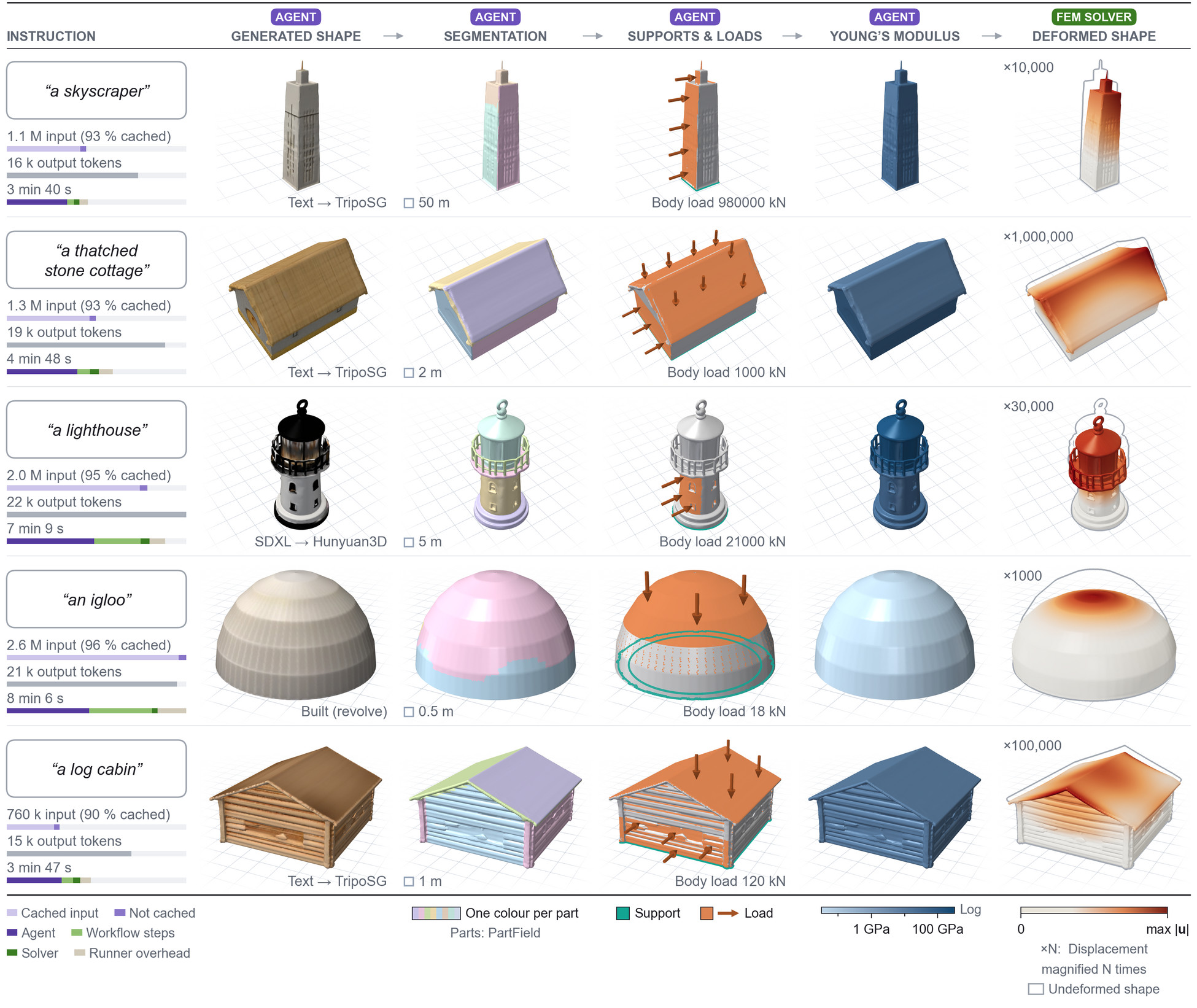}
    \caption{Settings and results on five buildings, by Claude Opus 5.5.}
\end{figure}

\begin{figure}[hp]
    \centering
    \includegraphics[width=0.7\linewidth]{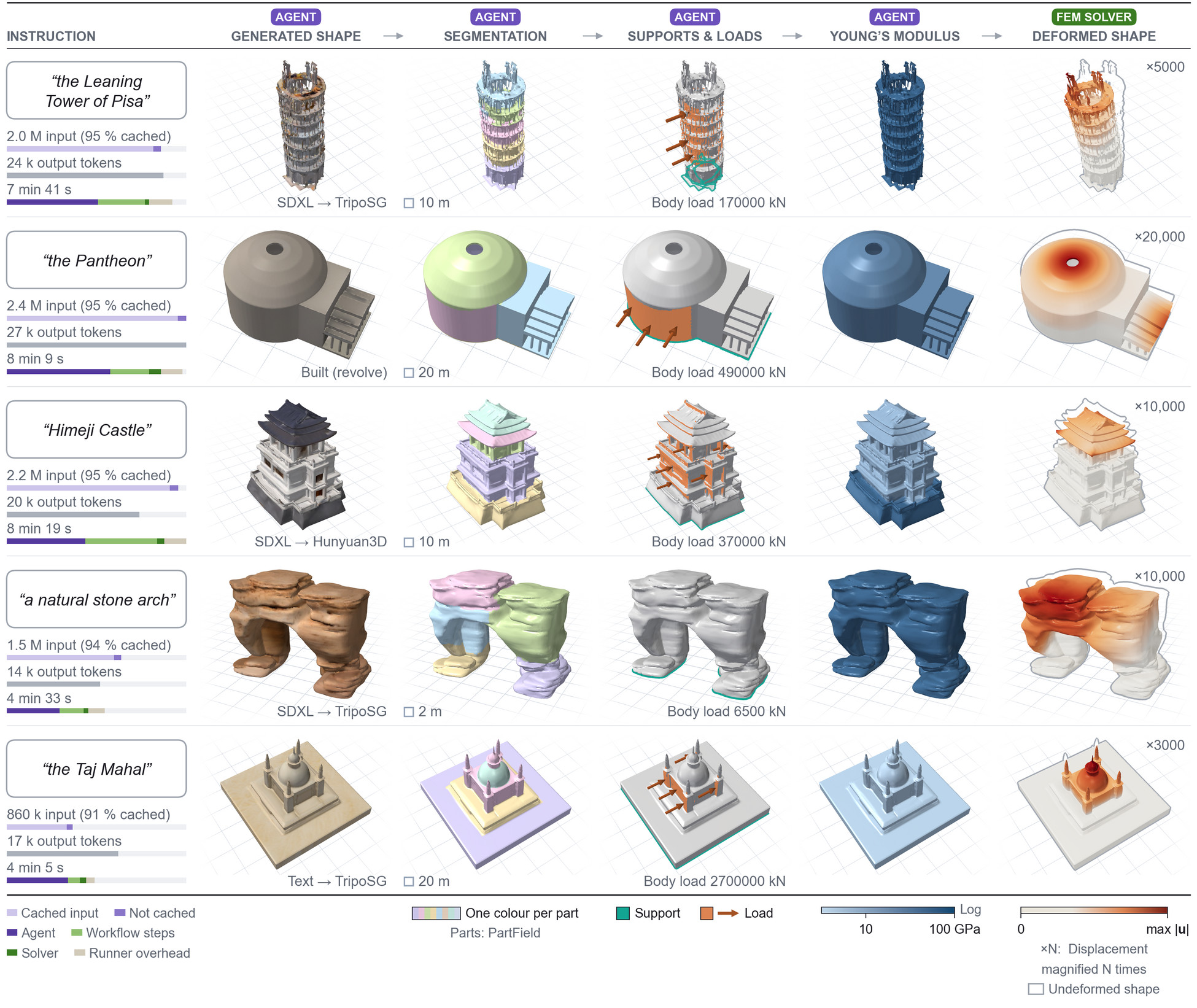}
    \caption{Settings and results on five landmarks, by Claude Opus 5.5.}
\end{figure}

\clearpage

\section{Comparison among Claude Haiku 5.5, Sonnet 5.5, and Opus 5.5}
\label{sec:si-model-compare}

\begin{figure}[hp]
    \centering
    \includegraphics[width=0.82\linewidth]{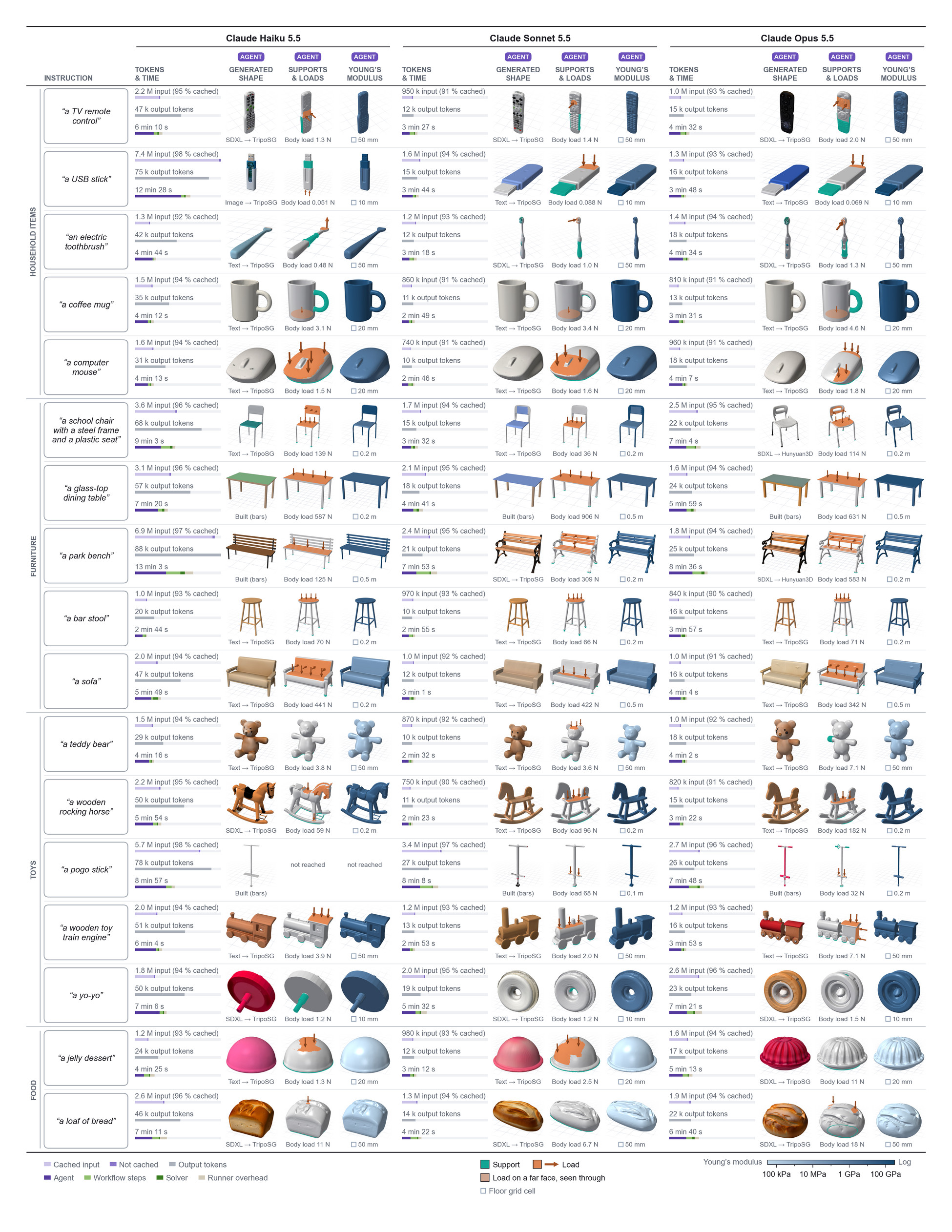}
    \caption{A side-by-side comparison among settings from Claude Haiku, Sonnet, and Opus. }
\end{figure}

\begin{figure}[hp]
    \centering
    \includegraphics[width=0.82\linewidth]{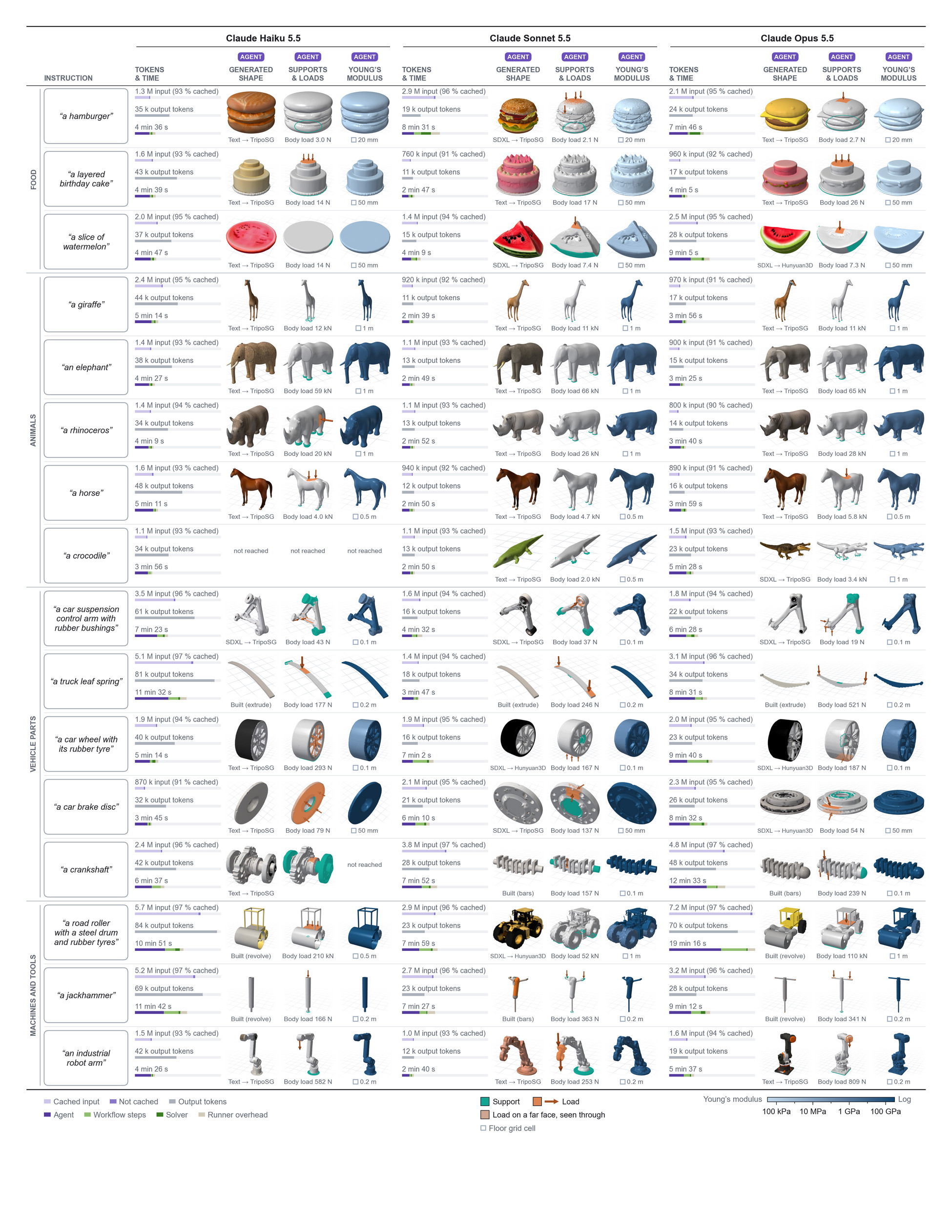}
    \caption{(Continued) A side-by-side comparison among settings from Claude Haiku, Sonnet, and Opus. }
\end{figure}

\begin{figure}[hp]
    \centering
    \includegraphics[width=0.82\linewidth]{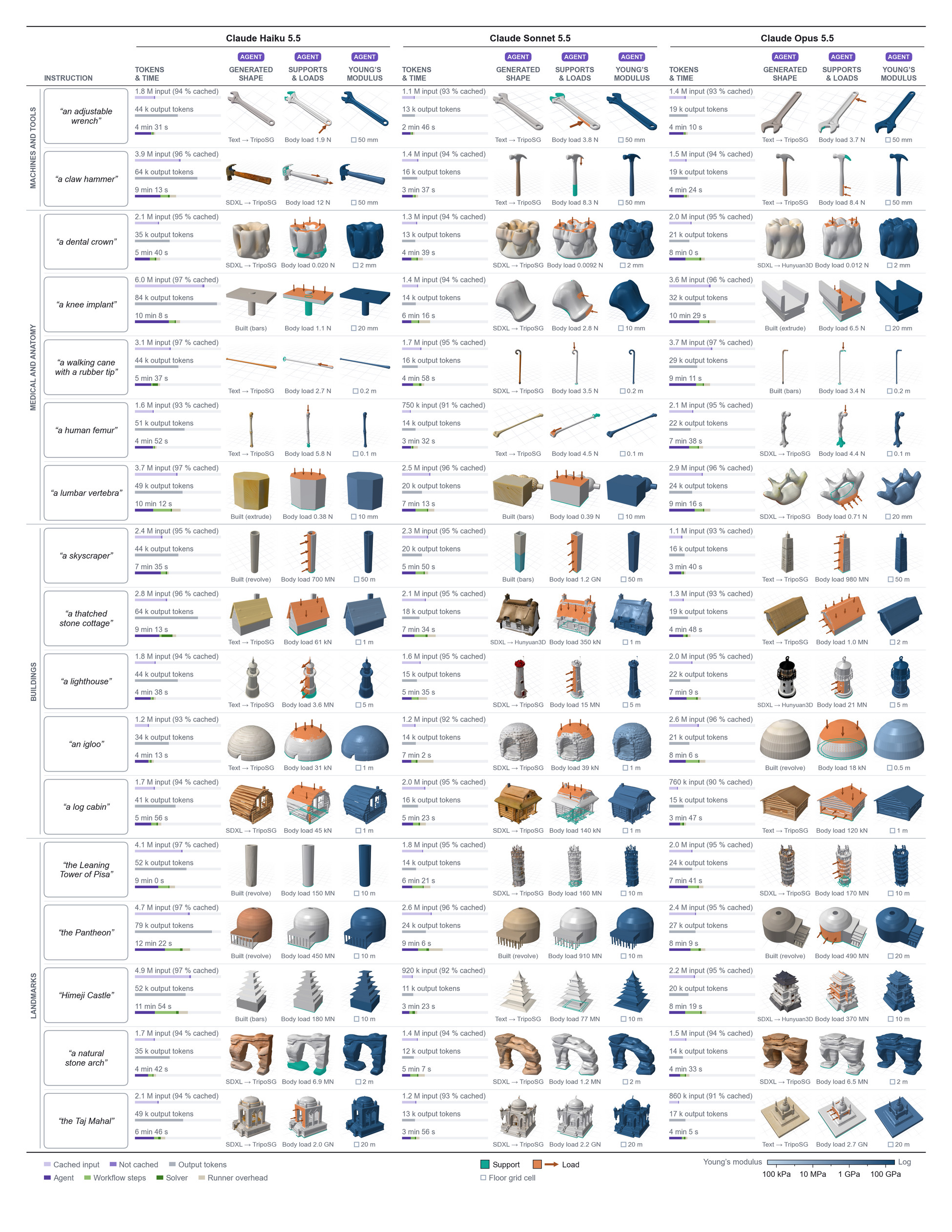}
    \caption{(Continued) A side-by-side comparison among settings from Claude Haiku, Sonnet, and Opus. }
\end{figure}

\clearpage

\section{Evaluation of simulations by different Claude models}
\label{sec:si-model-eval}

Two panels, each of three isolated reviewer sessions, revisit each simulation by each model, judge whether each criterion for geometry quality, boundary conditions quality, and material setting quality is met, partially met, or failed, and give a grade of high, moderate, low, or N/A for geometry, boundary condition, and material respectively. The sessions of one panel run Claude Fable~5.1 and those of the other run Astra~6 (GPT-6-Astra), all at the high reasoning-effort setting, following the LLM-as-a-judge approach~\citeS{gu2026survey}. The reviewers grade each session from figures like the side-by-side comparison figures in Section~\ref{sec:si-model-compare}, with the operator's name hidden, and from the operator's report and the session files. No human engineer graded the sessions. Between the majority votes of the two panels, Cohen's kappa is 0.25 for geometry, 0.23 for boundary conditions and 0.19 for materials. Evaluations by the Astra~6 panel are listed in Table~\ref{tab:S_model_evaluation_astra}, and those by the Fable~5.1 panel in Table~\ref{tab:S_model_evaluation}, adhering to the following rubric:

{\itshape
\begin{quote}
    Score every criterion met, partial or failed using the anchors. Then derive the category rating mechanically:
    \begin{itemize}
        \item \textbf{High}: every criterion met.
        \item \textbf{Moderate}: no criterion failed, and one or two partial.
        \item \textbf{Low}: any criterion failed, or three or more partial.
        \item \textbf{N/A}: only when the run never produced that setting. For example, no body was ever accepted, so there are no BCs and no materials. A setting that exists but whose solve failed is still rated.
    \end{itemize}

Judge sizes, loads and stiffnesses by order of magnitude against typical real values, using your own engineering knowledge. The anchors below are guides and do not form an exhaustive table.

\noindent \textbf{Geometry: is the simulated body recognizable, realistic, physical and complete?}

Judge only the body that was simulated (the Supports \& loads and Deformed columns); ignore an earlier attempt.

\begin{itemize}
    \item Recognizable:
    \begin{itemize}
        \item Met: identifiable as the object without reading the caption
        \item Partial: identifiable only once you know the caption
        \item Failed: not identifiable even with the caption
    \end{itemize}
    \item Realistic form:
    \begin{itemize}
        \item Met: proportions and the defining features look like a typical real one (a wrench's jaws and handle, a femur's head, shaft and condyles)
        \item Partial: one defining feature simplified or missing, or proportions clearly off; a hand-built proxy that keeps the load-bearing form
        \item Failed: a primitive stand-in (box, bar, blob) or a different object
    \end{itemize}
    \item Physical:
    \begin{itemize}
        \item Met: one connected solid with plausible member thicknesses; real-world size within 2× of typical (check the scale bar and the brief)
        \item Partial: a visible generation artefact (lumps, spurious tabs or slabs, fused gaps, holes) or size off 2-5×
        \item Failed: floating or disconnected pieces carrying load, a non-solid sheet, or size off more than 5×
    \end{itemize}
    \item Complete:
    \begin{itemize}
        \item Met: every major part present
        \item Partial: a minor part missing or truncated
        \item Failed: a major part missing (legs, wheels, a whole section)
    \end{itemize}
\end{itemize}

\noindent \textbf{BC: are the supports and loads where they would really be, with realistic magnitude and direction?}

First identify the scenario from the brief and reply: what holds the object, and what loads it.

\begin{itemize}
    \item Support regions:
    \begin{itemize}
        \item Met: where the object rests, is held, mounted or embedded in that scenario
        \item Partial: the right kind of place but the wrong extent or position, or a support missing or added
        \item Failed: on a surface that is free in reality, or a scenario that does not exist (the object held by a part nothing touches)
    \end{itemize}
    \item Load regions:
    \begin{itemize}
        \item Met: where the load acts in that scenario (the seat, the striking face, the top bun)
        \item Partial: the right kind of place but the wrong extent or position
        \item Failed: on a surface nothing loads in that scenario
    \end{itemize}
    \item Load magnitude and direction:
    \begin{itemize}
        \item Met: within 3$\times$ of a typical value for the scenario, and the direction is how it acts. Self-weight, if used, should match the body's mass
        \item Partial: 3-30$\times$ off a typical value, or direction off by more than 30°
        \item Failed: more than 30$\times$ off, or a physically impossible direction (a floor pulling, a press pulling away)
    \end{itemize}
    \item Held components:
    \begin{itemize}
        \item Met: held components fit the contact type: bonded, bolted, embedded or gripped $\rightarrow$ all components; a resting contact $\rightarrow$ at least the normal component (tangential holds standing in for friction or rigid-body removal are fine)
        \item Partial: over-restraint that materially changes this load case's response (a clamped end where free sliding or rotation governs), or under-restraint patched by an unphysical hold
        \item Failed: holds that contradict the contact, such as a support that must pull to resist the load where the real contact can only push
    \end{itemize}
\end{itemize}

\noindent \textbf{Material: does each part get a material that matches what the object is?}

\begin{itemize}
    \item Part mapping:
    \begin{itemize}
        \item Met: parts that are clearly different real materials get different materials, and the map follows the part boundaries; an object that is really one material is uniform
        \item Partial: a multi-material object treated as uniform where the other materials sit only in parts off the main load path; or a minor part misassigned; or a boundary visibly in the wrong place
        \item Failed: uniform where the main load path crosses clearly different materials (a rubber tyre given the rim's steel), or a major part assigned another part's material
    \end{itemize}
    \item Material class:
    \begin{itemize}
        \item Met: each material is the right class for its part under the chosen reading (metal, wood, rigid polymer, elastomer, foam, glass, stone, bone, food, soft tissue)
        \item Partial: a secondary part has the wrong class
        \item Failed: the main body has the wrong class
    \end{itemize}
    \item Stiffness value:
    \begin{itemize}
        \item Met: Young's modulus within 3$\times$ of typical for the named material
        \item Partial: 3-10$\times$ off
        \item Failed: more than 10$\times$ off
    \end{itemize}
\end{itemize}

\end{quote}
}

Under both panels (Tables~\ref{tab:S_model_evaluation_astra} and~\ref{tab:S_model_evaluation}), Opus~5.5 received the fewest low grades for geometry and Haiku~5.5 the most, and, by a smaller margin, Haiku~5.5 received the most low grades for the supports and loads. Apart from geometry, Sonnet~5.5 received grades close to those of Opus~5.5, with no consistent difference between them for the supports and loads or the materials, at less than half the cost (US\$26.22 against US\$65.29) and in less time (3\,h 56\,min against 5\,h 23\,min). Haiku~5.5 cost US\$7.44 and took 5\,h 39\,min, but did not complete 4 sessions. Each operator ran each object once, so differences of a few sessions may lie within the variation between repeated runs. Both panels graded the geometry, the supports and loads, and the materials as high in 10, 20 and 15 sessions of Haiku~5.5, in 15, 26 and 18 sessions of Sonnet~5.5, and in 15, 24 and 19 sessions of Opus~5.5. The grades that both panels gave as low show where the operators failed. Of the 12 materials graded low by both panels, 10 were in objects with soft parts, such as foam cushions, rubber tires and bushings. Both panels also graded the geometry low in 7 sessions of Haiku~5.5, against 3 for Sonnet~5.5 and 1 for Opus~5.5, and the supports and loads low in 3 sessions of Haiku~5.5 and in none of the others. The Fable~5.1 panel, from the same model family as the operators, graded higher than the GPT-6-Astra panel in most sessions where the two differed, and the two gave the same grade in 47--59\% of the sessions graded by both (Cohen's kappa 0.19--0.25, slight to fair agreement) and opposite grades in 2--15\% (Tables~\ref{tab:S_model_evaluation_astra} and~\ref{tab:S_model_evaluation}).

\begin{landscape}
\begingroup
\tiny
\let\footnotesize\tiny 
\setlength{\tabcolsep}{2pt}
\setlength{\LTleft}{-70pt plus 1fill}
\setlength{\LTright}{-70pt plus 1fill}
\setlength{\LTcapwidth}{\linewidth}
\begin{longtable}{@{}l *{3}{*{3}{c} c}@{}}
  \caption{Evaluation of simulations by different Claude models by the Astra~6 panel. Each cell gives the grades of reviewers 1, 2 and 3 and, after the arrow, their majority vote. $^{*}$ marks a cell where all three differed and the median is taken. N/A: the session never produced that setting. Haiku~5.5 did not complete 4 sessions: those of the crocodile, the pogo stick and the crankshaft, which have N/A cells, and that of the walking cane, marked Failed, whose solve failed under a load close to the buckling load. Cost: the operator's own model calls at API prices, excluding the AI subagent calls.}
  \label{tab:S_model_evaluation_astra} \\
  \toprule
  Object & \multicolumn{4}{c}{Haiku} & \multicolumn{4}{c}{Sonnet} & \multicolumn{4}{c}{Opus} \\
  \cmidrule(lr){2-5}\cmidrule(lr){6-9}\cmidrule(lr){10-13}
  & Geometry & BC & Material & Cost (USD) & Geometry & BC & Material & Cost (USD) & Geometry & BC & Material & Cost (USD) \\
  \midrule
  \endfirsthead
  \multicolumn{13}{l}{\tablename~\thetable{} (continued)} \\
  \toprule
  Object & \multicolumn{4}{c}{Haiku} & \multicolumn{4}{c}{Sonnet} & \multicolumn{4}{c}{Opus} \\
  \cmidrule(lr){2-5}\cmidrule(lr){6-9}\cmidrule(lr){10-13}
  & Geometry & BC & Material & Cost (USD) & Geometry & BC & Material & Cost (USD) & Geometry & BC & Material & Cost (USD) \\
  \midrule
  \endhead
  \midrule
  \multicolumn{13}{@{}r@{}}{\textit{Continued on next page}} \\
  \endfoot
  \bottomrule
  \endlastfoot
  a TV remote control                                  & M/M/M$\rightarrow$M & M/M/M$\rightarrow$M & L/L/M$\rightarrow$L & \$0.13     & M/M/M$\rightarrow$M & M/M/M$\rightarrow$M & L/L/L$\rightarrow$L & \$0.41     & M/M/M$\rightarrow$M & M/M/M$\rightarrow$M & L/L/L$\rightarrow$L & \$0.85 \\*
  a USB stick                                          & M/M/M$\rightarrow$M & M/M/H$\rightarrow$M & M/H/M$\rightarrow$M & \$0.40     & M/M/M$\rightarrow$M & H/H/H$\rightarrow$H & M/H/M$\rightarrow$M & \$0.52     & M/M/M$\rightarrow$M & H/H/H$\rightarrow$H & H/H/M$\rightarrow$H & \$1.00 \\*
  an electric toothbrush                               & L/M/M$\rightarrow$M & L/L/M$\rightarrow$L & M/M/M$\rightarrow$M & \$0.06     & M/M/M$\rightarrow$M & H/H/H$\rightarrow$H & M/M/M$\rightarrow$M & \$0.44     & H/M/M$\rightarrow$M & H/H/H$\rightarrow$H & M/M/M$\rightarrow$M & \$1.05 \\*
  a coffee mug                                         & H/H/H$\rightarrow$H & M/M/H$\rightarrow$M & H/H/H$\rightarrow$H & \$0.04     & H/H/H$\rightarrow$H & H/H/H$\rightarrow$H & H/H/H$\rightarrow$H & \$0.39     & H/H/H$\rightarrow$H & M/M/M$\rightarrow$M & H/H/H$\rightarrow$H & \$0.79 \\*
  a computer mouse                                     & M/M/M$\rightarrow$M & L/L/L$\rightarrow$L & M/M/M$\rightarrow$M & \$0.04     & M/M/M$\rightarrow$M & M/M/L$\rightarrow$M & M/M/M$\rightarrow$M & \$0.34     & M/M/M$\rightarrow$M & M/M/L$\rightarrow$M & M/M/M$\rightarrow$M & \$0.95 \\
  \midrule
  a school chair with a steel frame and a plastic seat & M/M/M$\rightarrow$M & H/H/H$\rightarrow$H & H/H/H$\rightarrow$H & \$0.24     & M/M/M$\rightarrow$M & H/M/M$\rightarrow$M & M/M/M$\rightarrow$M & \$0.53     & M/M/M$\rightarrow$M & H/H/H$\rightarrow$H & H/M/M$\rightarrow$M & \$1.50 \\*
  a glass-top dining table                             & M/M/M$\rightarrow$M & M/M/H$\rightarrow$M & M/M/M$\rightarrow$M & \$0.19     & M/M/M$\rightarrow$M & H/H/H$\rightarrow$H & L/L/L$\rightarrow$L & \$0.62     & H/H/H$\rightarrow$H & H/H/H$\rightarrow$H & H/H/H$\rightarrow$H & \$1.28 \\*
  a park bench                                         & M/M/M$\rightarrow$M & M/M/M$\rightarrow$M & H/L/H$\rightarrow$H & \$0.47     & H/H/H$\rightarrow$H & H/H/H$\rightarrow$H & H/H/H$\rightarrow$H & \$0.73     & H/H/H$\rightarrow$H & H/H/H$\rightarrow$H & H/H/H$\rightarrow$H & \$1.39 \\*
  a bar stool                                          & H/H/H$\rightarrow$H & H/H/H$\rightarrow$H & H/H/H$\rightarrow$H & \$0.03     & H/H/H$\rightarrow$H & H/H/H$\rightarrow$H & H/H/H$\rightarrow$H & \$0.37     & H/H/H$\rightarrow$H & H/H/M$\rightarrow$H & H/H/H$\rightarrow$H & \$0.88 \\*
  a sofa                                               & M/H/M$\rightarrow$M & H/H/H$\rightarrow$H & L/L/L$\rightarrow$L & \$0.13     & M/M/M$\rightarrow$M & M/M/M$\rightarrow$M & L/L/L$\rightarrow$L & \$0.41     & M/H/M$\rightarrow$M & H/M/M$\rightarrow$M & L/L/L$\rightarrow$L & \$0.93 \\
  \midrule
  a teddy bear                                         & H/H/H$\rightarrow$H & H/H/H$\rightarrow$H & H/M/M$\rightarrow$M & \$0.04     & H/H/H$\rightarrow$H & H/H/H$\rightarrow$H & H/M/M$\rightarrow$M & \$0.36     & H/H/H$\rightarrow$H & H/H/H$\rightarrow$H & H/M/M$\rightarrow$M & \$0.94 \\*
  a wooden rocking horse                               & M/M/M$\rightarrow$M & L/L/L$\rightarrow$L & H/H/M$\rightarrow$H & \$0.12     & H/H/H$\rightarrow$H & M/M/M$\rightarrow$M & H/H/H$\rightarrow$H & \$0.36     & H/H/H$\rightarrow$H & H/M/H$\rightarrow$H & H/H/H$\rightarrow$H & \$0.82 \\*
  a pogo stick                                         & L/L/L$\rightarrow$L & N/A & N/A & \$0.31     & L/L/L$\rightarrow$L & H/H/H$\rightarrow$H & L/L/L$\rightarrow$L & \$0.90     & L/L/L$\rightarrow$L & M/M/M$\rightarrow$M & L/L/L$\rightarrow$L & \$1.62 \\*
  a wooden toy train engine                            & H/H/M$\rightarrow$H & M/H/M$\rightarrow$M & H/H/H$\rightarrow$H & \$0.12     & H/H/M$\rightarrow$H & H/M/M$\rightarrow$M & H/H/H$\rightarrow$H & \$0.43     & M/M/M$\rightarrow$M & M/L/L$\rightarrow$L & H/H/M$\rightarrow$H & \$0.96 \\*
  a yo-yo                                              & L/L/L$\rightarrow$L & L/M/L$\rightarrow$L & H/H/H$\rightarrow$H & \$0.11     & L/L/L$\rightarrow$L & L/L/L$\rightarrow$L & H/H/H$\rightarrow$H & \$0.63     & M/M/M$\rightarrow$M & M/M/M$\rightarrow$M & M/H/H$\rightarrow$H & \$1.52 \\
  \midrule
  a jelly dessert                                      & H/H/H$\rightarrow$H & H/H/H$\rightarrow$H & H/H/H$\rightarrow$H & \$0.03     & M/H/H$\rightarrow$H & M/M/M$\rightarrow$M & H/H/H$\rightarrow$H & \$0.39     & H/H/H$\rightarrow$H & H/H/H$\rightarrow$H & H/H/H$\rightarrow$H & \$1.09 \\*
  a loaf of bread                                      & H/H/H$\rightarrow$H & H/H/H$\rightarrow$H & L/L/M$\rightarrow$L & \$0.12     & H/H/H$\rightarrow$H & H/H/H$\rightarrow$H & L/L/M$\rightarrow$L & \$0.47     & H/H/H$\rightarrow$H & M/M/M$\rightarrow$M & L/L/M$\rightarrow$L & \$1.32 \\*
  a hamburger                                          & L/M/M$\rightarrow$M & H/H/H$\rightarrow$H & L/L/L$\rightarrow$L & \$0.05     & M/M/M$\rightarrow$M & M/M/H$\rightarrow$M & M/H/H$\rightarrow$H & \$0.74     & H/M/M$\rightarrow$M & H/H/H$\rightarrow$H & H/H/H$\rightarrow$H & \$1.40 \\*
  a layered birthday cake                              & H/H/M$\rightarrow$H & M/M/M$\rightarrow$M & L/L/L$\rightarrow$L & \$0.08     & H/H/M$\rightarrow$H & H/H/H$\rightarrow$H & L/L/L$\rightarrow$L & \$0.36     & H/H/M$\rightarrow$H & M/M/M$\rightarrow$M & L/L/L$\rightarrow$L & \$0.92 \\*
  a slice of watermelon                                & M/M/M$\rightarrow$M & H/H/H$\rightarrow$H & L/L/L$\rightarrow$L & \$0.05     & M/H/H$\rightarrow$H & M/M/M$\rightarrow$M & H/H/H$\rightarrow$H & \$0.48     & M/M/M$\rightarrow$M & M/M/M$\rightarrow$M & M/L/L$\rightarrow$L & \$1.62 \\
  \midrule
  a giraffe                                            & H/H/H$\rightarrow$H & H/H/H$\rightarrow$H & L/L/L$\rightarrow$L & \$0.11     & H/H/H$\rightarrow$H & H/H/H$\rightarrow$H & L/L/L$\rightarrow$L & \$0.38     & H/H/H$\rightarrow$H & H/H/H$\rightarrow$H & L/L/L$\rightarrow$L & \$0.94 \\*
  an elephant                                          & H/H/H$\rightarrow$H & H/H/H$\rightarrow$H & L/L/L$\rightarrow$L & \$0.08     & H/H/H$\rightarrow$H & H/H/H$\rightarrow$H & L/L/L$\rightarrow$L & \$0.43     & H/H/H$\rightarrow$H & H/H/H$\rightarrow$H & L/L/L$\rightarrow$L & \$0.86 \\*
  a rhinoceros                                         & H/H/H$\rightarrow$H & M/M/M$\rightarrow$M & H/H/H$\rightarrow$H & \$0.04     & H/H/H$\rightarrow$H & H/H/H$\rightarrow$H & L/L/L$\rightarrow$L & \$0.43     & H/H/H$\rightarrow$H & H/H/H$\rightarrow$H & L/L/L$\rightarrow$L & \$0.82 \\*
  a horse                                              & H/H/H$\rightarrow$H & H/H/H$\rightarrow$H & L/L/L$\rightarrow$L & \$0.11     & H/H/H$\rightarrow$H & H/H/H$\rightarrow$H & L/L/L$\rightarrow$L & \$0.40     & H/H/H$\rightarrow$H & H/H/H$\rightarrow$H & L/L/L$\rightarrow$L & \$0.88 \\*
  a crocodile                                          & N/A & N/A & N/A & \$0.04     & M/M/M$\rightarrow$M & H/H/H$\rightarrow$H & L/L/L$\rightarrow$L & \$0.42     & M/M/M$\rightarrow$M & M/H/H$\rightarrow$H & L/L/L$\rightarrow$L & \$1.23 \\
  \midrule
  a car suspension control arm with rubber bushings    & H/H/H$\rightarrow$H & M/L/L$\rightarrow$L & L/L/L$\rightarrow$L & \$0.22     & M/M/M$\rightarrow$M & M/M/M$\rightarrow$M & L/L/L$\rightarrow$L & \$0.54     & M/H/M$\rightarrow$M & M/L/L$\rightarrow$L & L/L/L$\rightarrow$L & \$1.34 \\*
  a truck leaf spring                                  & L/L/M$\rightarrow$L & M/M/M$\rightarrow$M & H/H/H$\rightarrow$H & \$0.37     & L/L/M$\rightarrow$L & H/H/M$\rightarrow$H & H/H/H$\rightarrow$H & \$0.51     & M/M/M$\rightarrow$M & M/H/H$\rightarrow$H & H/H/H$\rightarrow$H & \$1.89 \\*
  a car wheel with its rubber tyre                     & M/M/M$\rightarrow$M & M/M/M$\rightarrow$M & M/L/L$\rightarrow$L & \$0.10     & M/M/M$\rightarrow$M & M/M/M$\rightarrow$M & L/L/L$\rightarrow$L & \$0.59     & M/M/M$\rightarrow$M & H/M/M$\rightarrow$M & L/L/L$\rightarrow$L & \$1.36 \\*
  a car brake disc                                     & L/L/L$\rightarrow$L & M/M/M$\rightarrow$M & H/H/H$\rightarrow$H & \$0.03     & M/M/M$\rightarrow$M & H/H/H$\rightarrow$H & H/H/H$\rightarrow$H & \$0.66     & M/M/M$\rightarrow$M & M/M/M$\rightarrow$M & H/H/H$\rightarrow$H & \$1.52 \\*
  a crankshaft                                         & L/L/L$\rightarrow$L & L/L/L$\rightarrow$L & N/A & \$0.07     & L/M/M$\rightarrow$M & L/L/L$\rightarrow$L & H/H/H$\rightarrow$H & \$0.96     & M/L/M$\rightarrow$M & M/L/L$\rightarrow$L & H/H/H$\rightarrow$H & \$2.67 \\
  \midrule
  a road roller with a steel drum and rubber tyres     & L/L/L$\rightarrow$L & M/M/M$\rightarrow$M & L/L/L$\rightarrow$L & \$0.36     & L/L/L$\rightarrow$L & M/L/L$\rightarrow$L & L/L/L$\rightarrow$L & \$0.80     & M/M/M$\rightarrow$M & M/M/M$\rightarrow$M & L/L/L$\rightarrow$L & \$3.74 \\*
  a jackhammer                                         & L/L/L$\rightarrow$L & M/M/L$\rightarrow$M & H/H/H$\rightarrow$H & \$0.35     & M/M/M$\rightarrow$M & H/H/H$\rightarrow$H & H/H/L$\rightarrow$H & \$0.76     & L/M/M$\rightarrow$M & H/H/H$\rightarrow$H & H/H/H$\rightarrow$H & \$1.78 \\*
  an industrial robot arm                              & M/M/M$\rightarrow$M & M/H/M$\rightarrow$M & L/M/L$\rightarrow$L & \$0.08     & M/M/M$\rightarrow$M & M/M/M$\rightarrow$M & L/M/M$\rightarrow$M & \$0.40     & M/M/M$\rightarrow$M & H/M/M$\rightarrow$M & L/M/M$\rightarrow$M & \$1.15 \\*
  an adjustable wrench                                 & L/M/M$\rightarrow$M & M/M/M$\rightarrow$M & H/H/H$\rightarrow$H & \$0.08     & M/M/M$\rightarrow$M & M/M/M$\rightarrow$M & H/H/H$\rightarrow$H & \$0.43     & M/M/M$\rightarrow$M & H/H/H$\rightarrow$H & H/H/H$\rightarrow$H & \$1.12 \\*
  a claw hammer                                        & M/H/M$\rightarrow$M & L/L/L$\rightarrow$L & H/H/H$\rightarrow$H & \$0.25     & M/M/M$\rightarrow$M & M/M/M$\rightarrow$M & H/H/H$\rightarrow$H & \$0.51     & M/M/M$\rightarrow$M & L/L/M$\rightarrow$L & H/H/H$\rightarrow$H & \$1.11 \\
  \midrule
  a dental crown                                       & L/M/M$\rightarrow$M & M/M/M$\rightarrow$M & H/H/H$\rightarrow$H & \$0.08     & L/M/M$\rightarrow$M & M/M/M$\rightarrow$M & L/H/H$\rightarrow$H & \$0.47     & M/M/M$\rightarrow$M & M/M/M$\rightarrow$M & L/H/H$\rightarrow$H & \$1.29 \\*
  a knee implant                                       & L/M/L$\rightarrow$L & M/H/H$\rightarrow$H & H/H/H$\rightarrow$H & \$0.42     & L/L/L$\rightarrow$L & H/M/H$\rightarrow$H & H/H/H$\rightarrow$H & \$0.49     & L/L/L$\rightarrow$L & H/H/H$\rightarrow$H & L/L/L$\rightarrow$L & \$2.01 \\*
  a walking cane with a rubber tip                     & M/L/L$\rightarrow$L & H/M/H$\rightarrow$H & M/M/M$\rightarrow$M & \$0.13 (Failed)     & H/H/H$\rightarrow$H & H/H/H$\rightarrow$H & M/M/M$\rightarrow$M & \$0.55     & H/H/H$\rightarrow$H & H/H/H$\rightarrow$H & M/M/M$\rightarrow$M & \$1.95 \\*
  a human femur                                        & L/L/L$\rightarrow$L & M/M/M$\rightarrow$M & L/L/L$\rightarrow$L & \$0.10     & M/M/M$\rightarrow$M & M/M/M$\rightarrow$M & L/L/L$\rightarrow$L & \$0.38     & M/M/M$\rightarrow$M & H/L/L$\rightarrow$L & L/L/L$\rightarrow$L & \$1.34 \\*
  a lumbar vertebra                                    & L/L/L$\rightarrow$L & H/H/H$\rightarrow$H & L/L/L$\rightarrow$L & \$0.17     & L/L/M$\rightarrow$L & M/H/H$\rightarrow$H & L/L/L$\rightarrow$L & \$0.71     & H/H/H$\rightarrow$H & M/M/M$\rightarrow$M & L/L/L$\rightarrow$L & \$1.60 \\
  \midrule
  a skyscraper                                         & L/L/L$\rightarrow$L & M/L/M$\rightarrow$M & H/M/M$\rightarrow$M & \$0.10     & L/L/L$\rightarrow$L & H/H/H$\rightarrow$H & H/M/M$\rightarrow$M & \$0.67     & M/M/M$\rightarrow$M & H/M/M$\rightarrow$M & M/M/M$\rightarrow$M & \$0.95 \\*
  a thatched stone cottage                             & L/L/M$\rightarrow$L & H/H/M$\rightarrow$H & L/L/L$\rightarrow$L & \$0.21     & M/M/M$\rightarrow$M & M/M/L$\rightarrow$M & L/L/L$\rightarrow$L & \$0.63     & M/M/M$\rightarrow$M & M/M/M$\rightarrow$M & L/L/L$\rightarrow$L & \$1.09 \\*
  a lighthouse                                         & M/M/M$\rightarrow$M & M/M/H$\rightarrow$M & M/M/M$\rightarrow$M & \$0.10     & L/L/M$\rightarrow$L & M/M/M$\rightarrow$M & M/M/M$\rightarrow$M & \$0.52     & M/M/M$\rightarrow$M & M/M/M$\rightarrow$M & M/M/M$\rightarrow$M & \$1.34 \\*
  an igloo                                             & M/M/M$\rightarrow$M & H/H/M$\rightarrow$H & M/M/M$\rightarrow$M & \$0.04     & M/M/M$\rightarrow$M & H/H/H$\rightarrow$H & H/H/H$\rightarrow$H & \$0.47     & L/L/L$\rightarrow$L & H/H/H$\rightarrow$H & H/H/H$\rightarrow$H & \$1.44 \\*
  a log cabin                                          & M/M/H$\rightarrow$M & M/M/M$\rightarrow$M & M/M/M$\rightarrow$M & \$0.09     & M/M/M$\rightarrow$M & M/M/M$\rightarrow$M & M/M/M$\rightarrow$M & \$0.60     & M/L/M$\rightarrow$M & M/M/M$\rightarrow$M & H/H/M$\rightarrow$H & \$0.83 \\
  \midrule
  the Leaning Tower of Pisa                            & L/L/L$\rightarrow$L & H/H/H$\rightarrow$H & L/L/H$\rightarrow$L & \$0.21     & L/L/M$\rightarrow$L & H/H/H$\rightarrow$H & M/L/M$\rightarrow$M & \$0.55     & L/L/M$\rightarrow$L & M/M/H$\rightarrow$M & M/L/H$\rightarrow$M$^{*}$ & \$1.36 \\*
  the Pantheon                                         & M/M/M$\rightarrow$M & H/H/H$\rightarrow$H & L/L/L$\rightarrow$L & \$0.31     & M/M/M$\rightarrow$M & H/H/H$\rightarrow$H & L/L/L$\rightarrow$L & \$0.77     & M/M/M$\rightarrow$M & M/M/H$\rightarrow$M & L/L/L$\rightarrow$L & \$1.53 \\*
  Himeji Castle                                        & L/L/L$\rightarrow$L & H/H/H$\rightarrow$H & M/L/M$\rightarrow$M & \$0.26     & L/L/L$\rightarrow$L & H/M/M$\rightarrow$M & L/L/L$\rightarrow$L & \$0.39     & L/L/M$\rightarrow$L & H/M/H$\rightarrow$H & M/L/M$\rightarrow$M & \$1.39 \\*
  a natural stone arch                                 & M/M/H$\rightarrow$M & M/H/H$\rightarrow$H & H/H/H$\rightarrow$H & \$0.07     & H/H/H$\rightarrow$H & H/H/H$\rightarrow$H & H/H/H$\rightarrow$H & \$0.46     & H/M/H$\rightarrow$H & H/H/H$\rightarrow$H & H/H/H$\rightarrow$H & \$1.03 \\*
  the Taj Mahal                                        & L/L/L$\rightarrow$L & M/M/H$\rightarrow$M & L/L/L$\rightarrow$L & \$0.12     & L/L/L$\rightarrow$L & H/H/H$\rightarrow$H & L/L/L$\rightarrow$L & \$0.46     & M/L/L$\rightarrow$L & M/H/H$\rightarrow$H & M/L/M$\rightarrow$M & \$0.89 \\
  \midrule
  Total: count of H/M/L or cost of all 50 objects & 12/21/16 (1 N/A) & 20/21/7 (2 N/A) & 17/11/19 (3 N/A) & \$7.44     & 16/23/11 & 27/20/3 & 19/11/20 & \$26.22    & 16/28/6 & 24/21/5 & 20/11/19 & \$65.29 \\*
  Total: time of all 50 objects                               & \multicolumn{4}{c}{5 h 39 min (4 failed in total)}                                                     & \multicolumn{4}{c}{3 h 56 min}                                                     & \multicolumn{4}{c}{5 h 23 min} \\
\end{longtable}
\endgroup
\end{landscape}

\begin{landscape}
\begingroup
\tiny
\let\footnotesize\tiny 
\setlength{\tabcolsep}{2pt}
\setlength{\LTleft}{-70pt plus 1fill}
\setlength{\LTright}{-70pt plus 1fill}
\setlength{\LTcapwidth}{\linewidth}
\begin{longtable}{@{}l *{3}{*{3}{c}}@{}}
  \caption{Evaluation of simulations by different Claude models by the Fable~5.1 panel. Each cell gives the grades of reviewers 1, 2 and 3 and, after the arrow, their majority vote. N/A: the session never produced that setting. The 4 Haiku~5.5 sessions that did not complete are those listed in Table~\ref{tab:S_model_evaluation_astra}.}
  \label{tab:S_model_evaluation} \\
  \toprule
  Object & \multicolumn{3}{c}{Haiku} & \multicolumn{3}{c}{Sonnet} & \multicolumn{3}{c}{Opus} \\
  \cmidrule(lr){2-4}\cmidrule(lr){5-7}\cmidrule(lr){8-10}
  & Geometry & BC & Material & Geometry & BC & Material & Geometry & BC & Material \\
  \midrule
  \endfirsthead
  \multicolumn{10}{l}{\tablename~\thetable{} (continued)} \\
  \toprule
  Object & \multicolumn{3}{c}{Haiku} & \multicolumn{3}{c}{Sonnet} & \multicolumn{3}{c}{Opus} \\
  \cmidrule(lr){2-4}\cmidrule(lr){5-7}\cmidrule(lr){8-10}
  & Geometry & BC & Material & Geometry & BC & Material & Geometry & BC & Material \\
  \midrule
  \endhead
  \midrule
  \multicolumn{10}{@{}r@{}}{\textit{Continued on next page}} \\
  \endfoot
  \bottomrule
  \endlastfoot
  a TV remote control                                  & H/H/H$\rightarrow$H & M/M/M$\rightarrow$M & H/H/H$\rightarrow$H & H/H/H$\rightarrow$H & M/M/M$\rightarrow$M & M/H/H$\rightarrow$H & H/H/H$\rightarrow$H & H/H/M$\rightarrow$H & M/H/H$\rightarrow$H \\*
  a USB stick                                          & H/H/H$\rightarrow$H & M/H/H$\rightarrow$H & H/H/H$\rightarrow$H & H/H/H$\rightarrow$H & H/M/H$\rightarrow$H & H/H/H$\rightarrow$H & H/H/H$\rightarrow$H & H/M/H$\rightarrow$H & H/H/H$\rightarrow$H \\*
  an electric toothbrush                               & M/L/M$\rightarrow$M & L/L/L$\rightarrow$L & H/H/H$\rightarrow$H & H/H/H$\rightarrow$H & H/H/H$\rightarrow$H & H/H/H$\rightarrow$H & H/H/H$\rightarrow$H & H/H/H$\rightarrow$H & H/H/H$\rightarrow$H \\*
  a coffee mug                                         & H/H/H$\rightarrow$H & H/H/H$\rightarrow$H & H/H/H$\rightarrow$H & H/H/H$\rightarrow$H & M/H/H$\rightarrow$H & H/H/H$\rightarrow$H & H/H/H$\rightarrow$H & H/H/H$\rightarrow$H & H/H/H$\rightarrow$H \\*
  a computer mouse                                     & H/H/H$\rightarrow$H & L/M/L$\rightarrow$L & H/H/H$\rightarrow$H & H/H/H$\rightarrow$H & M/M/M$\rightarrow$M & H/H/H$\rightarrow$H & H/H/H$\rightarrow$H & M/M/M$\rightarrow$M & H/H/H$\rightarrow$H \\
  \midrule
  a school chair with a steel frame and a plastic seat & H/H/H$\rightarrow$H & H/H/H$\rightarrow$H & M/M/M$\rightarrow$M & H/H/H$\rightarrow$H & H/H/H$\rightarrow$H & H/H/H$\rightarrow$H & H/H/H$\rightarrow$H & H/H/H$\rightarrow$H & H/H/M$\rightarrow$H \\*
  a glass-top dining table                             & M/M/M$\rightarrow$M & H/H/H$\rightarrow$H & M/M/M$\rightarrow$M & M/M/M$\rightarrow$M & H/H/H$\rightarrow$H & M/M/M$\rightarrow$M & M/M/M$\rightarrow$M & H/H/H$\rightarrow$H & H/H/H$\rightarrow$H \\*
  a park bench                                         & M/M/M$\rightarrow$M & H/H/H$\rightarrow$H & H/H/H$\rightarrow$H & H/H/H$\rightarrow$H & H/H/H$\rightarrow$H & H/H/H$\rightarrow$H & H/H/H$\rightarrow$H & H/H/H$\rightarrow$H & H/H/H$\rightarrow$H \\*
  a bar stool                                          & H/H/H$\rightarrow$H & H/H/H$\rightarrow$H & H/H/H$\rightarrow$H & H/H/H$\rightarrow$H & H/H/H$\rightarrow$H & H/H/H$\rightarrow$H & H/H/H$\rightarrow$H & H/H/H$\rightarrow$H & H/H/H$\rightarrow$H \\*
  a sofa                                               & H/H/H$\rightarrow$H & H/H/H$\rightarrow$H & L/L/L$\rightarrow$L & H/H/H$\rightarrow$H & H/H/H$\rightarrow$H & L/L/L$\rightarrow$L & H/H/H$\rightarrow$H & H/H/H$\rightarrow$H & L/L/L$\rightarrow$L \\
  \midrule
  a teddy bear                                         & H/H/H$\rightarrow$H & H/H/H$\rightarrow$H & H/H/H$\rightarrow$H & H/H/H$\rightarrow$H & H/H/H$\rightarrow$H & H/H/H$\rightarrow$H & H/H/H$\rightarrow$H & H/H/H$\rightarrow$H & M/H/H$\rightarrow$H \\*
  a wooden rocking horse                               & H/H/H$\rightarrow$H & H/H/H$\rightarrow$H & H/H/H$\rightarrow$H & H/H/H$\rightarrow$H & H/H/H$\rightarrow$H & H/H/H$\rightarrow$H & H/H/H$\rightarrow$H & H/H/H$\rightarrow$H & H/H/H$\rightarrow$H \\*
  a pogo stick                                         & L/L/L$\rightarrow$L & N/A                   & N/A                   & M/M/L$\rightarrow$M & H/H/H$\rightarrow$H & H/M/H$\rightarrow$H & M/M/L$\rightarrow$M & M/M/M$\rightarrow$M & L/L/L$\rightarrow$L \\*
  a wooden toy train engine                            & H/H/H$\rightarrow$H & H/M/H$\rightarrow$H & H/H/H$\rightarrow$H & H/H/H$\rightarrow$H & H/H/H$\rightarrow$H & H/H/H$\rightarrow$H & H/H/M$\rightarrow$H & H/M/M$\rightarrow$M & H/H/H$\rightarrow$H \\*
  a yo-yo                                              & L/L/L$\rightarrow$L & H/H/H$\rightarrow$H & H/H/H$\rightarrow$H & M/L/L$\rightarrow$L & M/M/H$\rightarrow$M & H/H/H$\rightarrow$H & M/M/M$\rightarrow$M & H/H/H$\rightarrow$H & H/H/H$\rightarrow$H \\
  \midrule
  a jelly dessert                                      & M/H/M$\rightarrow$M & M/H/H$\rightarrow$H & M/M/M$\rightarrow$M & M/H/M$\rightarrow$M & M/M/M$\rightarrow$M & M/M/M$\rightarrow$M & H/H/H$\rightarrow$H & H/H/H$\rightarrow$H & M/M/M$\rightarrow$M \\*
  a loaf of bread                                      & H/H/H$\rightarrow$H & H/H/H$\rightarrow$H & H/H/H$\rightarrow$H & H/H/H$\rightarrow$H & H/H/H$\rightarrow$H & H/H/H$\rightarrow$H & H/H/H$\rightarrow$H & H/H/H$\rightarrow$H & H/H/H$\rightarrow$H \\*
  a hamburger                                          & M/M/M$\rightarrow$M & H/H/H$\rightarrow$H & M/M/M$\rightarrow$M & H/M/M$\rightarrow$M & H/H/H$\rightarrow$H & H/H/H$\rightarrow$H & M/M/H$\rightarrow$M & H/H/H$\rightarrow$H & H/H/H$\rightarrow$H \\*
  a layered birthday cake                              & H/H/H$\rightarrow$H & M/M/M$\rightarrow$M & H/H/H$\rightarrow$H & H/H/H$\rightarrow$H & H/H/H$\rightarrow$H & H/H/H$\rightarrow$H & H/H/H$\rightarrow$H & M/H/M$\rightarrow$M & H/H/H$\rightarrow$H \\*
  a slice of watermelon                                & H/H/M$\rightarrow$H & H/H/H$\rightarrow$H & M/M/M$\rightarrow$M & H/H/H$\rightarrow$H & H/M/H$\rightarrow$H & H/H/H$\rightarrow$H & H/H/H$\rightarrow$H & H/H/H$\rightarrow$H & H/H/H$\rightarrow$H \\
  \midrule
  a giraffe                                            & H/H/H$\rightarrow$H & H/H/H$\rightarrow$H & M/M/M$\rightarrow$M & H/H/H$\rightarrow$H & H/H/H$\rightarrow$H & M/M/M$\rightarrow$M & H/H/H$\rightarrow$H & H/H/H$\rightarrow$H & M/M/M$\rightarrow$M \\*
  an elephant                                          & H/H/H$\rightarrow$H & H/H/H$\rightarrow$H & M/M/M$\rightarrow$M & H/H/H$\rightarrow$H & H/H/H$\rightarrow$H & M/M/M$\rightarrow$M & H/H/H$\rightarrow$H & H/H/H$\rightarrow$H & M/M/M$\rightarrow$M \\*
  a rhinoceros                                         & H/H/H$\rightarrow$H & M/H/H$\rightarrow$H & H/H/H$\rightarrow$H & H/H/H$\rightarrow$H & H/H/H$\rightarrow$H & M/M/M$\rightarrow$M & H/H/H$\rightarrow$H & H/H/H$\rightarrow$H & M/M/M$\rightarrow$M \\*
  a horse                                              & H/H/H$\rightarrow$H & H/H/H$\rightarrow$H & M/M/M$\rightarrow$M & H/H/H$\rightarrow$H & H/H/H$\rightarrow$H & M/M/M$\rightarrow$M & H/H/H$\rightarrow$H & H/H/H$\rightarrow$H & M/M/M$\rightarrow$M \\*
  a crocodile                                          & N/A                   & N/A                   & N/A                   & H/H/H$\rightarrow$H & H/H/H$\rightarrow$H & M/M/M$\rightarrow$M & M/H/M$\rightarrow$M & H/H/H$\rightarrow$H & M/M/M$\rightarrow$M \\
  \midrule
  a car suspension control arm with rubber bushings    & M/M/M$\rightarrow$M & H/M/H$\rightarrow$H & L/L/L$\rightarrow$L & M/M/M$\rightarrow$M & H/M/H$\rightarrow$H & M/M/M$\rightarrow$M & H/H/H$\rightarrow$H & H/M/H$\rightarrow$H & L/L/L$\rightarrow$L \\*
  a truck leaf spring                                  & L/M/M$\rightarrow$M & H/H/M$\rightarrow$H & H/H/H$\rightarrow$H & L/M/M$\rightarrow$M & H/H/H$\rightarrow$H & H/H/H$\rightarrow$H & M/H/H$\rightarrow$H & H/H/H$\rightarrow$H & H/H/H$\rightarrow$H \\*
  a car wheel with its rubber tyre                     & H/H/M$\rightarrow$H & M/M/M$\rightarrow$M & M/M/M$\rightarrow$M & H/H/M$\rightarrow$H & H/M/H$\rightarrow$H & L/L/L$\rightarrow$L & H/H/M$\rightarrow$H & H/H/H$\rightarrow$H & L/L/L$\rightarrow$L \\*
  a car brake disc                                     & M/H/M$\rightarrow$M & M/M/M$\rightarrow$M & H/H/H$\rightarrow$H & M/M/M$\rightarrow$M & H/H/H$\rightarrow$H & H/H/H$\rightarrow$H & M/H/H$\rightarrow$H & H/M/M$\rightarrow$M & H/H/H$\rightarrow$H \\*
  a crankshaft                                         & L/L/L$\rightarrow$L & M/M/M$\rightarrow$M & N/A                   & M/M/M$\rightarrow$M & M/M/H$\rightarrow$M & H/H/H$\rightarrow$H & M/M/L$\rightarrow$M & M/M/M$\rightarrow$M & H/H/H$\rightarrow$H \\
  \midrule
  a road roller with a steel drum and rubber tyres     & M/M/M$\rightarrow$M & H/H/H$\rightarrow$H & L/L/L$\rightarrow$L & L/L/L$\rightarrow$L & H/H/H$\rightarrow$H & L/L/L$\rightarrow$L & L/M/M$\rightarrow$M & H/H/H$\rightarrow$H & L/L/L$\rightarrow$L \\*
  a jackhammer                                         & L/L/L$\rightarrow$L & M/H/M$\rightarrow$M & H/H/H$\rightarrow$H & M/M/M$\rightarrow$M & H/H/H$\rightarrow$H & H/H/H$\rightarrow$H & M/M/M$\rightarrow$M & H/H/H$\rightarrow$H & H/H/H$\rightarrow$H \\*
  an industrial robot arm                              & H/H/H$\rightarrow$H & H/H/H$\rightarrow$H & H/H/H$\rightarrow$H & M/M/M$\rightarrow$M & M/M/M$\rightarrow$M & H/H/H$\rightarrow$H & H/H/H$\rightarrow$H & H/H/H$\rightarrow$H & H/H/H$\rightarrow$H \\*
  an adjustable wrench                                 & M/H/H$\rightarrow$H & H/H/H$\rightarrow$H & H/H/H$\rightarrow$H & H/H/H$\rightarrow$H & H/H/H$\rightarrow$H & H/H/H$\rightarrow$H & M/H/H$\rightarrow$H & H/H/H$\rightarrow$H & H/H/H$\rightarrow$H \\*
  a claw hammer                                        & H/H/H$\rightarrow$H & L/L/M$\rightarrow$L & H/H/H$\rightarrow$H & H/H/H$\rightarrow$H & L/M/M$\rightarrow$M & H/M/H$\rightarrow$H & H/H/H$\rightarrow$H & H/H/H$\rightarrow$H & H/H/H$\rightarrow$H \\
  \midrule
  a dental crown                                       & M/M/M$\rightarrow$M & M/H/H$\rightarrow$H & H/H/H$\rightarrow$H & M/M/M$\rightarrow$M & H/H/H$\rightarrow$H & H/H/H$\rightarrow$H & M/M/M$\rightarrow$M & H/H/H$\rightarrow$H & H/H/H$\rightarrow$H \\*
  a knee implant                                       & M/M/M$\rightarrow$M & H/H/H$\rightarrow$H & H/H/H$\rightarrow$H & M/L/M$\rightarrow$M & H/M/M$\rightarrow$M & H/H/H$\rightarrow$H & L/L/L$\rightarrow$L & H/H/H$\rightarrow$H & L/M/M$\rightarrow$M \\*
  a walking cane with a rubber tip                     & M/M/M$\rightarrow$M & H/H/H$\rightarrow$H & L/M/M$\rightarrow$M & H/H/H$\rightarrow$H & H/H/H$\rightarrow$H & L/M/M$\rightarrow$M & M/H/H$\rightarrow$H & H/H/H$\rightarrow$H & L/M/M$\rightarrow$M \\*
  a human femur                                        & M/M/M$\rightarrow$M & H/H/H$\rightarrow$H & H/M/H$\rightarrow$H & M/M/M$\rightarrow$M & H/H/H$\rightarrow$H & H/M/H$\rightarrow$H & M/M/M$\rightarrow$M & H/H/H$\rightarrow$H & H/M/H$\rightarrow$H \\*
  a lumbar vertebra                                    & L/L/L$\rightarrow$L & H/H/H$\rightarrow$H & H/H/H$\rightarrow$H & L/L/L$\rightarrow$L & H/H/H$\rightarrow$H & M/M/M$\rightarrow$M & H/H/H$\rightarrow$H & H/H/H$\rightarrow$H & H/H/H$\rightarrow$H \\
  \midrule
  a skyscraper                                         & M/L/M$\rightarrow$M & H/H/H$\rightarrow$H & H/H/H$\rightarrow$H & M/L/M$\rightarrow$M & H/H/H$\rightarrow$H & H/H/H$\rightarrow$H & H/H/H$\rightarrow$H & H/H/H$\rightarrow$H & H/H/H$\rightarrow$H \\*
  a thatched stone cottage                             & M/M/M$\rightarrow$M & H/H/M$\rightarrow$H & H/H/H$\rightarrow$H & H/H/H$\rightarrow$H & H/H/H$\rightarrow$H & M/M/L$\rightarrow$M & H/H/H$\rightarrow$H & H/H/H$\rightarrow$H & H/M/M$\rightarrow$M \\*
  a lighthouse                                         & H/H/H$\rightarrow$H & H/M/H$\rightarrow$H & H/H/M$\rightarrow$H & M/H/H$\rightarrow$H & H/H/H$\rightarrow$H & H/H/H$\rightarrow$H & M/M/M$\rightarrow$M & H/H/H$\rightarrow$H & H/H/H$\rightarrow$H \\*
  an igloo                                             & H/M/H$\rightarrow$H & H/H/H$\rightarrow$H & M/M/L$\rightarrow$M & H/M/H$\rightarrow$H & H/H/H$\rightarrow$H & H/H/H$\rightarrow$H & M/L/M$\rightarrow$M & H/H/H$\rightarrow$H & H/H/H$\rightarrow$H \\*
  a log cabin                                          & H/H/H$\rightarrow$H & H/H/H$\rightarrow$H & H/H/H$\rightarrow$H & H/H/H$\rightarrow$H & H/H/H$\rightarrow$H & H/H/H$\rightarrow$H & H/H/H$\rightarrow$H & H/H/H$\rightarrow$H & H/H/H$\rightarrow$H \\
  \midrule
  the Leaning Tower of Pisa                            & L/L/L$\rightarrow$L & H/H/H$\rightarrow$H & H/H/H$\rightarrow$H & M/M/M$\rightarrow$M & H/H/H$\rightarrow$H & H/H/H$\rightarrow$H & M/M/M$\rightarrow$M & H/H/H$\rightarrow$H & H/H/H$\rightarrow$H \\*
  the Pantheon                                         & M/M/M$\rightarrow$M & H/H/H$\rightarrow$H & H/H/H$\rightarrow$H & M/M/M$\rightarrow$M & H/H/H$\rightarrow$H & H/H/H$\rightarrow$H & M/M/M$\rightarrow$M & H/H/H$\rightarrow$H & H/H/H$\rightarrow$H \\*
  Himeji Castle                                        & L/L/L$\rightarrow$L & H/H/H$\rightarrow$H & H/H/H$\rightarrow$H & L/M/M$\rightarrow$M & H/H/H$\rightarrow$H & L/L/L$\rightarrow$L & M/M/H$\rightarrow$M & H/H/H$\rightarrow$H & H/H/H$\rightarrow$H \\*
  a natural stone arch                                 & H/H/H$\rightarrow$H & H/H/H$\rightarrow$H & H/H/H$\rightarrow$H & H/H/H$\rightarrow$H & H/H/H$\rightarrow$H & H/H/H$\rightarrow$H & H/H/H$\rightarrow$H & H/H/H$\rightarrow$H & H/H/H$\rightarrow$H \\*
  the Taj Mahal                                        & H/H/M$\rightarrow$H & H/H/H$\rightarrow$H & M/H/M$\rightarrow$M & H/H/H$\rightarrow$H & H/H/H$\rightarrow$H & M/H/M$\rightarrow$M & M/M/H$\rightarrow$M & H/H/H$\rightarrow$H & H/H/H$\rightarrow$H \\
  \midrule
  Total: count of H/M/L of all 50 objects & 26/16/7 (1 N/A)       & 39/6/3 (2 N/A)        & 32/12/3 (3 N/A)       & 30/17/3               & 42/8/0                & 34/12/4               & 33/16/1               & 44/6/0                & 36/9/5 \\
\end{longtable}
\endgroup
\end{landscape}

\newpage
\clearpage

\section{Comparison of the diffuse Nitsche method with body-fitted finite elements on 50 generated objects}
\label{si:error-comparison}

The diffuse Nitsche method solves on an octree that does not conform to the surface and imposes boundary conditions weakly on the diffused zero level set. Therefore, to see how far its solutions lie from those of a conventional finite element analysis of the same problems, each of the 50 sessions run by the Opus 5.5 operator (Sec.~\ref{si:opus-results}) was solved again with the diffuse Nitsche method and with body-fitted linear tetrahedral finite elements. Tetrahedra are used because a mesh that conforms to a generated surface is far easier to build from tetrahedra than from hexahedra~\citeS{schneider2022large}. Linear tetrahedra, however, are stiffer than trilinear hexahedra in bending and lock more readily~\citeS{schneider2022large}, so part of the discrepancy can come from the linear tetrahedral solution. Neither solution is a converged reference, so the reported numbers are discrepancies between two approaches to one problem. Tables~\ref{tab:si-bf-dofs-d7}--\ref{tab:si-bf-cost-d8} give, for each object at final octree levels 7 and 8, the degrees of freedom, the discrepancies and the computational cost.

\paragraph{Shared settings.} Both methods use each AI operator session's surface, boundary condition regions, tractions and body force, the neo-Hookean material, Newton's method and double precision. Three settings differ from the sessions. First, the diffuse Nitsche solve runs at two final octree levels, $l_{\mathrm{final}}=7$ and $8$. Second, Newton's method stops at a relative residual of $10^{-6}$ instead of the default $10^{-2}$. Third, a session with per-part materials uses, for both methods, one uniform material with the part-volume-weighted mean $E$, $\nu$ and $\rho$ of its parts.

\paragraph{Body-fitted problem settings.} The surface is tetrahedralized with TetGen \citeS{hang2015tetgen}, and each boundary triangle of the tetrahedral mesh inherits the support or load label of the surface triangle it lies on. The problem is posed the way the diffuse Nitsche method poses it, so that the discrepancy reflects the discretization rather than the way the boundary conditions are imposed. (i) Every node within a distance $\delta=\max(1.5\varepsilon,0.75\,h_{\min})$ of a support is held in that support's constrained displacement components, with $\varepsilon$ the phase-field width and $h_{\min}$ the finest cell size, which reproduces the band in which the Nitsche penalty acts. (ii) Each traction is applied to the facets of its region in proportion to how closely each facet faces the region's mean normal, so that the total force equals the force applied by the diffuse Nitsche method after its ``Load correction'' (see the paragraph above in Supplementary Methods).

\paragraph{Body-fitted meshes and retries.} Three body-fitted meshes are built at each level. The \textit{matched DOFs} mesh is uniform, with an element size chosen so that its vertex count matches the node count of the trimmed octree. The \textit{matched surface} mesh is uniform with element size $h_{\min}$. The \textit{graded} mesh has the surface of the \textit{matched surface} mesh and an interior that coarsens like the octree, again with a vertex count matching the octree's node count. The sizes are found by a search over a few trials that keeps the closest node count, so a vertex count can miss its target count (for example, the Leaning Tower of Pisa in Table~\ref{tab:si-bf-dofs-d7}). Before TetGen, the surface is remeshed to the target element size. If the remeshed surface intersects itself, it is remeshed again closer to the original surface, and if it still intersects itself, the original surface is used without remeshing. If TetGen fails, it is called again with less strict options.

\paragraph{Solvers and retries.} Every solve first uses JAX-AMG on the GPU. If it fails, the diffuse Nitsche solve is retried along a fixed sequence of solver paths: (1) four times the Krylov iteration limit, (2) a second algebraic multigrid implementation (pyamgx), (3) the Nitsche penalty $\beta$ capped at $500E$, for a session with a uniform material only, (4) the system scaled to a unit diagonal, (5) direct linear solves on the CPU, (6) path 5 with 10 load steps and line search, and (7) path 6 at a relative residual tolerance of $10^{-2}$. Paths 1 to 4 follow the operator skill, whereas paths 5 to 7 were added for this comparison. A body-fitted solve takes paths 1, 2 and 4 on the GPU and then direct linear solves on the CPU. Large body-fitted meshes with $450\,000$ nodes or more go to the CPU directly. Four body-fitted solves that the CPU could not finish within its time and memory limits were solved on the GPU by a separate, lightweight Newton solver, which assembles the tangent stiffness matrix chunk by chunk and solves it with JAX-AMG.

\paragraph{Timing.} Times are wall-clock seconds and exclude just-in-time (JIT) compilation. Each table reports the time of the attempt that succeeded. The number in parentheses in the tables adds the time spent in failed attempts. The preparation time of the diffuse Nitsche method includes the boundary conditions, signed distance field, octree, trimming and solver setup. The preparation time of body-fitted solves includes surface remeshing, TetGen, boundary mapping and solver setup. All times are on one NVIDIA RTX A6000 GPU and an AMD Ryzen Threadripper 3970X CPU.

\paragraph{Error measures.} Integrals of error measures are taken over the body $\Omega$ enclosed by the surface. At each quadrature point of the body-fitted mesh, the body-fitted displacement $\mathbf{u}_{\mathrm{bf}}$ is evaluated in its tetrahedron, and the diffuse Nitsche displacement $\mathbf{u}_{\mathrm{N}}$ is interpolated trilinearly in the octree cell that contains the point. With the difference $\mathbf{e}=\mathbf{u}_{\mathrm{N}}-\mathbf{u}_{\mathrm{bf}}$, the displacement error is
\begin{equation}
  \eta=\frac{\lVert\mathbf{e}-\mathbf{r}^\ast\rVert_{L^2(\Omega)}}{\lVert\mathbf{u}_{\mathrm{bf}}\rVert_{L^2(\Omega)}},
  \label{eq:si-bf-etarig}
\end{equation}
where $\mathbf{r}^\ast$ is the rigid-body motion, a translation plus a rotation, that best fits $\mathbf{e}$ in the least-squares sense. Minor differences in the solution of hinge-like structures, such as the ankle of the crocodile, can leave a strain-free rigid motion between the two displacement solutions, which is expected as the diffuse Nitsche approach discretizes such thin hinges differently from a body-fitted mesh. Removing $\mathbf{r}^\ast$ discounts such rigid body motion. The $H^1$ error $\eta^{H^1}$ is defined in the same way, with the same $\mathbf{r}^\ast$ removed, but in the $H^1$ norm, which additionally measures the difference in displacement gradients. The compliance ratio $\ell_{\mathrm{N}}/\ell_{\mathrm{bf}}$ compares the work of the applied loads on the two displacements. A compliance ratio below one means that the diffuse Nitsche solution is stiffer. The stress ratios $s_{95}$ and $s_{99}$ are the diffuse-Nitsche-to-body-fitted ratios of 95th and 99th percentiles of the von Mises stress.

\paragraph{Missing and marked entries.} The surfaces generated for the car brake disc, Himeji Castle and the park bench self-intersect, so these three objects have no body-fitted entries. At level 8, the diffuse Nitsche solve of the loaf of bread converged only on path 7. It is marked with $^{\dagger}$ and excluded from the medians.

\paragraph{Flags.} Some objects have features close to the resolution limit of either method. Each object is therefore flagged for thin walls (T, an area-median wall thickness below $4h_{\min}$), narrow gaps (G, gaps narrower than $h_{\min}$ on more than $1\%$ of the surface), a load near a support (L, a load region within $3h_{\min}$ of a support), or a small or partially clamped support (S, a support piece of area below $50h_{\min}^2$, or a clamp of only some displacement components on a section thinner than $4h_{\min}$).

\begin{landscape}
\begin{table}[p]
\centering
\footnotesize
\setlength{\tabcolsep}{3.5pt}
\renewcommand{\arraystretch}{0.85}
\caption{Degrees of freedom (thousands) at final octree level 7: diffuse Nitsche (three per free node of the trimmed octree) and the three body-fitted meshes (three per vertex). Matched DOFs: a uniform mesh whose vertex count targets the octree's node count. Matched surface: a uniform mesh of element size $h_{\min}$. Graded: element size $h_{\min}$ at the surface and coarser inside, like the octree. Flags (Sec.~\ref{si:error-comparison}): T: thin walls, G: narrow gaps, L: a load near a support, S: a small or partially clamped support. ---: no body-fitted mesh, because the session surface self-intersects. }\label{tab:si-bf-dofs-d7}
\smallskip
\begin{tabular}{@{}>{\raggedright\arraybackslash}p{3.7cm}crrrr@{\hspace{1.6em}}>{\raggedright\arraybackslash}p{3.7cm}crrrr@{}}
\toprule
 &  &  & \multicolumn{3}{c}{Body-fitted} &  &  &  & \multicolumn{3}{c}{Body-fitted} \\
\cmidrule(lr){4-6} \cmidrule(l){10-12}
\multicolumn{1}{@{}l}{Object} & Flags & \begin{tabular}[b]{@{}r@{}}Diffuse\\Nitsche\end{tabular} & \begin{tabular}[b]{@{}r@{}}Matched\\DOFs\end{tabular} & \begin{tabular}[b]{@{}r@{}}Matched\\surface\end{tabular} & \begin{tabular}[b]{@{}r@{}}Graded\end{tabular} & Object & Flags & \begin{tabular}[b]{@{}r@{}}Diffuse\\Nitsche\end{tabular} & \begin{tabular}[b]{@{}r@{}}Matched\\DOFs\end{tabular} & \begin{tabular}[b]{@{}r@{}}Matched\\surface\end{tabular} & \begin{tabular}[b]{@{}r@{}}Graded\end{tabular} \\
\midrule
a TV remote control &  & 49.4 & 49.5 & 57.0 & 49.0 & a car suspension control arm with rubber bushings &  & 82.1 & 81.7 & 76.9 & 80.1 \\
a USB stick &  & 43.1 & 43.2 & 50.3 & 42.3 & a truck leaf spring &  & 16.0 & 16.0 & 7.00 & 15.2 \\
an electric toothbrush &  & 17.2 & 17.2 & 11.0 & 17.2 & a car wheel with its rubber tyre & G & 308 & 279 & 708 & 318 \\
a coffee mug & T & 260 & 260 & 249 & 256 & a car brake disc & S & 184 & --- & --- & --- \\
a computer mouse &  & 108 & 108 & 237 & 108 & a crankshaft & G & 65.3 & 64.8 & 68.9 & 64.6 \\
a school chair with a steel frame and a plastic seat & T, S & 83.2 & 59.6 & 33.0 & 82.7 & a road roller with a steel drum and rubber tyres & T, G, S & 101 & 101 & 64.1 & 101 \\
a glass-top dining table & T, S & 73.8 & 72.7 & 53.4 & 71.7 & a jackhammer & S & 21.9 & 21.3 & 12.7 & 21.4 \\
a park bench & T, S & 94.4 & --- & --- & --- & an industrial robot arm & G & 90.3 & 91.0 & 97.8 & 87.1 \\
a bar stool & S & 86.1 & 84.4 & 46.5 & 83.9 & an adjustable wrench & T, S & 18.7 & 18.3 & 8.36 & 18.8 \\
a sofa & S & 132 & 133 & 195 & 131 & a claw hammer &  & 27.7 & 27.2 & 18.4 & 27.2 \\
a teddy bear &  & 116 & 116 & 192 & 116 & a dental crown &  & 357 & 370 & 1300 & 364 \\
a wooden rocking horse & T & 98.4 & 97.1 & 63.0 & 95.7 & a knee implant & G & 291 & 296 & 654 & 287 \\
a pogo stick & S & 14.3 & 14.6 & 3.39 & 13.8 & a walking cane with a rubber tip & T, L, S & 8.27 & 8.42 & 1.99 & 8.09 \\
a wooden toy train engine & G, L, S & 103 & 104 & 149 & 102 & a human femur &  & 36.6 & 35.7 & 28.4 & 35.5 \\
a yo-yo & G & 339 & 371 & 761 & 337 & a lumbar vertebra &  & 99.6 & 99.3 & 142 & 99.3 \\
a jelly dessert &  & 200 & 202 & 609 & 200 & a skyscraper & L & 52.8 & 53.0 & 80.8 & 51.8 \\
a loaf of bread &  & 235 & 238 & 843 & 247 & a thatched stone cottage & L & 219 & 221 & 720 & 218 \\
a hamburger & G & 237 & 238 & 837 & 248 & a lighthouse &  & 99.7 & 91.0 & 160 & 97.0 \\
a layered birthday cake &  & 196 & 197 & 570 & 195 & an igloo &  & 245 & 245 & 292 & 244 \\
a slice of watermelon &  & 113 & 113 & 245 & 113 & a log cabin & L & 264 & 270 & 814 & 263 \\
a giraffe & S & 36.4 & 35.3 & 23.9 & 35.1 & the Leaning Tower of Pisa & T, G, S & 86.0 & 182 & 228 & 127 \\
an elephant & S & 95.8 & 96.3 & 127 & 95.5 & the Pantheon & L & 206 & 207 & 256 & 206 \\
a rhinoceros & S & 79.1 & 79.8 & 114 & 78.5 & Himeji Castle & G & 313 & --- & --- & --- \\
a horse & S & 86.1 & 85.9 & 95.8 & 85.3 & a natural stone arch & G & 193 & 196 & 353 & 192 \\
a crocodile & S & 40.3 & 39.6 & 33.7 & 38.6 & the Taj Mahal &  & 209 & 213 & 422 & 207 \\
\midrule
Median (all objects) &  & 99.0 & 97.1 & 127 & 97.0 & Median (unflagged) &  & 104 & 104 & 176 & 103 \\
\hspace{1em}$n$ &  & 50 & 47 & 47 & 47 & \hspace{1em}$n$ &  & 18 & 18 & 18 & 18 \\
\bottomrule
\end{tabular}
\end{table}
\end{landscape}

\begin{landscape}
\footnotesize
\setlength{\tabcolsep}{2.5pt}
\ifdefined\SIbfColUnit\else\newlength{\SIbfColUnit}\fi
\setlength{\SIbfColUnit}{\dimexpr(\linewidth-4.6cm-1.05cm-32\tabcolsep)*100/1500\relax}
\setlength{\LTleft}{\fill}\setlength{\LTright}{\fill}
\setlength{\LTcapwidth}{\linewidth}
\begin{longtable}{@{}>{\raggedright\arraybackslash}p{4.6cm}>{\centering\arraybackslash}p{1.05cm}>{\raggedleft\arraybackslash}p{1\SIbfColUnit}>{\raggedleft\arraybackslash}p{1\SIbfColUnit}>{\raggedleft\arraybackslash}p{1\SIbfColUnit}>{\raggedleft\arraybackslash}p{1\SIbfColUnit}>{\raggedleft\arraybackslash}p{1\SIbfColUnit}>{\raggedleft\arraybackslash}p{1\SIbfColUnit}>{\raggedleft\arraybackslash}p{1\SIbfColUnit}>{\raggedleft\arraybackslash}p{1\SIbfColUnit}>{\raggedleft\arraybackslash}p{1\SIbfColUnit}>{\raggedleft\arraybackslash}p{1\SIbfColUnit}>{\raggedleft\arraybackslash}p{1\SIbfColUnit}>{\raggedleft\arraybackslash}p{1\SIbfColUnit}>{\raggedleft\arraybackslash}p{1\SIbfColUnit}>{\raggedleft\arraybackslash}p{1\SIbfColUnit}>{\raggedleft\arraybackslash}p{1\SIbfColUnit}@{}}
\caption{Discrepancy between the diffuse Nitsche solution and each body-fitted solution at final octree level 7. $\eta$ and $\eta^{H^1}$: relative $L^2$ and $H^1$ norms of the displacement difference after removal of its best-fit rigid motion (\%). $\ell_{\mathrm N}/\ell_{\mathrm{bf}}$: compliance ratio. $s_{95}$ and $s_{99}$: diffuse-Nitsche-to-body-fitted ratios of the 95th and 99th percentiles of the von Mises stress. Gray rows: $\eta$ above 10\% against at least one body-fitted mesh. Flags (Sec.~\ref{si:error-comparison}): T: thin walls, G: narrow gaps, L: a load near a support, S: a small or partially clamped support. ---: no body-fitted mesh, because the session surface self-intersects.}\label{tab:si-bf-accuracy-d7}\\
\toprule
 &  & \multicolumn{15}{c}{Body-fitted} \\
\cmidrule(l){3-17}
 &  & \multicolumn{5}{c}{Matched DOFs} & \multicolumn{5}{c}{Matched surface} & \multicolumn{5}{c}{Graded} \\
\cmidrule(lr){3-7} \cmidrule(lr){8-12} \cmidrule(l){13-17}
\multicolumn{1}{@{}l}{Object} & Flags & \begin{tabular}[b]{@{}r@{}}$\eta$\\(\%)\end{tabular} & \begin{tabular}[b]{@{}r@{}}$\eta^{H^1}$\\(\%)\end{tabular} & $\ell_{\mathrm N}/\ell_{\mathrm{bf}}$ & $s_{95}$ & $s_{99}$ & \begin{tabular}[b]{@{}r@{}}$\eta$\\(\%)\end{tabular} & \begin{tabular}[b]{@{}r@{}}$\eta^{H^1}$\\(\%)\end{tabular} & $\ell_{\mathrm N}/\ell_{\mathrm{bf}}$ & $s_{95}$ & $s_{99}$ & \begin{tabular}[b]{@{}r@{}}$\eta$\\(\%)\end{tabular} & \begin{tabular}[b]{@{}r@{}}$\eta^{H^1}$\\(\%)\end{tabular} & $\ell_{\mathrm N}/\ell_{\mathrm{bf}}$ & $s_{95}$ & $s_{99}$ \\
\midrule
\endfirsthead
\caption[]{Discrepancy at final octree level 7 (continued).}\\
\toprule
 &  & \multicolumn{15}{c}{Body-fitted} \\
\cmidrule(l){3-17}
 &  & \multicolumn{5}{c}{Matched DOFs} & \multicolumn{5}{c}{Matched surface} & \multicolumn{5}{c}{Graded} \\
\cmidrule(lr){3-7} \cmidrule(lr){8-12} \cmidrule(l){13-17}
\multicolumn{1}{@{}l}{Object} & Flags & \begin{tabular}[b]{@{}r@{}}$\eta$\\(\%)\end{tabular} & \begin{tabular}[b]{@{}r@{}}$\eta^{H^1}$\\(\%)\end{tabular} & $\ell_{\mathrm N}/\ell_{\mathrm{bf}}$ & $s_{95}$ & $s_{99}$ & \begin{tabular}[b]{@{}r@{}}$\eta$\\(\%)\end{tabular} & \begin{tabular}[b]{@{}r@{}}$\eta^{H^1}$\\(\%)\end{tabular} & $\ell_{\mathrm N}/\ell_{\mathrm{bf}}$ & $s_{95}$ & $s_{99}$ & \begin{tabular}[b]{@{}r@{}}$\eta$\\(\%)\end{tabular} & \begin{tabular}[b]{@{}r@{}}$\eta^{H^1}$\\(\%)\end{tabular} & $\ell_{\mathrm N}/\ell_{\mathrm{bf}}$ & $s_{95}$ & $s_{99}$ \\
\midrule
\endhead
\midrule
\multicolumn{17}{@{}r@{}}{\textit{Continued on next page}}\\
\endfoot
\bottomrule
\endlastfoot
a TV remote control &  & 3.7 & 7.0 & 0.909 & 0.924 & 0.893 & 2.9 & 5.8 & 0.935 & 0.942 & 0.910 & 2.8 & 5.7 & 0.937 & 0.943 & 0.912 \\
a USB stick &  & 1.4 & 4.4 & 0.871 & 0.886 & 0.841 & 1.1 & 3.8 & 0.886 & 0.904 & 0.847 & 1.0 & 3.6 & 0.886 & 0.907 & 0.851 \\
\rowcolor{gray!15}an electric toothbrush &  & 15 & 22 & 0.773 & 0.828 & 0.755 & 8.4 & 13 & 0.876 & 0.858 & 0.823 & 12 & 17 & 0.824 & 0.826 & 0.761 \\
a coffee mug & T & 4.7 & 10 & 0.869 & 0.885 & 0.842 & 4.6 & 10 & 0.872 & 0.887 & 0.840 & 4.8 & 10 & 0.868 & 0.884 & 0.837 \\
a computer mouse &  & 7.0 & 21 & 0.935 & 0.972 & 0.937 & 6.4 & 21 & 0.945 & 0.974 & 0.933 & 6.1 & 21 & 0.949 & 0.977 & 0.932 \\
\rowcolor{gray!15}a school chair with a steel frame and a plastic seat & T, S & 34 & 46 & 0.545 & 0.618 & 0.542 & 32 & 42 & 0.599 & 0.649 & 0.546 & 35 & 47 & 0.533 & 0.605 & 0.513 \\
\rowcolor{gray!15}a glass-top dining table & T, S & 16 & 28 & 0.852 & 0.893 & 0.953 & 12 & 21 & 0.907 & 0.940 & 1.003 & 15 & 26 & 0.870 & 0.909 & 0.969 \\
a park bench & T, S & --- & --- & --- & --- & --- & --- & --- & --- & --- & --- & --- & --- & --- & --- & --- \\
\rowcolor{gray!15}a bar stool & S & 15 & 30 & 0.862 & 0.911 & 0.818 & 12 & 27 & 0.865 & 0.901 & 0.831 & 15 & 31 & 0.845 & 0.888 & 0.803 \\
a sofa & S & 1.9 & 9.2 & 0.937 & 0.908 & 0.920 & 2.1 & 9.7 & 0.939 & 0.904 & 0.918 & 2.0 & 9.6 & 0.943 & 0.912 & 0.911 \\
a teddy bear &  & 1.5 & 4.2 & 0.831 & 0.878 & 0.893 & 1.6 & 4.3 & 0.837 & 0.927 & 0.877 & 1.6 & 4.3 & 0.839 & 0.928 & 0.876 \\
a wooden rocking horse & T & 9.1 & 16 & 0.857 & 0.883 & 0.888 & 8.9 & 15 & 0.857 & 0.887 & 0.881 & 10 & 17 & 0.844 & 0.873 & 0.873 \\
\rowcolor{gray!15}a pogo stick & S & 20 & 71 & 0.864 & 0.782 & 1.068 & 9.5 & 35 & 0.881 & 0.766 & 1.017 & 9.1 & 35 & 0.845 & 0.732 & 1.014 \\
a wooden toy train engine & G, L, S & 8.4 & 42 & 0.883 & 0.950 & 0.823 & 8.4 & 42 & 0.886 & 0.952 & 0.822 & 8.0 & 42 & 0.886 & 0.949 & 0.825 \\
\rowcolor{gray!15}a yo-yo & G & 21 & 61 & 0.461 & 0.676 & 0.565 & 21 & 62 & 0.454 & 0.669 & 0.554 & 21 & 61 & 0.460 & 0.673 & 0.561 \\
a jelly dessert &  & 2.7 & 23 & 0.957 & 0.958 & 0.963 & 3.2 & 22 & 0.963 & 0.948 & 0.960 & 2.6 & 23 & 0.969 & 0.962 & 0.974 \\
a loaf of bread &  & 4.8 & 23 & 0.886 & 0.924 & 0.863 & 4.5 & 23 & 0.902 & 0.938 & 0.877 & 4.2 & 23 & 0.907 & 0.944 & 0.879 \\
a hamburger & G & 7.1 & 23 & 0.911 & 0.937 & 0.859 & 7.6 & 24 & 0.908 & 0.927 & 0.843 & 5.8 & 24 & 0.930 & 0.944 & 0.869 \\
a layered birthday cake &  & 3.0 & 19 & 0.976 & 0.969 & 0.979 & 2.8 & 20 & 0.985 & 0.961 & 0.978 & 2.4 & 21 & 0.989 & 0.962 & 0.977 \\
a slice of watermelon &  & 4.0 & 21 & 0.923 & 0.951 & 0.897 & 3.7 & 20 & 0.926 & 0.949 & 0.886 & 3.4 & 20 & 0.931 & 0.953 & 0.889 \\
a giraffe & S & 2.7 & 7.5 & 1.016 & 0.946 & 1.180 & 2.1 & 6.1 & 1.030 & 0.939 & 1.146 & 2.5 & 7.1 & 1.019 & 0.932 & 1.152 \\
an elephant & S & 1.1 & 4.1 & 0.981 & 0.974 & 0.981 & 1.2 & 4.2 & 0.984 & 0.979 & 0.984 & 1.2 & 4.2 & 0.984 & 0.976 & 0.986 \\
a rhinoceros & S & 2.2 & 8.3 & 0.945 & 0.940 & 0.976 & 2.3 & 8.4 & 0.952 & 0.947 & 0.976 & 2.3 & 8.4 & 0.954 & 0.954 & 0.977 \\
\rowcolor{gray!15}a horse & S & 28 & 62 & 0.861 & 0.915 & 0.942 & 27 & 59 & 0.859 & 0.909 & 0.942 & 27 & 59 & 0.848 & 0.910 & 0.942 \\
a crocodile & S & 8.0 & 16 & 0.634 & 0.759 & 0.602 & 8.0 & 16 & 0.637 & 0.747 & 0.598 & 8.1 & 17 & 0.627 & 0.741 & 0.608 \\
a car suspension control arm with rubber bushings &  & 7.5 & 23 & 0.912 & 0.934 & 0.937 & 4.6 & 16 & 0.893 & 0.916 & 0.909 & 4.6 & 16 & 0.891 & 0.915 & 0.908 \\
\rowcolor{gray!15}a truck leaf spring &  & 24 & 37 & 0.710 & 0.693 & 0.642 & 17 & 27 & 0.793 & 0.742 & 0.727 & 22 & 33 & 0.736 & 0.689 & 0.672 \\
a car wheel with its rubber tyre & G & 7.4 & 16 & 0.822 & 0.901 & 0.816 & 8.3 & 18 & 0.799 & 0.882 & 0.795 & 7.7 & 17 & 0.816 & 0.897 & 0.811 \\
a car brake disc & S & --- & --- & --- & --- & --- & --- & --- & --- & --- & --- & --- & --- & --- & --- & --- \\
\rowcolor{gray!15}a crankshaft & G & 12 & 31 & 0.469 & 0.511 & 0.445 & 12 & 30 & 0.494 & 0.532 & 0.475 & 12 & 30 & 0.494 & 0.532 & 0.469 \\
\rowcolor{gray!15}a road roller with a steel drum and rubber tyres & T, G, S & 51 & 57 & 0.372 & 0.550 & 0.807 & 47 & 54 & 0.399 & 0.571 & 0.841 & 49 & 56 & 0.384 & 0.558 & 0.825 \\
\rowcolor{gray!15}a jackhammer & S & 6.6 & 24 & 0.547 & 0.856 & 0.539 & 11 & 32 & 0.605 & 0.821 & 0.542 & 13 & 38 & 0.549 & 0.814 & 0.494 \\
\rowcolor{gray!15}an industrial robot arm & G & 23 & 54 & 0.424 & 0.649 & 0.573 & 23 & 54 & 0.417 & 0.640 & 0.564 & 23 & 54 & 0.418 & 0.641 & 0.565 \\
an adjustable wrench & T, S & 2.1 & 7.1 & 1.077 & 1.099 & 1.092 & 2.6 & 8.8 & 1.094 & 1.112 & 1.105 & 2.5 & 8.7 & 1.083 & 1.103 & 1.091 \\
a claw hammer &  & 1.6 & 5.9 & 0.879 & 0.874 & 0.868 & 1.8 & 6.5 & 0.872 & 0.866 & 0.855 & 2.0 & 7.1 & 0.862 & 0.860 & 0.844 \\
a dental crown &  & 7.7 & 24 & 0.943 & 0.866 & 0.811 & 7.7 & 25 & 0.945 & 0.860 & 0.807 & 7.5 & 25 & 0.946 & 0.864 & 0.808 \\
\rowcolor{gray!15}a knee implant & G & 80 & 99 & 0.403 & 0.437 & 0.340 & 80 & 99 & 0.398 & 0.432 & 0.337 & 80 & 99 & 0.406 & 0.438 & 0.341 \\
\rowcolor{gray!15}a walking cane with a rubber tip & T, L, S & 24 & 39 & 0.921 & 0.807 & 0.741 & 18 & 30 & 0.938 & 0.862 & 0.741 & 21 & 35 & 0.925 & 0.837 & 0.703 \\
a human femur &  & 0.64 & 2.2 & 0.974 & 0.973 & 0.963 & 1.2 & 3.2 & 0.956 & 0.955 & 0.953 & 1.4 & 3.7 & 0.948 & 0.941 & 0.936 \\
a lumbar vertebra &  & 6.9 & 23 & 0.933 & 0.898 & 0.885 & 6.7 & 23 & 0.946 & 0.902 & 0.889 & 6.6 & 23 & 0.946 & 0.903 & 0.890 \\
a skyscraper & L & 0.35 & 0.37 & 0.992 & 0.981 & 0.979 & 0.34 & 0.37 & 1.004 & 0.980 & 0.983 & 0.32 & 0.34 & 1.003 & 0.981 & 0.982 \\
a thatched stone cottage & L & 0.73 & 4.9 & 0.983 & 0.995 & 1.006 & 0.62 & 5.3 & 1.012 & 0.990 & 0.987 & 0.59 & 5.4 & 1.009 & 0.992 & 0.992 \\
a lighthouse &  & 1.4 & 2.2 & 0.985 & 0.989 & 0.980 & 1.8 & 2.6 & 0.995 & 0.988 & 0.986 & 1.8 & 2.6 & 0.995 & 0.988 & 0.988 \\
an igloo &  & 2.4 & 9.1 & 0.986 & 0.965 & 0.949 & 2.4 & 9.2 & 0.988 & 0.965 & 0.948 & 2.4 & 9.3 & 0.988 & 0.965 & 0.949 \\
a log cabin & L & 2.3 & 12 & 0.916 & 0.942 & 0.886 & 2.9 & 15 & 0.909 & 0.938 & 0.883 & 2.7 & 14 & 0.913 & 0.944 & 0.883 \\
the Leaning Tower of Pisa & T, G, S & 4.4 & 4.9 & 0.946 & 0.958 & 0.962 & 4.3 & 4.9 & 0.942 & 0.952 & 0.959 & 4.3 & 4.8 & 0.968 & 1.010 & 0.977 \\
the Pantheon & L & 2.3 & 2.5 & 0.998 & 0.981 & 0.937 & 2.3 & 2.5 & 1.002 & 0.978 & 0.951 & 2.2 & 2.4 & 1.001 & 0.978 & 0.944 \\
Himeji Castle & G & --- & --- & --- & --- & --- & --- & --- & --- & --- & --- & --- & --- & --- & --- & --- \\*
a natural stone arch & G & 9.0 & 11 & 0.686 & 0.798 & 0.700 & 9.1 & 11 & 0.685 & 0.796 & 0.692 & 9.0 & 11 & 0.687 & 0.796 & 0.694 \\*
the Taj Mahal &  & 2.1 & 2.4 & 0.976 & 0.978 & 0.972 & 1.8 & 2.2 & 1.029 & 0.981 & 0.970 & 1.6 & 2.0 & 1.013 & 0.985 & 0.964 \\*
\noalign{\penalty10000\vskip\aboverulesep\hrule height\lightrulewidth\penalty10000\vskip\belowrulesep}
Median (all objects) &  & 6.6 & 19 & 0.909 & 0.911 & 0.888 & 4.6 & 16 & 0.907 & 0.916 & 0.883 & 4.8 & 17 & 0.907 & 0.912 & 0.883 \\*
\hspace{1em}$n$ &  & 47 & 47 & 47 & 47 & 47 & 47 & 47 & 47 & 47 & 47 & 47 & 47 & 47 & 47 & 47 \\*
\addlinespace
Median (unflagged) &  & 3.4 & 20 & 0.928 & 0.929 & 0.895 & 3.0 & 14 & 0.940 & 0.940 & 0.899 & 2.7 & 17 & 0.941 & 0.942 & 0.899 \\*
\hspace{1em}$n$ &  & 18 & 18 & 18 & 18 & 18 & 18 & 18 & 18 & 18 & 18 & 18 & 18 & 18 & 18 & 18 \\
\end{longtable}
\end{landscape}

\begin{landscape}
\footnotesize
\setlength{\tabcolsep}{2pt}
\ifdefined\SIbfColUnit\else\newlength{\SIbfColUnit}\fi
\setlength{\SIbfColUnit}{\dimexpr(\linewidth-3.0cm-1.05cm-40\tabcolsep)*100/2041\relax}
\setlength{\LTleft}{\fill}\setlength{\LTright}{\fill}
\setlength{\LTcapwidth}{\linewidth}
\begin{longtable}{@{}>{\raggedright\arraybackslash}p{3.0cm}>{\centering\arraybackslash}p{1.05cm}>{\centering\arraybackslash}p{1.03\SIbfColUnit}>{\raggedleft\arraybackslash}p{1.0\SIbfColUnit}>{\raggedleft\arraybackslash}p{1.3\SIbfColUnit}>{\raggedleft\arraybackslash}p{1.0\SIbfColUnit}>{\centering\arraybackslash}p{1.03\SIbfColUnit}>{\centering\arraybackslash}p{1.03\SIbfColUnit}>{\raggedleft\arraybackslash}p{1.0\SIbfColUnit}>{\raggedleft\arraybackslash}p{1.3\SIbfColUnit}>{\raggedleft\arraybackslash}p{1.0\SIbfColUnit}>{\centering\arraybackslash}p{1.03\SIbfColUnit}>{\centering\arraybackslash}p{1.03\SIbfColUnit}>{\raggedleft\arraybackslash}p{1.0\SIbfColUnit}>{\raggedleft\arraybackslash}p{1.3\SIbfColUnit}>{\raggedleft\arraybackslash}p{1.0\SIbfColUnit}>{\centering\arraybackslash}p{1.03\SIbfColUnit}>{\centering\arraybackslash}p{1.03\SIbfColUnit}>{\raggedleft\arraybackslash}p{1.0\SIbfColUnit}>{\raggedleft\arraybackslash}p{1.3\SIbfColUnit}>{\raggedleft\arraybackslash}p{1.0\SIbfColUnit}@{}}
\caption{Computational cost at final octree level 7: wall-clock times (in seconds) of the successful attempt. In parentheses: the total over all attempts, shown when failed attempts add more than 0.5\%. Solver retries: the number of attempts after the first solve. Mesh retries: failed TetGen calls plus rejected remeshes of the surface that the kept mesh was built from. Solve to $10^{-6}$: Newton's method to a relative residual of $10^{-6}$. Solve to $10^{-2}$: the time that the same solve takes to first reach a relative residual of $10^{-2}$. \textsuperscript{C}: solved with direct linear solves on the CPU. \textsuperscript{R}: solved by the lightweight GPU Newton solver. Flags (Sec.~\ref{si:error-comparison}): T: thin walls, G: narrow gaps, L: a load near a support, S: a small or partially clamped support. ---: no body-fitted mesh, because the session surface self-intersects. }\label{tab:si-bf-cost-d7}\\
\toprule
 &  &  &  &  &  & \multicolumn{15}{c}{Body-fitted} \\
\cmidrule(l){7-21}
 &  & \multicolumn{4}{c}{Diffuse Nitsche} & \multicolumn{5}{c}{Matched DOFs} & \multicolumn{5}{c}{Matched surface} & \multicolumn{5}{c}{Graded} \\
\cmidrule(lr){3-6} \cmidrule(lr){7-11} \cmidrule(lr){12-16} \cmidrule(l){17-21}
\multicolumn{1}{@{}l}{Object} & Flags & \begin{tabular}[b]{@{}c@{}}Solver\\retries\end{tabular} & \begin{tabular}[b]{@{}r@{}}Prep.\\(s)\end{tabular} & \begin{tabular}[b]{@{}r@{}}Solve\\to\\$10^{-6}$\\(s)\end{tabular} & \begin{tabular}[b]{@{}r@{}}Solve\\to\\$10^{-2}$\\(s)\end{tabular} & \begin{tabular}[b]{@{}c@{}}Mesh\\retries\end{tabular} & \begin{tabular}[b]{@{}c@{}}Solver\\retries\end{tabular} & \begin{tabular}[b]{@{}r@{}}Prep.\\(s)\end{tabular} & \begin{tabular}[b]{@{}r@{}}Solve\\to\\$10^{-6}$\\(s)\end{tabular} & \begin{tabular}[b]{@{}r@{}}Solve\\to\\$10^{-2}$\\(s)\end{tabular} & \begin{tabular}[b]{@{}c@{}}Mesh\\retries\end{tabular} & \begin{tabular}[b]{@{}c@{}}Solver\\retries\end{tabular} & \begin{tabular}[b]{@{}r@{}}Prep.\\(s)\end{tabular} & \begin{tabular}[b]{@{}r@{}}Solve\\to\\$10^{-6}$\\(s)\end{tabular} & \begin{tabular}[b]{@{}r@{}}Solve\\to\\$10^{-2}$\\(s)\end{tabular} & \begin{tabular}[b]{@{}c@{}}Mesh\\retries\end{tabular} & \begin{tabular}[b]{@{}c@{}}Solver\\retries\end{tabular} & \begin{tabular}[b]{@{}r@{}}Prep.\\(s)\end{tabular} & \begin{tabular}[b]{@{}r@{}}Solve\\to\\$10^{-6}$\\(s)\end{tabular} & \begin{tabular}[b]{@{}r@{}}Solve\\to\\$10^{-2}$\\(s)\end{tabular} \\
\midrule
\endfirsthead
\caption[]{Computational cost at final octree level 7 (continued).}\\
\toprule
 &  &  &  &  &  & \multicolumn{15}{c}{Body-fitted} \\
\cmidrule(l){7-21}
 &  & \multicolumn{4}{c}{Diffuse Nitsche} & \multicolumn{5}{c}{Matched DOFs} & \multicolumn{5}{c}{Matched surface} & \multicolumn{5}{c}{Graded} \\
\cmidrule(lr){3-6} \cmidrule(lr){7-11} \cmidrule(lr){12-16} \cmidrule(l){17-21}
\multicolumn{1}{@{}l}{Object} & Flags & \begin{tabular}[b]{@{}c@{}}Solver\\retries\end{tabular} & \begin{tabular}[b]{@{}r@{}}Prep.\\(s)\end{tabular} & \begin{tabular}[b]{@{}r@{}}Solve\\to\\$10^{-6}$\\(s)\end{tabular} & \begin{tabular}[b]{@{}r@{}}Solve\\to\\$10^{-2}$\\(s)\end{tabular} & \begin{tabular}[b]{@{}c@{}}Mesh\\retries\end{tabular} & \begin{tabular}[b]{@{}c@{}}Solver\\retries\end{tabular} & \begin{tabular}[b]{@{}r@{}}Prep.\\(s)\end{tabular} & \begin{tabular}[b]{@{}r@{}}Solve\\to\\$10^{-6}$\\(s)\end{tabular} & \begin{tabular}[b]{@{}r@{}}Solve\\to\\$10^{-2}$\\(s)\end{tabular} & \begin{tabular}[b]{@{}c@{}}Mesh\\retries\end{tabular} & \begin{tabular}[b]{@{}c@{}}Solver\\retries\end{tabular} & \begin{tabular}[b]{@{}r@{}}Prep.\\(s)\end{tabular} & \begin{tabular}[b]{@{}r@{}}Solve\\to\\$10^{-6}$\\(s)\end{tabular} & \begin{tabular}[b]{@{}r@{}}Solve\\to\\$10^{-2}$\\(s)\end{tabular} & \begin{tabular}[b]{@{}c@{}}Mesh\\retries\end{tabular} & \begin{tabular}[b]{@{}c@{}}Solver\\retries\end{tabular} & \begin{tabular}[b]{@{}r@{}}Prep.\\(s)\end{tabular} & \begin{tabular}[b]{@{}r@{}}Solve\\to\\$10^{-6}$\\(s)\end{tabular} & \begin{tabular}[b]{@{}r@{}}Solve\\to\\$10^{-2}$\\(s)\end{tabular} \\
\midrule
\endhead
\midrule
\multicolumn{21}{@{}r@{}}{\textit{Continued on next page}}\\
\endfoot
\bottomrule
\endlastfoot
a TV remote control &  & 0 & 0.336 & 0.725 & 0.725 & 0 & 0 & 4.46 & 0.654 & 0.654 & 0 & 0 & 4.71 & 0.924 & 0.704 & 0 & 0 & 4.58 & 0.646 & 0.646 \\
a USB stick &  & 0 & 0.305 & 0.868 & 0.603 & 0 & 0 & 4.26 & 0.844 & 0.644 & 0 & 0 & 4.36 & 0.963 & 0.710 & 0 & 0 & 4.47 & 1.09 & 0.630 \\
an electric toothbrush &  & 0 & 0.172 & 0.939 & 0.785 & 0 & 0 & 3.63 & 2.17 & 1.44 & 0 & 0 & 3.55 & 1.60 & 1.15 & 0 & 0 & 3.83 & 1.87 & 1.50 \\
a coffee mug & T & 0 & 1.40 & 4.13 & 2.22 & 0 & 0 & 14.5 & 4.36 & 1.65 & 0 & 0 & 14.1 & 4.14 & 1.58 & 0 & 0 & 16.3 & 4.25 & 1.63 \\
a computer mouse &  & 0 & 0.585 & 0.731 & 0.411 & 0 & 0 & 6.67 & 0.800 & 0.525 & 0 & 0 & 11.4 & 1.50 & 0.950 & 0 & 0 & 7.39 & 1.08 & 0.747 \\
a school chair with a steel frame and a plastic seat & T, S & 0 & 0.491 & 3.60 & 2.51 & 0 & 1 & 6.55 & 2.96 (4.13) & 1.79 & 0 & 1 & 5.78 & 2.54 (3.53) & 1.52 & 0 & 1 & 9.40 & 4.55 (5.91) & 2.29 \\
a glass-top dining table & T, S & 0 & 0.677 & 2.77 & 1.80 & 0 & 0 & 8.37 & 2.25 & 1.61 & 0 & 0 & 16.7 & 1.83 & 1.31 & 0 & 0 & 15.1 & 1.82 & 1.26 \\
a park bench & T, S & 0 & 0.695 & 3.40 & 2.27 & --- & --- & --- & --- & --- & --- & --- & --- & --- & --- & --- & --- & --- & --- & --- \\
a bar stool & S & 0 & 0.498 & 2.64 & 1.72 & 0 & 0 & 6.47 & 2.92 & 1.77 & 0 & 1 & 5.42 & 1.93 (3.04) & 1.34 & 0 & 1 & 7.54 & 2.73 (4.11) & 1.88 \\
a sofa & S & 0 & 2.33 & 5.79 & 3.07 & 0 & 0 & 7.97 & 1.42 & 0.870 & 0 & 0 & 10.5 & 1.98 & 1.21 & 0 & 0 & 9.10 & 1.39 & 0.867 \\
a teddy bear &  & 0 & 0.603 & 3.08 & 1.89 & 0 & 0 & 7.30 & 2.82 & 2.30 & 0 & 0 & 11.4 & 4.37 & 3.57 & 0 & 0 & 8.26 & 2.81 & 2.30 \\
a wooden rocking horse & T & 0 & 0.573 & 1.42 & 1.42 & 0 & 0 & 6.79 & 2.17 & 1.51 & 0 & 0 & 5.79 & 1.58 & 1.12 & 0 & 0 & 7.87 & 2.18 & 1.51 \\
a pogo stick & S & 0 & 0.173 & 0.766 & 0.489 & 0 & 0 & 3.67 & 2.17 & 0.704 & 0 & 0 & 3.55 & 0.991 & 0.478 & 0 & 0 & 3.87 & 1.77 & 0.712 \\
a wooden toy train engine & G, L, S & 0 & 0.713 & 1.10 & 0.632 & 0 & 0 & 12.0 & 0.988 & 0.645 & 0 & 0 & 19.8 & 1.58 & 1.06 & 0 & 0 & 14.7 & 1.28 & 0.854 \\
a yo-yo & G & 0 & 1.70 & 3.10 & 1.63 & 0 & 0 & 18.4 & 4.67 & 1.93 & 0 & 0 & 33.8 & 10.3 & 7.27 & 0 & 0 & 19.3 & 3.15 & 1.85 \\
a jelly dessert &  & 0 & 2.16 & 6.21 & 5.25 & 0 & 0 & 25.1 & 2.21 & 1.68 & 0 & 0 & 58.3 & 8.80 & 7.05 & 0 & 0 & 17.5 & 3.98 & 3.25 \\
a loaf of bread &  & 0 & 1.17 & 24.5 & 1.52 & 0 & 0 & 13.2 & 11.5 & 1.62 & 0 & 0 & 36.2 & 40.1 & 5.47 & 0 & 0 & 15.1 & 11.4 & 1.60 \\
a hamburger & G & 0 & 1.18 & 21.8 & 1.53 & 0 & 0 & 13.0 & 11.8 & 1.58 & 0 & 0 & 34.7 & 43.9 & 5.33 & 0 & 0 & 14.9 & 11.7 & 1.58 \\
a layered birthday cake &  & 0 & 1.60 & 6.03 & 4.34 & 0 & 0 & 11.3 & 4.37 & 2.99 & 0 & 0 & 36.3 & 11.0 & 6.83 & 0 & 0 & 26.8 & 3.27 & 2.26 \\
a slice of watermelon &  & 0 & 0.643 & 1.41 & 0.749 & 0 & 0 & 6.86 & 1.02 & 0.577 & 0 & 0 & 11.6 & 2.00 & 1.08 & 0 & 0 & 7.95 & 0.992 & 0.563 \\
a giraffe & S & 0 & 0.260 & 1.26 & 0.865 & 0 & 0 & 4.22 & 1.58 & 1.12 & 0 & 0 & 4.08 & 1.35 & 0.956 & 0 & 0 & 4.65 & 2.08 & 1.14 \\
an elephant & S & 0 & 0.508 & 1.48 & 0.788 & 0 & 0 & 6.41 & 1.34 & 0.736 & 0 & 0 & 7.35 & 1.61 & 0.850 & 0 & 0 & 7.13 & 1.31 & 0.698 \\
a rhinoceros & S & 0 & 0.459 & 0.864 & 0.477 & 0 & 0 & 5.66 & 1.40 & 0.614 & 0 & 0 & 6.62 & 1.79 & 0.775 & 0 & 0 & 6.08 & 1.00 & 0.616 \\
a horse & S & 0 & 0.668 & 2.79 & 1.77 & 0 & 0 & 14.6 & 3.78 & 2.04 & 0 & 0 & 16.5 & 3.16 & 2.21 & 0 & 0 & 7.76 & 4.31 & 1.83 \\
a crocodile & S & 0 & 0.435 & 0.894 & 0.894 & 0 & 0 & 10.6 & 1.85 & 1.32 & 0 & 0 & 10.5 & 1.70 & 1.22 & 0 & 0 & 11.8 & 1.86 & 1.32 \\
a car suspension control arm with rubber bushings &  & 0 & 0.547 & 1.65 & 1.16 & 0 & 0 & 6.03 & 1.67 & 1.20 & 0 & 0 & 5.96 & 1.52 & 1.08 & 0 & 0 & 6.95 & 1.59 & 1.13 \\
a truck leaf spring &  & 0 & 0.188 & 0.528 & 0.349 & 0 & 0 & 3.75 & 0.999 & 0.799 & 0 & 0 & 3.70 & 0.713 & 0.410 & 0 & 0 & 3.94 & 0.956 & 0.760 \\
a car wheel with its rubber tyre & G & 0 & 1.52 & 4.58 & 2.49 & 0 & 0 & 12.3 & 3.07 & 1.67 & 0 & 0 & 23.4 & 7.89 & 4.20 & 0 & 0 & 13.0 & 3.52 & 1.92 \\
a car brake disc & S & 0 & 1.02 & 1.76 & 0.987 & --- & --- & --- & --- & --- & --- & --- & --- & --- & --- & --- & --- & --- & --- & --- \\
a crankshaft & G & 0 & 0.462 & 0.759 & 0.759 & 0 & 0 & 5.66 & 1.17 & 0.846 & 0 & 0 & 5.72 & 0.864 & 0.864 & 0 & 0 & 5.96 & 1.20 & 0.874 \\
a road roller with a steel drum and rubber tyres & T, G, S & 0 & 1.07 & 1.78 & 0.920 & 0 & 0 & 21.5 & 3.28 & 2.35 & 0 & 0 & 32.5 & 3.04 & 2.16 & 0 & 0 & 35.3 & 5.19 & 3.96 \\
a jackhammer & S & 0 & 0.200 & 2.04 & 0.795 & 0 & 3 & 3.97 & 2.46\textsuperscript{C} (33.5) & 1.70\textsuperscript{C} & 0 & 3 & 3.85 & 1.52\textsuperscript{C} (22.9) & 1.07\textsuperscript{C} & 0 & 4 & 4.22 & 2.48\textsuperscript{C} (44.8) & 1.71\textsuperscript{C} \\
an industrial robot arm & G & 0 & 0.537 & 1.66 & 1.10 & 0 & 0 & 6.78 & 2.23 & 1.23 & 0 & 0 & 6.93 & 2.37 & 1.31 & 0 & 0 & 7.65 & 2.16 & 1.19 \\
an adjustable wrench & T, S & 0 & 0.188 & 0.694 & 0.495 & 0 & 0 & 3.79 & 1.57 & 0.652 & 0 & 0 & 3.54 & 0.917 & 0.462 & 0 & 0 & 3.95 & 1.28 & 0.616 \\
a claw hammer &  & 0 & 0.227 & 1.08 & 0.565 & 0 & 0 & 3.91 & 1.18 & 0.670 & 0 & 0 & 3.78 & 0.993 & 0.589 & 0 & 0 & 4.17 & 1.52 & 0.739 \\
a dental crown &  & 0 & 1.55 & 2.47 & 1.30 & 0 & 0 & 21.9 & 3.10 & 2.03 & 0 & 3 & 76.1 & 148\textsuperscript{C} & 139\textsuperscript{C} & 0 & 0 & 18.5 & 2.21 & 1.36 \\
a knee implant & G & 0 & 1.49 & 3.18 & 1.61 & 0 & 0 & 14.3 & 3.07 & 1.76 & 0 & 0 & 26.0 & 6.92 & 3.92 & 0 & 0 & 12.0 & 2.79 & 1.60 \\
a walking cane with a rubber tip & T, L, S & 0 & 0.142 & 0.417 & 0.275 & 0 & 0 & 3.48 & 0.654 & 0.482 & 0 & 0 & 3.44 & 0.455 & 0.341 & 0 & 0 & 3.67 & 0.667 & 0.490 \\
a human femur &  & 0 & 0.257 & 1.08 & 0.659 & 0 & 0 & 4.22 & 1.57 & 0.904 & 0 & 0 & 4.11 & 1.65 & 0.851 & 0 & 0 & 4.71 & 1.89 & 0.943 \\
a lumbar vertebra &  & 0 & 0.897 & 1.42 & 0.709 & 0 & 0 & 16.8 & 1.04 & 0.634 & 0 & 0 & 41.4 & 1.55 & 0.955 & 0 & 0 & 7.80 & 0.991 & 0.603 \\
a skyscraper & L & 0 & 0.324 & 0.528 & 0.307 & 0 & 0 & 4.53 & 0.670 & 0.432 & 0 & 0 & 5.28 & 0.834 & 0.523 & 0 & 0 & 4.88 & 0.650 & 0.421 \\
a thatched stone cottage & L & 0 & 1.08 & 2.43 & 1.29 & 0 & 0 & 11.7 & 1.72 & 0.947 & 0 & 0 & 30.3 & 5.48 & 2.94 & 0 & 0 & 13.0 & 1.67 & 0.928 \\
a lighthouse &  & 0 & 0.671 & 0.979 & 0.584 & 0 & 0 & 7.31 & 1.00 & 0.615 & 0 & 0 & 10.8 & 1.59 & 0.984 & 0 & 0 & 8.38 & 1.06 & 0.659 \\
an igloo &  & 0 & 1.41 & 1.84 & 0.943 & 0 & 0 & 12.1 & 1.68 & 1.05 & 0 & 0 & 13.9 & 1.95 & 1.21 & 0 & 0 & 13.7 & 1.62 & 1.01 \\
a log cabin & L & 0 & 1.29 & 1.83 & 0.977 & 0 & 0 & 14.5 & 1.95 & 1.20 & 0 & 0 & 34.3 & 6.09 & 3.63 & 0 & 0 & 16.1 & 1.88 & 1.15 \\
the Leaning Tower of Pisa & T, G, S & 0 & 0.593 & 0.993 & 0.555 & 2 & 1 & 22.2 & 8.48 (10.4) & 3.21 & 2 & 1 & 24.6 & 6.40 (8.66) & 1.64 & 8 & 1 & 21.6 (48.4) & 4.89 (6.53) & 1.99 \\
the Pantheon & L & 0 & 1.06 & 1.61 & 0.891 & 0 & 0 & 11.3 & 1.50 & 0.957 & 0 & 0 & 13.4 & 1.80 & 1.14 & 0 & 0 & 12.8 & 1.44 & 0.915 \\
Himeji Castle & G & 0 & 1.52 & 3.82 & 2.04 & --- & --- & --- & --- & --- & --- & --- & --- & --- & --- & --- & --- & --- & --- & --- \\*
a natural stone arch & G & 0 & 1.45 & 2.05 & 1.15 & 0 & 0 & 33.0 & 2.15 & 0.970 & 0 & 0 & 36.7 & 5.99 & 3.10 & 0 & 0 & 12.4 & 1.75 & 1.10 \\*
the Taj Mahal &  & 0 & 1.03 & 2.55 & 1.36 & 0 & 0 & 11.7 & 2.03 & 1.13 & 0 & 0 & 18.8 & 3.85 & 2.15 & 0 & 0 & 13.7 & 1.88 & 1.05 \\*
\noalign{\penalty10000\vskip\aboverulesep\hrule height\lightrulewidth\penalty10000\vskip\belowrulesep}
Median (all objects) &  & 0 & 0.655 & 1.71 & 0.960 & 0 & 0 & 7.31 & 1.95 & 1.13 & 0 & 0 & 11.4 & 1.83 & 1.21 & 0 & 0 & 8.26 & 1.86 & 1.13 \\*
\hspace{1em}$n$ &  & 50 & 50 & 50 & 50 & 47 & 47 & 47 & 47 & 47 & 47 & 47 & 47 & 47 & 47 & 47 & 47 & 47 & 47 & 47 \\*
\addlinespace
Median (unflagged) &  & 0 & 0.623 & 1.42 & 0.767 & 0 & 0 & 7.08 & 1.62 & 0.975 & 0 & 0 & 11.4 & 1.62 & 1.08 & 0 & 0 & 7.87 & 1.60 & 0.976 \\*
\hspace{1em}$n$ &  & 18 & 18 & 18 & 18 & 18 & 18 & 18 & 18 & 18 & 18 & 18 & 18 & 18 & 18 & 18 & 18 & 18 & 18 & 18 \\
\end{longtable}
\end{landscape}

\begin{landscape}
\begin{table}[p]
\centering
\footnotesize
\setlength{\tabcolsep}{3.5pt}
\renewcommand{\arraystretch}{0.85}
\caption{Degrees of freedom (thousands) at final octree level 8: diffuse Nitsche (three per free node of the trimmed octree) and the three body-fitted meshes (three per vertex). Matched DOFs: a uniform mesh whose vertex count targets the octree's node count. Matched surface: a uniform mesh of element size $h_{\min}$. Graded: element size $h_{\min}$ at the surface and coarser inside, like the octree. Flags (Sec.~\ref{si:error-comparison}): T: thin walls, G: narrow gaps, L: a load near a support, S: a small or partially clamped support. ---: no body-fitted mesh, because the session surface self-intersects. }\label{tab:si-bf-dofs-d8}
\smallskip
\begin{tabular}{@{}>{\raggedright\arraybackslash}p{3.7cm}crrrr@{\hspace{1.6em}}>{\raggedright\arraybackslash}p{3.7cm}crrrr@{}}
\toprule
 &  &  & \multicolumn{3}{c}{Body-fitted} &  &  &  & \multicolumn{3}{c}{Body-fitted} \\
\cmidrule(lr){4-6} \cmidrule(l){10-12}
\multicolumn{1}{@{}l}{Object} & Flags & \begin{tabular}[b]{@{}r@{}}Diffuse\\Nitsche\end{tabular} & \begin{tabular}[b]{@{}r@{}}Matched\\DOFs\end{tabular} & \begin{tabular}[b]{@{}r@{}}Matched\\surface\end{tabular} & \begin{tabular}[b]{@{}r@{}}Graded\end{tabular} & Object & Flags & \begin{tabular}[b]{@{}r@{}}Diffuse\\Nitsche\end{tabular} & \begin{tabular}[b]{@{}r@{}}Matched\\DOFs\end{tabular} & \begin{tabular}[b]{@{}r@{}}Matched\\surface\end{tabular} & \begin{tabular}[b]{@{}r@{}}Graded\end{tabular} \\
\midrule
a TV remote control &  & 209 & 210 & 443 & 208 & a car suspension control arm with rubber bushings &  & 353 & 358 & 588 & 350 \\
a USB stick &  & 193 & 193 & 378 & 193 & a truck leaf spring &  & 58.0 & 56.8 & 54.3 & 57.6 \\
an electric toothbrush &  & 66.7 & 68.1 & 74.3 & 65.8 & a car wheel with its rubber tyre & G & 1410 & 1460 & 3580 & 1390 \\
a coffee mug & T & 1260 & 1270 & 1970 & 1250 & a car brake disc & S & 912 & --- & --- & --- \\
a computer mouse &  & 445 & 446 & 1890 & 443 & a crankshaft & G & 299 & 309 & 552 & 296 \\
a school chair with a steel frame and a plastic seat & T, S & 336 & 344 & 453 & 335 & a road roller with a steel drum and rubber tyres & T, G, S & 454 & 464 & 501 & 437 \\
a glass-top dining table & T, S & 293 & 217 & 193 & 293 & a jackhammer & S & 87.6 & 90.8 & 105 & 84.2 \\
a park bench & T, S & 400 & --- & --- & --- & an industrial robot arm & G & 398 & 402 & 723 & 392 \\
a bar stool & S & 353 & 351 & 348 & 348 & an adjustable wrench & T, S & 76.9 & 44.9 & 55.2 & 76.4 \\
a sofa & S & 544 & 569 & 1510 & 541 & a claw hammer &  & 113 & 119 & 144 & 111 \\
a teddy bear &  & 486 & 453 & 1550 & 482 & a dental crown &  & 1480 & 1470 & 9040 & 1470 \\
a wooden rocking horse & T & 412 & 423 & 476 & 407 & a knee implant & G & 1310 & 1280 & 4950 & 1300 \\
a pogo stick & S & 48.0 & 47.0 & 30.6 & 46.6 & a walking cane with a rubber tip & T, L, S & 30.6 & 29.2 & 18.2 & 30.2 \\
a wooden toy train engine & G, L, S & 438 & 455 & 1700 & 454 & a human femur &  & 153 & 155 & 214 & 150 \\
a yo-yo & G & 1560 & 1510 & 4490 & 1550 & a lumbar vertebra &  & 418 & 423 & 1130 & 415 \\
a jelly dessert &  & 831 & 829 & 4890 & 826 & a skyscraper & L & 218 & 207 & 620 & 216 \\
a loaf of bread &  & 983 & 984 & 6730 & 979 & a thatched stone cottage & L & 897 & 902 & 5710 & 894 \\
a hamburger & G & 1000 & 999 & 6690 & 997 & a lighthouse &  & 437 & 434 & 1480 & 443 \\
a layered birthday cake &  & 841 & 846 & 4520 & 840 & an igloo &  & 1070 & 1080 & 2470 & 1070 \\
a slice of watermelon &  & 473 & 460 & 2000 & 477 & a log cabin & L & 1130 & 1140 & 6420 & 1120 \\
a giraffe & S & 145 & 152 & 181 & 141 & the Leaning Tower of Pisa & T, G, S & 422 & 433 & 709 & 341 \\
an elephant & S & 404 & 412 & 1010 & 399 & the Pantheon & L & 935 & 968 & 2190 & 965 \\
a rhinoceros & S & 333 & 351 & 904 & 330 & Himeji Castle & G & 1420 & --- & --- & --- \\
a horse & S & 354 & 358 & 761 & 349 & a natural stone arch & G & 825 & 835 & 2790 & 822 \\
a crocodile & S & 169 & 170 & 244 & 167 & the Taj Mahal &  & 880 & 862 & 3320 & 902 \\
\midrule
Median (all objects) &  & 420 & 423 & 904 & 407 & Median (unflagged) &  & 441 & 440 & 1510 & 443 \\
\hspace{1em}$n$ &  & 50 & 47 & 47 & 47 & \hspace{1em}$n$ &  & 18 & 18 & 18 & 18 \\
\bottomrule
\end{tabular}
\end{table}
\end{landscape}

\begin{landscape}
\footnotesize
\setlength{\tabcolsep}{2.5pt}
\ifdefined\SIbfColUnit\else\newlength{\SIbfColUnit}\fi
\setlength{\SIbfColUnit}{\dimexpr(\linewidth-4.6cm-1.05cm-32\tabcolsep)*100/1500\relax}
\setlength{\LTleft}{\fill}\setlength{\LTright}{\fill}
\setlength{\LTcapwidth}{\linewidth}
\begin{longtable}{@{}>{\raggedright\arraybackslash}p{4.6cm}>{\centering\arraybackslash}p{1.05cm}>{\raggedleft\arraybackslash}p{1\SIbfColUnit}>{\raggedleft\arraybackslash}p{1\SIbfColUnit}>{\raggedleft\arraybackslash}p{1\SIbfColUnit}>{\raggedleft\arraybackslash}p{1\SIbfColUnit}>{\raggedleft\arraybackslash}p{1\SIbfColUnit}>{\raggedleft\arraybackslash}p{1\SIbfColUnit}>{\raggedleft\arraybackslash}p{1\SIbfColUnit}>{\raggedleft\arraybackslash}p{1\SIbfColUnit}>{\raggedleft\arraybackslash}p{1\SIbfColUnit}>{\raggedleft\arraybackslash}p{1\SIbfColUnit}>{\raggedleft\arraybackslash}p{1\SIbfColUnit}>{\raggedleft\arraybackslash}p{1\SIbfColUnit}>{\raggedleft\arraybackslash}p{1\SIbfColUnit}>{\raggedleft\arraybackslash}p{1\SIbfColUnit}>{\raggedleft\arraybackslash}p{1\SIbfColUnit}@{}}
\caption{Discrepancy between the diffuse Nitsche solution and each body-fitted solution at final octree level 8. $\eta$ and $\eta^{H^1}$: relative $L^2$ and $H^1$ norms of the displacement difference after removal of its best-fit rigid motion (\%). $\ell_{\mathrm N}/\ell_{\mathrm{bf}}$: compliance ratio. $s_{95}$ and $s_{99}$: diffuse-Nitsche-to-body-fitted ratios of the 95th and 99th percentiles of the von Mises stress. Gray rows: $\eta$ above 10\% against at least one body-fitted mesh. $^{\dagger}$: diffuse Nitsche solve converged only to a relative residual of $10^{-2}$, excluded from the medians. Flags (Sec.~\ref{si:error-comparison}): T: thin walls, G: narrow gaps, L: a load near a support, S: a small or partially clamped support. ---: no body-fitted mesh, because the session surface self-intersects.}\label{tab:si-bf-accuracy-d8}\\
\toprule
 &  & \multicolumn{15}{c}{Body-fitted} \\
\cmidrule(l){3-17}
 &  & \multicolumn{5}{c}{Matched DOFs} & \multicolumn{5}{c}{Matched surface} & \multicolumn{5}{c}{Graded} \\
\cmidrule(lr){3-7} \cmidrule(lr){8-12} \cmidrule(l){13-17}
\multicolumn{1}{@{}l}{Object} & Flags & \begin{tabular}[b]{@{}r@{}}$\eta$\\(\%)\end{tabular} & \begin{tabular}[b]{@{}r@{}}$\eta^{H^1}$\\(\%)\end{tabular} & $\ell_{\mathrm N}/\ell_{\mathrm{bf}}$ & $s_{95}$ & $s_{99}$ & \begin{tabular}[b]{@{}r@{}}$\eta$\\(\%)\end{tabular} & \begin{tabular}[b]{@{}r@{}}$\eta^{H^1}$\\(\%)\end{tabular} & $\ell_{\mathrm N}/\ell_{\mathrm{bf}}$ & $s_{95}$ & $s_{99}$ & \begin{tabular}[b]{@{}r@{}}$\eta$\\(\%)\end{tabular} & \begin{tabular}[b]{@{}r@{}}$\eta^{H^1}$\\(\%)\end{tabular} & $\ell_{\mathrm N}/\ell_{\mathrm{bf}}$ & $s_{95}$ & $s_{99}$ \\
\midrule
\endfirsthead
\caption[]{Discrepancy at final octree level 8 (continued).}\\
\toprule
 &  & \multicolumn{15}{c}{Body-fitted} \\
\cmidrule(l){3-17}
 &  & \multicolumn{5}{c}{Matched DOFs} & \multicolumn{5}{c}{Matched surface} & \multicolumn{5}{c}{Graded} \\
\cmidrule(lr){3-7} \cmidrule(lr){8-12} \cmidrule(l){13-17}
\multicolumn{1}{@{}l}{Object} & Flags & \begin{tabular}[b]{@{}r@{}}$\eta$\\(\%)\end{tabular} & \begin{tabular}[b]{@{}r@{}}$\eta^{H^1}$\\(\%)\end{tabular} & $\ell_{\mathrm N}/\ell_{\mathrm{bf}}$ & $s_{95}$ & $s_{99}$ & \begin{tabular}[b]{@{}r@{}}$\eta$\\(\%)\end{tabular} & \begin{tabular}[b]{@{}r@{}}$\eta^{H^1}$\\(\%)\end{tabular} & $\ell_{\mathrm N}/\ell_{\mathrm{bf}}$ & $s_{95}$ & $s_{99}$ & \begin{tabular}[b]{@{}r@{}}$\eta$\\(\%)\end{tabular} & \begin{tabular}[b]{@{}r@{}}$\eta^{H^1}$\\(\%)\end{tabular} & $\ell_{\mathrm N}/\ell_{\mathrm{bf}}$ & $s_{95}$ & $s_{99}$ \\
\midrule
\endhead
\midrule
\multicolumn{17}{@{}r@{}}{\textit{Continued on next page}}\\
\endfoot
\bottomrule
\endlastfoot
a TV remote control &  & 2.6 & 5.1 & 0.929 & 0.948 & 0.940 & 2.8 & 5.3 & 0.925 & 0.942 & 0.933 & 2.4 & 4.8 & 0.934 & 0.950 & 0.949 \\
a USB stick &  & 0.43 & 2.0 & 0.938 & 0.965 & 0.924 & 0.49 & 2.1 & 0.934 & 0.963 & 0.915 & 0.33 & 1.9 & 0.949 & 0.972 & 0.931 \\
an electric toothbrush &  & 6.1 & 10 & 1.091 & 0.950 & 1.014 & 5.2 & 8.5 & 1.077 & 0.943 & 1.025 & 2.9 & 4.9 & 1.042 & 0.942 & 0.977 \\
a coffee mug & T & 1.7 & 4.3 & 0.943 & 0.954 & 0.937 & 2.0 & 4.7 & 0.938 & 0.951 & 0.933 & 1.8 & 4.3 & 0.944 & 0.952 & 0.938 \\
a computer mouse &  & 3.4 & 16 & 0.975 & 0.997 & 0.970 & 3.1 & 14 & 0.976 & 0.999 & 0.957 & 2.8 & 15 & 0.980 & 1.003 & 0.960 \\
a school chair with a steel frame and a plastic seat & T, S & 4.2 & 5.7 & 0.932 & 0.934 & 0.917 & 5.5 & 7.5 & 0.908 & 0.917 & 0.902 & 4.3 & 5.9 & 0.926 & 0.926 & 0.917 \\
\rowcolor{gray!15}a glass-top dining table & T, S & 6.2 & 11 & 0.951 & 0.975 & 0.929 & 5.1 & 9.2 & 0.960 & 0.981 & 0.932 & 10 & 18 & 0.921 & 0.962 & 0.917 \\
a park bench & T, S & --- & --- & --- & --- & --- & --- & --- & --- & --- & --- & --- & --- & --- & --- & --- \\
a bar stool & S & 5.3 & 10 & 0.949 & 0.954 & 0.933 & 5.2 & 10 & 0.949 & 0.954 & 0.931 & 5.6 & 11 & 0.945 & 0.951 & 0.926 \\
a sofa & S & 0.58 & 5.3 & 0.997 & 0.973 & 0.963 & 0.71 & 5.3 & 0.991 & 0.966 & 0.960 & 0.49 & 5.2 & 0.998 & 0.981 & 0.964 \\
a teddy bear &  & 0.58 & 1.8 & 0.926 & 0.936 & 0.958 & 0.60 & 1.8 & 0.929 & 0.985 & 0.951 & 0.55 & 1.7 & 0.935 & 0.988 & 0.948 \\
a wooden rocking horse & T & 2.7 & 5.3 & 0.960 & 0.966 & 0.969 & 2.9 & 5.6 & 0.956 & 0.965 & 0.967 & 2.7 & 5.3 & 0.960 & 0.966 & 0.969 \\
\rowcolor{gray!15}a pogo stick & S & 16 & 46 & 0.957 & 0.876 & 1.042 & 16 & 46 & 0.965 & 0.886 & 1.047 & 16 & 46 & 0.957 & 0.881 & 1.042 \\
\rowcolor{gray!15}a wooden toy train engine & G, L, S & 11 & 40 & 0.901 & 0.971 & 0.874 & 10 & 41 & 0.905 & 0.975 & 0.869 & 9.8 & 40 & 0.914 & 0.978 & 0.876 \\
\rowcolor{gray!15}a yo-yo & G & 16 & 48 & 0.606 & 0.761 & 0.674 & 16 & 48 & 0.596 & 0.751 & 0.663 & 16 & 48 & 0.601 & 0.754 & 0.667 \\
a jelly dessert &  & 1.8 & 18 & 0.976 & 0.971 & 0.972 & 2.2 & 17 & 0.974 & 0.960 & 0.955 & 1.6 & 18 & 0.982 & 0.976 & 0.973 \\
a loaf of bread$^{\dagger}$ &  & 2.0 & 16 & 0.972 & 0.998 & 0.965 & 2.7 & 16 & 0.935 & 0.956 & 0.927 & 2.3 & 17 & 0.943 & 0.966 & 0.930 \\
a hamburger & G & 5.3 & 19 & 0.939 & 0.963 & 0.917 & 5.6 & 19 & 0.933 & 0.954 & 0.905 & 3.8 & 19 & 0.955 & 0.973 & 0.935 \\
a layered birthday cake &  & 1.4 & 13 & 0.990 & 0.981 & 0.994 & 1.6 & 14 & 0.991 & 0.974 & 0.987 & 1.3 & 15 & 0.992 & 0.976 & 0.986 \\
a slice of watermelon &  & 2.7 & 15 & 0.947 & 0.968 & 0.935 & 1.9 & 14 & 0.960 & 0.971 & 0.933 & 1.5 & 15 & 0.967 & 0.978 & 0.938 \\
a giraffe & S & 0.84 & 3.0 & 1.048 & 0.988 & 1.034 & 0.89 & 3.1 & 1.051 & 0.988 & 1.032 & 0.87 & 2.9 & 1.032 & 0.984 & 1.029 \\
an elephant & S & 0.42 & 2.5 & 0.993 & 0.997 & 1.003 & 0.49 & 2.6 & 0.994 & 0.997 & 1.002 & 0.46 & 2.7 & 0.996 & 0.998 & 1.002 \\
a rhinoceros & S & 0.67 & 4.8 & 0.991 & 0.984 & 1.005 & 0.75 & 4.7 & 0.992 & 0.997 & 1.003 & 0.70 & 4.8 & 0.993 & 0.993 & 1.003 \\
\rowcolor{gray!15}a horse & S & 26 & 54 & 0.967 & 0.968 & 0.981 & 26 & 54 & 0.949 & 0.966 & 0.969 & 26 & 54 & 0.951 & 0.966 & 0.969 \\
a crocodile & S & 1.9 & 5.6 & 0.905 & 0.901 & 0.912 & 2.3 & 5.9 & 0.891 & 0.894 & 0.898 & 2.2 & 5.9 & 0.891 & 0.901 & 0.891 \\
a car suspension control arm with rubber bushings &  & 4.2 & 11 & 0.968 & 0.974 & 0.979 & 4.1 & 11 & 0.964 & 0.972 & 0.975 & 4.2 & 11 & 0.969 & 0.975 & 0.980 \\
\rowcolor{gray!15}a truck leaf spring &  & 14 & 24 & 0.831 & 0.856 & 0.727 & 14 & 24 & 0.832 & 0.857 & 0.729 & 15 & 25 & 0.824 & 0.859 & 0.718 \\
a car wheel with its rubber tyre & G & 2.8 & 7.5 & 0.924 & 0.950 & 0.925 & 3.1 & 8.1 & 0.918 & 0.946 & 0.917 & 2.9 & 7.9 & 0.925 & 0.950 & 0.923 \\
a car brake disc & S & --- & --- & --- & --- & --- & --- & --- & --- & --- & --- & --- & --- & --- & --- & --- \\
a crankshaft & G & 5.4 & 14 & 0.769 & 0.797 & 0.753 & 4.7 & 12 & 0.801 & 0.820 & 0.783 & 4.6 & 12 & 0.806 & 0.824 & 0.785 \\
\rowcolor{gray!15}a road roller with a steel drum and rubber tyres & T, G, S & 37 & 42 & 0.493 & 0.654 & 0.592 & 37 & 42 & 0.491 & 0.655 & 0.592 & 37 & 42 & 0.489 & 0.653 & 0.589 \\
\rowcolor{gray!15}a jackhammer & S & 10 & 16 & 0.841 & 0.983 & 0.836 & 11 & 16 & 0.834 & 0.984 & 0.829 & 11 & 17 & 0.831 & 0.984 & 0.820 \\
\rowcolor{gray!15}an industrial robot arm & G & 20 & 46 & 0.556 & 0.737 & 0.607 & 20 & 47 & 0.549 & 0.733 & 0.600 & 20 & 47 & 0.553 & 0.736 & 0.605 \\
an adjustable wrench & T, S & 0.27 & 2.2 & 1.007 & 1.008 & 1.002 & 0.27 & 2.2 & 1.007 & 1.006 & 1.004 & 0.31 & 2.3 & 1.005 & 1.004 & 1.001 \\
a claw hammer &  & 0.42 & 1.8 & 0.951 & 0.955 & 0.968 & 0.45 & 1.9 & 0.949 & 0.953 & 0.962 & 0.39 & 1.7 & 0.953 & 0.955 & 0.971 \\
a dental crown &  & 4.5 & 16 & 0.966 & 0.920 & 0.886 & 4.6 & 16 & 0.967 & 0.917 & 0.884 & 4.5 & 16 & 0.968 & 0.920 & 0.884 \\
\rowcolor{gray!15}a knee implant & G & 65 & 80 & 0.456 & 0.487 & 0.428 & 65 & 81 & 0.451 & 0.480 & 0.422 & 65 & 80 & 0.458 & 0.486 & 0.427 \\
a walking cane with a rubber tip & T, L, S & 7.6 & 13 & 0.958 & 0.899 & 0.836 & 6.4 & 11 & 0.967 & 0.913 & 0.868 & 9.0 & 15 & 0.937 & 0.889 & 0.830 \\
a human femur &  & 0.18 & 0.94 & 0.992 & 0.989 & 0.981 & 0.26 & 1.0 & 0.990 & 0.985 & 0.979 & 0.10 & 0.89 & 0.995 & 0.992 & 0.983 \\
a lumbar vertebra &  & 4.1 & 16 & 0.961 & 0.941 & 0.934 & 3.9 & 16 & 0.965 & 0.939 & 0.933 & 3.8 & 16 & 0.966 & 0.941 & 0.934 \\
a skyscraper & L & 0.13 & 0.16 & 0.994 & 0.994 & 0.992 & 0.10 & 0.14 & 1.003 & 0.993 & 0.990 & 0.081 & 0.13 & 1.003 & 0.992 & 0.994 \\
a thatched stone cottage & L & 0.70 & 4.1 & 0.985 & 0.984 & 0.982 & 0.53 & 4.3 & 0.998 & 0.986 & 0.975 & 0.47 & 4.4 & 0.997 & 0.987 & 0.976 \\
a lighthouse &  & 0.90 & 1.4 & 0.994 & 0.997 & 0.991 & 1.6 & 2.2 & 0.996 & 0.991 & 0.992 & 1.3 & 1.9 & 0.996 & 0.992 & 0.992 \\
an igloo &  & 2.5 & 7.9 & 0.971 & 0.965 & 0.952 & 1.5 & 6.4 & 0.991 & 0.981 & 0.977 & 1.5 & 6.8 & 0.989 & 0.981 & 0.977 \\
a log cabin & L & 1.5 & 8.6 & 0.945 & 0.954 & 0.965 & 2.0 & 11 & 0.933 & 0.954 & 0.936 & 1.8 & 11 & 0.939 & 0.960 & 0.935 \\
the Leaning Tower of Pisa & T, G, S & 3.7 & 4.2 & 0.964 & 0.961 & 0.958 & 3.7 & 4.2 & 0.965 & 0.958 & 0.956 & 4.6 & 5.4 & 0.969 & 0.965 & 0.969 \\
the Pantheon & L & 0.96 & 1.1 & 0.992 & 0.984 & 0.973 & 1.2 & 1.3 & 1.004 & 0.991 & 0.981 & 0.96 & 1.1 & 1.002 & 0.988 & 0.985 \\
Himeji Castle & G & --- & --- & --- & --- & --- & --- & --- & --- & --- & --- & --- & --- & --- & --- & --- \\*
a natural stone arch & G & 6.7 & 8.3 & 0.776 & 0.867 & 0.797 & 6.7 & 8.4 & 0.773 & 0.864 & 0.792 & 6.7 & 8.3 & 0.776 & 0.865 & 0.796 \\*
the Taj Mahal &  & 1.2 & 1.4 & 0.980 & 0.987 & 0.989 & 0.95 & 1.3 & 1.013 & 0.992 & 0.989 & 0.77 & 1.2 & 1.003 & 0.996 & 0.990 \\*
\noalign{\penalty10000\vskip\aboverulesep\hrule height\lightrulewidth\penalty10000\vskip\belowrulesep}
Median (all objects) &  & 2.7 & 8.4 & 0.958 & 0.964 & 0.955 & 3.0 & 8.4 & 0.960 & 0.961 & 0.944 & 2.8 & 8.1 & 0.956 & 0.966 & 0.948 \\*
\hspace{1em}$n$ &  & 46 & 46 & 46 & 46 & 46 & 46 & 46 & 46 & 46 & 46 & 46 & 46 & 46 & 46 & 46 \\*
\addlinespace
Median (unflagged) &  & 2.5 & 10 & 0.968 & 0.965 & 0.968 & 1.9 & 8.5 & 0.967 & 0.971 & 0.957 & 1.5 & 6.8 & 0.969 & 0.976 & 0.971 \\*
\hspace{1em}$n$ &  & 17 & 17 & 17 & 17 & 17 & 17 & 17 & 17 & 17 & 17 & 17 & 17 & 17 & 17 & 17 \\
\end{longtable}
\end{landscape}

\begin{landscape}
\footnotesize
\setlength{\tabcolsep}{2pt}
\ifdefined\SIbfColUnit\else\newlength{\SIbfColUnit}\fi
\setlength{\SIbfColUnit}{\dimexpr(\linewidth-3.0cm-1.05cm-40\tabcolsep)*100/2041\relax}
\setlength{\LTleft}{\fill}\setlength{\LTright}{\fill}
\setlength{\LTcapwidth}{\linewidth}
\begin{longtable}{@{}>{\raggedright\arraybackslash}p{3.0cm}>{\centering\arraybackslash}p{1.05cm}>{\centering\arraybackslash}p{1.03\SIbfColUnit}>{\raggedleft\arraybackslash}p{1.15\SIbfColUnit}>{\raggedleft\arraybackslash}p{1.15\SIbfColUnit}>{\raggedleft\arraybackslash}p{1.0\SIbfColUnit}>{\centering\arraybackslash}p{1.03\SIbfColUnit}>{\centering\arraybackslash}p{1.03\SIbfColUnit}>{\raggedleft\arraybackslash}p{1.0\SIbfColUnit}>{\raggedleft\arraybackslash}p{1.3\SIbfColUnit}>{\raggedleft\arraybackslash}p{1.0\SIbfColUnit}>{\centering\arraybackslash}p{1.03\SIbfColUnit}>{\centering\arraybackslash}p{1.03\SIbfColUnit}>{\raggedleft\arraybackslash}p{1.0\SIbfColUnit}>{\raggedleft\arraybackslash}p{1.3\SIbfColUnit}>{\raggedleft\arraybackslash}p{1.0\SIbfColUnit}>{\centering\arraybackslash}p{1.03\SIbfColUnit}>{\centering\arraybackslash}p{1.03\SIbfColUnit}>{\raggedleft\arraybackslash}p{1.0\SIbfColUnit}>{\raggedleft\arraybackslash}p{1.3\SIbfColUnit}>{\raggedleft\arraybackslash}p{1.0\SIbfColUnit}@{}}
\caption{Computational cost at final octree level 8: wall-clock times (in seconds) of the successful attempt. In parentheses: the total over all attempts, shown when failed attempts add more than 0.5\%. Solver retries: the number of attempts after the first solve. Mesh retries: failed TetGen calls plus rejected remeshes of the surface that the kept mesh was built from. Solve to $10^{-6}$: Newton's method to a relative residual of $10^{-6}$. Solve to $10^{-2}$: the time that the same solve takes to first reach a relative residual of $10^{-2}$. \textsuperscript{C}: solved with direct linear solves on the CPU. \textsuperscript{R}: solved by the lightweight GPU Newton solver. $^{\dagger}$: diffuse Nitsche solve converged only to a relative residual of $10^{-2}$, excluded from the medians. Flags (Sec.~\ref{si:error-comparison}): T: thin walls, G: narrow gaps, L: a load near a support, S: a small or partially clamped support. ---: no body-fitted mesh, because the session surface self-intersects. }\label{tab:si-bf-cost-d8}\\
\toprule
 &  &  &  &  &  & \multicolumn{15}{c}{Body-fitted} \\
\cmidrule(l){7-21}
 &  & \multicolumn{4}{c}{Diffuse Nitsche} & \multicolumn{5}{c}{Matched DOFs} & \multicolumn{5}{c}{Matched surface} & \multicolumn{5}{c}{Graded} \\
\cmidrule(lr){3-6} \cmidrule(lr){7-11} \cmidrule(lr){12-16} \cmidrule(l){17-21}
\multicolumn{1}{@{}l}{Object} & Flags & \begin{tabular}[b]{@{}c@{}}Solver\\retries\end{tabular} & \begin{tabular}[b]{@{}r@{}}Prep.\\(s)\end{tabular} & \begin{tabular}[b]{@{}r@{}}Solve\\to\\$10^{-6}$\\(s)\end{tabular} & \begin{tabular}[b]{@{}r@{}}Solve\\to\\$10^{-2}$\\(s)\end{tabular} & \begin{tabular}[b]{@{}c@{}}Mesh\\retries\end{tabular} & \begin{tabular}[b]{@{}c@{}}Solver\\retries\end{tabular} & \begin{tabular}[b]{@{}r@{}}Prep.\\(s)\end{tabular} & \begin{tabular}[b]{@{}r@{}}Solve\\to\\$10^{-6}$\\(s)\end{tabular} & \begin{tabular}[b]{@{}r@{}}Solve\\to\\$10^{-2}$\\(s)\end{tabular} & \begin{tabular}[b]{@{}c@{}}Mesh\\retries\end{tabular} & \begin{tabular}[b]{@{}c@{}}Solver\\retries\end{tabular} & \begin{tabular}[b]{@{}r@{}}Prep.\\(s)\end{tabular} & \begin{tabular}[b]{@{}r@{}}Solve\\to\\$10^{-6}$\\(s)\end{tabular} & \begin{tabular}[b]{@{}r@{}}Solve\\to\\$10^{-2}$\\(s)\end{tabular} & \begin{tabular}[b]{@{}c@{}}Mesh\\retries\end{tabular} & \begin{tabular}[b]{@{}c@{}}Solver\\retries\end{tabular} & \begin{tabular}[b]{@{}r@{}}Prep.\\(s)\end{tabular} & \begin{tabular}[b]{@{}r@{}}Solve\\to\\$10^{-6}$\\(s)\end{tabular} & \begin{tabular}[b]{@{}r@{}}Solve\\to\\$10^{-2}$\\(s)\end{tabular} \\
\midrule
\endfirsthead
\caption[]{Computational cost at final octree level 8 (continued).}\\
\toprule
 &  &  &  &  &  & \multicolumn{15}{c}{Body-fitted} \\
\cmidrule(l){7-21}
 &  & \multicolumn{4}{c}{Diffuse Nitsche} & \multicolumn{5}{c}{Matched DOFs} & \multicolumn{5}{c}{Matched surface} & \multicolumn{5}{c}{Graded} \\
\cmidrule(lr){3-6} \cmidrule(lr){7-11} \cmidrule(lr){12-16} \cmidrule(l){17-21}
\multicolumn{1}{@{}l}{Object} & Flags & \begin{tabular}[b]{@{}c@{}}Solver\\retries\end{tabular} & \begin{tabular}[b]{@{}r@{}}Prep.\\(s)\end{tabular} & \begin{tabular}[b]{@{}r@{}}Solve\\to\\$10^{-6}$\\(s)\end{tabular} & \begin{tabular}[b]{@{}r@{}}Solve\\to\\$10^{-2}$\\(s)\end{tabular} & \begin{tabular}[b]{@{}c@{}}Mesh\\retries\end{tabular} & \begin{tabular}[b]{@{}c@{}}Solver\\retries\end{tabular} & \begin{tabular}[b]{@{}r@{}}Prep.\\(s)\end{tabular} & \begin{tabular}[b]{@{}r@{}}Solve\\to\\$10^{-6}$\\(s)\end{tabular} & \begin{tabular}[b]{@{}r@{}}Solve\\to\\$10^{-2}$\\(s)\end{tabular} & \begin{tabular}[b]{@{}c@{}}Mesh\\retries\end{tabular} & \begin{tabular}[b]{@{}c@{}}Solver\\retries\end{tabular} & \begin{tabular}[b]{@{}r@{}}Prep.\\(s)\end{tabular} & \begin{tabular}[b]{@{}r@{}}Solve\\to\\$10^{-6}$\\(s)\end{tabular} & \begin{tabular}[b]{@{}r@{}}Solve\\to\\$10^{-2}$\\(s)\end{tabular} & \begin{tabular}[b]{@{}c@{}}Mesh\\retries\end{tabular} & \begin{tabular}[b]{@{}c@{}}Solver\\retries\end{tabular} & \begin{tabular}[b]{@{}r@{}}Prep.\\(s)\end{tabular} & \begin{tabular}[b]{@{}r@{}}Solve\\to\\$10^{-6}$\\(s)\end{tabular} & \begin{tabular}[b]{@{}r@{}}Solve\\to\\$10^{-2}$\\(s)\end{tabular} \\
\midrule
\endhead
\midrule
\multicolumn{21}{@{}r@{}}{\textit{Continued on next page}}\\
\endfoot
\bottomrule
\endlastfoot
a TV remote control &  & 0 & 1.15 & 2.66 & 2.66 & 0 & 0 & 8.74 & 2.63 & 1.96 & 0 & 0 & 17.2 & 5.58 & 4.07 & 0 & 0 & 9.77 & 2.56 & 1.90 \\
a USB stick &  & 0 & 1.08 & 3.97 & 2.02 & 0 & 0 & 8.01 & 2.69 & 1.91 & 0 & 0 & 15.0 & 6.96 & 3.75 & 0 & 0 & 9.17 & 3.44 & 1.87 \\
an electric toothbrush &  & 0 & 0.505 & 3.22 & 2.17 & 0 & 0 & 4.29 & 5.26 & 3.62 & 0 & 0 & 4.49 & 4.53 & 3.17 & 0 & 0 & 4.68 & 9.73 & 8.05 \\
a coffee mug & T & 0 & 6.25 & 29.8 & 22.1 & 0 & 0 & 73.0 & 24.4 & 17.5 & 0 & 0 & 77.4 & 127\textsuperscript{C} & 114\textsuperscript{C} & 0 & 0 & 80.1 & 23.8 & 16.9 \\
a computer mouse &  & 0 & 2.36 & 2.86 & 1.51 & 0 & 0 & 18.7 & 2.75 & 1.71 & 0 & 0 & 72.3 & 437\textsuperscript{C} & 423\textsuperscript{C} & 0 & 0 & 55.6 & 3.58 & 2.41 \\
a school chair with a steel frame and a plastic seat & T, S & 1 & 1.79 (3.52) & 14.1 (17.6) & 7.12 & 0 & 1 & 10.9 & 11.5 (14.5) & 6.94 & 0 & 1 & 14.2 & 20.6 (24.5) & 8.96 & 0 & 1 & 38.5 & 13.7 (16.7) & 6.90 \\
a glass-top dining table & T, S & 0 & 1.55 & 8.85 & 5.90 & 0 & 1 & 7.89 & 4.16 (6.31) & 2.75 & 0 & 1 & 6.75 & 4.42 (6.48) & 2.86 & 0 & 1 & 33.3 & 6.10 (8.82) & 4.09 \\
a park bench & T, S & 1 & 2.38 (4.78) & 12.4 (16.6) & 8.28 & --- & --- & --- & --- & --- & --- & --- & --- & --- & --- & --- & --- & --- & --- & --- \\
a bar stool & S & 1 & 1.78 (3.52) & 10.4 (14.0) & 6.73 & 0 & 1 & 14.5 & 7.91 (11.3) & 5.30 & 0 & 1 & 14.2 & 7.81 (11.2) & 5.23 & 0 & 1 & 24.6 & 10.4 (13.7) & 5.24 \\
a sofa & S & 0 & 2.64 & 9.56 & 3.51 & 0 & 0 & 25.2 & 6.31 & 3.74 & 0 & 0 & 58.4 & 158\textsuperscript{C} & 147\textsuperscript{C} & 0 & 0 & 30.5 & 19.6 & 10.8 \\
a teddy bear &  & 0 & 2.36 & 12.7 & 7.67 & 0 & 0 & 22.3 & 12.1 & 9.81 & 0 & 0 & 59.8 & 1743\textsuperscript{C} & 1436\textsuperscript{C} & 0 & 0 & 37.1 & 13.4 & 10.9 \\
a wooden rocking horse & T & 0 & 2.10 & 6.55 & 6.55 & 0 & 1 & 16.5 & 10.2 (14.3) & 6.93 & 0 & 1 & 18.7 & 11.1 (15.7) & 7.55 & 0 & 1 & 27.0 & 9.56 (13.4) & 6.60 \\
a pogo stick & S & 1 & 0.400 (0.793) & 3.20 (5.77) & 0.748 & 0 & 1 & 3.79 & 5.98 (7.12) & 0.800 & 0 & 0 & 3.50 & 1.97 & 0.620 & 0 & 1 & 4.32 & 4.79 (5.90) & 0.792 \\
a wooden toy train engine & G, L, S & 0 & 2.95 & 4.40 & 2.40 & 0 & 0 & 44.2 & 4.20 & 2.68 & 0 & 0 & 248 & 263\textsuperscript{C} & 245\textsuperscript{C} & 0 & 0 & 29.8 & 6.07 & 4.10 \\
a yo-yo & G & 0 & 7.45 & 15.4 & 7.92 & 0 & 0 & 137 & 270\textsuperscript{C} & 248\textsuperscript{C} & 0 & 0 & 181 & 874\textsuperscript{C} & 824\textsuperscript{C} & 0 & 0 & 161 & 101\textsuperscript{C} & 91.2\textsuperscript{C} \\
a jelly dessert &  & 0 & 3.97 & 12.6 & 9.61 & 0 & 0 & 54.7 & 11.7 & 9.51 & 0 & 1 & 378 & 71.5\textsuperscript{R} (2284) & 50.1\textsuperscript{R} & 0 & 0 & 75.6 & 10.1 & 6.23 \\
a loaf of bread &  & 6 & 4.74 (32.6) & 1991\textsuperscript{C}$^{\dagger}$ (8765) & 170\textsuperscript{C} & 0 & 0 & 40.2 & 47.4 & 6.32 & 0 & 1 & 281 & 365\textsuperscript{R} (2549) & 40.7\textsuperscript{R} & 0 & 0 & 53.4 & 59.9 & 7.17 \\
a hamburger & G & 0 & 4.96 & 110 & 6.10 & 0 & 0 & 88.5 & 53.3 & 6.53 & 0 & 1 & 347 & 338\textsuperscript{R} (1473) & 38.8\textsuperscript{R} & 0 & 0 & 51.0 & 51.5 & 6.26 \\
a layered birthday cake &  & 0 & 3.80 & 16.1 & 10.3 & 0 & 0 & 67.6 & 9.95 & 6.15 & 0 & 1 & 408 & 67.7\textsuperscript{R} (1259) & 36.4\textsuperscript{R} & 0 & 0 & 82.2 & 9.36 & 5.82 \\
a slice of watermelon &  & 0 & 2.59 & 5.44 & 5.44 & 0 & 0 & 35.5 & 3.77 & 2.01 & 2 & 0 & 164 (472) & 911\textsuperscript{C} & 885\textsuperscript{C} & 0 & 0 & 96.3 & 8.66 & 6.99 \\
a giraffe & S & 0 & 0.945 & 4.56 & 3.20 & 0 & 1 & 6.94 & 6.27 (8.20) & 2.53 & 0 & 1 & 8.15 & 7.31 (9.42) & 2.95 & 0 & 1 & 10.4 & 8.48 (10.3) & 3.15 \\
an elephant & S & 0 & 1.99 & 6.39 & 3.20 & 0 & 0 & 15.4 & 7.72 & 2.76 & 0 & 0 & 38.6 & 21.2 & 15.2 & 0 & 0 & 22.3 & 7.60 & 2.69 \\
a rhinoceros & S & 0 & 1.71 & 3.63 & 1.90 & 0 & 0 & 13.6 & 5.49 & 3.83 & 0 & 0 & 34.2 & 15.3 & 10.6 & 0 & 0 & 17.1 & 5.16 & 2.08 \\
a horse & S & 1 & 1.86 (3.74) & 21.0 (32.1) & 7.20 & 0 & 1 & 13.9 & 13.4 (16.8) & 5.42 & 0 & 1 & 68.6 & 51.6 (59.0) & 11.5 & 0 & 1 & 22.1 & 18.3 (21.7) & 5.27 \\
a crocodile & S & 0 & 1.18 & 5.45 & 3.74 & 0 & 1 & 8.80 & 6.37 (8.86) & 3.49 & 0 & 1 & 25.2 & 7.84 (11.2) & 3.96 & 0 & 1 & 11.4 & 6.03 (8.06) & 3.13 \\
a car suspension control arm with rubber bushings &  & 0 & 1.90 & 7.12 & 5.13 & 0 & 0 & 13.6 & 10.5 & 4.62 & 0 & 0 & 22.7 & 18.0 & 7.91 & 0 & 0 & 25.6 & 8.33 & 4.53 \\
a truck leaf spring &  & 0 & 0.452 & 1.89 & 1.43 & 0 & 0 & 3.85 & 1.94 & 1.51 & 0 & 0 & 3.83 & 1.91 & 1.47 & 0 & 0 & 4.19 & 2.02 & 1.57 \\
a car wheel with its rubber tyre & G & 0 & 6.78 & 23.1 & 11.7 & 0 & 0 & 125 & 244\textsuperscript{C} & 210\textsuperscript{C} & 0 & 0 & 409 & 558\textsuperscript{C} & 500\textsuperscript{C} & 0 & 0 & 68.2 & 57.1\textsuperscript{C} & 44.0\textsuperscript{C} \\
a car brake disc & S & 0 & 4.85 & 9.57 & 5.31 & --- & --- & --- & --- & --- & --- & --- & --- & --- & --- & --- & --- & --- & --- & --- \\
a crankshaft & G & 0 & 1.57 & 4.82 & 3.29 & 0 & 0 & 13.5 & 4.57 & 3.22 & 0 & 0 & 19.7 & 8.41 & 5.86 & 0 & 0 & 11.9 & 4.20 & 2.96 \\
a road roller with a steel drum and rubber tyres & T, G, S & 0 & 3.66 & 15.4 & 4.85 & 0 & 1 & 16.8 & 13.3 (17.8) & 6.67 & 0 & 1 & 65.2 & 10.6 (17.2) & 7.18 & 0 & 1 & 38.1 & 9.14 (13.1) & 5.97 \\
a jackhammer & S & 1 & 0.571 (1.15) & 34.2 (35.5) & 25.9 & 0 & 3 & 4.77 & 9.35\textsuperscript{C} (48.4) & 6.25\textsuperscript{C} & 0 & 3 & 5.12 & 11.3\textsuperscript{C} (49.4) & 7.61\textsuperscript{C} & 0 & 3 & 6.46 & 8.39\textsuperscript{C} (56.4) & 5.61\textsuperscript{C} \\
an industrial robot arm & G & 0 & 2.08 & 7.78 & 5.34 & 0 & 0 & 15.0 & 11.8 & 5.09 & 0 & 0 & 28.1 & 22.7 & 14.0 & 0 & 0 & 27.0 & 9.17 & 4.87 \\
an adjustable wrench & T, S & 0 & 0.573 & 2.71 & 1.45 & 0 & 0 & 3.81 & 1.84 & 1.02 & 0 & 0 & 4.02 & 4.45 & 1.63 & 0 & 0 & 5.26 & 3.26 & 1.46 \\
a claw hammer &  & 0 & 0.739 & 3.34 & 1.76 & 0 & 0 & 5.81 & 3.82 & 2.39 & 0 & 0 & 6.69 & 4.49 & 1.98 & 0 & 0 & 7.12 & 3.62 & 1.62 \\
a dental crown &  & 0 & 6.58 & 9.94 & 5.15 & 0 & 0 & 53.1 & 193\textsuperscript{C} & 182\textsuperscript{C} & 0 & 0 & 387 & 1496\textsuperscript{C} & 1088\textsuperscript{C} & 0 & 0 & 72.1 & 71.0\textsuperscript{C} & 62.0\textsuperscript{C} \\
a knee implant & G & 0 & 5.96 & 15.2 & 7.41 & 0 & 3 & 46.8 & 114\textsuperscript{C} & 105\textsuperscript{C} & 0 & 0 & 190 & 794\textsuperscript{C} & 575\textsuperscript{C} & 0 & 0 & 76.6 & 14.2 & 7.83 \\
a walking cane with a rubber tip & T, L, S & 0 & 0.315 & 0.947 & 0.621 & 0 & 0 & 3.56 & 1.24 & 0.855 & 0 & 0 & 3.42 & 0.969 & 0.686 & 0 & 0 & 3.81 & 1.25 & 0.872 \\
a human femur &  & 0 & 0.871 & 4.00 & 2.37 & 0 & 0 & 6.96 & 4.05 & 1.91 & 0 & 0 & 9.19 & 5.56 & 3.61 & 0 & 0 & 9.19 & 3.99 & 2.57 \\
a lumbar vertebra &  & 0 & 2.17 & 5.10 & 1.83 & 0 & 0 & 16.1 & 5.19 & 3.73 & 0 & 0 & 43.5 & 14.8 & 5.98 & 0 & 0 & 31.3 & 5.05 & 3.63 \\
a skyscraper & L & 0 & 1.12 & 1.95 & 1.05 & 0 & 0 & 9.90 & 1.75 & 1.07 & 0 & 0 & 23.8 & 5.34 & 3.12 & 0 & 0 & 10.3 & 1.84 & 1.13 \\
a thatched stone cottage & L & 0 & 4.16 & 9.93 & 5.12 & 0 & 0 & 36.5 & 6.89 & 3.69 & 0 & 0 & 232 & 2796\textsuperscript{C} & 1487\textsuperscript{C} & 0 & 0 & 46.5 & 6.53 & 3.50 \\
a lighthouse &  & 0 & 2.19 & 4.20 & 2.24 & 0 & 0 & 43.0 & 4.76 & 2.92 & 0 & 0 & 57.6 & 158\textsuperscript{C} & 148\textsuperscript{C} & 0 & 0 & 23.3 & 5.06 & 2.97 \\
an igloo &  & 0 & 5.45 & 7.61 & 3.97 & 0 & 0 & 67.6 & 9.33 & 6.11 & 0 & 0 & 129 & 271\textsuperscript{C} & 243\textsuperscript{C} & 0 & 0 & 72.6 & 6.88 & 4.14 \\
a log cabin & L & 0 & 5.40 & 8.28 & 4.39 & 0 & 0 & 51.7 & 8.79 & 5.22 & 0 & 0 & 265 & 1759\textsuperscript{C} & 1178\textsuperscript{C} & 0 & 0 & 56.7 & 10.3 & 5.94 \\
the Leaning Tower of Pisa & T, G, S & 0 & 2.17 & 4.73 & 2.62 & 0 & 1 & 15.2 & 12.1 (16.1) & 3.00 & 0 & 1 & 26.4 & 20.3 (26.7) & 10.2 & 8 & 1 & 13.5 (82.2) & 21.1 (24.0) & 4.62 \\
the Pantheon & L & 0 & 4.46 & 7.60 & 4.23 & 0 & 0 & 35.3 & 8.60 & 5.63 & 0 & 0 & 78.1 & 225\textsuperscript{C} & 210\textsuperscript{C} & 0 & 0 & 86.4 & 7.87 & 4.82 \\
Himeji Castle & G & 0 & 6.76 & 19.7 & 10.7 & --- & --- & --- & --- & --- & --- & --- & --- & --- & --- & --- & --- & --- & --- & --- \\*
a natural stone arch & G & 0 & 4.47 & 7.31 & 3.89 & 1 & 0 & 43.2 & 9.53 & 6.77 & 0 & 0 & 155 & 490\textsuperscript{C} & 470\textsuperscript{C} & 0 & 0 & 93.4 & 7.18 & 4.42 \\*
the Taj Mahal &  & 0 & 3.93 & 11.3 & 6.04 & 0 & 0 & 162 & 11.2 & 5.72 & 0 & 0 & 124 & 1228\textsuperscript{C} & 1164\textsuperscript{C} & 0 & 0 & 269 & 14.6 & 7.86 \\*
\noalign{\penalty10000\vskip\aboverulesep\hrule height\lightrulewidth\penalty10000\vskip\belowrulesep}
Median (all objects) &  & 0 & 2.18 & 7.60 & 4.62 & 0 & 0 & 16.1 & 7.91 & 3.83 & 0 & 0 & 43.5 & 21.2 & 11.5 & 0 & 0 & 29.8 & 8.39 & 4.62 \\*
\hspace{1em}$n$ &  & 50 & 50 & 49 & 50 & 47 & 47 & 47 & 47 & 47 & 47 & 47 & 47 & 47 & 47 & 47 & 47 & 47 & 47 & 47 \\*
\addlinespace
Median (unflagged) &  & 0 & 2.28 & 5.10 & 3.31 & 0 & 0 & 20.5 & 5.23 & 3.68 & 0 & 0 & 58.7 & 69.6 & 38.5 & 0 & 0 & 34.2 & 7.60 & 4.34 \\*
\hspace{1em}$n$ &  & 18 & 18 & 17 & 18 & 18 & 18 & 18 & 18 & 18 & 18 & 18 & 18 & 18 & 18 & 18 & 18 & 18 & 18 & 18 \\
\end{longtable}
\end{landscape}

\newpage
\clearpage

\section{Web user interface}
\label{sec:si-webui}

A recording of operating the web user interface is available at \url{https://drive.google.com/file/d/11mT9NLNuYj64azS-_JYqRVnaQDN8s4Af/view?usp=drive_link}

\begin{figure}[h]
    \centering
    \includegraphics[width=0.95\linewidth]{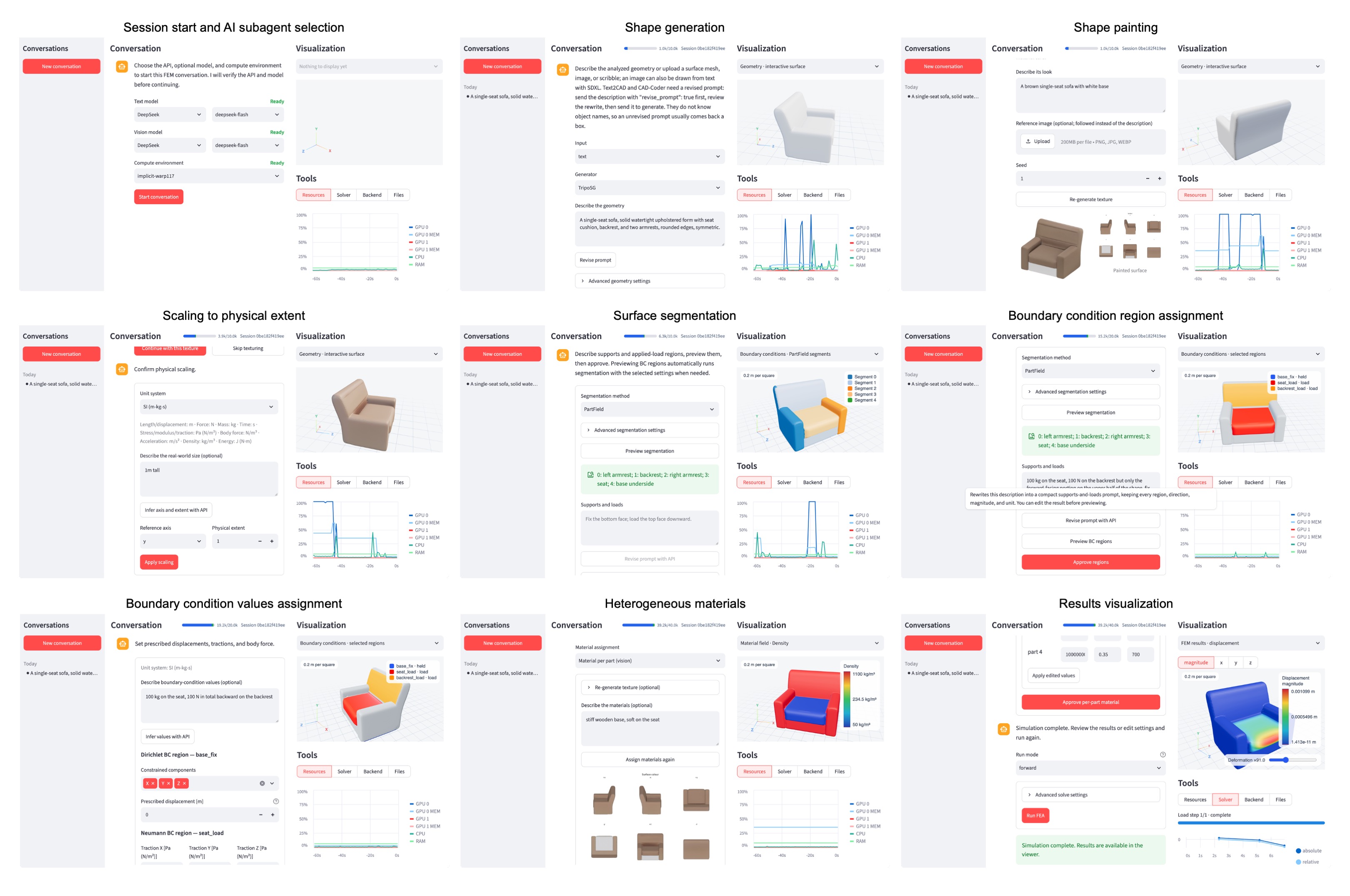}
    \caption{Screenshots of the web user interface at different workflow stages.}
    \label{fig:si-webui}
\end{figure}

\clearpage
\setcounter{NAT@ctr}{0}
\begingroup\makeatletter\def\@extra@b@citeb{@S}\makeatother

\endgroup


\begin{thebibliography}{10}
\expandafter\ifx\csname url\endcsname\relax
  \def\url#1{\texttt{#1}}\fi
\expandafter\ifx\csname urlprefix\endcsname\relax\def\urlprefix{URL }\fi
\expandafter\ifx\csname href\endcsname\relax
  \def\href#1#2{#2} \def\path#1{#1}\fi

\bibitem{jacobson2013robust}
A.~Jacobson, L.~Kavan, O.~Sorkine-Hornung, Robust inside-outside segmentation using generalized winding numbers, ACM Transactions on Graphics (TOG) 32~(4) (2013) 1--12.

\bibitem{li2025triposg}
Y.~Li, Z.-X. Zou, Z.~Liu, D.~Wang, Y.~Liang, Z.~Yu, X.~Liu, Y.-C. Guo, D.~Liang, W.~Ouyang, et~al., Triposg: High-fidelity 3d shape synthesis using large-scale rectified flow models, IEEE Transactions on Pattern Analysis and Machine Intelligence (2025).

\bibitem{liu2022flow}
X.~Liu, C.~Gong, Q.~Liu, Flow straight and fast: Learning to generate and transfer data with rectified flow, arXiv preprint arXiv:2209.03003 (2022).

\bibitem{khan2024text2cad}
M.~S. Khan, S.~Sinha, T.~U. Sheikh, D.~Stricker, S.~A. Ali, M.~Z. Afzal, Text2cad: Generating sequential cad designs from beginner-to-expert level text prompts, Advances in Neural Information Processing Systems 37 (2024) 7552--7579.

\bibitem{devlin2019bert}
J.~Devlin, M.-W. Chang, K.~Lee, K.~Toutanova, Bert: Pre-training of deep bidirectional transformers for language understanding, in: Proceedings of the 2019 conference of the North American chapter of the association for computational linguistics: human language technologies, volume 1 (long and short papers), 2019, pp. 4171--4186.

\bibitem{guan2026cad}
Y.~Guan, X.~Wang, X.~Xing, J.~Zhang, D.~Xu, Q.~Yu, Cad-coder: Text-to-cad generation with chain-of-thought and geometric reward, Advances in Neural Information Processing Systems 38 (2026) 59765--59789.

\bibitem{hunyuan3d2025hunyuan3d}
T.~H. Team, Hunyuan3d 2.1: From images to high-fidelity 3d assets with production-ready pbr material (2025).
\newblock \href {http://arxiv.org/abs/2506.15442} {\path{arXiv:2506.15442}}.

\bibitem{lai2025unleashingvecsetdiffusionmodel}
Z.~Lai, Y.~Zhao, Z.~Zhao, H.~Liu, F.~Wang, H.~Shi, X.~Yang, Q.~Lin, J.~Huang, Y.~Liu, J.~Jiang, C.~Guo, X.~Yue, \href{https://arxiv.org/abs/2503.16302}{Unleashing vecset diffusion model for fast shape generation} (2025).
\newblock \href {http://arxiv.org/abs/2503.16302} {\path{arXiv:2503.16302}}.
\newline\urlprefix\url{https://arxiv.org/abs/2503.16302}

\bibitem{podell2024sdxl}
D.~Podell, Z.~English, K.~Lacey, A.~Blattmann, T.~Dockhorn, J.~M{\"u}ller, J.~Penna, R.~Rombach, Sdxl: Improving latent diffusion models for high-resolution image synthesis, in: International Conference on Learning Representations, Vol. 2024, 2024, pp. 1862--1874.

\bibitem{ho2022classifier}
J.~Ho, T.~Salimans, Classifier-free diffusion guidance, arXiv preprint arXiv:2207.12598 (2022).

\bibitem{zhang2023adding}
L.~Zhang, A.~Rao, M.~Agrawala, Adding conditional control to text-to-image diffusion models, in: 2023 IEEE/CVF International Conference on Computer Vision (ICCV), IEEE, 2023, pp. 3813--3824.

\bibitem{mescheder2019occupancy}
L.~Mescheder, M.~Oechsle, M.~Niemeyer, S.~Nowozin, A.~Geiger, Occupancy networks: Learning 3d reconstruction in function space, in: 2019 IEEE/CVF Conference on Computer Vision and Pattern Recognition (CVPR), IEEE, 2019, pp. 4455--4465.

\bibitem{sundar2008bottom}
H.~Sundar, R.~S. Sampath, G.~Biros, Bottom-up construction and 2: 1 balance refinement of linear octrees in parallel, SIAM Journal on Scientific Computing 30~(5) (2008) 2675--2708.

\bibitem{burstedde2011p4est}
C.~Burstedde, L.~C. Wilcox, O.~Ghattas, p4est: Scalable algorithms for parallel adaptive mesh refinement on forests of octrees, SIAM Journal on Scientific Computing 33~(3) (2011) 1103--1133.

\bibitem{lorensen1987marching}
W.~E. Lorensen, H.~E. Cline, \href{https://doi.org/10.1145/37401.37422}{Marching cubes: A high resolution 3d surface construction algorithm}, in: Proceedings of the 14th Annual Conference on Computer Graphics and Interactive Techniques, SIGGRAPH '87, Association for Computing Machinery, New York, NY, USA, 1987, p. 163–169.
\newblock \href {https://doi.org/10.1145/37401.37422} {\path{doi:10.1145/37401.37422}}.
\newline\urlprefix\url{https://doi.org/10.1145/37401.37422}

\bibitem{liu2025partfield}
M.~Liu, M.~A. Uy, D.~Xiang, H.~Su, S.~Fidler, N.~Sharp, J.~Gao, Partfield: Learning 3d feature fields for part segmentation and beyond, in: 2025 IEEE/CVF International Conference on Computer Vision (ICCV), IEEE, 2025, pp. 9704--9715.

\bibitem{ward1963hierarchical}
J.~H. Ward~Jr, Hierarchical grouping to optimize an objective function, Journal of the American statistical association 58~(301) (1963) 236--244.

\bibitem{shapira2008consistent}
L.~Shapira, A.~Shamir, D.~Cohen-Or, Consistent mesh partitioning and skeletonisation using the shape diameter function, The Visual Computer 24~(4) (2008) 249--259.

\bibitem{alliez2016cgal}
P.~Alliez, A.~Fabri, Cgal: the computational geometry algorithms library (2016).

\bibitem{boykov2001fast}
Y.~Boykov, O.~Veksler, R.~Zabih, Fast approximate energy minimization via graph cuts, IEEE Transactions on pattern analysis and machine intelligence 23~(11) (2001) 1222--1239.

\bibitem{li2009solving}
X.~Li, J.~Lowengrub, A.~R{\"a}tz, A.~Voigt, Solving pdes in complex geometries: a diffuse domain approach, Communications in mathematical sciences 7~(1) (2009) 81.

\bibitem{nguyen2018diffuse}
L.~H. Nguyen, S.~K. Stoter, M.~Ruess, M.~A. Sanchez~Uribe, D.~Schillinger, The diffuse nitsche method: Dirichlet constraints on phase-field boundaries, International Journal for Numerical Methods in Engineering 113~(4) (2018) 601--633.

\bibitem{bonet2008nonlinear}
J.~Bonet, R.~D. Wood, Nonlinear continuum mechanics for finite element analysis, Cambridge university press, 2008.

\bibitem{dagli2025vomp}
R.~Dagli, D.~Xiang, V.~Modi, C.~Loop, C.~F. Tsang, A.~H. Chen, A.~Hu, G.~State, D.~I. Levin, M.~Shugrina, Vomp: Predicting volumetric mechanical property fields, arXiv preprint arXiv:2510.22975 (2025).

\bibitem{xu2026deepseek}
A.~Xu, B.~Li, B.~Lin, B.~Xue, B.~Xian, B.~Xu, B.~Wu, B.~Zhang, B.~Deng, C.~Yu, et~al., Deepseek-v4. 1-flash: Pushing the limits of kv cache compression, arXiv preprint arXiv:2609.19969 (2026).

\bibitem{nitsche1971variationsprinzip}
J.~Nitsche, {\"U}ber ein variationsprinzip zur l{\"o}sung von dirichlet-problemen bei verwendung von teilr{\"a}umen, die keinen randbedingungen unterworfen sind, in: Abhandlungen aus dem mathematischen Seminar der Universit{\"a}t Hamburg, Vol.~36, Springer, 1971, pp. 9--15.

\bibitem{embar2010imposing}
A.~Embar, J.~Dolbow, I.~Harari, Imposing dirichlet boundary conditions with nitsche's method and spline-based finite elements, International journal for numerical methods in engineering 83~(7) (2010) 877--898.

\bibitem{parvizian2007finite}
J.~Parvizian, A.~D{\"u}ster, E.~Rank, Finite cell method: h-and p-extension for embedded domain problems in solid mechanics, Computational Mechanics 41~(1) (2007) 121--133.

\bibitem{naumov2015amgx}
M.~Naumov, M.~Arsaev, P.~Castonguay, J.~Cohen, J.~Demouth, J.~Eaton, S.~Layton, N.~Markovskiy, I.~Reguly, N.~Sakharnykh, et~al., Amgx: A library for gpu accelerated algebraic multigrid and preconditioned iterative methods, SIAM Journal on Scientific Computing 37~(5) (2015) S602--S626.

\bibitem{liu2026jax}
Y.~Liu, X.~Fan, J.-X. Wang, Jax-amg: A gpu-accelerated differentiable sparse linear solver library for jax, arXiv preprint arXiv:2606.09001 (2026).

\bibitem{allaire2004structural}
G.~Allaire, F.~Jouve, A.-M. Toader, Structural optimization using sensitivity analysis and a level-set method, Journal of computational physics 194~(1) (2004) 363--393.

\bibitem{wang2003level}
M.~Y. Wang, X.~Wang, D.~Guo, A level set method for structural topology optimization, Computer methods in applied mechanics and engineering 192~(1-2) (2003) 227--246.

\bibitem{bendsoe1989optimal}
M.~P. Bends{\o}e, Optimal shape design as a material distribution problem, Structural optimization 1~(4) (1989) 193--202.

\bibitem{fuchs1977optimal}
H.~Fuchs, Z.~M. Kedem, S.~P. Uselton, Optimal surface reconstruction from planar contours, Communications of the ACM 20~(10) (1977) 693--702.

\bibitem{taubin1995signal}
G.~Taubin, A signal processing approach to fair surface design, in: Proceedings of the 22nd annual conference on Computer graphics and interactive techniques, 1995, pp. 351--358.

\bibitem{luo2026towards}
X.-L. Luo, S.~Yang, J.~Tan, R.~F. Shepherd, N.~Cohen, N.~Bouklas, Towards end-to-end optimization in multimaterial 3d printing, arXiv preprint arXiv:2607.13174 (2026).

\bibitem{xia2019stable}
Q.~Xia, T.~Shi, L.~Xia, Stable hole nucleation in level set based topology optimization by using the material removal scheme of beso, Computer Methods in Applied Mechanics and Engineering 343 (2019) 438--452.

\bibitem{wang2024agent}
J.~Wang, Z.~Duan, Agent ai with langgraph: A modular framework for enhancing machine translation using large language models, arXiv preprint arXiv:2412.03801 (2024).

\bibitem{paszke2019pytorch}
A.~Paszke, S.~Gross, F.~Massa, A.~Lerer, J.~Bradbury, G.~Chanan, T.~Killeen, Z.~Lin, N.~Gimelshein, L.~Antiga, et~al., Pytorch: An imperative style, high-performance deep learning library, Advances in neural information processing systems 32 (2019).

\bibitem{macklin2022warp}
M.~Macklin, Warp: A high-performance python framework for gpu simulation and graphics, \url{https://github.com/nvidia/warp}, {NVIDIA} {GPU} Technology Conference ({GTC}) (2022).

\bibitem{gu2026survey}
J.~Gu, X.~Jiang, Z.~Shi, H.~Tan, X.~Zhai, C.~Xu, W.~Li, Y.~Shen, S.~Ma, H.~Liu, et~al., A survey on llm-as-a-judge, The Innovation 7~(6) (2026).

\bibitem{schneider2022large}
T.~Schneider, Y.~Hu, X.~Gao, J.~Dumas, D.~Zorin, D.~Panozzo, A large-scale comparison of tetrahedral and hexahedral elements for solving elliptic pdes with the finite element method, ACM Transactions on Graphics (TOG) 41~(3) (2022) 1--14.

\bibitem{hang2015tetgen}
S.~Hang, Tetgen, a delaunay-based quality tetrahedral mesh generator, ACM Trans. Math. Softw 41~(2) (2015) 11.

\end{thebibliography}


\begin{thebibliography}{10}
\expandafter\ifx\csname url\endcsname\relax
  \def\url#1{\texttt{#1}}\fi
\expandafter\ifx\csname urlprefix\endcsname\relax\def\urlprefix{URL }\fi
\expandafter\ifx\csname href\endcsname\relax
  \def\href#1#2{#2} \def\path#1{#1}\fi

\bibitem{owen2026survey}
S.~Owen, N.~Brown, N.~Chrisochoides, R.~Garimella, X.~Gu, F.~Ledoux, N.~Lei, R.~Quadros, N.~Ray, N.~Winovich, et~al., A survey of ai methods for geometry preparation and mesh generation in engineering simulation, Engineering with Computers 42~(5) (2026) 171.

\bibitem{sigmund2013topology}
O.~Sigmund, K.~Maute, Topology optimization approaches: A comparative review, Structural and multidisciplinary optimization 48~(6) (2013) 1031--1055.

\bibitem{allaire2004structural}
G.~Allaire, F.~Jouve, A.-M. Toader, Structural optimization using sensitivity analysis and a level-set method, Journal of computational physics 194~(1) (2004) 363--393.

\bibitem{wang2003level}
M.~Y. Wang, X.~Wang, D.~Guo, A level set method for structural topology optimization, Computer methods in applied mechanics and engineering 192~(1-2) (2003) 227--246.

\bibitem{papadopoulos2021computing}
I.~P. Papadopoulos, P.~E. Farrell, T.~M. Surowiec, Computing multiple solutions of topology optimization problems, SIAM Journal on Scientific Computing 43~(3) (2021) A1555--A1582.

\bibitem{boiko2023autonomous}
D.~A. Boiko, R.~MacKnight, B.~Kline, G.~Gomes, Autonomous chemical research with large language models, Nature 624~(7992) (2023) 570--578.

\bibitem{szymanski2023autonomous}
N.~J. Szymanski, B.~Rendy, Y.~Fei, R.~E. Kumar, T.~He, D.~Milsted, M.~J. McDermott, M.~Gallant, E.~D. Cubuk, A.~Merchant, et~al., An autonomous laboratory for the accelerated synthesis of inorganic materials, Nature 624~(7990) (2023) 86--91.

\bibitem{ni2024mechagents}
B.~Ni, M.~J. Buehler, Mechagents: Large language model multi-agent collaborations can solve mechanics problems, generate new data, and integrate knowledge, Extreme Mechanics Letters 67 (2024) 102131.

\bibitem{mudur2025feabench}
N.~Mudur, H.~Cui, S.~Venugopalan, P.~Raccuglia, M.~P. Brenner, P.~Norgaard, Feabench: Evaluating language models on multiphysics reasoning ability, arXiv preprint arXiv:2504.06260 (2025).

\bibitem{qi2025feagpt}
Y.~Qi, R.~Xu, X.~Chu, Feagpt: an end-to-end agentic-ai for finite element analysis, arXiv preprint arXiv:2510.21993 (2025).

\bibitem{li2025triposg}
Y.~Li, Z.-X. Zou, Z.~Liu, D.~Wang, Y.~Liang, Z.~Yu, X.~Liu, Y.-C. Guo, D.~Liang, W.~Ouyang, et~al., Triposg: High-fidelity 3d shape synthesis using large-scale rectified flow models, IEEE Transactions on Pattern Analysis and Machine Intelligence (2025).

\bibitem{hunyuan3d2025hunyuan3d}
T.~H. Team, Hunyuan3d 2.1: From images to high-fidelity 3d assets with production-ready pbr material (2025).
\newblock \href {http://arxiv.org/abs/2506.15442} {\path{arXiv:2506.15442}}.

\bibitem{khan2024text2cad}
M.~S. Khan, S.~Sinha, T.~U. Sheikh, D.~Stricker, S.~A. Ali, M.~Z. Afzal, Text2cad: Generating sequential cad designs from beginner-to-expert level text prompts, Advances in Neural Information Processing Systems 37 (2024) 7552--7579.

\bibitem{guan2026cad}
Y.~Guan, X.~Wang, X.~Xing, J.~Zhang, D.~Xu, Q.~Yu, Cad-coder: Text-to-cad generation with chain-of-thought and geometric reward, Advances in Neural Information Processing Systems 38 (2026) 59765--59789.

\bibitem{cao2026physx}
Z.~Cao, Z.~Chen, L.~Pan, Z.~Liu, Physx-3d: Physical-grounded 3d asset generation, Advances in Neural Information Processing Systems 38 (2026) 93771--93784.

\bibitem{chen2024atlas3d}
Y.~Chen, T.~Xie, Z.~Zong, X.~Li, F.~Gao, Y.~Yang, Y.~N. Wu, C.~Jiang, Atlas3d: Physically constrained self-supporting text-to-3d for simulation and fabrication, Advances in Neural Information Processing Systems 37 (2024) 102501--102530.

\bibitem{ranawaka2026simfoundry}
N.~Ranawaka, J.~Wong, W.-L. Pai, W.-T. Chu, T.~Dai, M.~Moghani, H.~Yin, Y.~Jiang, W.~Durbano, B.~Huynh, et~al., Simfoundry: Modular and automated scene generation for policy learning and evaluation, arXiv preprint arXiv:2606.28276 (2026).

\bibitem{chen2026diaggen}
G.~Chen, Y.~Qu, Q.~Xia, P.~Jing, Y.~Cheng, B.~Ma, P.~Yang, B.~Zhou, Z.~Li, S.~Suri, et~al., Diaggen: Agentic generation of deformable assets with sim-based diagnostics for robotic simulation, arXiv preprint arXiv:2609.23103 (2026).

\bibitem{guo2024physically}
M.~Guo, B.~Wang, P.~Ma, T.~Zhang, C.~E. Owens, C.~Gan, J.~B. Tenenbaum, K.~He, W.~Matusik, Physically compatible 3d object modeling from a single image, Advances in Neural Information Processing Systems 37 (2024) 119260--119282.

\bibitem{zhan2025physiopt}
X.~Zhan, C.~Jambon, E.~Thompson, K.~Ng, M.~Konakovi{\'c}~Lukovi{\'c}, Physiopt: Physics-driven shape optimization for 3d generative models, in: Proceedings of the SIGGRAPH Asia 2025 Conference Papers, 2025, pp. 1--11.

\bibitem{hu2020fast}
Y.~Hu, T.~Schneider, B.~Wang, D.~Zorin, D.~Panozzo, Fast tetrahedral meshing in the wild, ACM Transactions on Graphics (ToG) 39~(4) (2020) 117--1.

\bibitem{schneider2022large}
T.~Schneider, Y.~Hu, X.~Gao, J.~Dumas, D.~Zorin, D.~Panozzo, A large-scale comparison of tetrahedral and hexahedral elements for solving elliptic pdes with the finite element method, ACM Transactions on Graphics (TOG) 41~(3) (2022) 1--14.

\bibitem{nguyen2018diffuse}
L.~H. Nguyen, S.~K. Stoter, M.~Ruess, M.~A. Sanchez~Uribe, D.~Schillinger, The diffuse nitsche method: Dirichlet constraints on phase-field boundaries, International Journal for Numerical Methods in Engineering 113~(4) (2018) 601--633.

\bibitem{macklin2022warp}
M.~Macklin, Warp: A high-performance python framework for gpu simulation and graphics, \url{https://github.com/nvidia/warp}, {NVIDIA} {GPU} Technology Conference ({GTC}) (2022).

\bibitem{podell2024sdxl}
D.~Podell, Z.~English, K.~Lacey, A.~Blattmann, T.~Dockhorn, J.~M{\"u}ller, J.~Penna, R.~Rombach, Sdxl: Improving latent diffusion models for high-resolution image synthesis, in: International Conference on Learning Representations, Vol. 2024, 2024, pp. 1862--1874.

\bibitem{liu2025partfield}
M.~Liu, M.~A. Uy, D.~Xiang, H.~Su, S.~Fidler, N.~Sharp, J.~Gao, Partfield: Learning 3d feature fields for part segmentation and beyond, in: 2025 IEEE/CVF International Conference on Computer Vision (ICCV), IEEE, 2025, pp. 9704--9715.

\bibitem{jacobson2013robust}
A.~Jacobson, L.~Kavan, O.~Sorkine-Hornung, Robust inside-outside segmentation using generalized winding numbers, ACM Transactions on Graphics (TOG) 32~(4) (2013) 1--12.

\bibitem{xu2026deepseek}
A.~Xu, B.~Li, B.~Lin, B.~Xue, B.~Xian, B.~Xu, B.~Wu, B.~Zhang, B.~Deng, C.~Yu, et~al., Deepseek-v4. 1-flash: Pushing the limits of kv cache compression, arXiv preprint arXiv:2609.19969 (2026).

\bibitem{yang2021deep}
Z.~Yang, C.-H. Yu, M.~J. Buehler, Deep learning model to predict complex stress and strain fields in hierarchical composites, Science Advances 7~(15) (2021) eabd7416.

\bibitem{hong2025deepjeb}
S.~Hong, Y.~Kwon, D.~Shin, J.~Park, N.~Kang, Deepjeb: 3d deep learning-based synthetic jet engine bracket dataset, Journal of Mechanical Design 147~(4) (2025) 041703.

\bibitem{luo2026towards}
X.-L. Luo, S.~Yang, J.~Tan, R.~F. Shepherd, N.~Cohen, N.~Bouklas, Towards end-to-end optimization in multimaterial 3d printing, arXiv preprint arXiv:2607.13174 (2026).

\bibitem{fuhg2023modular}
J.~N. Fuhg, C.~M. Hamel, K.~Johnson, R.~Jones, N.~Bouklas, Modular machine learning-based elastoplasticity: Generalization in the context of limited data, Computer Methods in Applied Mechanics and Engineering 407 (2023) 115930.

\bibitem{fuhg2025review}
J.~N. Fuhg, G.~Anantha~Padmanabha, N.~Bouklas, B.~Bahmani, W.~Sun, N.~N. Vlassis, M.~Flaschel, P.~Carrara, L.~De~Lorenzis, A review on data-driven constitutive laws for solids: Jn fuhg et al., Archives of Computational Methods in Engineering 32~(3) (2025) 1841--1883.

\bibitem{lu2021learning}
L.~Lu, P.~Jin, G.~Pang, Z.~Zhang, G.~E. Karniadakis, Learning nonlinear operators via deeponet based on the universal approximation theorem of operators, Nature machine intelligence 3~(3) (2021) 218--229.

\bibitem{karniadakis2021physics}
G.~E. Karniadakis, I.~G. Kevrekidis, L.~Lu, P.~Perdikaris, S.~Wang, L.~Yang, Physics-informed machine learning, Nature Reviews Physics 3~(6) (2021) 422--440.

\bibitem{lejeune2020mechanical}
E.~Lejeune, Mechanical mnist: A benchmark dataset for mechanical metamodels, Extreme Mechanics Letters 36 (2020) 100659.

\bibitem{hendriks2025wallpaper}
F.~Hendriks, V.~Menkovski, M.~Do{\v{s}}k{\'a}{\v{r}}, M.~G. Geers, K.~Verbeek, O.~Roko{\v{s}}, Wallpaper group-based mechanical metamaterials: Dataset including mechanical responses, Scientific Data 12~(1) (2025) 1880.

\bibitem{hwangbo2019learning}
J.~Hwangbo, J.~Lee, A.~Dosovitskiy, D.~Bellicoso, V.~Tsounis, V.~Koltun, M.~Hutter, Learning agile and dynamic motor skills for legged robots, Science robotics 4~(26) (2019) eaau5872.

\bibitem{yin2021modeling}
H.~Yin, A.~Varava, D.~Kragic, Modeling, learning, perception, and control methods for deformable object manipulation, Science Robotics 6~(54) (2021) eabd8803.

\bibitem{niederer2021scaling}
S.~A. Niederer, M.~S. Sacks, M.~Girolami, K.~Willcox, Scaling digital twins from the artisanal to the industrial, Nature Computational Science 1~(5) (2021) 313--320.

\bibitem{hafner2025mastering}
D.~Hafner, J.~Pasukonis, J.~Ba, T.~Lillicrap, Mastering diverse control tasks through world models, Nature 640~(8059) (2025) 647--653.

\bibitem{motamed2026generative}
S.~Motamed, L.~Culp, K.~Swersky, P.~Jaini, R.~Geirhos, Do generative video models understand physical principles?, in: 2026 IEEE/CVF Winter Conference on Applications of Computer Vision (WACV), IEEE, 2026, pp. 948--958.

\bibitem{paszke2019pytorch}
A.~Paszke, S.~Gross, F.~Massa, A.~Lerer, J.~Bradbury, G.~Chanan, T.~Killeen, Z.~Lin, N.~Gimelshein, L.~Antiga, et~al., Pytorch: An imperative style, high-performance deep learning library, Advances in neural information processing systems 32 (2019).

\bibitem{liu2022flow}
X.~Liu, C.~Gong, Q.~Liu, Flow straight and fast: Learning to generate and transfer data with rectified flow, arXiv preprint arXiv:2209.03003 (2022).

\bibitem{mescheder2019occupancy}
L.~Mescheder, M.~Oechsle, M.~Niemeyer, S.~Nowozin, A.~Geiger, Occupancy networks: Learning 3d reconstruction in function space, in: 2019 IEEE/CVF Conference on Computer Vision and Pattern Recognition (CVPR), IEEE, 2019, pp. 4455--4465.

\bibitem{sundar2008bottom}
H.~Sundar, R.~S. Sampath, G.~Biros, Bottom-up construction and 2: 1 balance refinement of linear octrees in parallel, SIAM Journal on Scientific Computing 30~(5) (2008) 2675--2708.

\bibitem{burstedde2011p4est}
C.~Burstedde, L.~C. Wilcox, O.~Ghattas, p4est: Scalable algorithms for parallel adaptive mesh refinement on forests of octrees, SIAM Journal on Scientific Computing 33~(3) (2011) 1103--1133.

\bibitem{lorensen1987marching}
W.~E. Lorensen, H.~E. Cline, \href{https://doi.org/10.1145/37401.37422}{Marching cubes: A high resolution 3d surface construction algorithm}, in: Proceedings of the 14th Annual Conference on Computer Graphics and Interactive Techniques, SIGGRAPH '87, Association for Computing Machinery, New York, NY, USA, 1987, p. 163–169.
\newblock \href {https://doi.org/10.1145/37401.37422} {\path{doi:10.1145/37401.37422}}.
\newline\urlprefix\url{https://doi.org/10.1145/37401.37422}

\bibitem{ward1963hierarchical}
J.~H. Ward~Jr, Hierarchical grouping to optimize an objective function, Journal of the American statistical association 58~(301) (1963) 236--244.

\bibitem{li2009solving}
X.~Li, J.~Lowengrub, A.~R{\"a}tz, A.~Voigt, Solving pdes in complex geometries: a diffuse domain approach, Communications in mathematical sciences 7~(1) (2009) 81.

\bibitem{bonet2008nonlinear}
J.~Bonet, R.~D. Wood, Nonlinear continuum mechanics for finite element analysis, Cambridge university press, 2008.

\bibitem{nitsche1971variationsprinzip}
J.~Nitsche, {\"U}ber ein variationsprinzip zur l{\"o}sung von dirichlet-problemen bei verwendung von teilr{\"a}umen, die keinen randbedingungen unterworfen sind, in: Abhandlungen aus dem mathematischen Seminar der Universit{\"a}t Hamburg, Vol.~36, Springer, 1971, pp. 9--15.

\bibitem{embar2010imposing}
A.~Embar, J.~Dolbow, I.~Harari, Imposing dirichlet boundary conditions with nitsche's method and spline-based finite elements, International journal for numerical methods in engineering 83~(7) (2010) 877--898.

\bibitem{naumov2015amgx}
M.~Naumov, M.~Arsaev, P.~Castonguay, J.~Cohen, J.~Demouth, J.~Eaton, S.~Layton, N.~Markovskiy, I.~Reguly, N.~Sakharnykh, et~al., Amgx: A library for gpu accelerated algebraic multigrid and preconditioned iterative methods, SIAM Journal on Scientific Computing 37~(5) (2015) S602--S626.

\bibitem{liu2026jax}
Y.~Liu, X.~Fan, J.-X. Wang, Jax-amg: A gpu-accelerated differentiable sparse linear solver library for jax, arXiv preprint arXiv:2606.09001 (2026).

\bibitem{bendsoe1989optimal}
M.~P. Bends{\o}e, Optimal shape design as a material distribution problem, Structural optimization 1~(4) (1989) 193--202.

\bibitem{fuchs1977optimal}
H.~Fuchs, Z.~M. Kedem, S.~P. Uselton, Optimal surface reconstruction from planar contours, Communications of the ACM 20~(10) (1977) 693--702.

\bibitem{taubin1995signal}
G.~Taubin, A signal processing approach to fair surface design, in: Proceedings of the 22nd annual conference on Computer graphics and interactive techniques, 1995, pp. 351--358.

\bibitem{xia2019stable}
Q.~Xia, T.~Shi, L.~Xia, Stable hole nucleation in level set based topology optimization by using the material removal scheme of beso, Computer Methods in Applied Mechanics and Engineering 343 (2019) 438--452.

\bibitem{wang2024agent}
J.~Wang, Z.~Duan, Agent ai with langgraph: A modular framework for enhancing machine translation using large language models, arXiv preprint arXiv:2412.03801 (2024).

\end{thebibliography}
\end{document}